\documentclass[12pt,twoside,titlepage,a4paper]{scrreprt}
\usepackage{amsthm,amssymb}
\usepackage[intlimits,sumlimits,namelimits]{amsmath}
\usepackage{multind}
\usepackage{graphicx}
\usepackage{caption}
\makeindex{ind}
\makeindex{sym}
\makeatletter 
\renewcommand{\printindex}[2] 
               {\if@twocolumn 
                  \@restonecolfalse 
                \else 
                  \@restonecoltrue 
                \fi 
                \twocolumn[\@makeschapterhead{#2}]%
                \@mkboth{\MakeUppercase#2}%
                        {\MakeUppercase#2}%
                \addcontentsline{toc}{chapter}{\numberline{}#2}%
                \thispagestyle{plain}\parindent\z@ 
                \@input{#1.ind}} 
 
\makeatother 
\usepackage{accents}
\usepackage[english,german]{babel}
\usepackage{a4}
\usepackage{cancel}

\theoremstyle{definition}
\newtheorem{Def}{Definition}[section]

\theoremstyle{plain}
\newtheorem{Lem}[Def]{Lemma}
\newtheorem{Prop}[Def]{Proposition}
\newtheorem{Thm}[Def]{Theorem}
\newtheorem{Coro}[Def]{Corollary}

\newtheorem{ThmDef}[Def]{Theorem and Definition}
\newtheorem{LemDef}[Def]{Lemma and Definition}

\theoremstyle{remark}
\newtheorem*{Rem}{Remark}
\newtheorem*{Rems}{Remarks}

\newtheorem*{Pf}{Proof}
\newtheorem*{Thm*}{}

\newcommand{\R}{\mathbb{R}}

\newcommand{\N}{\mathbb{N}}
\newcommand{\Z}{\mathbb{Z}}
\newcommand{\g}{\operatorname{g}}

\newcommand{\T}{\operatorname{T}}
\newcommand{\Ric}{\operatorname{Ric}}
\newcommand{\Rm}{\operatorname{Rm}}
\newcommand{\Scal}{\operatorname{R}}
\newcommand{\tr}{\operatorname{tr}}
\newcommand{\grad}{\operatorname{grad}}
\newcommand{\rot}{\operatorname{rot}}
\newcommand{\diver}{\operatorname{div}}
\newcommand{\supp}{\operatorname{supp}}
\newcommand{\id}{\operatorname{id}}
\newcommand{\dist}{\operatorname{dist}}
\newcommand{\inter}{\operatorname{int}}

\newcommand{\myg}[1]{{}^{#1}\!g}
\newcommand{\mygam}[1]{{}^{#1}\!\gamma}
\newcommand{\mylap}[1]{{}^{#1}\!\triangle}
\newcommand{\my}[2]{{}^{#1}{#2}}
\newcommand{\inv}[2]{{}^{#1}\!g{^{#2}}}
\newcommand{\der}[1]{\accentset{#1}{;}}
\newcommand{\free}{\accentset{\,\circ}{h}}
\newcommand*{\dt}[1]{%
  \accentset{\mbox{\large\bfseries .}}{#1}}
\newcommand*{\ddt}[1]{%
  \accentset{\mbox{\large\bfseries .\hspace{-0.25ex}.}}{#1}}

\newcounter{todocounter}
 
\newcommand{\schild}{Schwarz\-schild }
\newcommand{\mytitle}{\sc The Newtonian Limit of Geometrostatics}

\makeindex{ind}
\makeindex{sym}

\begin{titlepage}
\title{\mytitle}

\author{Carla Cederbaum\\[12pt]}

\subject{Dissertation}
\date{}

\end{titlepage}

\begin{document}
\maketitle


\pagestyle{headings}
{\pagenumbering{roman}
\setcounter{page}{1}

\begin{center}\begin{minipage}{14cm}
\section*{Summary}
This thesis compares Newton's classical gravitational theory (NG) to Einstein's theory of General Relativity (GR).
In particular, we study the so-called Newtonian limit of General Relativity: the question if and in what sense Newtonian Gravity arises as a limit of General Relativity for speeds that are small when compared to the speed of light. 
This question is relevant for consistency reasons but there are also practical reasons for its significance. 
These arise as the Newtonian theory of gravitation is still being used for astrophysical, astronomical, and technical computations and observations, today. Furthermore,
a deeper understanding of the Newtonian limit can simplify and improve relativistic modeling, numerical simulations, and physical interpretation.

We analyze the Newtonian limit mathematically in the language of frame theory. This theory was introduced in the 1980s by J\"urgen Ehlers \cite{Ehl5}.
It allows for a uniform description of the Newtonian (coordinate variant) and relativistic (coordinate invariant) theories.\\

Here, we specifically use Ehlers' frame theory to investigate the Newtonian limit of physical properties of a relativistic system like its mass and center of mass.
Our analysis focuses on static isolated relativistic systems with compactly supported matter. 
We introduce the name {\it geometrostatics} for the study of these systems to emphasize the significance of geometry in our approach and to discriminate it from the more general field of geometrodynamics.

By establishing and reinforcing analogies to the Newtonian setting, our analysis of the Newtonian limit also deepens our understanding of geometrostatics itself.
Through a conformal transformation, geometrostatics is closely connected to {\it pseudo-Newtonian gravity}, a theory which proves useful for the consideration of the Newtonian limit of geometrostatics.
At the same time, it allows us to translate many Newtonian concepts into geometrostatics. For example, we formulate a Second Pseudo-Newtonian Law of Motion, characterize equipotential surfaces,
and answer uniqueness questions concerning geometric and physical properties of geometrostatic systems.\\

Using the framework of geometrostatics and pseudo-Newtonian gravity, we furthermore introduce new quasi-local definitions and explicit formulas for the mass and center of mass of a physical system.
We relate these to the asymptotic behavior of the geometrostatic variables (cf.~e.g.~\cite{Beig2,KM}) as well as to the established notions of mass and center of mass in General Relativity (cf.~e.g.~\cite{HY,Bartnik}).
The new notions differ from the established ones in their local character: they are not only determined asymptotically but can be computed exactly in the immediate vicinity of the matter or black holes. Thus, they 
provide a new tool for the analysis of static relativistic systems. Simultaneously, they facilitate the proof of convergence of mass and center of mass in the Newtonian limit. This proof is presented in the last chapter of this thesis.
\end{minipage}\end{center}
\cleardoublepage

\vspace*{1ex}
\begin{figure}[h]\begin{center}
\includegraphics[scale=1.2]{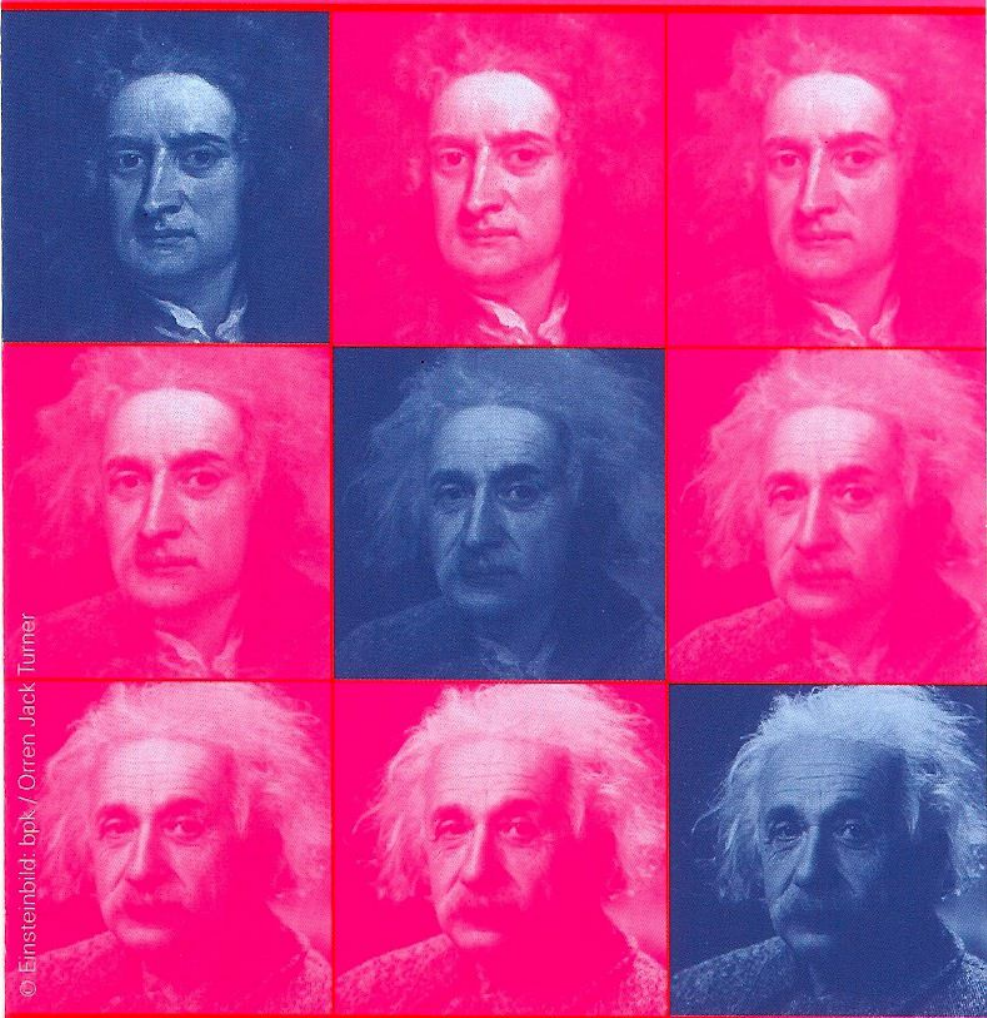}
\caption*{{\large\sc The Newtonian Limit in a Nutshell}\footnotemark}\end{center}\end{figure}
\footnotetext{\copyright~Einsteinbild: bpk / Orren Jack Turner; layout: Sascha Rieger, Milde Marketing}
\cleardoublepage

\selectlanguage{english}

\tableofcontents{}
\cleardoublepage}

\setcounter{chapter}{-1}
\pagenumbering{arabic}
\setcounter{page}{1}
\chapter{Introduction}
Gravity is one of the four fundamental physical forces in our universe. It gives rise to the orbits of the planets around the sun and to many other equally omnipresent phenomena. Starting with Isaac Newton in the 18th century, natural philosophers and physicists have theorized on and experimented with gravity and its astronomical consequences throughout history. Two major gravitational theories have evolved that are still relevant today. One of them is the so-called Newtonian theory of gravity or ``Newtonian gravity (NG)''\index{ind}{NG}\index{ind}{Newtonian gravity} for short. It builds upon the Newtonian laws of motion and characterizes gravitation as a force field acting instantly and at a distance. In modern language, it is often formulated in terms of a (Newtonian) potential $U$\index{ind}{potential ! Newtonian}\index{ind}{potential}\index{sym}{$U$} that satisfies a Poisson equation $\triangle U=4\pi G\rho$ relating it to the matter density $\rho$ via the gravitational constant $G$ on the Euclidean space $\R^3$. We have this formulation of Newtonian gravity at the back of our minds throughout this thesis.

The other major theory of gravitation that remained important and is moreover still vividly researched on today is Albert Einstein's theory of general relativity or ``general relativity (GR)'' for short dating back to the early years of the twentieth century. It takes a very different approach eliminating the concept of force altogether and unifying space and time into a curved Lorentzian $4$-manifold $(L^4,g)$ called spacetime. The ``Einstein equations'' $\Ric-\frac{1}{2}\Scal g=\frac{8\pi G}{c^4}T$ then relate the curvatures $\Ric$ and $\Scal$ of the Lorentzian metric to an ``energy-momentum'' or ``matter tensor field'' $T$, also coupling them via the gravitational constant $G$. In addition, however, another constant called $c$ enters the game. $c$ denotes the speed of light; it gave rise to the birth of the theory of (special) relativity in the first place and holds responsible for the coupling of space and time in GR.

In this thesis, we are interested in shedding light on the relationship of these two theories. This relationship has been discussed and studied by numerous scientists and philosophers and from many different perspectives. Before we turn to its physical and mathematical aspects, let us have a quick look at some of its philosophical facets following J\"urgen Ehlers in \cite{Ehl5,Ehl4,Ehl6}. First of all, the relationship between the theories NG and GR is one of the most important examples of how a ``finalized'' theory is replaced (or not so?) by a more comprehensive one in science. It can thus add to the theory of science debate which tries to settle the issue of whether such change comes by revolution or rather occurs in small steps. Secondly, and maybe even deeper, the two gravitational theories provide a good example for studying the epistemological question whether progress in science can at all be made or whether what we as scientists consider to be an ``instant of progress'' is just a ``change of taste or opinion''.

Coming back to the physical aspects of the relationship of NG and GR, it is important to realize that many if not most empirical observations and measurements which have been designed to test GR have been pursued in weakly coupled relatively isolated systems moving slowly with respect to each other. Moreover, predictions and measurement methodology are often devised and calculated within a Newtonian framework of thought and in a perturbative approach starting with Newtonian concepts and equations. This kind of approach in fact implicitly presumes that GR ``goes over to NG as $c\to\infty$''. This, however, is not a prerequisite but a claim which is intuitively convincing, suggestive, and desirable for the purpose of perturbative methods but needs to be proven rigorously. In the last thirty years, mathematicians and physicists have put a lot of effort into formalizing and proving that these perturbative approaches (which go by the names of ``Newtonian limit'' ($c\to\infty$) and ``post-Newtonian expansions''(power series in $1/c^2$ or $1/c$)) are actually justified and that there are reasonable examples for their usefulness (cf.~e.~g.~Alan D.~Rendall \cite{Rend3}, Todd Oliynyk \cite{Oly}, Martin Lottermoser \cite{Lott1}, and J\"urgen Ehlers \cite{Lott1} as well as references therein). Their arguments rely on a mathematical frame theory devised and named after J\"urgen Ehlers, cf.~\cite{Ehl6}. This frame theory allows to not only formally compare Newtonian and relativistic equations or specific components of relevant tensors but empowers researchers to phrase questions on convergence of full solutions to those equations and therewith formalizes the intuitive idea of ``$c\to\infty$''.
 
In this thesis, we use the framework provided by J\"urgen Ehlers to study the Newtonian limit of physical properties like mass and center of mass of given solutions to GR. To the best of the author's knowledge, this question has not yet obtained much attention. We focus on the special case of static isolated relativistic systems with compactly supported matter, a setting which we call ``geometrostatic'' to underline the geometric approach we take and in accordance with the name ``geometrodynamics'' that GR is frequently given when its geometric character is to be emphasized. Static isolated systems model individual stellar bodies or groups of stars and black holes that are not changing in time (``static'') and are not influenced from the outside world (``isolated'').

Mathematically speaking, staticity is modeled by a special timelike Killing vector field in the spacetime while isolatedness is modeled by asymptotic flatness of the relevant mathematical objects (a Riemannian $3$-metric as well as additional geometric quantities). The assumption that the matter has compact support is made for two reasons: The first one is that many analytical arguments simplify technically in that case and much of literature is dedicated to this situation. The second motivation for choosing to consider systems with compactly supported matter, only, is that it is intuitively appropriate from a physical point of view as one would not expect stars to be infinitely extended. 

In order to prove that the mass and center of mass of a geometrostatic system converge to the mass and center of mass of the Newtonian system that constitutes its Newtonian limit, we execute a number of steps. First of all, after summarizing and contextualizing well-known results on static and isolated relativistic systems and introducing the concept of geometrostatic systems in Chapters \ref{chap:prelims} through \ref{chap:static}, we prove in Theorem \ref{thm:asym-CoM} that every such system possesses a well-defined asymptotic (total) center of mass. This center of mass is completely equivalent to other notions of center of mass that have been suggested in different contexts e.~g.~by Gerhard Huisken and Shing-Tung Yau \cite{HY}, by Richard Arnowitt, Stanley Deser, and Charles W.~Misner \cite{ADM}, and by Lan-Hsuan Huang \cite{Huang2}, cf.~Theorem \ref{thm:centers}.

Our analysis builds upon earlier results on the asymptotics of static isolated relativistic systems, e.~g.~by Robert Beig and Walter Simon \cite{Beig2,BS1} and by Daniel Kennefick and Niall \'O Murchadha \cite{KM} and on considerations of the total mass by again Richard Arnowitt, Stanley Deser, and Charles W.~Misner \cite{ADM} and by Robert Bartnik \cite{Bartnik}, among others. It is formulated by means of weighted Sobolev spaces and relies on the faster fall trick \ref{higherreg} which we prove in Chapter \ref{chap:prelims}.

Using similar techniques as for the definition of the asymptotic center of mass, we prove uniqueness results for the Riemannian $3$-metric and the ``lapse function'' constituting a geometrostatic system (cf.~Section \ref{sec:asym}). These uniqueness assertions clarify the relationship between the $3$-metric and the lapse function. They point towards similarities as well as towards differences between geometrostatics and static Newtonian gravity.

In a second step which we pursue in Section \ref{sec:pseudo}, we recast geometrostatics into a language which is more similar to the Newtonian setting and which we thus call ``pseudo-Newtonian gravity (pNG)''. This recasting happens through a rescaling of the lapse function and by the aid of a conformal change of the $3$-metric. Although the term ``pseudo-Newtonian'' might be new, the rescaling and conformal change procedure is well-known in the literature\footnote{and I would like to thank Bernd Schmidt for suggesting its use to me.}. After translating the afore-made definitions and results into pseudo-Newtonian gravity, we introduce quasi-local notions of mass and center of mass of a pseudo-Newtonian system in Sections \ref{sec:mpseudo} and \ref{sec:CoMpseudo}. These notions are closely analogous to the Newtonian concepts of mass and center of mass. We prove that they asymptotically agree with the asymptotically defined ADM-mass and asymptotic center of mass, cf.~Theorems \ref{ADM=pseudo} and \ref{asym=pseudo}. Moreover, other than the afore-mentioned asymptotic concepts, the pseudo-Newtonian mass and center of mass can be read off in the vicinity of the matter just as one would intuitively expect from everyday experience (and from NG).

In Chapter \ref{chap:geo}, we continue to pursue the approach of gaining insight into geometrostatics and pseudo-Newtonian gravity by analogy to Newtonian gravity. In this spirit, we give a simple proof of a static (toy) version of the positive mass theorem originally and much more generally proven by Richard Schoen and Shing-Tung Yau \cite{SchoenYau}. Afterwards, we define a concept of force acting on a test particle in a geometrostatic system and prove a pseudo-Newtonian version of Newton's second law of motion for it. In Section \ref{sec:equi}, we get closer to the heart of Newtonian gravity and present a variational proof of a relativistic equivalent (``surfaces of equilibrium'') of the well-known Newtonian fact that test particles constrained to an equipotential surface do not accelerate. Combining this result with an important result by Yvonne Choquet-Bruhat \cite{CHB} stating that the Einstein equations can be reformulated as a well-posed initial value problem, we reprove one of the analytical uniqueness results asserted in Chapter \ref{chap:static} from a more geometric and at the same time more physical perspective.  Finally, in Section \ref{sec:photon}, as a by-product of our improved understanding of geometrostatics and of surfaces of equilibrium in particular, we prove uniqueness of so-called photon spheres -- a construct intimately related to black holes -- mimicking a proof by Werner Israel \cite{Israel} where he proves uniqueness of black holes.

Coming back to our main thesis on the Newtonian limit of mass and center of mass in Chapter \ref{chap:FT}, we first review the foundations of frame theory. Afterwards, we generalize the concept of staticity from general relativity to frame theory and present, refine, and adapt the concept of Newtonian limit to the context of static isolated systems. The quasi-local pseudo-Newtonian notions of mass and center of mass introduced in Chapter \ref{chap:pseudo} then allow us to prove the main result of this final chapter: Convergence of mass and center of mass under the Newtonian limit, cf.~Theorems \ref{thm:NLm} and \ref{thm:NLCoM}, respectively.

I would like to thank my advisor Gerhard Huisken for many inspiring discussions and helpful advice. I give my sincere thanks to both Gerhard Huisken and J\"urgen Ehlers for suggesting this fascinating topic to me. Last but not least, I thank my friends, colleagues, family, and my Wissenschaftssommer team and visitors for answering and asking innumerable questions.

\chapter{Mathematical Preliminaries}\label{chap:prelims}
Before we begin with a short introduction into general relativity and in particular into geometrostatics, we need to introduce some notational conventions and to collect and present some results from differential geometry and from geometric analysis which we will do in this chapter. In Section \ref{sec:notation}, we fix notations. In Section \ref{sec:subs}, we quickly discuss well-known results on submanifold geometry. In Section \ref{sec:LieKill}, we would like to remind the reader of the concept of Lie derivatives and present a slight generalization of this concept as a preparation for our discussion of the Newtonian limit within the framework of Ehlers' frame theory in Chapter \ref{chap:FT}. Finally, in Section \ref{sec:sobo}, we quote a number of facts on the analytic tool of weighted Sobolev spaces and prove a faster fall-off trick.

\section{Notation and Conventions}\label{sec:notation}
Geometrically speaking, general relativity is a theory of 4-dimensional Lorentzian manifolds (``spacetimes'')\index{ind}{spacetime}. These 4-manifolds are frequently split into a ``time direction'' and a collection of $3$-dimensional spacelike hypersurfaces with induced Riemannian metrics, the ``spatial slices''. This makes it necessary to simultaneously deal with $3$- and $4$-dimensional curvature and connection fields. Moreover, asymptotic flatness conditions that will be introduced in Chapter \ref{chap:iso} make it desirable to allude to connection and curvature fields of different metrics like Euclidean and Schwarzschildian ones, simultaneously. Thus, in order to notationally distinguish between the curvature tensors and connection fields induced from different metrics, we will in all these cases use a label like the dimension of the manifold or an individual letter like ``E'' for ``Euclidean'' or ``S'' for ``Schwarzschildian'' attached to all objects from the left.

To fix notation, assume that $(M^n,\myg{A})$ is an $n$-dimensional pseudo-Riemannian manifold with label ``A'' and that $\{X_i\}_{i=1,\dots,n}$ is a local frame for $M^n$. Then let $({\myg{A}}_{ij})$ stand for the matrix $(\myg{A}(X_i,X_j))$ and $(\inv{A}{ij})$ for its inverse matrix representing the induced metric on the dual bundle $T^\ast\!M^n$\index{sym}{$T^\ast M^n$}\index{ind}{cotangent bundle}. We will continuously use $({\myg{A}}_{ij})$ and $(\inv{A}{ij})$ to pull indices up and to push them down\index{ind}{musical operations} -- but we would like to draw the reader's attention to the fact that we will not do so in Section \ref{sec:FT}, where we discuss J\"urgen Ehlers' frame theory in which the position of an index is of physical significance.

Coming back to notational issues, let $\{\omega^j\}_{j=1,\dots,n}$ denote the dual frame to $\{X_i\}$ and let $T$ be any $(s,t)$-tensor on $M^n$. We will use the expression $T_{i_1 \dots i_s}^{j_1\dots j_t}$ as a shorthand standing for $T(X_{i_1},\dots,X_{i_s},\omega^{j_1},\dots,\omega^{j_t})$. Moreover, if $\myg{A}$ is Riemannian, we will denote the induced norm of a tensor $T$ by $\lvert T \rvert_A:=\left(T_{i_1,\dots i_s}^{j_1 \dots j_t}\, T_{k_1 \dots k_s}^{l_1 \dots l_t}\, \myg{A}_{j_1l_1}\dots\myg{A}_{j_t l_t}\,\inv{A}{i_1k_1}\dots\inv{A}{i_s k_s} \right)\!^{1/2}$\index{ind}{norm}\index{sym}{$\lvert . \rvert_A$} where we have used Einstein's summation convention\index{ind}{Einstein's summation convention} as we will continue to do throughout the text.

Coming back to an arbitrary pseudo-Riemannian manifold $(M^n,\myg{A})$, we will use the symbol $\my{A}{\nabla}$\index{sym}{$\my{A}{\nabla}$} to refer to the induced {\it Levi-Civit\`a connection}\index{ind}{Levi-Civit\`a connection}\index{ind}{connection ! Levi-Civit\`a}. The {\it Christoffel symbols}\index{ind}{Christoffel symbols} corresponding to $\my{A}{\nabla}$ will be denoted as $\my{A}{\Gamma}^k_{ij}$\index{sym}{$\my{A}{\Gamma}$} such that $\my{A}{\Gamma}^k_{ij}=\omega^k(\my{A}{\nabla}\!_{X_i} X_j)$. If $\myg{A}$ is Riemannian, then $d\mu_{A}$\index{sym}{$d\mu_{A}$} or $d\sigma_{A}$\index{sym}{$d\sigma_{A}$} denote the induced area or surface measure on $M^n$. In general, $\!\my{A}{\tr}$ will stand for the trace\index{ind}{trace}\index{sym}{$\my{A}{\tr}$} of a tensor with respect to $\myg{A}$, $\my{A}{\diver}$ will denote the induced {\it divergence operator} acting on vector fields\index{ind}{divergence}\index{sym}{$\my{A}{\diver}$},  $\my{A}{\grad} f$ the induced {\it gradient (vector)}\index{ind}{gradient}\index{sym}{$\my{A}{\grad}$}, and $\my{A}{\nabla}^2 f$ the {\it Hessian}\index{ind}{Hessian}\index{sym}{$\my{A}{\nabla}^2$} of a real-valued function $f$. If $\myg{A}$ is Riemannian, $\mylap{A} f$ will denote the {\it Laplacian}\index{ind}{Laplacian}\index{sym}{$\mylap{A}$} while if it is Lorentzian, we will use the symbol $\my{A}{\Box}f$ to denote the {\it wave operator}\index{ind}{wave operator}\index{sym}{$\my{A}{\Box}$} acting on a real-valued function $f$. Finally, $\my{A}{\Rm}$ will denote the induced {\it Riemannian curvature endomorphism}\index{ind}{Riemannian curvature endomorphism ! of a pseudo-Riemannian metric}\index{ind}{curvature ! endomorphism}\index{sym}{$\my{A}{\Rm}$} with sign convention such that
\begin{equation}\label{Rm}
{\my{A}{\Rm}_{ijk}}^{l} X_l=\my{A}{\nabla}\!_{X_i} \my{A}{\nabla}\!_{X_j} X_k -\my{A}{\nabla}\!_{X_j} \my{A}{\nabla}\!_{X_i} X_k -\my{A}{\nabla}\!_{[X_i,X_j]} X_k\in\Gamma(T\!M),
\end{equation}
where $[\cdot,\cdot]$ denotes the Lie bracket\index{sym}{$[\cdot,\cdot]$}\index{ind}{Lie bracket} on $M^n$, $T\!M$ will denote the tangent bundle over $M$\index{sym}{$TM$} and $\Gamma(B)$\index{sym}{$\Gamma(B)$} denotes the set of all smooth sections of a bundle $B$. If, in particular, $\{X_i\}$ is a coordinate frame, this has the coordinate version 
\begin{equation}\label{Rmcoord}
{\my{A}{\Rm}_{ijk}}^{l}=\my{A}{\Gamma}_{jk,i}^l-\my{A}{\Gamma}_{ik,j}^l+\my{A}{\Gamma}_{jk}^s\my{A}{\Gamma}_{is}^l-\my{A}{\Gamma}_{ik}^s\my{A}{\Gamma}_{js}^l. 
\end{equation}
For the {\it Ricci curvature tensor}\index{ind}{Ricci curvature tensor}\index{ind}{curvature ! Ricci}\index{sym}{$\my{A}{\Ric}$}, we will use the sign convention that
\begin{equation}\label{Ric}
\my{A}{\Ric}_{ij}= \,\inv{A}{kl}\,\myg{A}_{sj}\,{\my{A}{\Rm}_{ikl}}^{s}=-\,{\my{A}{\Rm}_{ikj}}^{k} 
\end{equation}
while the {\it scalar curvature}\index{ind}{scalar curvature}\index{ind}{curvature ! scalar}\index{sym}{$\my{A}{\Scal}$} is given by $\my{A}{\Scal}=\inv{A}{ij}\,\my{A}{\Ric}_{ij}$ as usual. These curvature tensors have special properties in low dimensions which we will exploit in this thesis. For example, it is well-known that -- in $3$ dimensions --, the Ricci tensor determines the Riemannian curvature endomorphism through the formula
\begin{equation}\label{formula:3Rm}
\my{3}{\Rm}_{ijkl}=\my{3}{\Ric}_{il}\myg{3}_{jk}-\my{3}{\Ric}_{ik}\myg{3}_{jl}-\my{3}{\Ric}_{jl}\myg{3}_{ik}+\my{3}{\Ric}_{jk}\myg{3}_{il}-\frac{1}{2}\my{3}{\Scal}\,(\myg{3}_{il}\myg{3}_{jk}-\myg{3}_{ik}\myg{3}_{jl}),
\end{equation}
which will be of great help to us in Sections \ref{sec:pm} and \ref{sec:photon}. In $2$ dimensions, there is a different well-known theorem which we will frequently refer to: If we denote the {\it Gau{\ss}-curvature}\index{ind}{Gau{\ss}-curvature} of a $2$-dimensional Riemannian manifold by $K$\index{sym}{$K$}, then $2K=\my{2}{\Scal}$ and the Gau{\ss}-Bonnet theorem\index{ind}{Gau{\ss}-Bonnet theorem} tells us that
\begin{equation}\label{GB}
 \int_\Sigma K \,d\sigma = 4\pi
\end{equation}
holds on (topological) $2$-spheres $\Sigma$ with intrinsic geometrically induced measure $d\sigma$.

In this thesis, we will mainly encounter manifolds of dimensions $2$ (``surfaces'')\index{ind}{surface}, $3$ (``spatial slices'')\index{ind}{spatial slice}\index{ind}{slice}, and $4$ (``spacetimes'')\index{ind}{spacetime}. All appearing manifolds are tacitly assumed to be connected. From now on, indices on surfaces will be denoted by upper case Latin letters\index{sym}{$I,J,K$} running from $1$ to $2$, those on $3$-dimensional manifolds by lower case Latin letters\index{sym}{$i,j,k$} running from $1$ to $3$, and spacetime indices by lower case Greek letters\index{sym}{$\alpha,\beta,\gamma$} running from $0$ to $3$.

In most of the physics and part of the mathematics literature, the $4$-dimensional Lorent\-zian metric of
general relativity is usually denoted by $ds^2$\index{sym}{$ds^2$}. We will stick to this convention, here, but use the label ``$4$'' on all derived quantities in order to indicate their $4$-dimensional nature. In particular, we will denote the matrix components of $ds^2$ by $\left(\myg{4}_{\alpha\beta}\right)$. Lorentzian metrics will have signature $(-,+,+,+)$. For the consideration of the center of mass in Chapter \ref{chap:mCoM} and for the asymptotic considerations in Chapters \ref{chap:iso} through \ref{chap:mCoM}, we will also need the notion of harmonic and wave harmonic coordinates.
\begin{Def}[Harmonic Coordinates]
Let $(M^n,g)$ be a smooth Riemannian (Lorent\-zian) manifold. We call a system of coordinates $\{x^i\}$ on an open subset $U\subset M^n$ {\it harmonic}\index{ind}{coordinates ! harmonic}\index{ind}{harmonic ! coordinates} ({\it wave harmonic}\index{ind}{coordinates ! wave harmonic}) if the coordinate functions satisfy the Laplace (wave) equation $\mylap{g}x^i=0$ ($\my{g}{\Box}x^i=0$) in all of $U$.
\end{Def}

Sometimes, in particular in Section \ref{sec:FT}, it will be more convenient to use abstract index notation\index{ind}{abstract index notation} (cf.~e.~g.~\cite{Rendall}) which makes the contra- and covariant type of a given tensor explicit. For example, $g_{ij}$ then does not refer to the matrix element of $g$ at position $(i,j)$ but to the metric tensor $g$ understood that it is a (2,0)-tensor a priori. When we are using abstract index notation on a pseudo-Riemannian manifold $(M^n,\myg{A})$,  $\der{A}$\index{sym}{$\der{A}$} will be used to denote covariant differentiation, and index (anti-)symmetrization will be indicated through (square brackets) parenthesis. We will frequently exchange classical tensorial notation referring for example to the metric as $g$ for abstract index notation (calling it $g_{ij}$) and use whichever is better suited. We will however draw the reader's attention to any possible source of confusion arising from this.

\section{Submanifold Geometry}\label{sec:subs}
As already indicated above, we will frequently be dealing with submanifolds of pseudo-Riemannian manifolds in this thesis. They appear both as spacelike slices in (Lorentzian) spacetimes and as ($2$-dimensional) surfaces in these spacelike slices which will help us to define and study the center of mass of an isolated gravitational system in Chapters \ref{chap:iso} and \ref{chap:mCoM}. Let us therefore shortly review some concepts and formulae relating curvatures of a submanifold to those of the ambient manifold.

To this end, let $(M^{n+1},g)$ be a pseudo-Riemannian manifold hosting an embedded submanifold $\Sigma^n\subset M^{n+1}$. Assume that $g$ has signature $(\sigma,+,\dots,+)$ with $\sigma\in\{+,-\}$ and that $\Sigma$ is spacelike if $\sigma=-$. Let $\nu$\index{sym}{$\nu$} denote a $g$-unit normal vector field for $\Sigma$ tacitly assumed to point {\it outwards}\index{ind}{outwards}\index{ind}{unit normal ! outward} if there is a suitable interpretation of this term as, for example, there is in the setting of surfaces embedded into asymptotically flat manifolds as discussed in Chapters \ref{chap:iso}, \ref{chap:static}, and \ref{chap:mCoM}. Observe that $g(\nu,\nu)=\sigma \cdot 1$ and fix the sign of the {\it second fundamental form}\index{ind}{second fundamental form} $h$\index{sym}{$h$} of $\Sigma$ by
\begin{equation}\label{2nd}
h_{IJ}:=g(\my{n+1}{\nabla}_{X_I} \nu,X_J)=-g(\my{n+1}{\nabla}_{X_I} X_J,\nu)
\end{equation}
on any local frame $\{X_I\}_{I=1,\dots,n}$ of $\Sigma$. From this, we can read off the identity
\begin{equation}\label{nabla}
\my{n+1}{\nabla}\!_{X_I}X_J=\my{n}{\nabla}\!_{X_I} X_J -\sigma\, h_{IJ}\,\nu. 
\end{equation}

As usual, the $\myg{n}$-trace of $h$ is called the {\it (scalar) mean curvature}\index{ind}{mean curvature} and is denoted by $H=\inv{n}{IJ}\,h_{IJ}$\index{sym}{$H$}. We use the expression $\free$\index{sym}{$\free$} to denote the trace-free part $\free_{IJ}:=h_{IJ}-\frac{1}{n}Hg_{IJ}$ of $h$. Motivated by the fact that round spheres in Euclidean spaces have vanishing trace-free part of their second fundamental forms, we call $\Sigma$ {\it extrinsically round}\index{ind}{extrinsically round} if it satisfies $\free=0$.

The following equations are well-known consequences of the above definitions, cf. e.~g. \cite{Lee}. The first one is the {\it Gau{\ss}-equation}\index{ind}{Gau{\ss}-equation}
\begin{equation}\label{Gauss1}
\my{n+1}{\Rm}_{IJKL}=\my{n}{\Rm}_{IJKL}-\sigma\,h_{IL}h_{JK}+\sigma\,h_{IK}h_{JL}.
\end{equation}
Taking the $n$-trace of this equation over $J$ and $K$, we obtain
\begin{equation}\label{Gauss2}
\my{n+1}{\Ric}_{IL}-\sigma\,\,\my{n+1}{\Rm}(X_I,\nu,\nu,X_L)=\my{n}{\Ric}_{IL}-\sigma\,H\,h_{IL}+\sigma\,(h^2)_{IL},
\end{equation}
with $(h^2)_{IL}:={h_I}^{K}\,h_{KL}$\index{sym}{$h^2$}. Again taking the $n$-trace, this leads to
\begin{equation}\label{Gauss3}
\my{n+1}{\Scal}-2\sigma\;\my{n+1}{\Ric}(\nu,\nu)= \my{n}{\Scal}-\sigma\,H^2+\sigma\lvert h\rvert^2
\end{equation}
where we have used the well-known symmetries of the Riemannian curvature tensor. Furthermore, we have the {\it Codazzi equation}\index{ind}{Codazzi equation}
\begin{equation}\label{Codazzi}
\myg{n+1}(\,\my{n+1}{\Rm}(X_K,X_I,\nu),X_J)=\left(\my{n}{\nabla}\!_{X_K}h\right)_{IJ}-\left(\my{n}{\nabla}\!_{X_I}h\right)_{KJ}.
\end{equation}

Moreover, if $f:M^{n+1}\to\R$ is a smooth function and if $\sigma=1$, Equation \eqref{nabla} leads to the helpful equality\begin{equation}\label{surflap}
\mylap{n+1}f=\mylap{n}f+\my{n}{\nabla}^2f(\nu,\nu)+H\nu(f).
\end{equation}

\subsection*{Foliations and the Global Frobenius Theorem}\label{subsec:Frobenius}
In this thesis, submanifolds will frequently combine to a so-called {\it foliation}\index{ind}{foliation} of a manifold $M^n$. Following \cite{Lee2}, we understand a ($k$-dimensional) foliation of $M^n$ to be a collection of disjoint, connected, immersed $k$-dimensional submanifolds of $M^n$ (called the {\it leaves}\index{ind}{leaf} of the foliation) whose union is $M^n$ and such that in a neighborhood $U$ of each point $p\in M^n$, there is a smooth chart $\phi:U\to\R^n$ with the property that $\phi(U)=U_k\times U_{n-k}$ with $U_k, U_{n-k}$ open subsets of $\R^k, \R^{n-k}$, respectively, and such that each leaf of the foliation intersects $U$ in either the empty set or a countable union of $k$-dimensional slices of the form $\phi^{-1}\left(\{x\in\phi(U)\,\vert\,x^i=c^i\mbox{ for }i=k+1,\dots,n\}\right)$ for constants $c^i\in\R$.

The question of whether a given manifold $M^n$ is foliated by submanifolds possessing a specific property like e.~g.~constant mean curvature in the case of hypersurfaces in a pseudo-Riemannian manifold plays a central role in the geometric analysis of manifolds. It is closely related to the concepts of ``tangent distributions'' and ``integral submanifolds'' by Frobenius' theorem. Here, a (smooth $k$-dimensional) {\it tangent distribution}\index{ind}{distribution}\index{ind}{distribution ! tangent} $D$ is a smooth subbundle of the tangent bundle $T\!M^n$. A tangent distribution is called {\it involutive}\index{ind}{distribution ! involutive} if it is closed under the Lie bracket operation or in other words if the Lie bracket of any two smooth sections $X,Y\in\Gamma(D)$ satisfies $\left[X,Y\right]\in\Gamma(D)$. An immersed submanifold $\Sigma^k\subset M^n$ is called an {\it integral submanifold}\index{ind}{submanifold ! integral} of a tangent distribution $D$ if $D=T\Sigma^k$. While it is straightforward to see that the tangent space of any immersed submanifold is a smooth involutive tangent distribution, the converse is the content of a theorem by Ferdinand Georg Frobenius.
\begin{Thm}[Global Frobenius Theorem]\index{ind}{Frobenius' theorem}\label{Frob}
Let $D$ be a smooth involutive tangent distribution on a smooth manifold $M^n$. Then the collection of all maximal connected integral submanifolds of $D$ forms a foliation of $M^n$ and $D$ is called {\bf integrable}\index{ind}{tangent distribution ! integrable}.
\end{Thm}
We will apply this theorem in Chapter \ref{chap:static} in order to better understand the geometric structure of static spacetimes. It will also play a role in Section \ref{sec:NLStatic} where we study related questions in frame theory.

\section{Lie Derivatives, Connections, and Killing Vectors}\label{sec:LieKill}
As we intend to consider the Newtonian limit of static spacetimes -- or in other words of spacetimes that are invariant under translation and reflection of time --, we will have to adapt the concept of Killing vector fields and thus the idea of Lie derivatives to the setting of frame theory, the formal unification of Newtonian gravity and general relativity which allows for this limit, cf.~Chapter \ref{chap:FT}. In particular, we need to generalize the well-known concept of Lie derivatives\index{ind}{Lie derivative} of tensor fields to include the possibility of Lie derivating connections. Before we present the generalization to connections, let us shortly review the concept of Lie derivatives in the conventional setting. Recall that the Lie derivative\index{ind}{Lie derivative ! of a vector field}\index{sym}{$\mathfrak{L}_X$} of a vector field $Y$ in direction of another vector field $X$ is defined as $$\mathfrak{L}_XY(p)=\lim_{t\to 0}\frac{\theta_t^\ast Y\vert_{\theta_t(p)}-Y\vert_p}{t}$$ where $\theta_t$ denotes the flow of $X$ and the asterisk stands for the induced pull-back\index{ind}{pull-back}\index{sym}{$\theta^\ast$}. Now, it is well-known that this notion of Lie derivative of a vector field can be generalized to general tensor fields in this straightforward fashion\index{ind}{Lie derivative ! of a tensor field} $$\mathfrak{L}_X T(p)=\lim_{t\to 0}\frac{\theta_t^\ast T\vert_{\theta_t(p)}-T\vert_p}{t}$$ such that $\mathfrak{L}_X T$ is a tensor field of the same rank as $T$. There is a Leibniz rule\index{ind}{Leibniz rule ! for Lie derivatives} saying that
\begin{eqnarray}\nonumber
X\left(T(Y_1,\dots,Y_s,\omega^1,\dots,\omega^t)\right)&=&\left(\mathfrak{L}_XT\right)(Y_1,\dots,Y_s,\omega^1,\dots,\omega^t)\\\label{Lie product rule}
&&+\sum_{i=1,\dots s}T(Y_1,\dots,\mathfrak{L}_X Y_i,\dots,Y_s,\omega^1,\dots,\omega^t)\\\nonumber
&&+\sum_{j=1,\dots t}T(Y_1,\dots,Y_s,\omega^1,\dots,\mathfrak{L}_X\omega^j,\dots,\omega^t).
\end{eqnarray}
where $Y_i\in\Gamma(T\!M)$ and $\omega^j\in\Gamma(T^\ast\! M)$ are arbitrary fields.

To put us in a position where we can define Lie derivatives of connections, recall that an {\it affine connection}\index{ind}{affine connection}\index{ind}{connection ! affine}\index{sym}{$\nabla$} $\nabla:\Gamma(T\!M)\times\Gamma(T\!M)\to\Gamma(T\!M)$ on a smooth manifold $M^n$ is a map that is linear over $C^\infty(M)$ in its first and linear over $\R$ in its second argument and that satisfies the Leibniz rule\index{ind}{Leibniz rule ! for covariant derivatives} $$\nabla\!_X(fY)=f\nabla\!_XY+X(f)Y$$ for any vector fields $X, Y\in\Gamma(T\!M)$ and any $f\in C^\infty(M)$. Its {\it torsion field}\index{ind}{torsion field} $\mathfrak{T}$\index{sym}{$\mathfrak{T}$} is defined by
\begin{equation}\label{torsion}
\mathfrak{T}(X,Y)=\nabla\!_X Y -\nabla\!_Y X-[X,Y]
\end{equation}
for $X, Y\in\Gamma(T\!M)$. $\nabla$ is called {\it torsion free}\index{ind}{connection ! torsion free} or {\it symmetric}\index{ind}{connection ! symmetric} if its torsion tensor vanishes. In the following, we will only consider torsion free connections such as the Levi-Civit\`a connection of a pseudo-Riemannian metric or the Cartan connection\index{ind}{Cartan connection} introduced in Chapter \ref{chap:FT}. The {\it Riemannian curvature endomorphism}\index{ind}{Riemannian curvature endomorphism ! of a connection}\index{ind}{curvature ! endomorphism}\index{sym}{$\my{A}{\Rm}$} of a symmetric connection $\nabla$ is consistently defined by
\begin{equation}\label{Rieconnection}
\Rm(X,Y,Z):=\nabla\!_{X} \nabla\!_{Y} Z -\nabla\!_{Y} \nabla\!_{X} Z -\nabla\!_{[X,Y]} Z\in\Gamma(T\!M)
\end{equation}
for all $X,Y,Z\in\Gamma(T\!M)$. Its components are again denoted by ${\Rm_{ijk}}^{l}$.
$\Rm$ is obviously antisymmetric in its first two indices and satisfies the first Bianchi identity\index{ind}{Bianchi identity ! first} $\Rm(X,Y,Z)+\Rm(Y,Z,X)+\Rm(Z,X,Y)=0$ as $\nabla$ is torsion free. Let us now introduce the concept of a pull-back of a connection. Let $\theta_\ast$ denote the push-forward\index{ind}{push-forward}\index{sym}{$\theta_\ast$} through the flow $\theta$ of a vector field $X$.
\begin{ThmDef}[Pullback of Connection]\label{Pullback of Gamma}
The {\it pull-back}\index{ind}{pull-back ! of a connection} of a symmetric connection $\nabla:\Gamma(T\!M)\times\Gamma(T\!M)\to\Gamma(T\!M)$ along a diffeomorphism $\theta:M\to M$ is the symmetric connection $\theta^\ast\nabla$ given by
\begin{equation}\label{pullback}
(\theta^\ast\nabla)_X Y:=(\theta^{-1})_\ast\left(\nabla\!_{\theta_\ast X}(\theta_\ast Y)\right)
\end{equation} for any $X,Y\in\Gamma(T\!M)$.
\end{ThmDef}
\begin{Rem}
The connection $\theta^\ast\nabla$ is well-defined. If we denote the components of $\nabla$ in a frame $\{X_i\}$ by $\Gamma_{ij}^k$, then $(\theta^\ast\nabla)_{ij}^k=(\theta_\ast)_i^r\,(\theta_\ast)_j^s\left[((\theta^{-1})_\ast)^k_t\,\Gamma_{rs}^t-((\theta^{-1})_\ast)^k_{r,s}\right].$
\end{Rem}
\begin{Pf}
For fixed $X,Y\in\Gamma(T\!M)$, the right hand side of \eqref{pullback} clearly is a smooth vector field. It depends $\R$-linearly on $X$ and $Y$ by linearity of the differentials $\theta_\ast$ and $(\theta^{-1})_\ast$. Moreover, $$(\theta^\ast\nabla)_{fX} Y=(\theta^{-1})_\ast\left((f\circ\theta)\nabla\!_{\theta_\ast X}(\theta_\ast Y)\right)=f(\theta^\ast\nabla)_{X} Y$$ and by the Leibniz rule for $\nabla$,
\begin{eqnarray*}
(\theta^\ast\nabla)_{X} (fY)&=&(\theta^{-1})_\ast \left((f\circ\theta) \nabla\!_{\theta_\ast X}(\theta_\ast Y)+(\theta_\ast X)(f)\,\theta_\ast Y\right)\\
&=&f(\theta^\ast\nabla)_{X} (Y)
\end{eqnarray*}
so that $\theta^\ast\nabla$ is indeed a connection. From $$(\theta^\ast\nabla)_X Y-(\theta^\ast\nabla)_Y X=(\theta^{-1})_\ast\left(\nabla\!_{\theta_\ast X}(\theta_\ast Y)-\nabla\!_{\theta_\ast Y}(\theta_\ast X)\right)=(\theta^{-1})_\ast \left[\theta_\ast X,\theta_\ast Y\right]=\left[X,Y\right]$$ we deduce that $\theta^\ast\nabla$ has vanishing torsion. Its coordinate expression is a trivial consequence of the above.
\qed\end{Pf}

It is thus possible to make the following definition.
\begin{Def}[Lie Derivative of Connection]\label{Lie}
The {\it Lie derivative}\index{ind}{Lie derivative ! of a connection}\index{ind}{Lie derivative}\index{sym}{$\mathfrak{L}_{X}$} of a connection $\nabla:\Gamma(T\!M)\times\Gamma(T\!M)\to\Gamma(T\!M)$ along a vector field $X$ with flow $(\theta_t)$ is the $(2,1)$-tensor field $\mathfrak{L}_X\!\nabla$\index{sym}{$\mathfrak{L}_X\nabla$} given by
\begin{equation}
\mathfrak{L}_X\!\nabla:=\lim_{t\to 0}\frac{(\theta_t^\ast\nabla)-\nabla}{t}.
\end{equation}
\end{Def}
\begin{Prop}
The tensor field $\mathfrak{L}_X\nabla$ is well-defined, symmetric, and satisfies
\begin{eqnarray}\label{Leib1}
\mathfrak{L}_X(\nabla\!_Y Z)&=&(\mathfrak{L}_X\!\nabla)(Y,Z)+\nabla_{\mathfrak{L}_X Y}Z+\nabla\!_Y(\mathfrak{L}_X Z)\mbox{ and}\\\label{Leib2}
(\mathfrak{L}_X\!\nabla)(Y,Z)&=&\nabla\!_Y\nabla\!_Z X+\Rm(X,Y,Z)-\nabla\!_{\nabla\!_Y Z}X.
\end{eqnarray}
\end{Prop}
\begin{Pf}
We have seen in the above remark that $\theta_t^\ast\nabla$ is a symmetric connection so that its difference from $\nabla$ and hence $\mathfrak{L}_X\nabla$ must be a tensor field as the limit $t\to 0$ exists by smoothness. To prove Equation \eqref{Leib1}, let us argue that
\begin{eqnarray*}
(\mathfrak{L}_X\!\nabla)(Y,Z)&=&\lim_{t\to0}\frac{(\theta^\ast\nabla)_Y Z-\nabla\!_Y Z}{t}\\
&=&\lim_{t\to0}\frac{\nabla\!_{\theta_{t\ast}Y}(\theta_{t\ast}Z)-\theta_{t\ast}(\nabla\!_Y Z)}{t}\\
&\stackrel{\pm0}{=}&\lim_{t\to0}\nabla\!_{\theta_{t\ast}Y}(\mathfrak{L}_{-X}Z)+\nabla\!_{\mathfrak{L}_{-X}Y}Z+\mathfrak{L}_X(\nabla\!_Y Z)\\
&=&-\nabla\!_Y(\mathfrak{L}_X Z)-\nabla\!_{\mathfrak{L}_X Y}Z+\mathfrak{L}_X(\nabla\!_Y Z)
\end{eqnarray*}
where we have used the linearity properties of $\nabla$ and the fact that $\theta_{t\ast}\vert_{t=0}=\id_{T\!M}$. Equation \eqref{Leib2} holds because
\begin{eqnarray*}
(\mathfrak{L}_X\!\nabla)(Y,Z)&\stackrel{\eqref{Leib1}}{=}&\mathfrak{L}_X(\nabla\!_Y Z)-\nabla\!_{\mathfrak{L}_X Y}Z-\nabla\!_Y(\mathfrak{L}_X Z)\\&{=}&\nabla\!_X\nabla\!_Y Z-\nabla\!_{\nabla\!_Y Z} X-\nabla\!_{\mathfrak{L}_X Y}Z-\nabla\!_Y\nabla\!_X Z+\nabla\!_Y\nabla\!_Z X\\
&=&\Rm(X,Y,Z)-\nabla\!_{\nabla\!_Y Z}X+\nabla\!_Y\nabla\!_Z X
\end{eqnarray*}
by $\mathfrak{L}_X Y=\left[X,Y\right]$ (recall that $\nabla$ is symmetric). Finally, $\mathfrak{L}_X\!\nabla$ is symmetric as
$$(\mathfrak{L}_X\!\nabla)(Y,Z)-(\mathfrak{L}_X\!\nabla)(Z,Y)\stackrel{\eqref{Leib2}}{=}\Rm(X,Y,Z)+\Rm(Y,Z,X)+\Rm(Z,X,Y)\stackrel{\mbox{\scriptsize Bianchi}}{=}0.\qed$$
\end{Pf}

Now recall that a {\it Killing vector field}\index{ind}{Killing vector field} in a pseudo-Riemannian manifold $(M,g)$ is a vector field $X$ satisfying $\mathfrak{L}_X g=0$. It will be important for the generalization of the concept of Killing vector fields to the setting of frame theory to be aware of the following fact.
\begin{Prop}\label{Killing}
Let $(M^n,g)$ be a pseudo-Riemannian manifold possessing a Killing vector field $X$. Then if $\nabla$ is the induced Levi-Civit\`a connection, $g^{-1}$ is the induced metric on $T^\ast M$, and $\Rm, \Ric, \Scal$ are the induced curvature tensors, one finds $$\mathfrak{L}_X g^{-1}=\mathfrak{L}_X\!\nabla=\mathfrak{L}_X\!\Rm=\mathfrak{L}_X\!\Ric=\mathfrak{L}_X\!\Scal=0.$$
\end{Prop}
\begin{Pf}
A direct computation gives $\mathfrak{L}_X(\omega^\sharp)=(\mathfrak{L}_X\omega)^\sharp-(\mathfrak{L}_Xg)(\omega^\sharp,\cdot)=(\mathfrak{L}_X\omega)^\sharp$ for every $\omega\in\Gamma(T^\ast\!M)$ as $X$ is Killing. Thus, for any $\omega,\tau\in\Gamma(T^\ast\!M)$
\begin{eqnarray*}
(\mathfrak{L}_Xg^{-1})(\omega,\tau)&=&X(g^{-1}(\omega,\tau))-g^{-1}(\mathfrak{L}_X\omega,\tau)-g^{-1}(\omega,\mathfrak{L}_X\tau)\\
&=&X(g(\omega^\sharp,\tau^\sharp))-g^{-1}(\mathfrak{L}_X\omega,\tau)-g^{-1}(\omega,\mathfrak{L}_X\tau)\\
&\stackrel{X\mbox{{\scriptsize Killing}}}{=}&0.
\end{eqnarray*}
For the proof of $\mathfrak{L}_X\!\nabla=0$, we make use of the fact that $\nabla$ is the Levi-Civit\`a connection of $g$. As the Riemannian curvature tensor stems from the Levi-Civit\'a connection of a pseudo-Riemannian metric, it possesses the additional symmetry \index{ind}{Riemannian curvature endomorphism ! symmetry} $g(\Rm(Y,Z,W),A)=g(\Rm(W,A,Y),Z)$ for any $Y,Z,W,A\in\Gamma(T\!M)$. Thus, by the Killing equation for $X$, 
$g(\nabla\!_YX,Z)+g(\nabla\!_Z X,Y)=0$, we deduce 
\begin{eqnarray*}
g((\mathfrak{L}_X\!\nabla)(Y,Y),Z)&\stackrel{\eqref{Leib2}}{=}&g(\nabla\!_Y\nabla\!_YX-\nabla\!_{\nabla\!_YY}X+\Rm(X,Y,Y),Z)\\
&\stackrel{\nabla\,\mbox{\scriptsize Levi-Civit\`a}}{=}&Y(g(\nabla\!_YX,Z))-g(\nabla\!_YX,\nabla\!_YZ)-g(\nabla\!_{\nabla\!_YY}X,Z)\\
&&+g(\Rm(Y,Z,X),Y)\\
&\stackrel{X\,\mbox{{\scriptsize Killing}}}{=}&\cancel{-Y(g(\nabla\!_ZX,Y))}-g(\nabla\!_YX,\nabla\!_YZ)+\cancel{g(\nabla\!_ZX,\nabla\!_YY)}\\
&&+g(\cancel{\nabla\!_Y\nabla\!_ZX}-\nabla\!_Z\nabla\!_YX,Y)+g(\nabla\!_Y X,\mathfrak{L}_YZ)\\
&\stackrel{\nabla\,\mbox{\scriptsize Levi-Civit\`a}}{=}&-g(\nabla\!_YX,\nabla\!_ZY)-Z(\cancelto{0}{g(\nabla\!_YX,Y)})+g(\nabla\!_Y X,\nabla\!_ZY)\\
&=&0
\end{eqnarray*}
for any $Y,Z\in\Gamma(T\!M)$, so that symmetry of $\mathfrak{L}_X\!\nabla$ implies $\mathfrak{L}_X\!\nabla=0$ by polarization. To show $\mathfrak{L}_X\!\Rm=0$, let $\{X_i\}_{i=1,\dots,n}$ be a coordinate frame on $M^n$ (i.~e.~satisfying $\left[X_i,X_j\right]=0$ for all $i,j=1,\dots,n$) with $X=X_1$. Such a frame exists by ODE theory, cf.~e.~g.~Proposition 1.53 in \cite{Warner}. Then
\begin{eqnarray*}
(\mathfrak{L}_X\!\Rm)(X_i,X_j,X_k)&\stackrel{\mbox{\scriptsize Leibniz rule}}{=}&\mathfrak{L}_X(\Rm(X_i,X_j,X_k))\\
&=&\mathfrak{L}_X(\nabla\!_i\nabla\!_jX_k-\nabla\!_j\nabla\!_iX_k)\\
&\stackrel{\mathfrak{L}_X\!\nabla=0=\left[X,X_i\right]}{=}&\nabla\!_i\mathfrak{L}_X(\nabla\!_jX_k)-\nabla\!_j\mathfrak{L}_X(\nabla\!_iX_k)\\
&\stackrel{\mbox{\scriptsize iterate}}{=}&0.
\end{eqnarray*}
By the Leibniz rule for Lie derivatives \eqref{Lie product rule}, $\Ric$ and $\Scal$ also have vanishing Lie derivatives in direction $X$.
\qed\end{Pf}

\section{Weighted Sobolev Spaces}\label{sec:sobo}
In this section, we collect some well-known results on weighted Sobolev spaces defined on $\R^n$ and subsets thereof. We follow the exposition in \cite{Bartnik}, but adapt the results slightly so that it fits with our notation and conventions. Although we only apply these results in dimension $n=3$, we state the theorems for arbitrary $n\geq 3$ for convenience of the reader. We use the following notation: If $R>0$, $B_R:=\overline{B_R(0)}\subset\R^n$\index{sym}{$B_R$} denotes the closed ball of radius $R$ around $0$ and $E_R:=\R^n\setminus B_R$ denotes the {\it associated (open) exterior region}\index{ind}{exterior region}\index{sym}{$E_R$}. Abusing notation, we will denote $\R^n\setminus\{0\}$ by $E_0$ and also call it an exterior region. We refer to a fixed Cartesian coordinate system $(x^i)_{i=1,\dots,n}$ on $\R^n$ and denote the Cartesian radius by $r:=\vert x\vert:=\sqrt{(x^1)^2+\dots+(x^n)^2}$. The Euclidean metric on $\R^n$ will be denoted by $\delta$ and $g$ will denote an arbitrary smooth Riemannian metric on $\R^n$ or subsets thereof. Using the {\it weight functions}\index{ind}{weight function} $\sigma:\R^n\setminus\{0\}\to\R:x\mapsto\sqrt{1+r^2}$ and $r\vert_{E_0}$, we define the so-called ``weighted Lebesgue'' and ``weighted Sobolev spaces''.
\begin{Def}\label{Lebesgue}
Let $k\in\N$,  $1\leq q<\infty$, and $\varepsilon\in\R$. Let $d\mu_{\delta}$\index{sym}{$d\mu_{\delta}$} denote the ordinary Lebesgue measure on $\R^n$. The {\it weighted Lebesgue space}\index{ind}{weighted Lebesgue space} $L^q_\varepsilon:=L^q_\varepsilon(\R^n)$\index{sym}{$L^q_\varepsilon$}\index{sym}{$L^q_\varepsilon(\R^n)$} is defined as the space of all measurable functions $u\in L^q_{loc}(\R^n)$ such that the norm
\begin{equation*}
\vert\vert u\vert\vert_{q,\varepsilon}:=\left(\,\int_{\R^n}\vert u\vert^q\sigma^{-\varepsilon q-n}\,d\mu_{\delta}\right)^{1/q}
\end{equation*}
is finite. Similarly, the {\it weighted Sobolev space}\index{ind}{weighted Sobolev space} $W^{k,q}_\varepsilon:=W^{k,q}_\varepsilon(\R^n)$\index{sym}{$W^{k,q}_\varepsilon$}\index{sym}{$W^{k,q}_\varepsilon(\R^n)$} is given as the subspace of $L^q_\varepsilon(\R^n)$ consisting of all those functions $u:\R^n\to\R$ which have weak (partial) derivatives of order $j\leq k$ in $L^q_{\varepsilon-j}(\R^n)$. The weighted Sobolev norm on this space is defined as
\begin{equation*}
\vert\vert u\vert\vert_{k,q,\varepsilon}:=\sum_{j=0}^k\vert\vert D^j u\vert\vert_{q,\varepsilon-j}.
\end{equation*}
Similarly, the {\it weighted Lebesgue space on the exterior region $E_0=\R^n\setminus\{0\}$}\index{ind}{weighted Lebesgue space}, $\dt{L}^q_\varepsilon:=L^q_\varepsilon(E_0)$\index{sym}{$L^q_\varepsilon(E_R)$}\index{sym}{$\dt{L}^q_\varepsilon$}, is defined as the space of all measurable functions $u\in L^q_{loc}(E_0)$ such that the norm
\begin{equation*}
\vert\vert u\vert\vert_{q,\varepsilon}^\bullet:=\left(\,\int_{E_0}\vert u\vert^q r^{-\varepsilon q-n}\,d\mu_{\delta}\right)^{1/q}
\end{equation*}
is finite. The {\it weighted Sobolev space on the exterior region $E_0$}\index{ind}{weighted Sobolev space}, $\dt{W}^{k,q}_\varepsilon:=W^{k,q}_\varepsilon(E_0)$\index{sym}{$W^{k,q}_\varepsilon(E_0)$}\index{sym}{$\dt{W}^{k,q}_\varepsilon$}, is defined as the subspace of $L^q_\varepsilon(E_0)$ consisting of all those functions $u:\R^n\to\R$ which have weak (partial) derivatives of order $j\leq k$ in $L^q_{\varepsilon-j}(E_0)$. The weighted Sobolev norm on this space is then given by
\begin{equation*}
\vert\vert u\vert\vert_{k,q,\varepsilon}^\bullet:=\sum_{j=0}^k\vert\vert D^j u\vert\vert_{q,\varepsilon-j}^\bullet.
\end{equation*}
\end{Def}
Observe that the weight function $r$ used on $E_0=\R^n\setminus\{0\}$ differs from the one used on all of $\R^n$, namely from $\sigma$, but that they are asymptotically identical at infinity. Note furthermore that $C_c^\infty(\R^n)$, $C_c^\infty(E_0)$ are dense in $W^{k,q}_\varepsilon$, $\dt{W}^{k,q}_\varepsilon$, respectively, and that $L^q_\varepsilon=L^q(\R^n)$ for $\varepsilon=-n/q$.

The following theorems correspond to Theorem 1.2, Theorem 1.3, Lemma 1.4, Theorem 1.7, and Corollary 1.9 in \cite{Bartnik}. They will help us prove the faster fall-off trick Theorem \ref{higherreg} and will be helpful for our center of mass considerations in Chapter \ref{chap:static}.
\begin{Thm}\label{Sobprop}
Let $n,k\in\N$, $n\geq 2$. The following claims hold true:
\begin{enumerate}
\item If $1\leq q_1\leq q_2<\infty$ and $\varepsilon_1>\varepsilon_2$, then $L^{q_2}_{\varepsilon_2}$ continuously embeds into $L^{q_1}_{\varepsilon_1}$ via inclusion.
\item If $1\leq q_1,q_2,q<\infty$ satisfy $1/q=1/{q_1}+1/{q_2}$ and if in addition $\varepsilon=\varepsilon_1+\varepsilon_2$, then for all $u\in L^{q_1}_{\varepsilon_1}$, $v\in L^{q_1}_{\varepsilon_1}$ we have the {\bf weighted H\"older inequality}\index{ind}{weighted H\"older inequality}
\begin{equation}\label{Hoelder}
\vert\vert uv\vert\vert_{q,\varepsilon}\leq\vert\vert u\vert\vert_{q_1,\varepsilon_1}\vert\vert v\vert\vert_{q_2,\varepsilon_2}.
\end{equation}
\item If $1\leq q_2\leq q_1\leq nq_2/(n-kq_2)<\infty$ and $n-kq_2>0$, then there is a constant $C>0$ such that any $u\in W^{k,q_2}_\varepsilon$ satisfies the {\bf weighted Sobolev inequality}\index{ind}{weighted Sobolev inequality}
\begin{equation}\label{B:1.2}
\vert\vert u \vert\vert_{nq_2/(n-kq_2),\varepsilon}\leq C\vert\vert u \vert\vert_{k,q_1,\varepsilon}
\end{equation}
and in fact $\vert u(x)\vert r^{-\varepsilon}\to0$ as $r\to\infty$.
\item Let $C^0_\varepsilon$ consist of all continuous functions $u:\R^n\to\R$ such that $$\vert\vert u\vert\vert_{C^0_\varepsilon}:=\sup_{x\in\R^n}\{\sigma^{-\varepsilon}(x)\vert u(x)\vert\}<\infty.$$
If $1\leq q<\infty$ and $0<k-n/q\leq1$, there is a constant $C>0$ such that for any $u\in W^{k,q}_{\varepsilon}$, we have $$\vert\vert u\vert\vert_{C^0_{\varepsilon}}\leq C\vert\vert u\vert\vert_{k,q,\varepsilon}.$$
\end{enumerate}
\end{Thm}
We will also frequently apply the following multiplication theorem which we quote from p.~153 in \cite{KM}.
\begin{Thm}[Multiplication Theorem]\label{multi}
For all $k_{1},k_{2},k\in\N$, $k_{1}+k_{2}> k+3/2$, and all $\varepsilon>\varepsilon_{1}+\varepsilon_{2}$, pointwise multiplication is a continuous bilinear map $$\cdot:W^{k_{1},2}_{\varepsilon_{1}}(\R^3)\times W^{k_{2},2}_{\varepsilon_{2}}(\R^3)\to W^{k,2}_{\varepsilon}(\R^3).$$
\end{Thm}

Before we continue to list facts on weighted Sobolev spaces, we quote the following definition from \cite{Bartnik}.
\begin{Def}
$\varepsilon\in\R$ is said to be an {\it exceptional}\index{ind}{weight parameter ! exceptional} weight parameter if it is an integer which satisfies $\varepsilon\neq-1,-2,\dots,n-3$; otherwise it is called {\it non-exceptional}\index{ind}{weight parameter ! non-exceptional}. We set $\overline{\varepsilon}:=\max\{\mu\mbox{ exceptional }\vert\,\mu <\varepsilon\}$ for any weight parameter $\varepsilon\in\R$.
\end{Def}
Observe that $\overline{\varepsilon}\in\Z$ and that precisely all $\varepsilon\in\Z$ are exceptional for the case $n=3$ which will be of most concern to us in geometrostatics. The exceptional values of the weight parameter correspond to the orders of growth of harmonic functions in $\R^n\setminus B_1$. The following theorem shows the relevance of these exceptional weight parameters. It refers to the Laplacian $\triangle$ which is induced from the flat metric $\delta$ and we quote it from \cite{McO}.

\begin{Thm}[Robert McOwen]\label{thm:Owen}
Suppose that $\varepsilon$ is non-exceptional, $1<q<\infty$ and $k\in\N_0$. Then the map $$\triangle:\dt{W}^{k+2,q}_\varepsilon\to\dt{W}^{k,q}_{\varepsilon-2}$$ is a bounded isomorphism with bounded inverse.
\end{Thm}

\subsection*{$\mathcal{O}$-notation}\label{subseq:O}
In order to simplify notation, we will use the $\mathcal{O}$-notation\index{sym}{$\mathcal{O}(r^\varepsilon)$} as an abbreviation for a precise statement in the language of weighted Sobolev spaces when the precise statement is straightforward. For example, we say that a smooth function $f:M^n\to\R$ on an asymptotically flat manifold $(M^n,g)$ with radial coordinate $r$ {\it lies in the class $\mathcal{O}(r^{-k})$ as $r\to\infty$}\index{sym}{$\mathcal{O}(\cdot)$} if there exists a constant $C>0$ such that $$\vert f(p)\vert\leq C r^{-k}$$ holds for all $p\in M^n$ or all $p$ in a specified neighborhood of infinity within $M^n$. In addition, we say that $f$ lies in the class $\mathcal{O}(r^{-k})$ {\it with $l$ derivatives} if $$\partial^\alpha\!f\in\mathcal{O}(r^{-k-\vert\alpha\vert})$$ for all multi-indices $\alpha$ with $\vert\alpha\vert\leq l$.\index{ind}{lie in the class $\mathcal{O}(\cdot)$}\index{ind}{lie in the class $\mathcal{O}(\cdot)$ ! with $l$ derivatives} Compare this to item number $4$ in Theorem \ref{Sobprop}.

\subsection{The Faster Fall-Off Trick}\label{subsec:faster}
The above results on differential operators asymptotic to $\triangle$ can and will be used to show faster fall-off for the solutions of the static metric equations of general relativity which we will do in Chapter \ref{chap:static}. To this end, we will now prove a more general faster fall-off trick making use of the above results and of the Kelvin transform. We refer the reader to the book \cite{AxBourRam} for an introductory exposition of the Kelvin transform.

We begin by giving the definition of the Kelvin transform. Let $n\geq3$ in all of this section.
\begin{Def}[Kelvin Transform]\label{def:Kel}
Set $\overline{\R^n}:=\R^n\cup\{\infty\}$ where $\infty$ is considered as the topological one point compactification. For any $x\in\overline{\R^n}$, we set
\begin{eqnarray*}
x^*:=\begin{cases}\frac{x}{\vert x\vert^2}&\mbox{ if }x\neq0,\infty\\
0&\mbox{ if }x=\infty\\
\infty&\mbox{ if }x=0\end{cases}.
\end{eqnarray*}
If $E\subset\overline{\R^n}$, we set $E^*:=\{x^*\,\vert\,x\in E\}\subset\overline{\R^n}$ and endow both $E$ and $E^\ast$ with the induced topology. We define the {\it Kelvin transform}\index{ind}{Kelvin transform}\index{sym}{$\mathcal{K}\left[\cdot\right]$} of a continuous function $u:E\to\R$ to be the function $\mathcal{K}\left[u\right]:E^*\to\R$ given by $$\mathcal{K}\left[u\right](x):=\vert x\vert^{2-n}u(x^*).$$
\end{Def}
The Kelvin transform is a higher dimensional analog of inversion in a sphere. It is a very useful technique for transporting the ``point at infinity'' of $\R^n$ to the interior and thus makes power series expansions at infinity and related harmonicity questions more approachable as we will see in the following lemmata.
\begin{Lem}\label{lem:Kelv}
The Kelvin transform is a linear transform from $C^0(E)$ to $C^0(E^*)$ that preserves uniform convergence on compact subsets of $\R^n\setminus\{0\}$. $\mathcal{K}$ is its own inverse in the sense that $\mathcal{K}\left[\mathcal{K}\left[u\right]\right]=u$ for all $u\in C^0(E)$. It maps positive homogeneous functions on $\R^n\setminus\{0\}$ of degree $k\in\Z$ to positive homogeneous functions on $\R^n\setminus\{0\}$ of degree $2-n-k$. Moreover, it maps
harmonic functions on $E$ onto harmonic functions on $E^*$ if $E\subset\R^n\setminus\{0\}$ is open.
\end{Lem}
\begin{Def}
Let $E\subset\R^n$ be compact and let $u\in C^0(\R^n\setminus E)$ be harmonic. We call $u$ {\it harmonic at $\infty$}\index{ind}{harmonic ! at infinity} provided that $\mathcal{K}\left[u\right]$ has a removable singularity at $0$, i.~e.~if there exists a harmonic function $\overline{u}\in C^0((\R^n\setminus E)^*\cup\{0\})$ with $\overline{u}\equiv\mathcal{K}\left[u\right]$ on $(\R^n\setminus E)^*$.
\end{Def}
\begin{Lem}\label{lem:remov}
Let $E\subset\R^n$ be compact. Then a harmonic function $u:\R^n\setminus E\to\R$ is harmonic at infinity if and only if $\lim_{x\to\infty}u(x)=0$.
\end{Lem}

Let us now turn our attention to the faster fall-off trick announced above. For easiness of notation, we change the name of weight parameter from $\varepsilon$ to $-\tau$ from now on.\index{sym}{$\tau$}
\begin{Thm}[Faster Fall-Off Trick]\label{higherreg}
Let $n,k\in\N_0$, $n\geq3$, $1\leq q<\infty$, $\tau>0$ such that $-\tau$ is non-exceptional, and let $f:\R^n\to\R$ be a smooth function satisfying
\begin{equation*}
f\in W_{-\tau+1}^{k+2,q}\mbox{ and }\, \triangle f\in W_{-\tau-2}^{k, q}.
\end{equation*}
Then there is a harmonic polynomial $p$ of degree $d\leq\lceil\tau\rceil$ such that for any radius $R>0$ and any smooth cut-off function $\eta:\R^n\to\R$ with support in $E_R$, $\eta(f-\mathcal{K}\left[p\right])\in W_{-\tau}^{k+2, q}\cap\dt{W}^{k+2,q}_{-\tau}$.
\end{Thm}
\begin{Rem}
The behavior of $f$ is of interest to us only in a neighborhood of infinity, where the faster fall-off trick gives us precise fall-off information. We can interpret this informally as saying that if $f=\mathcal{O}(r^{-l+1})$ and $\triangle f=\mathcal{O}(r^{-l-2})$ as $r\to\infty$, then $f-\mathcal{K}\left[p\right]=\mathcal{O}(r^{-l})$ as $r\to\infty$ for all decay orders $l\in\N$.
\end{Rem}
\begin{Pf}
Let $\eta$ be a smooth cut-off function with support in $E_R$ for some $R>0$ and without loss of generality $\eta\equiv1$ in $E_{2R}$. Set $g:=\eta f$ and $h:=\triangle g$ so that $h\in W_{-\tau-2}^{k,q}\cap C^\infty(\R^n)$ by the assumptions that $f, \eta$ be smooth, $\eta\equiv1$ in $E_{2R}$, and $\triangle f\in W^{k,q}_{-\tau-2}$. Since $g\equiv0$ in $B_R$, it also holds that $h\in \dt{W}^{k,q}_{-\tau-2}$. Robert McOwen's result \ref{thm:Owen} now tells us that there exists a unique solution $f_h\in \dt{W}^{k+2,q}_{-\tau}$ satisfying $\triangle f_h=h$ in $\R^n\setminus\{0\}$ as $-\tau$ is non-exceptional. This means that the two functions $f_h$ and $g$ both satisfy the Poisson equation with right hand side $h$ on $\R^n\setminus\{0\}$ and in consequence, $g-f_h$ is a harmonic function on $E_0=\R^n\setminus\{0\}$ with respect to the flat Laplacian.

By construction, we find that $g\in\dt{W}^{k,q}_{-\tau+1}$. $f_h\in\dt{W}^{k,q}_{-\tau+1}$ follows from the first item in Theorem \ref{Sobprop} and thus we also know that $g-f_h\in \dt{W}^{k,q}_{-\tau+1}$. Cutting $g-f_h$ off with the cut-off function $\eta$, we obtain a function $\eta(g-f_h)\in W^{k,q}_{-\tau+1}$. This helps us seeing that $\eta(g-f_h)$ and thus also $g-f_h$ falls off to $0$ as $r\to\infty$ by the weighted Sobolev inequality \eqref{B:1.2}. $(g-f_h)\vert_{E_0}$ is thus harmonic at $\infty$ by Lemma \ref{lem:remov} and thus its Kelvin transform $K:=\mathcal{K}[(g-f_h)\vert_{E_0}]:\R^n\to\R$ is harmonic and hence real analytic at $0$. We define a polynomial $p$ on $\R^n$ by $$p(y):=\sum_{\vert\alpha\vert=0}^{\vert\alpha\vert=\lceil\tau\rceil}\frac{\partial_\alpha K(0)\,y^\alpha}{\alpha!}$$ and deduce that $p$ is in fact a harmonic\footnote{cf.~pp.~22-24 in \cite{AxBourRam}.} polynomial of degree $d\leq\lceil\tau\rceil$. This implies that $K-p$ is a harmonic function on $\R^n$. $K$ then splits into
\begin{equation}\label{K} K(y)=p(y)+\sum_{\vert\alpha\vert=\lceil\tau\rceil+1}^\infty\frac{\partial_\alpha K(0)\,y^\alpha}{\alpha!}
\end{equation}
in a neighborhood $B_\varepsilon$ of $0$ by harmonic function theory, where the series converges absolutely and uniformly\footnote{cf.~again pp.~22-24 in \cite{AxBourRam}.}. We can therefore calculate an explicit expression for the harmonic function $\left(g-f_h-\mathcal{K}\left[ p\right]\right)\vert_{E_{\varepsilon^{-1}}}$ using the different assertions of Lemma \ref{lem:Kelv} and the definition of the Kelvin transform. For $x\in E_{\varepsilon^{-1}}$, we obtain:
\begin{eqnarray*}
\left(g-f_h-\mathcal{K}\left[ p\right]\right)(x)
&=&\mathcal{K}\left[(K-p)\right](x)\\
&\stackrel{\eqref{K}}{=}&\mathcal{K}\left[\sum_{\vert\alpha\vert=\lceil\tau\rceil+1}^\infty\frac{\partial_{\alpha} K(0)\,y^\alpha}{\alpha!}\right](x)\\Ü
&=&\sum_{\vert\alpha\vert=\lceil\tau\rceil+1}^\infty\frac{\partial_{\alpha} K(0)\,\mathcal{K}\left[y^\alpha\right](x)}{\alpha!}\\
&=&\sum_{\vert\alpha\vert=\lceil\tau\rceil+1}^\infty\frac{\partial_{\alpha} K(0)\,x^\alpha}{\vert x\vert^{n+2\vert\alpha\vert-2}\,\alpha!}.
\end{eqnarray*}
This explicit expression now allows us to estimate
\begin{eqnarray*}
\vert\left(g-f_h-\mathcal{K}\left[ p\right]\right)(x)\vert
&\leq&\lvert\sum_{\vert\alpha\vert=\lceil\tau\rceil+1}^\infty\frac{\partial_{\alpha} K(0)\,x^\alpha}{\vert x\vert^{n+2\vert\alpha\vert-2}\,\alpha!}\rvert\\
&\leq&\sum_{\vert\alpha\vert=\lceil\tau\rceil+1}^\infty\frac{\vert \partial_{\alpha} K(0)\vert}{\alpha!}\vert x\vert^{2-n-\vert\alpha\vert}\\
&\leq&\sum_{\vert\alpha\vert=\lceil\tau\rceil+1}^\infty\frac{\vert \partial_{\alpha} K(0)\vert}{\alpha!}\vert x\vert^{1-n-\lceil\tau\rceil}\varepsilon^{-\lceil\tau\rceil-1+\vert\alpha\vert}\\
&\leq&C\,\vert x\vert^{1-n-\lceil\tau\rceil}
\end{eqnarray*}
for all $x\in E_{\varepsilon^{-1}}$ as the power series converges absolutely in $E_{\varepsilon^{-1}}$. Similarly, one can show by induction that $$\vert \partial_{\beta}(g-f_h-\mathcal{K}\left[ p\right])(x)\vert\leq C\vert x\vert^{1-n-\lceil\tau\rceil-\vert\beta\vert}$$ for all $x\in E_{\varepsilon^{-1}}$ and all multi-indices $\beta$ with $\vert\beta\vert\leq k+2$. Since by assumption $f$ and $\eta$ (and hence $g=\eta f$) are smooth on $\R^n$ and since $g-f_h$ is harmonic and thus smooth on $\R^n\setminus\{0\}$, $f_h$ must be smooth on $\R^n\setminus\{0\}$. This implies that $\eta(g-f_h-\mathcal{K}\left[ p\right])$ is smooth on $\R^n$ and vanishes in a neighborhood of the origin so that -- together with the above estimates --, we can deduce that $$\eta(g-f_h-\mathcal{K}\left[ p\right])\in W^{k+2,q}_{-\tau}\cap\dt{W}^{k+2,q}_{-\tau}.$$ As $f_h\in\dt{W}^{k+2,q}_{-\tau}$ and thus also $\eta f_h\in W^{k+2,q}_{-\tau}\cap\dt{W}^{k+2,q}_{-\tau}$ by construction, it follows by linearity that $\eta(g-\mathcal{K}\left[ p\right])\in W^{k+2,q}_{-\tau}\cap\dt{W}^{k+2,q}_{-\tau}$. Now $g=\eta f$ differs from $\eta g$ only in the compact annulus $\overline{E_R}\setminus E_{2R}$ where $\eta, f,\mbox{ and }\mathcal{K}\!\left[ p\right]$ are smooth so that $\eta(g-f)\in W^{k+2,q}_{-\tau}\cap\dt{W}^{k+2,q}_{-\tau}$ and hence again by linearity $\eta(f-\mathcal{K}\left[ p\right])\in W^{k+2,q}_{-\tau}\cap\dt{W}^{k+2,q}_{-\tau}$ which proves the theorem.
\qed\end{Pf}
\chapter{Isolated Relativistic Systems}\label{chap:iso}
In this chapter, we will provide a short introduction into general relativity, focusing on isolated systems. It is structured as follows: In Section \ref{sec:sett}, we introduce the main variables and equations. In Section \ref{sec:3+1}, we shortly discuss the initial value formulation of general relativity (the so-called Cauchy problem) while in Section \ref{sec:flat}, we give a short overview over boundary conditions insuring isolatedness of the systems under consideration. In Sections \ref{sec:energy} and \ref{sec:HCoM}, we present definitions and a small number of results on the concepts of mass and center of mass of isolated systems, respectively. As one of the main goals of this thesis is to prove that mass and center of mass converge to Newtonian mass and center of mass in the Newtonian limit, these concepts lie at the heart of our considerations.

\section{Setting and Notation}\label{sec:sett}
When modeling a relativistic system in GR as a smooth\footnote{We will not discuss the regularity of general dynamical relativistic solutions as we will focus on static solutions in this thesis. For these, regularity issues will be discussed in Chapter \ref{chap:static}.} spacetime\footnote{We are not assuming any orientability or hyperbolicity conditions on spacetimes, a priori.}\index{ind}{spacetime} $(L^4,ds^2)$\index{sym}{$(L^4,ds^2)$}, one usually has to specify a matter model, cf.~e.~g.~\cite{Rendall}. This matter model is usually defined in terms of a matter Lagrangian depending on both the metric $ds^2$ and the appropriate matter fields. One can then derive a symmetric $(0,2)$-tensor field $T$ on $L^4$ by variation of the matter Lagrangian. $T$ is called the {\it energy-momentum tensor}\index{ind}{energy-momentum tensor}\index{sym}{$T$}\label{T} or {\it stress-energy tensor}\index{ind}{stress-energy tensor} of the system. If $V\in T_p L^4$ is a future-pointing timelike unit vector (or an {\it observer}\index{ind}{observer}) situated at the spacetime event $p$, one can introduce the observed \label{Tsplit}{\it mass density}\index{ind}{mass density} $\rho(p):=T_p(V,V)/c^2$\index{sym}{$\rho$}, where $c$ the {\it speed of light}\index{ind}{speed of light}\index{sym}{$c$}. The observed {\it momentum density}\index{ind}{momentum density} is given by $J_p(X):=-T_p(V,X)$ in the spatial direction $X\in T_pL^4$ with $ds^2_p(V,X)=0$\index{sym}{$J$}, and the observed {\it stress}\index{ind}{stress} is defined as $S_p(X_1,X_2):=T_p(X_1,X_2)$ in spatial directions $X_i\in T_pL^4$ with $ds^2_p(V,X_i)=0$, $i=1,2$\index{sym}{$S$}. The vector $P_p:=(T_p(V,\cdot))^\#$ can be interpreted as the $4$-momentum vector\index{ind}{momentum} observed by $V$. Note that we have defined $\rho$ to be a mass and not an energy density as this will be more suitable for tackling the Newtonian limit in Chapter \ref{chap:FT}.

Besides suitable {\it energy conditions}\index{ind}{energy conditions} like the weak energy condition $T(V,V)\geq 0$ for any future-directed causal (timelike or lightlike) vector field $V\in\Gamma(T\!L^4)$ that one usually expects to hold, the energy-momentum tensor is required to satisfy the {\it equation of motion}\index{ind}{equation of motion}\label{conserv} $\diver\T=0$ which is designed to assure (differential) energy conservation, cf.~e.~g.~\cite{Wald,SW}. Some sorts of matter, for example perfect fluids, will also be required to satisfy an appropriate equation of state. We will, however, not focus on specific matter models in this thesis as we are mainly interested in the behavior in the vacuum region outside a given matter distribution.

A {\it relativistic spacetime}\index{ind}{spacetime ! relativistic}\index{ind}{relativistic} or {\it relativistic system}\index{ind}{system ! relativistic} then is formally represented by a triple $(L^4,ds^2,T)$ consisting of a spacetime $(L^4,ds^2=g)$ and an energy-momentum tensor $T$ that satisfy the equation of motion and {\it Einstein's equation}\index{ind}{Einstein's equation} \begin{equation}\label{EEq}
\Ric-\frac{1}{2}\Scal g=\frac{8\pi G}{c^4}\T
\end{equation}
with $c$ again the {\it speed of light}\index{ind}{speed of light}\index{sym}{$c$} and $G$ the {\it (Newtonian) gravitational constant}\index{ind}{gravitational constant}\index{sym}{$G$}. Observe that Einstein's equation relates the metric $ds^2$ to the $(2,0)$-variant of $T$. This is of no significance in general relativity but will become important when we ``unify'' general relativity and Newtonian gravity in Chapter \ref{chap:FT}.

\section{3+1 Decomposition and Cauchy Problem}\label{sec:3+1}
It is often useful (and gives deep insight) to rewrite Einstein's equation as an initial value problem (``Cauchy problem'')\index{ind}{Cauchy problem}. Yvonne Choquet-Bruhat's famous theorem \ref{CHB} essentially states that this is possible, and we will apply this theorem in order to prove a uniqueness result for the lapse function on page \pageref{subsec:gunique}. Before we can cite the theorem, we have to introduce a few notions which we will now begin with. We follow the approaches taken in \cite{Wald,HE,CHB,Bartnik3}.

First of all, a spacetime is said to be {\it time orientable}\index{ind}{time orientable} if it possesses a smooth global timelike vector field (which automatically induces a time orientation). In this case, the {\it chronal future}\index{ind}{chronal future}\index{sym}{$I^+(q)$} of $q\in L^4$ is defined via $$I^+(q):=\{p\in L^4\,\vert\mbox{ There is a smooth timelike future directed curve from }q\mbox{ to }p.\}$$ A set $S\subset L^4$ is called {\it achronal}\index{ind}{achronal} if it satisfies $S\cap I^+(S)=\emptyset$, where $I^+(S):=\cup_{q\in S}I^+(q)$. Furthermore, the {\it future/past domain of dependence}\index{ind}{domain of dependence ! future}\index{ind}{domain of dependence ! past}\index{sym}{$D^+(S)$}\index{sym}{$D^-(S)$}\index{sym}{$D(S)$} of an achronal set $S\subset L^4$ is given by $$D^\pm(S):=\{p\in L^4\,\vert\mbox{ Every past/future inextensible curve through }p\mbox{ intersects }S.\},$$
where a smooth timelike curve $\kappa:\left(s_0,s_1\right)\to L^4$ is {\it past/future inextensible}\index{ind}{inextensible} if and only if it has no limit as $s\to s_0/s_1$, respectively. The {\it domain of dependence of $S$}\index{ind}{domain of dependence} then is the union $D(S):=D^+(S)\cup D^-(S)$\index{sym}{$D(S)$}. A closed achronal set $S\subset L^4$ which fulfills $D(S)=L^4$ is called a {\it Cauchy surface}\index{ind}{Cauchy surface} and indeed can be seen to be a 3-dimensional spatial submanifold\footnote{As the remainder of this thesis is not severely depending on this section, we will not take regularity questions into account.} of $L^4$. $L^4$ is said to be {\it globally hyperbolic}\index{ind}{globally hyperbolic}\footnote{This definition is unconventional but equivalent to the traditional one, cf.~p.~209 in \cite{Wald}. As we are not primarily concerned with the Cauchy problem, this just seems the least cumbersome approach.} if it possesses a Cauchy surface. It can be shown (cf. Theorem 8.3.14 in \cite{Wald} and references therein) that every globally hyperbolic spacetime possesses a global {\it time function}\index{ind}{time function} $t$ (i.~e.~a function having $\grad t$ past directed timelike), the level sets of which are smooth Cauchy surfaces. Thus every globally hyperbolic $L^4$ can be foliated by Cauchy surfaces and thus has the differential topology of $\R\times M^3$, where $M^3$ denotes any such Cauchy surface.

Now let $(L^4,ds^2)$ be a globally hyperbolic spacetime with time function $t$. We think of a point $p\in L^4$ as representing a {\it spacetime incident}\index{ind}{spacetime incident} $(t,x)$ with coordinate representation $(t(p),x^i(p))$ with respect to some coordinate system $\{x^i\}_{i=1,2,3}$ on an open neighborhood of $x\in M^3$ (which is then transported along the integral curves of $\grad t$). Clearly, there is a unique future-pointing timelike unit normal vector field $\nu\in\Gamma(T\!L^4)$ $ds^2$-orthogonal to the leaves $\{t\}\times M^3$. We can therefore uniquely decompose the coordinate vector field $\partial_t\in\Gamma(T\!L^4)$ into its normal and tangential components, meaning that $$\partial_t=cN\nu+X$$ with $N:I\times M^3\to\R$ the {\it lapse function}\index{ind}{lapse function}\index{sym}{$N$} and $X:I\times M^3\to T\!M^3$ the {\it shift vector field}\index{ind}{shift vector field} of the foliation. Moreover, $ds^2$ induces a Riemannian metric $\myg{3}(t)$\index{sym}{$\myg{3}$} on each of its (spacelike) submanifolds $\{t\}\times M^3$ which we sloppily understand to be a time-dependent metric on $M^3$. One can then see that $ds^2$ can be rewritten as
\begin{equation}\label{3-metric}
ds^2=-c^2N^2 dt^2+\myg{3}_{ij}(dx^i+X^idt)(dx^j+X^jdt).
\end{equation}
In this setting, the second fundamental form $h(t)$ of the slice $\{t\}\times M^3\subset (I\times M^3,ds^2)$ with respect to the chosen normal is often referred to as its {\it extrinsic curvature}\index{ind}{extrinsic curvature}\index{ind}{curvature ! extrinsic} (and is usually denoted by $K$\index{sym}{$K$} in the physics literature). Just as $\myg{3}$, $h$ is understood to be a time-dependent symmetric $(2,0)$-tensor field on $M^3$. It is straight forward from the definition of $h$ that
\begin{equation}\label{Lie-der}
 \mathcal{L}_{\nu} \myg{3}= 2h.
\end{equation}
Besides the Gau{\ss}- and Codazzi equations \eqref{Gauss1} through \eqref{Codazzi}, there is another submanifold equation induced by the decomposition of the normal $\nu$. It is called the {\it Mainardi identity}\index{ind}{Mainardi identity} and reads\footnote{Recall that the label $4$ corresponds to fields derived from $ds^2=\myg{4}$.}
\begin{equation}\label{Mainardi}
ds^2(\my{4}{\Rm}(X_i,\nu,\nu),X_j)=(1/N)\,\my{3}{\nabla}_{ij}^2N+h^2_{ij}-\mathcal{L}_{\nu} h_{ij}
\end{equation}
on any frame $\{X_i\}_{i=1,2,3}$ of $M^3$. As a consequence of these submanifold equations, the {\it Einstein tensor}\index{ind}{Einstein tensor}\index{sym}{$\mathcal{G}$} $\mathcal{G}:=\my{4}{\Ric}-\frac{1}{2}\,\my{4}{\Scal}\,ds^2$ can be decomposed into its normal and mixed parts
\begin{eqnarray}\label{energy preconstraint}
2\mathcal{G}(\nu,\nu)&=&\my{3}{\Scal}+H^2-|h|_{3}^2\\\label{momentum preconstraint}
\mathcal{G}(\nu,\partial_i)&=&\my{3}{\nabla}\!_j\,h_i^{\;j}-H_{,i}
\end{eqnarray}
while the remaining information leads to
\begin{equation}\label{predynamics}
\mathcal{L}_\nu h=\mathcal{G}-\frac{\my{4}{\tr}\,\mathcal{G}}{2}\myg{3}-\my{3}{\Ric}+2h^2-Hh+\frac{\my{3}{\nabla}^2 N}{N}.
\end{equation}

If combined with Einstein's equation \eqref{EEq}, equations \eqref{energy preconstraint} and \eqref{momentum preconstraint} lead to constraints on the geometry of the spatial hypersurface $M^3$, the so-called {\it energy or Hamiltonian}\index{ind}{energy constraint}\index{ind}{Hamiltonian constraint}\index{ind}{constraint ! Hamiltonian}\index{ind}{constraint ! energy} and {\it momentum constraints}\index{ind}{momentum constraints}\index{ind}{constraint ! momentum}
\begin{eqnarray}\label{energy constraint}
\my{3}{\Scal}+H^2-|h|_{3}^2&=&\frac{16\pi G}{c^4} T(\nu,\nu)\\\label{momentum constraint}
\my{3}{\nabla}\!_j\,h_i^{\:j}-H\!_{,i}&=&\frac{8\pi G}{c^4} T(\nu,\partial_i).
\end{eqnarray}
Equations \eqref{EEq} and \eqref{predynamics} lead to the dynamical equation
\begin{equation}\label{dynamics}
\mathcal{L}_\nu h=-\my{3}{\Ric}+2h^2-Hh+\frac{\my{3}{\nabla}^2 N}{N}+\frac{8\pi G}{c^4}(T\vert_{TM^3\times TM^3}-\frac{\my{4}{\tr}\, T}{2}\,{\myg{3}}).
\end{equation}

Before we proceed to the initial value formulation of Einstein's equation, we wish to put down explicitly its 3+1 version using the timelike unit normal $\nu$ as an observer, i.~e.~the {\it 3+1 decomposed Einstein equations}\index{ind}{Einstein equations ! decomposed}
\begin{eqnarray}\nonumber
\my{3}{\Scal}+H^2-|h|_{3}^2&=&\frac{16\pi G\rho}{c^2}\\\label{Einstmatt}
\my{3}{\nabla}\!_j\,h_i^{\:j}-H\!_{,i}&=&-\frac{8\pi G J_i}{c^4}\\\nonumber
\mathcal{L}_\nu h+\my{3}{\Ric}-2h^2-Hh-\frac{\my{3}{\nabla}^2 N}{N}&=&\frac{8\pi G}{c^4}(S-\frac{\my{3}{\tr}S}{2}\,{\myg{3}})+\frac{4\pi G\rho}{c^2}\,\myg{3}.
\end{eqnarray}

Following Yvonne Choquet-Bruhat (and James W. York, Jr.~in \cite{CHB}), we will now proceed to reformulate Einstein's equation \eqref{EEq} as an initial value problem\footnote{We follow the exposition in \cite{Ring}.}. To this end, let an {\it initial data set}\index{ind}{initial data set} be a triple $(M^3,\myg{3},h)$ with $(M^3,\myg{3})$ a 3-dimensional Riemannian manifold and $h$ a symmetric $(2,0)$-tensor field on $M^3$, satisfying the constraint equations \eqref{energy constraint} and \eqref{momentum constraint} in vacuum. A {\it development}\index{ind}{development} of an initial data set $(M^3,\myg{3},h)$ then is a triple $(L^4,ds^2,\Lambda)$ consisting of a spacetime $(L^4,ds^2)$ satisfying the vacuum Einstein equation \eqref{EEq} and an isometric embedding $\Lambda:(M^3,\myg{3})\to(L^4,ds^2)$ being such that the induced second fundamental form of $\Lambda(M^3)$ agrees with the push-forward of $h$ under $\Lambda$. A development $(L^4,ds^2,\Lambda)$ is called a {\it globally hyperbolic development}\index{ind}{development ! globally hyperbolic} if $\Lambda(M^3)$ is a Cauchy surface in $(L^4,ds^2)$ (which then automatically implies that $(L^4,ds^2)$ is a globally hyperbolic spacetime).

A development $(\overline{L}^4,\overline{ds^2},\overline{\Lambda})$ of an initial data set $(M^3,\myg{3},h)$ is said to be an {\it extension}\index{ind}{extension} of a development $(L^4,ds^2,\Lambda)$ if $(L^4,ds^2)$ can be isometrically embedded into $(\overline{L}^4,\overline{ds^2})$ through a time-orientation preserving diffeomorphism $D$ satisfying $D\circ\Lambda=\overline{\Lambda}$ or in sloppy terms ``with $(M^3,\myg{3})$ sitting inside $(L^4,ds^2)$ in the same way as inside $(\overline{L}^4,\overline{ds^2})$ when compared via $D$''. Two developments are considered identical if they are extensions of each other and a globally hyperbolic development is considered {\it maximal}\index{ind}{development ! maximal} if it extends any other globally hyperbolic development. The following theorems answer the question of well-posedness of the Einsteinian Initial Value Problem (the {\it Cauchy problem})\index{ind}{Cauchy problem} in the affirmative.

\begin{Thm}[Yvonne Choquet-Bruhat]\label{CHBloc}
Let $(M^3,\myg{3},h)$ be an initial data set for which there are two developments $\mathcal{D}_i=(L^4_i,ds^2_i,\Lambda_i)$ $(i=1,2)$ (not necessarily globally hyperbolic). Then there is a globally hyperbolic development $(L^4,ds^2,\Lambda)$ which is extended by both developments $\mathcal{D}_1,\mathcal{D}_2$.
\end{Thm}

\begin{Thm}[Yvonne Choquet-Bruhat \& Robert Geroch]\label{CHB}
Every initial data set admits a maximal globally hyperbolic development. This maximal globally hyperbolic development is unique up to isometry.
\end{Thm}
\begin{Rems}
There also exist versions of both of these theorems where initial data sets and developments need not be vacuum but satisfy the constraints and Einstein equations with respect to some matter tensor, respectively. In this thesis, we are only going to apply the first of these theorems (in vacuum) to prove a uniqueness property of static systems in Section \ref{sec:equi} and therefore prefer not to discuss to any detail what conditions the chosen matter model must obey in order for these theorems to be true also in the presence of matter. Details can be found e.~g.~in \cite{CHB,Wald} and references cited therein.
\end{Rems}

\section{Asymptotically Flat Ends and Their Properties}\label{sec:flat}
So far, we have not explicitly discussed boundary data or asymptotics of the Lorentzian metric of a spacetime. We will do so now. In this thesis, we focus on {\it isolated systems}\index{ind}{isolated systems}\index{ind}{system ! isolated} -- spacetimes modeling stars or black holes that do not interact with other systems and cannot be reached by intruding gravitational waves, either. Isolated systems are mathematically modeled by so-called {\it asymptotically flat}\index{ind}{asymptotically flat}\index{ind}{spacetime ! asymptotically flat} Lorentzian manifolds, i.~e.~by specifying the fall-off of the metric on the boundary ``at spatial infinity''.

More concretely, we assume that for a decomposed spacetime $(L^4,ds^2)$ with time function $t$, each of the diffeomorphic time slices $M^3=\{t=\mbox{const}\}$ can be decomposed into a (possibly empty) compact {\it interior} $K\subset M^3$ and a finite number of {\it ends}\index{ind}{end}\index{ind}{interior}, i.~e.~unbounded components of $M^3\setminus K$ in which both $\myg{3}(t)$ and $h(t)$ satisfy certain fall-off conditions ``at spatial infinity'' which we will describe below.

As we will see in Chapter \ref{chap:static}, static spacetimes can be decomposed canonically such that the second fundamental forms $h(t)$ vanish for all times $t$. We will therefore only discuss the asymptotics of $\myg{3}(t)$ in more detail. Moreover, although we will only need and apply the asymptotic flatness conditions in dimension $n=3$, we will state them for arbitrary $n\geq3$ for convenience of the reader.

Let $(M^n,g)$ now be a Riemannian manifold that can be decomposed into a compact $K\subset M^n$ and a finite number of unbounded ends. In order to ensure isolatedness, we require these ends to be {\it asymptotically flat}\index{ind}{end ! asymptotically flat} in the sense that they are each diffeomorphic to $\R^n\setminus B$ for some closed balls $B$ and such that the pushed forward metrics in these ends behave as those of the flat (Euclidean) metric on $\R^n\setminus B$ plus error terms which decay as $\mathcal{O}(r^{-1})$ as $r\to\infty$, combined with suitable decay conditions on the derivatives; for a more precise definition please see below. As above, $r$ is the radial coordinate $r:=\vert x\vert:=\sqrt{(x^1)^2+\dots+(x^n)^2}$ associated to the system of asymptotically flat coordinates given by the chosen diffeomorphism in the given end.

We remind the reader of the following notation introduced in Section \ref{sec:sobo}: If $R>0$, $B_R:=\overline{B_R(0)}\subset\R^n$\index{sym}{$B_R$} denotes the closed ball of radius $R$ around $0$ and $E_R:=\R^n\setminus B_R$ denotes the associated (open) exterior region\index{ind}{exterior region}\index{sym}{$E_R$}. Following Robert Bartnik \cite{Bartnik} and Daniel Kennefick and Niall \'O Murchadha \cite{KM}\footnote{Other than these authors, we allow the manifold to have several ends. We require the metric to be smooth as this will be the case anyway when we consider static systems later on. This allows us to drop the $g\in W^{k,q}_{loc}$ condition stated in \cite{Bartnik}. Moreover, we include orders $k>1$ for later ease of formulation, where for $k>2$, we extend Robert Bartnik's condition to suitable $q\leq3$.}, we make the following more precise definition.

\begin{Def}\label{def:asym}
Let $n,k\in\N$, $n\geq3$, $1\leq q<\infty$, and $\tau>0$ be such that $W^{k,q}_{-\tau}(E_{R})\hookrightarrow W^{1,\overline{q}}_{-\tau}(E_{R})$ for some $\overline{q}>n$ and all $R\geq 1$. A smooth Riemannian manifold $(E^n,g)$ is then called a {\it $(k,q,\tau)$-asymptotically flat end}\index{ind}{asymptotically flat}\index{ind}{end ! asymptotically flat} if it carries a {\it structure of infinity of type $(k,q,\tau)$}\index{ind}{structure of infinity}, i.~e.~if there is a radius $R\geq1$ and a smooth diffeomorphism $\Phi:E^n\to E_R$ such that
\begin{enumerate}
\item[(i)] there exists $\mu\geq1$ such that $\mu^{-1}\vert\xi\vert^2\leq(\Phi_\ast g)\vert_x(\xi,\xi)\leq\mu\vert\xi\vert^2$ for all $x\in E_{R}$, $\xi\in\R^n$ (uniform positive definiteness and uniform boundedness of $\Phi_*g$), and
\item[(ii)] $(\Phi_\ast g)_{ij}-\delta_{ij}\in W^{k,q}_{-\tau}(E_{R})$ for $i,j=1,\dots,n$\\
(asymptotic decay of {\it order}\index{ind}{order of decay} $k$ and {\it decay rate}\index{ind}{decay rate} $\tau$\index{sym}{$\tau$}).
\end{enumerate}
We will call $(E^n,g)$ an asymptotically flat end for short if $k,q,\tau$ are either clear from context or arbitrary. A smooth connected Riemannian manifold $(M^n,g)$ is then called {\it $(k,q,\tau)$-asymptotically flat}\index{ind}{asymptotically flat}\index{ind}{asymptotically flat ! manifold} if there is a (possibly empty) compact $K\subset M^n$ such that $M^n\setminus K$ is a disjoint union of finitely many $(k,q,\tau)$-asymptotically flat ends. In particular, asymptotically flat ends are asymptotically flat manifolds in their own right. However, if $K$ is non-empty, we additionally assume that $(M^n,g)$ is geodesically complete. Observe that the diffeomorphsims $\Phi$ define the announced {\it coordinates at infinity}\index{ind}{coordinates ! at infinity} (in the end $E^n$) or {\it asymptotically flat coordinates for $g$}\index{ind}{coordinates ! asymptotically flat} (also in the end $E^n$). 

Moreover, we say that a sequence of points $\{p_l\}_{l\in\N}\subset M^n$ {\it tends to infinity}\index{ind}{tend to infinity} as $l\to\infty$ if it is ultimately contained in one of the ends of $M^n$ and if $r(\Phi(p_l))\to\infty$ holds there. Conversely, a subset $S\subset M^n$ will be called {\it bounded away from infinity}\index{ind}{bounded away from infinity} if there is a constant $C>0$ such that $\vert\Phi(S\cap E^n)\vert\leq C$ for all ends $E^n$ of $M^n$ and corresponding diffeomorphisms $\Phi$. Finally, $S$ will be called a {\it standardized compact interior}\index{ind}{standardized compact interior} if it is relatively compact with respect to $M^n$, contains $K$, and if for each {\it standardized exterior}\index{ind}{standardized exterior} $E^n\setminus S$ of $(M^n,g)$ there is a radius $R_1\geq R$ such that $\Phi^{-1}(E_{R_1})=E^n\setminus S$. A standardized compact interior is called {\it non-trivial}\index{ind}{standardized compact interior ! non-trivial} if $R_1>R$ for at least one standardized exterior. 
\end{Def}

We will frequently replace a given compact interior $K$ by one of its standardized compact interiors when we are not interested in the behavior of the metric in a neighborhood of $K$. The notion of standardized compact interior then ensures that the standardized exteriors $(E^n\setminus S,\Phi\vert_{E^n\setminus S})$ again qualify as asymptotically flat ends so that with the notion of a standardized exterior we can formalize the idea of an ``end of an end'' having certain properties. In Sections \ref{sec:known}, \ref{sec:asym}, \ref{sec:CoMasym} we will see that static asymptotically flat solutions of the Einstein equations in fact automatically possess better fall-off properties.

\begin{Rem}
There also exist other notions of asymptotic flatness in the literature, for example the so-called Regge-Teitelboim conditions and asymptotic Schwarzschildian behavior, cf.~e.~g.~\cite{KM}, \cite{Huang2} as well as Section \ref{sec:CoMasym}. Other approaches to define asymptotic flatness include methods of conformal compactification, an approach initiated by Robert Geroch in \cite{Geroch3}, cf.~e.~g.~Abhay Ashtekar and Rolf Hansen's paper \cite{AshHan}.
\end{Rem}

Before we continue by introducing the notions of mass and center of mass of asymptotically flat manifolds, we need to shortly discuss specific geometric coordinate systems available on and useful for our study of asymptotically flat ends. They are called ``harmonic coordinates''. The following existence theorem for asymptotically flat harmonic coordinates has been established by Niall \'O Murchadha \cite{Mur} and Robert Bartnik \cite{Bartnik}, independently. We quote it from \cite{Bartnik}, here, adapting the statement to our notation.
\begin{Thm}[Niall \'O Murchadha, Robert Bartnik]\label{harmonic}
Let $(M^n,g)$ be a $(k,q,\tau)$-asympto\-tically flat manifold with one end $E^n$. Let the structure of infinity of $E^n$ be denoted by $\Phi:E^n\to E_R$, let $(x^i=\Phi^i)$ be the associated asymptotically flat coordinates, and fix $1<\eta<2$. Then there are functions $y^i\in L^q_\eta(M^n)$, $i=1,\dots,n$, such that $\mylap{g}y^i=0$ and $x^i-y^i\in W^{k+1,q}_{1-\tau}$. If $(M^n,g)$ is geodesically complete, these harmonic coordinates are unique up to a Euclidean motion\index{ind}{Euclidean motion} at infinity in the sense that for any other global harmonic and asymptotically flat system of coordinates $(z^i)$ there exist a vector $\vec{b}\in\R^3$ and an orthogonal matrix $O\in O(\R^n)$ such that $$z^{i}=O^i_{j}y^j+b^{i}.$$
\end{Thm}\index{sym}{$O(\R^n)$}

\section{Mass and Energy}\label{sec:energy}
Starting from the famous formula $E=mc^2$ discovered by Albert Einstein in his (special) theory of relativity \cite{Einstein}\index{ind}{mass}\index{ind}{energy}, ``mass'' $m$\index{sym}{$m$} and ``energy'' $E$\index{sym}{$E$} are usually treated as interchangeable concepts in the general theory of relativity (where the speed of light $c$ is usually set to $1$ by the choice of units). This is, however, interfering with our attempt of understanding the behavior of physical properties under the Newtonian limit ($c\to\infty$) as the coupling of energy and mass will certainly not persist in this limit. We therefore do not interchangeably use these terms but restrict our attention to the mass of a system. As we shall see in Chapter \ref{chap:FT}, the (suitably defined) mass will have a finite Newtonian limit which obviously implies that the Einsteinian energy $E$ diverges. 

In contrast to Newtonian gravity, where local (and thus also global) mass can straightforwardly be defined as
\begin{equation}\label{local mass}
m(\Omega)=\int_{\Omega}\!\rho\,\,dV
\end{equation}
in a region $\Omega\subseteq\R^3$ with (Newtonian) mass density $\rho$, it is not very well understood how to define how much mass/energy\index{ind}{mass}\index{ind}{energy} exists in a given region $\Omega\subseteq M^3$ of a spatial slice in general relativity. This difficulty is due to the existence of vacuum spacetimes (which in particular satisfy $\rho\equiv 0$) with positive total mass implying invalidity of \eqref{local mass} even globally: a relativistic phenomenon usually explained by the existence of extra ``gravitational energy'' and intimately related to the existence of gravitational waves and/or black holes.

As an attempt to circumvent this problem of local mass in general relativity, several concepts of ``quasi-local mass''\index{ind}{quasi-local mass}\index{ind}{mass ! quasi-local} have been proposed, for example by Robert Bartnik \cite{Bartnik2}, Robert Geroch \cite{Geroch}, Gerhard Huisken \cite{HuiskenIso}, Arthur Komar \cite{Komar}, Roger Penrose and Wolfgang Rindler \cite{Penrose1,Penrose2}, and Stephen Hawking \cite{Hawking} -- some of them only for spacetimes with certain symmetries --, cf.~\cite{Szabados} for a review. The adjective ``quasi-local'' indicates that a mass/energy is assigned to the 2-surface enclosing a region instead of to the region itself as one would classically expect.

When defining such a quasi-local mass, one has to make sure that it converges to the ``total mass'' (ADM-mass) of the system along a suitably chosen sequence of 2-surfaces. Demetrios Christodoulou and Shing-Tung Yau \cite{ChrisYau} and Robert Bartnik \cite{Bartnik4} proposed lists of additional properties a quasi-local notion of mass should possess. We will come back to these in Section \ref{sec:mpseudo}, where we will describe a new notion of quasi-local mass for static spacetimes. This notion is inspired by a Newtonian construction and will be of central importance for our study of the Newtonian limit of mass. Let us now recall the notion of ``total mass'' of asymptotically flat manifolds, the so-called ``ADM-mass''.

\subsection*{ADM-Mass}\label{subsec:ADM}
In 1961, Richard Arnowitt, Stanley Deser, and Charles W. Misner \cite{ADM} have suggested a notion of total mass of asymptotically flat manifolds. This notion is nowadays referred to as the {\it ADM-mass}\index{ind}{ADM-mass}\index{ind}{mass ! ADM}. It is closely related to the Hamiltonian formulation of general relativity where it appears as the surface or flux integral ``at infinity'' corresponding to a divergence term appearing in the variation\footnote{cf.~\cite{ADM,Bartnik}.}. Robert Bartnik \cite{Bartnik}, Piotr Chru\'sciel \cite{Chrus2}, and Niall \'O Murchadha \cite{Mur} have shown that the ADM-mass is a well-defined and geometrically invariant property of the Riemannian $3$-metric on an asymptotically flat slice. Their proofs rely on the use of asymptotically flat harmonic coordinates. We refer the reader to John M.~Lee and Thomas H.~Parker's survey article \cite{LP} for an overview and a list of references on asymptotically flat manifolds and their masses. The ADM-mass is defined as follows.

\begin{Def}[Formal]
The ADM-mass of an asymptotically flat end $(E^n,g)$ is formally defined as $$m_{ADM}(E^n,g):=\frac{c^2}{16\pi G}\lim_{r\to\infty}\int_{S^{n-1}_r}\sum_{i=1}^3\left(g_{ii,j}-g_{ij,i}\right)\nu^j\,d\sigma,$$ where $S^{n-1}_r$ are coordinate spheres in a given system of asymptotically flat coordinates for $g$, indices are pulled down with the flat metric in this system of coordinates, and $\nu$ and $\sigma$ are the normal vector to and surface element of $S^{n-1}_r$ w.~r.~t.~this flat background metric.
\end{Def}

We cite the following theorem from p.~682 in \cite{Bartnik}. Variants of it can be found in \cite{Chrus2,Mur}.

\begin{Thm}[Robert Bartnik]\label{thm:Bartnik}
Let $\eta>0$, $k\geq2$, and  $q>n$. Suppose that a complete Riemannian manifold $(M^n,g)$ has a structure of infinity $\Phi:M^n\setminus K\to E_R$ of type $(k,q,\eta)$ for some $R\geq1$ and $K\subset M^n$ compact, and suppose that the Ricci tensor of $(M^n,g)$ satisfies $$\Ric\in L^q_{-2-\tau}(M^n)\mbox{ for some non-exceptional }\tau>\eta.$$ Then if $\tau\geq(n-2)/2$, the ADM-mass exists and is unique. Moreover, it is zero if $\tau>n-2$. 
\end{Thm}

\begin{Rems}
In our setting, the metric is always going to be smooth (as a direct consequence of being static and solving the Einstein equations). The condition on the Ricci tensor therefore reduces to a pure fall-off condition. Moreover, although the theorem is only stated for manifolds with one end, it can be straightforwardly generalized to include complete asymptotically flat manifolds with a finite number of ends. The statement of the theorem then holds true in each end, individually.
\end{Rems}

In simple terms, Theorem \ref{thm:Bartnik} states that the ADM-mass is a geometric quantity. Observe that this is a necessary requirement for a concept of mass as the mass of an object should not depend on the observer (at infinity). From now on, we will therefore assume that all asymptotically flat manifolds fall off at least as fast as required by Theorem \ref{thm:Bartnik}. Besides being a geometric quantity (or differently put ``being generally covariant''), one of the most important other physical requirements of a concept of total mass is that it is non-negative. This has been proven by Richard M.~Schoen and Shing-Tung Yau in \cite{SchoenYau} in the ``positive mass theorem''\index{ind}{positive mass theorem}\footnote{We cite the positive mass theorem from \cite{Huis2}. where we have added the implicit assumption of geodesic completeness explicitly. As we are not going to apply the theorem, we do not specify exact fall-off conditions.}.

\begin{Thm}[Richard M.~Schoen and Shing-Tung Yau]\label{positive mass thm}
If $(M^3,g)$ is an asymptotically flat Riemannian $3$-manifold with non-negative scalar curvature, then the mass of each end is non-negative. If the manifold is geodesically complete and if the mass is zero in one end, then $(M^3,g)$ is isometric to flat space $(\R^3,\delta)$.
\end{Thm}

The positive mass theorem has been proved under many different sets of additional assumptions, cf.~\cite{SchoenYau} and \cite{Bekenstein} for an exposition. In particular, Piotr Chru\'sciel and Gregory Galloway \cite{ChruGall} proved a stationary version with a technique very different from the original one by Richard M.~Schoen and Shing-Tung Yau. To the author's knowledge, however, there is as yet no specific proof for the static realm. We will prove such a static version of the positive mass theorem in Section \ref{sec:pm} using again a different method which relies on the quasi-local concept of pseudo-Newtonian mass introduced in Chapter \ref{chap:mCoM}. This pseudo-Newtonian mass agrees with the ADM-mass at infinity and will also be useful to study the Newtonian limit of mass. It will help us to prove that the Newtonian limit of mass is the mass of the Newtonian limit along any family of static relativistic systems which has a Newtonian limit, cf.~Theorem \ref{thm:NLm}.

\section{The Center of Mass}\label{sec:HCoM}
Asymptotically flat manifolds model isolated systems like stars or galaxies. Therefore, if they have non-vanishing mass, one would expect that they can also be attributed a ``center of gravity''\index{ind}{center of gravity} or differently said a ``center of mass (CoM)''\index{ind}{center of mass}\index{ind}{CoM}. Intuitively, one might expect that this center corresponds to a point in the manifold (at least in case the manifold is geodesically complete and thus does not haphazardly miss this point). However, asymptotically flat manifolds can contain black holes and can moreover display a very involved topology in their interiors even if no black holes are present. In addition, they can possess multiple ends with different ADM-masses.

It thus seems more adequate to define individual centers of mass for each of the ends of an asymptotically flat manifold and to formulate these centers in terms of asymptotically flat coordinates instead of defining them as points in the manifold. Differently put, the center of mass of an end $E$ then does not lie in the manifold itself but in the linear space $\R^3$ which extends the image $\Phi(E)=E_R$ of the coordinate diffeomorphism $\Phi:E\to E_R$ mapping the end to an exterior domain $E_R\subset\R^3$. It will therefore depend on the specific system of coordinates chosen. Alternatively, one can interpret the center of mass as a point in the affine space corresponding to the tangent space to $M^3$ at the point at infinity.

Several definitions for such a center of mass\index{ind}{center of mass} have been put forward, namely (possibly among others) by Tullio Regge and Claudio Teitelboim \cite{RT}, by Gerhard Huisken and Shing-Tung Yau \cite{HY} with a generalization by Jan Metzger \cite{Jan2}, and by Lan-Hsuan Huang \cite{Huang1}. Some of these centers of mass are defined for general asymptotically flat ends while others are only coined for asymptotically flat solutions of the vacuum Einstein constraint, see below. We will give a very short overview over these definitions and their interrelations and would like to refer the interested reader to the references cited above and to the overview article \cite{CoPo} by Justin Corvino and Daniel Pollack for more information. The different concepts of center of mass are devised as follows\footnote{We cite these definitions from \cite{Huang1,Huang2,CoPo,CoWu,HY}. The definitions are given in dimension $3$ only, as some of them assume that the metric satisfies the $3$-dimensional vacuum Einstein constraints and others rely on the Gau{\ss}-Bonnet and $2$-dimensional Sobolev embedding theorems. Moreover, we have introduced the constants $c$ (speed of light) and $G$ (gravitational constant) for later convenience; in the quoted references, they are set to $1$ by choice of units as usual in the literature.}.

In analogy to the definition of the ADM-mass, Tullio Regge and Claudio Teitelboim define an ADM center of mass with the help of a surface integral ``at infinity''.
\begin{Def}[Formal]\label{def:zADM}
The {\it ADM center of mass} $\vec{z}_{ADM}(E^3,g)\in\R^3$\index{ind}{center of mass ! ADM}\index{sym}{$\vec{z}_{ADM}$} of an asymptotically flat end $(E^3,g)$ with non-vanishing ADM-mass $m$ is formally defined in components as 
$$z_{ADM}^k(E^3,g):=\frac{c^2}{16\pi mG}\lim_{r\to\infty}\int_{S^2_{r}}\left[\sum_{i=1}^3x^k(g_{ij,i}-g_{ii,j})\nu^j-\sum_{i=1}^3(g_{i}^k\nu^{i}-g_{ii}\,\nu^k)\right]d\sigma$$  where $S^{2}_r$ are coordinate spheres in a given system of asymptotically flat coordinates for $g$. Indices are pushed up and pulled down with the flat ``background'' metric in this system of coordinates, and $\nu$ and $\sigma$ are the normal vector to and surface element of $S^{2}_r$ with respect to this background metric.
\end{Def}

Observe that, just as for the ADM-mass, it is not a priori obvious that the above definition is a geometric quantity, i.~e.~independent of the chosen system of asymptotically flat coordinates. Neither is it obvious how fast the asymptotic decay has to be in order that the integral expression in the definition converges as $r\to\infty$. This issue is settled for solutions of the vacuum Hamiltonian constraint having a specific kind of fall-off.

\begin{Prop}[Tullio Regge and Claudio Teitelboim]\label{prop:RT}
Suppose $(E^3,g,h)$ is an asymptotically flat solution of the vacuum Hamiltonian constraint \eqref{energy constraint} with positive ADM-mass $m$ which satisfies the \emph{Regge-Teitelboim conditions}\index{ind}{Regge-Teitelboim conditions}
\begin{eqnarray}\nonumber
\partial^l g_{ij}=\mathcal{O}(r^{-l-\tau})&\mbox{ with }&0\leq l\leq 2\\\nonumber
\partial^l g_{ij}^{odd}=\mathcal{O}(r^{-1-l-\tau})&\mbox{ with }&0\leq l\leq 1\\[-0.5ex]\label{propdef:RT}
\\[-0.5ex]\nonumber
\partial^l h_{ij}=\mathcal{O}(r^{-1-l-\tau})&\mbox{ with }&0\leq l\leq 2\\\nonumber
\partial^l h^{even}_{ij}=\mathcal{O}(r^{-2-l-\tau})&\mbox{ with }&0\leq l\leq 1
\end{eqnarray}
for some $\tau\in\left(1/2,1\right]$, where $f^{odd}$ and $f^{even}$ denote the odd and even parts $f^{odd}(x)=f(x)-f(-x)$ and $f^{even}(x)=f(x)+f(-x)$ of a given function $f$, respectively. Then $\vec{z}_{ADM}(E^3,g)\in\R^3$ given by Definition \ref{def:zADM} is well-defined.
\end{Prop}

In the static setting in focus in this thesis, we have $h=0$ and the vacuum Hamiltonian constraint \eqref{energy constraint} reduces to $\Scal=0$ for the $3$-metric $g$. Moreover, we will recall in Section \ref{sec:known} that asymptotically flat static metrics automatically possess better ({\it asymptotically Schwarzschildian}\footnote{Metrics of this type are also sometimes referred to as {\it asymptotically spherically round}.}\index{ind}{asymptotically Schwarzschildian}\index{ind}{asymptotically spherically round}) fall-off behavior, a well-known result in static general relativity first proven by Daniel Kennefick and Niall \'O Murchadha. In other words, they assert that $$\partial^l g_{ij}=\left(1+\frac{mG}{2c^2r}\right)^4\delta_{ij}+\mathcal{O}(r^{-2-l-\tau}),$$ so that the Regge-Teitelboim conditions are automatically satisfied and thus the ADM center of mass is well-defined in the static realm.

A very different, geometric approach is taken by Gerhard Huisken and Shing-Tung Yau who define the center of mass of a general asymptotically flat end by an intricate construction using a foliation of the end by constant mean curvature (CMC) surfaces\index{ind}{CMC-surface}. They prove existence and uniqueness results on these CMC-surfaces as asymptotic roundness and convergence of the Euclidean centers of mass of these CMC-surfaces as $r\to\infty$ thus intrinsically and geometrically defining a unique center of mass of the end itself. More concretely said, they prove the following quantitative theorem.

\begin{Thm}[Gerhard Huisken and Shing-Tung Yau]\label{CMC-CoM}
Let $(E^3,g)$ be an asymptotically flat end with positive ADM-mass $m$. Assume furthermore that $(E^3,g)$ is asymptotically Schwarzschildian in the sense that there is an asymptotically flat system of coordinates such that $$g_{ij}=\left(1+\frac{mG}{2c^2r}\right)^4\delta_{ij}+P_{ij}\mbox{ with }\vert\partial^lP_{ij}\vert\leq C_{l+1}r^{-l-2}\mbox{ for } 0\leq l\leq 4$$ for some constants $C_{1},\dots,C_{5}$.
Then there is a constant $\rho_{0}>0$ depending only on $m$ and on $C_{0}:=\max\{1,m,C_{1},\dots C_{5}\}$, and a foliation $\{\Sigma_{\rho}\}_{\rho\geq\rho_{0}}$ of a standardized exterior of $E^3$ by strictly stable constant mean curvature spheres. Furthermore, there exist constants $D_{1},D_{2}$ depending only on $C_{0},m$ and not on $\rho$ such that radial coordinate $r$ satisfies $\vert r-\rho\vert\leq D_{1}$ and $\vert H_{\rho}-\frac{2}{\rho}+\frac{4mG}{c^2\rho^2}\vert\leq D_{2}\rho^{-3}$ on $\Sigma_{\rho}$, where $H_{\rho}$ denotes the mean curvature on $\Sigma_{\rho}$. Finally, there is a vector $\vec{z}_{CMC}:=\vec{z}_{HY}(E^3,g)\in\R^3$ depending only on the end $(E^3,g)$ such that the Euclidean centers of mass of the surfaces $\Sigma_{\rho}$, $$\vec{z}\,(\Sigma_{\rho}):=\frac{\int_{\Sigma_{\rho}}\vec{x}\,d\sigma}{\int_{\Sigma_{\rho}}\,d\sigma}$$ converge to $\vec{z}_{HY}(E^3,g)$ as $\rho\to\infty$, where $d\sigma$ is the surface element with respect to the flat (Euclidean) background. We call $\vec{z}_{HY}$ the constant mean curvature (CMC) or Huisken-Yau center of mass.\index{ind}{center of mass ! CMC}\index{ind}{center of mass ! Huisken-Yau}\index{sym}{$\vec{z}_{HY}$}\index{sym}{$\vec{z}_{CMC}$}

The surfaces $\Sigma_{\rho}$ are unique (in a certain class). They arise as solutions to the curvature flow $F^\rho:S^2\to E^3$, 
\begin{eqnarray*}
\frac{d}{dt}F^\rho(p,t)&=&(\frac{\int_{F^\rho(S^2,t)}H\,d\sigma}{\int_{F^\rho(S^2,t)}d\sigma}-H)\,\nu(p,t),\quad\mbox{ for all } t\geq 0, p\in S^2,\\
F^\rho(0,p)&=&F^\rho_{0}(p)\quad\quad\quad\quad\quad\quad\quad\quad\;\quad\quad\mbox{ for all }p\in S^2
\end{eqnarray*} where $H(p,t)$, $\mu$, and $\nu$ denote the mean curvature, the surface measure, and the outer unit normal of $F^\rho(S^2,t)$ at $p$ with respect to the background $3$-metric $g$.
\end{Thm}

Jan Metzger has generalized this theorem to general asymptotically flat ends $(E^3,g,h)$ also allowing for a symmetric tensor field $h$ thought of as a second fundamental form from a $3+1$-perspective. At the same time, he has significantly weakened the assumptions on the fall-off of $g$; namely, it suffices to assume $$\sup_{\R^3\setminus B_{\sigma(0)}}\left(r\,\vert g-\myg{S}\vert+r^2\,\vert\my{g}{\nabla}-\my{S}{\nabla}\vert+r^3\,\vert\my{g}{\Ric}-\my{S}{\Ric}\vert\right)<\eta$$ for some $\eta>0$ sufficiently small where $\vert\cdot\vert$ denotes the norm with respect to either $g$ or $\myg{S}$ and $\myg{S}$ denotes the Schwarzschild metric\footnote{cf.~page \pageref{subsec:schwarz}ff for a short introduction into Schwarzschild metrics..} of the same (positive) ADM-mass as $g$. Other authors like Rugang Ye \cite{Ye} and Lan-Hsuan Huang \cite{Huang1} have proven similar existence, uniqueness, and asymptotic roundness statements with different methods and under different asymptotic flatness assumptions; in particular, the latter work relaxes the asymptotic Schwarzschildian condition to the Regge-Teitelboim conditions provided the vacuum Einstein constraints \eqref{energy constraint} and \eqref{momentum constraint} are both satisfied. Anticipating again the well-known asymptotic Schwarzschildness of asymptotically flat static metrics (cf.~Theorem \ref{thm:KM}), Jan Metzger and Lan-Hsuan Huang's results will in fact apply to the geometrostatic setting we will study in the remainder of this thesis.

Having looked at two very different definitions of center of mass, it is natural to ask whether or in what circumstances these centers agree. This question is addressed in Justin Corvino and Haotian Wu's paper \cite{CoWu} where they prove that -- in case of sufficiently fast fall-off -- both centers coincide: $\vec{z}_{ADM}=\vec{z}_{HY}.$

\begin{Thm}[Justin Corvino and Haotian Wu]\label{thm:CoWu}
Consider an asymptotically flat end $(E^3,g)$ with positive ADM-mass $m$ which satisfies $$g_{ij}(x)=(1+\frac{mG}{2 r c^2}+\frac{\sum_{k=1}^3B^k x^k}{r^3})^4\delta_{ij}+P_{ij}\mbox{ with } \partial^l P_{ij}=\mathcal{O}(r^{-3-l})\mbox{ for }0\leq l\leq 5$$ in some asymptotically flat chart for some constants $B^k\in\R$. Then, in this chart, $$z_{HY}^k(E^3,g)=\frac{2c^2 B^k}{mG}=z_{ADM}^k(E^3,g)$$ holds for all $k=1,2,3$.
\end{Thm}

Observe that this theorem does not request that the vacuum Hamiltonian constraint is satisfied; instead, convergence of the surface integrals in the formal Definition \ref{def:zADM} is secured by the stronger fall-off assumption. In her paper \cite{Huang1}, Lan-Hsuan Huang proves a different version of this result for solutions of the full vacuum Einstein constraints. 

\begin{Thm}[Lan-Hsuan Huang]\label{thm:Huang1}
If $(E^3,g,h)$ is an asymptotically flat end solving the vacuum Einstein constraints with positive ADM-mass $m$ and satisfying the Regge-Teitelboim conditions, then there exists a foliation by surfaces $\Sigma_{\rho}$ with constant mean curvature $H_{\rho}=\frac{2}{\rho}+\mathcal{O}(\rho^{-1-\tau})$ in some standardized exterior of $E^3$. Each leaf $\Sigma_{\rho}$ is a $c_{0}\rho^{1-\tau}$-graph over $S^2_{\rho}(\vec{z}_{ADM}(E^3,g))\subset\R^3$ and is strictly stable.
\end{Thm}

Again, both of these theorems apply in the static setting though this time, this is not obvious from our previous understanding of the fall-off of static metrics. In fact, we will prove in Chapter \ref{chap:static} that asymptotically flat static metrics (with positive ADM-mass) can always be put into the form assumed in Theorems \ref{thm:CoWu} or \ref{thm:Huang1}, cf.~Theorem \ref{thm:asym-CoM}. We will also illustrate what systems of asymptotically flat coordinates bring the metric into this form. In the proof of Theorem \ref{thm:Huang1}, Lan-Hsuan proves the following formula which will become useful for our purposes in Chapter \ref{chap:mCoM}.
\begin{Prop}[Lan-Hsuan Huang]\label{prop:Huang-formula}
Under the conditions of Theorem \ref{thm:Huang1}, there is a radius $r_{0}\geq1$ such that $$\int_{S^2_{r}(\vec{p})}\left(x^k-p^k\right)\left(H-\frac{2}{r}\right)d\sigma=\frac{8\pi mG}{c^2}\left(p^k-z^k\!\!\!_{ADM}(E^3,g)\right)+\mathcal{O}(r^{-\tau})$$ holds as $r\to\infty$ for any vector $\vec{p}=(p^1,p^2,p^3)^t\in\R^3$, any radius $r\geq r_{0}$, and all $k=1,2,3$. In this formula, $H$ denotes the mean curvature of the surface $S^2_{r}(\vec{p})$ with respect to the metric $g$ and $d\sigma$ is the surface element with respect to the flat metric. The coordinates are chosen such that the assumed Regge-Teitelboim conditions are fulfilled.
\end{Prop}

In \cite{Huang2}, Lan-Hsuan Huang gives\footnote{She states to have studied this expression on suggestion of Richard Schoen.} another ``intrinsic'' definition of center of mass. In the same paper, she proves that this intrinsic center of mass $\vec{z}_{I}(E^3,g)$ agrees with $\vec{z}_{ADM}(E^3,g)$ under the conditions of Theorem \ref{thm:Huang1} and thus also with $\vec{z}_{HY}(E^3,g)$ in each suitably asymptotically flat end.
\begin{ThmDef}[Lan-Hsuan Huang]\label{def:Huang}
Under the conditions of Theorem \ref{thm:Huang1}, the {\bf intrinsic center of mass}\index{ind}{center of mass ! intrinsic} $\vec{z}_{I}\in\R^3$\index{sym}{$\vec{z}_{I}$} of an asymptotically flat end $(E^3,g,h)$ is defined as $$z_{I}^k(E^3,g)=\frac{c^2}{16\pi mG}\lim_{r\to\infty}\int_{S^2_{r}}\big(\Ric-\frac{1}{2}\Scal\,g\big)\Big(r^2g^{kj}\partial_j-2x^k x^j\partial_{j},\nu\Big)d\sigma,$$ where the Ricci and scalar curvatures, the normal $\nu$, and the surface element $d\sigma$ correspond to $g$. $x^j$, $r$, and $\partial_{j}$ are the coordinate functions, the coordinate radius and the partial derivative with respect to the given chart at infinity, respectively. The intrinsic center of mass of the end $(E^3,g,h)$ agrees with its ADM and HY centers.
\end{ThmDef}

\begin{Rem}
The field $Y^k:=r^2g^{kj}\partial_j-2x^k x^j\partial_{j}$ is a conformal Killing vector field of the Euclidean background metric in the coordinates described in the theorem.
\end{Rem}

In Chapter \ref{chap:mCoM}, we will present a new quasi-local notion of center of mass in the static setting and prove that it agrees with the centers of mass described above whenever the considered manifolds are static solutions of the Einstein equations. This new notion will allow us to show that the Newtonian limit of the center of mass converges to the Newtonian center of mass of the limit along any family of static relativistic systems which has a Newtonian limit, cf.~Theorem \ref{thm:NLCoM}.

\chapter{Geometrostatics}\label{chap:static}
In this chapter, we will introduce and discuss geometrostatic systems and the equations governing them. They model static asymptotically flat spacetimes with compactly supported matter as we will explain in Section \ref{sec:SME}. For convenience of the reader, we will then collect some well-known facts about geometrostatic systems (known as ``static solutions'' in the literature) in Section \ref{sec:known}. In Sections \ref{sec:CoMasym} and \ref{sec:asym}, we will analyze the fall-off behavior of these systems repeatedly using the faster fall-off trick introduced on page \pageref{subsec:faster}ff. We begin by defining staticity and deriving the static version of the Einstein equations, the ``static metric equations''.

\section{The Static Metric Equations}\label{sec:SME}
A spacetime $(L^4,ds^2)$ is called {\it static}\index{ind}{static}\index{ind}{spacetime ! static} if there exists a timelike Killing vector field $X$ that is {\it irrotational}\index{ind}{irrotational}\index{ind}{Killing vector field ! irrotational} or {\it hypersurface-orthogonal}\index{ind}{hypersurface-orthogonal}\index{ind}{Killing vector field ! hypersurface-orthogonal}, i.~e.~that satisfies $X_{\left[\alpha\right.}\!{\my{4}{\nabla}}_{\beta} X_{\left.\gamma\right]}=0$. Now let $D$ be the distribution\index{ind}{distribution} given by $$D:=\bigcup_{p\in L^4}\{Y\in T_pL^4\,\vert\,ds^2(Y,X)\vert_p=0\}\subset T\!L^4$$ and equipped with the smooth bundle structure inherited from the tangent bundle. Then $D$ is clearly a smooth tangent distribution. $D$ is involutive as
\begin{eqnarray*}
ds^2(\left[Y,Z\right],X)&=&ds^2(Y,\my{4}{\nabla}\!_ZX)-ds^2(Z,\my{4}{\nabla}\!_YX)\\
&=&6\,X_{\left[\alpha\right.}\!{\my{4}{\nabla}}_{\beta} X_{\left.\gamma\right]}Y^\alpha Z^\beta=0
\end{eqnarray*}
and hence by the global Frobenius theorem\index{ind}{Frobenius' theorem} \ref{Frob}, $L^4$ is foliated by maximal connected integral submanifolds. Let $M^3$ be any of these maximal connected submanifolds. Then $M^3$ is spacelike as its tangent bundle is the orthogonal complement of the timelike vector field $X$ and therefore $ds^2$ induces a Riemannian $3$-metric on $M^3$ which we denote by $\myg{3}$. Let $h$ denote the induced second fundamental form and $\nu:=X/\vert X\vert$ the associated unit normal vector field, where $\vert X\vert:=\sqrt{-ds^2(X,X)}$. Then symmetry of $h$ and the Killing equation for $X$ give us that for all vector fields $Y,Z\in\Gamma(T\!M^3)$
$$h(Y,Z)=\frac{2}{\vert X\vert}\,\my{4}{\nabla}\!_{\left(\alpha\right.} X_{\left.\beta\right)}Y^\alpha Z^\beta=0$$ so that $M^3$ is a totally geodesic submanifold. Observe that although $L^4$ possess this natural foliation, it needs not in general be a product manifold. And even if so, the metric $ds^2$ must not necessarily globally split into a product metric. In this thesis, however, we will assume\footnote{This assumption is conventional but usually hidden in the terminology that a static spacetime is given by a Riemannian $3$-metric and a lapse function both defined on a common $3$-manifold.} that any static spacetime is {\it standard static}\index{ind}{standard static} which means that it can be globally decomposed as
\begin{eqnarray}\label{L^4}
L^4&=&\R\times M^3\\\label{ds^2}
ds^2&=&-N^2c^2dt^2+\myg{3}
\end{eqnarray}
with $N:=\vert X\vert=\sqrt{-ds^2(X,X)}c^{-2}>0$\label{DefN}, $t$ a global time function, $X=\partial_{t}$, and $\myg{3}$ the $3$-metric induced on the spatial slices arising as the integral manifolds of the above distribution $D$. For a discussion of when a given static metric is standard static we refer the interested reader to Miguel S\'anchez' articles \cite{Sanchez,SS}. Now observe furthermore that even a standard static metric needs not in general be globally hyperbolic, cf.~\cite{Sanchez} and references cited therein. A very simple example for a non-globally hyperbolic standard static spacetime would be a restriction of the Minkowski spacetime $(\R^4,\eta)$ with $\eta=-c^2dt^2+\delta,$ $\delta$\index{sym}{$\delta$} the flat metric on $\R^3$, to the submanifold $L^4:=\R^4\setminus\left(\R\times B\right)$, where $B$ is a closed ball in $\R^3$. However, an asymptotically flat standard static spacetime $(M^3,\myg{3},N)$ is globally hyperbolic if and only if $(M^3,\myg{3})$ is geodesically complete. The reason for this is that the lapse function and $3$-metric are uniformly bounded by asymptotic flatness, cf.~Spiros Cotsakis \cite{Cotsakis}.

Having discussed the spacetime structure of static metrics let us now study the associated symmetry reduction of Einstein's equation \eqref{Einstmatt}. If $(L^4,ds^2)$ is a static spacetime and $T$ an energy-momentum tensor, the 3+1 decomposed Einstein equations applied to $Y,Z\in\Gamma(T\!M^3)$ reduce to
\begin{eqnarray}\label{SM1}
\my{3}{\Scal}&=&\frac{16\pi G\rho}{c^2}\\\label{SM2}
0&=&-\frac{8\pi G}{c^4}J(Y)\\\label{SM3}
\my{3}{\Ric}(Y,Z)-\frac{\my{3}{\nabla}^2 N(Y,Z)}{N}&=&\frac{8\pi G}{c^4}\left(S(Y,Z)-\frac{\my{3}{\tr}S}{2}\myg{3}(Y,Z)\right)\\\nonumber&&+\frac{4\pi G\rho}{c^2}\,\myg{3}(Y,Z).
\end{eqnarray}
In these equations, the mass density $\rho$, the momentum density $J$, and the stress tensor $S$ are defined as explained on p.~\pageref{Tsplit}. Taking the trace of Equation \eqref{SM3}, we see that 
\begin{equation}\label{lapN}
\mylap{3}N=\frac{4\pi G}{c^2}N\left(\rho+\frac{\my{3}{\tr}S}{c^2}\right).
\end{equation}
\eqref{lapN} and \eqref{SM3} imply \eqref{SM1} and combine to the {\it static metric equations}\index{ind}{static metric equations}\footnote{The equation $J=0$ is not usually included in the static metric equations but it must be added in order to obtain the full Einstein equations.}
\begin{eqnarray}\nonumber
\my{3}{\Ric}&=&\frac{\my{3}{\nabla}^2 N}{N}+\frac{8\pi G}{c^4}\left(S-\frac{\my{3}{\tr}S}{2}\myg{3}\right)+\frac{4\pi G\rho}{c^2}\,\myg{3}\\[-1ex]\label{SME}
&&\\[-1ex]\nonumber
\mylap{3}N&=&\frac{4\pi G}{c^2}N\left(\rho+\frac{\my{3}{\tr}S}{c^2}\right).
\end{eqnarray}
In vacuum ($T=0$), the static metric equations read
\begin{eqnarray}\nonumber
N\,\my{3}{\Ric}&=&\my{3}{\nabla}^2 N\\[-1ex]\label{SMEvac}
&&\\[-1ex]\nonumber
\mylap{3}N&=&0.
\end{eqnarray}

For standard static spacetimes $(L^4,ds^2)$, the lapse function $N$ and the $3$-metric $\myg{3}$ with respect to the described canonical 3+1 decomposition characterize $ds^2$ uniquely by \eqref{ds^2}. In what follows, we will therefore sloppily refer to $(M^3,\myg{3},N)$ as a static spacetime and/or as a solution to the static metric equations. We remark that this terminology implicitly requires that $N$ be positive.

In dynamical GR, it is well-known that the equation of motion $\my{4}{\diver}T=0$ is a direct consequence of Einstein's equation (cf.~p.~\pageref{conserv}). The following proposition states that this feature persists in the static setting.
\begin{Prop}
The equation of motion $\my{4}{\diver} T=0$ is an automatic consequence of the static metric equations \eqref{SME}.
\end{Prop}
\begin{Pf}
Suppose that $(M^3,\myg{3},N,\rho,S)$ satisfy the static metric equations \eqref{SME} with $T$ induced from $\rho,S$ and $J=0$ as usual (cf.~p.~\pageref{T}) and compute that $(\diver T)_t=0$ by staticity.
$(\my{4}{\diver} T)_i=(\my{3}{\diver} S+N^{-1}\,S(\my{3}{\grad}N,\cdot)+c^2\rho N^{-1}\,dN)_i$ holds for a general static spacetime. Taking the exterior derivative of both the trace of the first and of the second static metric equation as well as the covariant divergence of the first, we obtain the equations
\begin{eqnarray*}
\my{3}{\Scal}\,dN+Nd(\my{3}{\Scal})&=&d(\mylap{3}N)+\frac{4\pi G}{c^4}\left(\left[\rho\,c^2-\my{3}{\tr}S\right]dN+Nd\!\left[\rho\,c^2-\my{3}{\tr}S\right]\right)\\
d(\mylap{3} N)&=&\frac{4\pi G}{c^4}\left(\left[\rho\,c^2+\my{3}{\tr}S\right]dN+Nd\!\left[\rho\,c^2+\my{3}{\tr}S\right]\right)\\
N(\my{3}{\diver}\,\my{3}{\Ric})&=&d(\mylap{3}N)+\frac{8\pi G}{c^4}\left(S(\my{3}{\grad}N,\cdot)+N\,\my{3}{\diver}S\right)\\
&&+\frac{4\pi G}{c^4}\left(\left[\rho\,c^2-\my{3}{\tr}S\right]dN+Nd\!\left[\rho\,c^2-\my{3}{\tr}S\right]\right)
\end{eqnarray*}
where we have used the definition of Ricci curvature. Applying Schur's lemma (or in other words the contracted second Bianchi identity)\index{ind}{Bianchi identity ! second}\index{ind}{Schur's lemma} $\my{3}{\diver}\,\my{3}{\Ric}=d(\my{3}{\Scal})/2$, we obtain
\begin{equation*}
\my{3}{\Scal}\,dN=-\frac{16\pi G}{c^4}\left(S(\my{3}{\grad}N,\cdot)+N\,\my{3}{\diver}S\right)
\end{equation*}
by a substraction of twice the last equation from the first one. Using \eqref{SM1} which is a simple consequence of the static metric equations \eqref{SME}, we therefore recover $\my{4}{\diver}T=0$.
\qed\end{Pf}

Asymptotically flat solutions of the static metric equations are the main objects to be studied in this thesis. Let us therefore continue by giving the definition of ``geometrostatic systems'', a term collecting a set of appropriate assumptions.

\begin{Def}[Geometrostatic Systems]\label{def:geomstatic}
Let $\mathcal{S}=(M^3,\myg{3},N,\rho,S)$ be a solution of the static metric equations \eqref{SME} and let $k\in\N$, $k\geq3$, and $\tau\geq1/2$ such that $-\tau$ is non-exceptional (i.~e.~$\tau\notin\Z$). We call $\mathcal{S}$ a {\it $(k,\tau)$-geometrostatic system}\index{ind}{geometrostatic system}\index{ind}{system ! geometrostatic} if, in addition, the following conditions hold:
\begin{enumerate}
\item[(i)] $(M^3,\myg{3})$ is a $(k,q=2,\tau)$-asymptotically flat manifold.
\item[(ii)] $N>0$ in $M^3$ and $N(p)\to1$ as $p\to\infty$ in each end of $M^3$.
\item[(iii)] $\rho\geq0$, $S\geq0$, and the supports of $\rho$ and $S$ are bounded away from infinity.
\end{enumerate}
As before, we will call $\mathcal{S}$ a geometrostatic system for short if $k$ and $\tau$ are either clear from context or arbitrary. Moreover, if we are concerned with a vacuum solution (i.~e.~if $\rho=0$, $S=0$), we call $(M^3,\myg{3},N)$ a {\it vacuum geometrostatic system}\index{ind}{vacuum} for simplicity.
\end{Def}

\begin{Rems}
$\tau\geq1/2$ and $k\geq3$ ensure that the ADM-masses of all ends are well-defined by Theorem \ref{thm:Bartnik} and the embedding theorems\footnote{To be concrete, observe that in particular $W^{3,2}_{-\tau}(E_{R})\hookrightarrow W^{2,\overline{q}}_{-\tau}(E_{R})$ for $\overline{q}=6>3$ by the Sobolev inequality \eqref{B:1.2}.} stated in Section \ref{sec:sobo}. The restrictions $q=2$ and $k\geq3$ relate to better a priori decay results for $g$ and $N$ which have been obtained by Daniel Kennefick and Niall \'O Murchadha, cf.~Theorem \ref{thm:KM}. Clearly, the static spacetime $(L^4,ds^2)$ constructed from a geometrostatic system $(M^3,\myg{3},N,\rho,S)$ via \eqref{L^4} and \eqref{ds^2} is a standard static spacetime.
\end{Rems}

The following lemma will be very useful in the sequel. It is implicit in many papers on static metrics and the metric $\gamma$ appearing in it will be very important in Chapters \ref{chap:pseudo} and \ref{chap:FT} where it will be called the ``pseudo-Newtonian metric'' corresponding to the given geometrostatic system.
\begin{Lem}\label{waveharm}\label{confharm}
Let $(M^3,g,N,\rho,S)$ be a geometrostatic system and let $(x^{i})$ be local coordinates for $M^3$. Then $(x^i)$ are wave harmonic\index{ind}{wave harmonic} with respect to the induced Lorentzian metric $ds^2$ (i.~e.~satisfying $\my{ds^2}{\square}\,x^{i}=0$) if and only if they satisfy $$\mylap{3}x^i=-\frac{N_{,j}}{N}\,\inv{3}{ji}.$$ Equivalently put, $(x^{i})$ are wave harmonic with respect to $ds^2$ if and only if they are harmonic with respect to the conformally transformed metric $\gamma:=N^2\,\myg{3}$.\index{sym}{$\gamma$}\index{ind}{conformally transformed metric}
\end{Lem}
\begin{Pf} Straightforward computation.\qed\end{Pf}

\subsection*{Example: Schwarzschild Solutions}\label{subsec:schwarz}
The most important example for an asymptotically flat solution of the vacuum static metric equations is the family of (spatial) {\it Schwarzschild metrics} named after their discoverer Karl Schwarzschild \cite{Schwarz}:\index{ind}{Schwarzschild ! metric}\index{ind}{metric ! Schwarzschild}\index{sym}{$\myg{m,S}$}\index{sym}{$\myg{S}$}
\begin{equation}\label{schwarz}
\myg{m,S}:= \left(1+\frac{mG}{2rc^2}\right)^4\delta
\end{equation}
on $\R^3\setminus\{0\}$ where $r$ denotes the radial coordinate on $\R^3\setminus\{0\}$, $m\in\R$ is called the ``mass parameter'' of the family, and $\delta$\index{sym}{$\delta$} denotes the Euclidean metric\index{ind}{Euclidean metric}\index{ind}{metric ! Euclidean} on $\R^3\setminus\{0\}$. Schwarzschild metrics model the (spatial) exterior of rotationally symmetric static stars or blackholes in general relativity.  They are rotationally symmetric, conformal to the flat metric $\delta$, and asymptotically flat with decay rate $0<\tau<1$ with respect to Euclidean coordinates on $\R^3\setminus\{0\}$. We have $\myg{0,S}=\delta$ and $m$ coincides with the ADM-mass of $\myg{m,S}$. Together with the {\it Schwarzschild lapse functions}\index{ind}{Schwarzschild ! lapse function}\index{sym}{$\my{m,S}{N}$}\index{sym}{$\my{S}{N}$}
\begin{equation}\label{Nschwarz}
\my{m,S}\!{N}:\R^3\setminus\{0\}\to\R:p\mapsto\left(1-\frac{mG}{2r(p)c^2}\right)\left(1+\frac{mG}{2r(p)c^2}\right)^{-1},
\end{equation}
$\myg{m,S}$ solves the vacuum static metric equations on its domain of definition. As moreover $N\to1$ at infinity, $(\R^3\setminus\{0\},\myg{m,S},\my{m,S}\!{N},0,0)$ is a geometrostatic system as defined above. For later use, let us introduce the abbreviation
\begin{equation}\label{M}
 M:=\frac{mG}{c^2}
\end{equation}\index{sym}{$M$}
so that the Schwarzschild metrics obtain the more familiar form $\myg{m,S}=\left(1+\frac{M}{2r}\right)^4\delta$. When the mass $m$ is implicitly understood, we also write $\myg{S}$ and $\my{S}\!{N}$ instead of the lengthy but more precise terms $\myg{m,S}$ and $\my{m,S}\!{N}$, respectively.

The coordinates used in \eqref{schwarz} are called {\it isotropic coordinates}\index{ind}{coordinates ! isotropic}. They are very useful for computations relying on rotational symmetry but they are not wave harmonic with respect to the corresponding $4$-metric $\my{m,S}\!{ds^2}=-\my{m,S}\!{N}^2c^2dt^2+\myg{m,S}$. As our considerations of the center of mass of an asymptotically flat manifold will rely on such wave harmonic systems of coordinates, we will now introduce such a system for the Schwarzschild metrics. If $(x^i)$ are the isotropic coordinates for a Schwarzschild metric $\myg{S}$, set
\begin{eqnarray}\label{schwarzharm}
y^i:&=&\left(1+\frac{M^2}{4r^2}\right)x^i\\\nonumber
s:&=&\vert(y^i)\vert=\sqrt{(y^1)^2+(y^2)^2+(y^3)^2}
\end{eqnarray}
which implies
\begin{equation}\label{(s,r)}
s(r)=\left(1+\frac{M^2}{4r^2}\right)r\quad\mbox{and}\quad r(s)=\frac{s}{2}\left(1+\sqrt{1-\frac{M^2}{s^2}}\right).
\end{equation}
A straightforward computation shows that $(y^i)$ is indeed a smooth system of coordinates on $\R^3\setminus\overline{B_{M/2}(0)}$ with values in $\R^3\setminus\overline{B_M(0)}$. The transformation reads
\begin{equation}\label{schwarztrafo}
\frac{\partial y^i}{\partial x^j}=\left(1+\frac{M^2}{4r^2}\right)\delta^i_j-\frac{M^2x^ix_j}{2r^4},
\end{equation}
where $\delta^i_j$ denotes the Kronecker delta symbol. With the abbreviations $\varphi(r):=1+M/2r$ and $\psi(s):=(1+M^2/4r(s)^2)^{-1}=(1+\sqrt{1-M^2/s^2)})/2$, one finds that $r(s)=\psi(s)s$ and
\begin{eqnarray}\nonumber
\partial_{y^i}&=&\psi\partial_{x^i}+\psi'\frac{y^ky_i}{s}\partial_{x^k},\\
\myg{S}(\partial_{y^i},\partial_{y^j})&=&(\varphi\circ r)^4\left(\psi^2\delta_{ij}+2s\psi\psi'\,\frac{y_iy_j}{s^2}+s^2(\psi')^2\frac{y_iy_j}{s^2}\right)\\\label{SN}
\my{S}\!{N}(s)&=&(1-\frac{M}{s})^{1/2}(1+\frac{M}{s})^{-1/2}.
\end{eqnarray}
Moreover, for later convenience, let us summarize the explicit expressions for the Christoffel symbols, curvature tensors, and derivatives of the lapse for the Schwarzschild family both in isotropic
\begin{eqnarray}
\my{S}{\Gamma}_{ij}^k&=&-\frac{M}{r^2\varphi}\left(\frac{x_i}{r}\delta_j^k+\frac{x_j}{r}\delta_i^k-\frac{x^k}{r}\delta_{ij}\right)\\
\my{S}{\Ric}_{ij}&=&\frac{M}{r^3\varphi^2}\left(\delta_{ij}-\frac{3x_ix_j}{r^2}\right)\\\label{N,i}
\my{S}\!{N}_{,i}&=&\frac{Mx_i}{r^3\varphi^2}
\end{eqnarray}
and in wave harmonic coordinates
\begin{eqnarray}\nonumber
\my{S}{\Gamma}_{ij}^k&=&-\frac{M}{s^2\psi^2\varphi\circ r}\left[(\frac{y_i}{s}\delta_j^k+\frac{y_j}{s}\delta_i^k-\frac{y^k}{s}\delta_{ij})\psi+s\psi'(\frac{y_i}{s}\delta^k_j+\frac{y_j}{s}\delta^k_i)+\frac{M^2y^k\delta_{ij}}{2s^3}\right.\\
&&\quad+\left.\frac{(\psi')^2y_iy_jy^k}{s\psi}-\frac{M^2y_iy_jy^k}{s^5}\left(1+s^2(\psi')^2+\frac{s\psi'}{\psi}-\frac{s^2(\psi')^2}{2\psi^3}\right)\right]\\
\my{S}{\Ric}_{ij}&=&\frac{M}{s^3\psi^3(\varphi\circ r)^2}\left(\psi^2\delta_{ij}-\frac{y_iy_j}{s^2}\left[3\psi^2+4s\psi\psi'+2s^2(\psi')^2\right]\right)\\\label{N,yi}
\my{S}\!{N}_{,i}&=&\frac{M\psi y_i}{s^3\psi^3(\varphi\circ r)^2}\left(\psi+s\psi'\right).
\end{eqnarray}

Having derived these expressions, it is straightforward to see that $\myg{S}$ is asymptotically flat with respect to both $(x^i)$ and $(y^i)$:
\begin{Lem}\label{schwarziso}
Let $m\geq0$ and let $M:=mG/c^2$ as before. Let $\myg{S}:=\myg{m,S}$ be given in isotropic coordinates with $r>M/2$. Then $\myg{S}$ is $(k,q,\tau)$-asymptotically flat in $E_{M/2}$ for all $k\in\N$, $1\leq q<\infty$, and $0\leq\tau<1$.
\end{Lem}
\begin{Pf}
As for $m=0$ the claim must trivially hold, we assume that $m>0$ without loss of generality. We can then immediately see that $\myg{S}\in W^{k,q}_{loc}(\R^3\setminus\overline{B_{M/2}})$ by smoothness and that $\myg{S}$ is uniformly positive and bounded as $1\leq\varphi\leq2$ for $r>M/2$. An induction over the order of (weak) differentiability combined with the chain rule and the well-known fact that $\int_{M/2}^\infty r^\kappa dr<\infty$ iff $\kappa<-1$ ensures that $\myg{S}_{ij}-\delta_{ij}\in W^{k,q}_{-\tau}(\R^3\setminus\overline{B_{M/2}(0)})$. Hence by Definition \ref{def:asym}, $\myg{S}$ is $(k,q,\tau)$-asymptotically flat with respect to its isotropic coordinates.
\qed\end{Pf}

\begin{Lem}\label{schwarzharmonic}
Let $m\geq0$ and let $M:=mG/c^2$ as before. Let $\myg{S}:=\myg{m,S}$ be given in the wave harmonic coordinates $(y^i)$ described in \eqref{schwarzharm} with $s>M^\ast>M$. Then $\myg{S}$ is $(k,q,\tau)$-asymptotically flat in $E_{M^\ast}$ for all $k\in\N$, $1\leq q<\infty$, and $0<\tau<1$.
\end{Lem}
\begin{Pf}
For $m=0$, we have that $x^i=y^i$ so that the claim follows from Lemma \ref{schwarziso}. Let us therefore assume that $m>0$ without loss of generality. Again, $\myg{S}\in W^{k,q}_{loc}(\R^3\setminus\overline{B_{M^\ast}})$ must hold by smoothness and $\myg{S}$ is uniformly positive and bounded as $1\leq\varphi\circ r\leq2$, $1/2<\psi<1$, and $0<s\psi'<M^2/2M^\ast\sqrt{(M^\ast)^2-M^2}$ for $s>M^\ast$. An argument similar to the one in Lemma \ref{schwarziso} using the decay behavior of $\psi(s)=(1+\sqrt{1-M^2/s^2})/2$ and its derivatives leads to $\myg{S}_{ij}-\delta_{ij}\in W^{k,q}_{-\tau}(\R^3\setminus\overline{B_{M^\ast}(0)})$ in wave harmonic coordinates. Hence by Definition \ref{def:asym}, $\myg{S}$ is $(k,q,\tau)$-asymptotically flat with respect to these coordinates.
\qed\end{Pf}

Using these lemmata, it is immediate that the Schwarzschild metrics and lapse functions constitute vacuum geometrostatic systems outside suitable balls.
\begin{Prop}\label{prop:schwarzgeo}
Let $m\geq0$, $k\in\N$, $k\geq3$, $1/2\leq\tau<1$, $M^\ast>M$, and set $M:=mG/c^2$ as before. Then the Schwarzschild systems $\mathcal{S}_{iso}:=(E_{M/2}(0),\myg{m,S},\my{m,S}\!N)$ and $\mathcal{S}_{har}:=(E_{M^*}(0),\myg{m,S},\my{m,S}\!N)$ are $(k,\tau)$-geometrostatic systems (with structures of infinity given by the isotropic coordinates and the wave harmonic coordinates, respectively). If $m\neq0$, the ADM center of mass vanishes in these wave harmonic coordinates.
\end{Prop}
\begin{Pf}
Lemmata \ref{schwarziso} and \ref{schwarzharmonic} assert the asymptotics of the metric in the respective systems of coordinates. The lapse function $\my{m,S}\!N$ satisfies $\my{m,S}\!N(p)\to1$ as $p\to\infty$ because $r(p)\to\infty$ and $r(s(p))\to\infty$ as $p\to\infty$, cf.~\eqref{(s,r)}. The embedding statements from Section \ref{sec:sobo} ensure property (ii). The center of mass claim follows from a direct computation.
\qed\end{Pf}

For later reference, we quickly summarize the asymptotic expansions of $\my{S}\!{N}$ and $\myg{S}_{ij}$ in wave harmonic coordinates using analyticity at infinity of the relevant expressions:
\begin{eqnarray}\label{Nfall}
\my{S}\!{N}&=&1-\frac{M}{s}+\frac{M^2}{2s^2}+\mathcal{O}(s^{-3})\\\label{gfall}
\myg{S}_{ij}&=&\left(1+\frac{2M}{s}+\frac{M^2}{s^2}\right)\delta_{ij}+\frac{2M^2y_{i}y_{j}}{s^4}+\mathcal{O}(s^{-3})
\end{eqnarray}

Let us close our introduction into the topic of Schwarzschild metrics by mentioning that although the spacetimes corresponding to the Schwarzschild $3$-metric and lapse are well-defined and satisfy the static vacuum Einstein equations even inside the (pointed) balls cut out in the above proposition, they cannot be understood as geometrostatic systems, there, as the lapse $N=\sqrt{-ds^2(X,X)}$ passes through zero at the so-called ``horizon''\index{ind}{horizon}, a surface which is located at $r=M/2$ in isotropic coordinates, cf.~the remark on page \pageref{rem:horizon}. For reasons that will become clear later\footnote{One of them being the fact that we will use $N^2$ as a factor for a conformal transform in Chapter \ref{chap:pseudo}.}, the condition $N>0$ will however be vital and cannot be dropped in the definition of geometrostatic systems.

\section{Well-Known Properties of Geometrostatic Systems}\label{sec:known}
(Standard) static spacetimes that satisfy the static metric equations \eqref{SME} (with a suitable matter model) have been studied in abundance. For convenience of the reader, we will now collect some of their well-known and very useful properties like regularity and fall-off behavior. Most of these properties will be stated in their vacuum\index{ind}{vacuum}\index{ind}{vacuum ! solution} versions, which means that $T=0$ or equivalently in the static setting that $S=0$ and $\rho=0$. In these versions, the theorems apply to the vacuum regions outside the compactly supported matter in the case of non-vacuum geometrostatic systems. We will state them in their original formulation for the purpose of recognition by the knowledgeable reader.

Let us begin with a regularity result. Surely, classical solutions $(M^3,\myg{3},N>0)$ of the vacuum static metric equations will have some assumed a priori regularity. The weakest possible assumption will probably be that $N$ and $\myg{3}$ be $C^2$ assuring existence and continuity of the curvature and derivative terms on which the equations rely. Assuming slightly more regularity, Henning M\"uller zum Hagen \cite{MzH} showed a remarkable automatic analyticity property of static vacuum metrics. Observe that this analyticity property is local and does not assume or imply specific boundary behavior on the spatial slice $M^3$.

\begin{Thm}[Henning M\"uller zum Hagen]\label{MzH}
Let $L^4$ be manifold of class $C^5$, $ds^2$ a Lorentzian metric on $L^4$ of class $C^3$ and assume that there is a irrotational timelike Killing vector field $X$ of class $C^4$ making $(L^4,ds^2)$ a static spacetime. Assume that $(L^4,ds^2)$ satisfies the vacuum static metric equations. Then $M^3$ is analytic and its analytic atlas is generated by wave harmonic coordinates\footnote{In \cite{MzH}, these coordinates are referred to as harmonic coordinates with respect to the conformally transformed metric $\gamma:=N^2\,\myg{3}$, cf.~Lemma \ref{confharm}.} charts. Moreover, the induced $3$-metric $\myg{3}$ is analytic with respect to any analytic chart.
\end{Thm}

Having this result in mind, we will not be concerned with interior regularity issues\footnote{It is well-known that singularities do arise even in static spacetimes -- e.~g.~in the Schwarzschild spacetime of page \pageref{subsec:schwarz}ff -- but these are not interior points of the manifold and therefore do not destroy interior regularity.} in the remainder of this thesis. In harmony with the positive mass theorem \ref{positive mass thm}, there can only be trivial geodesically complete vacuum solutions as Andr\'e Lichnerowicz (\cite{Lich}, chapter VIII) showed using the maximum principle for the Laplace equation and the fact that $M^3$ is $3$-dimensional (in terms of formula \eqref{formula:3Rm}).

\begin{Thm}[Andr\'e Lichnerowicz]\label{Lich}
If $(M^3,g,N)$ is a (geodesically) complete solution of the vacuum static metric equations with $N>0$ everywhere, then either $M^3$ is compact, $N$ is constant, and $g$ is flat or $M^3$ admits a {\it domain of infinity}\index{ind}{domain of infinity} in the sense of possessing arbitrarily remote points from a given point $p\in M^3$ (with respect to the Riemannian distance function $\dist$). If in that case there exists a constant $N_{0}\in\R^+$ such that $N(q)\to N_{0}$ as $\dist(p,q)\to\infty$, it follows that $N\equiv N_{0}$ and that $g$ is flat on all of $M^3$.
\end{Thm}

Michael T.~Anderson \cite{Anderson1} has generalized this result using geometric PDE techniques and again the $3$-dimensionality in form of formula \eqref{formula:3Rm} to get rid of the constant asymptotic behavior assumption.
\begin{Thm}[Michael T.~Anderson]\label{Anders}
If $(M^3,g,N)$ is a (geodesically) complete solution of the vacuum static metric equations with $N>0$ everywhere, then $N$ is constant and $g$ is flat. Moreover, $M^3$ is diffeomorphic to $\R^3$ or a quotient thereof.
\end{Thm}

Following Michael T.~Anderson in \cite{Anderson2}, a solution $(M^3,g,N)$ of the vacuum static metric equations is called {\it maximal}\index{ind}{maximal solution}\index{ind}{solution ! maximal}  if there is no solution $(\widetilde{M}\,\!^3,\widetilde{\g},\widetilde{N})$ of the vacuum static metric equations properly extending $(M^3,g,N)$ in the sense that $M^3\subsetneq\widetilde{M}\,\!^3$, $\widetilde{N}\vert_{M^3}=N$, and $\widetilde{\g}\vert_{T\!M^3\times T\!M^3}=g$. Furthermore, if $M^3$ is not complete we denote its metrical completion by $\overline{M}\,\!^3$ and its metrical boundary by $\partial{M}\,\!^3$ and extend the Riemannian distance function to $\overline{M}\,\!^3$ -- without assuming that $\overline{M}\,\!^3$ be a smooth manifold with boundary or for that matter that $g$ and $N$ can be smoothly extended to $\partial M^3$. $\partial M^3$ is called {\it pseudo-compact}\index{ind}{pseudo-compact} if there is a tubular neighborhood $U\subset\overline{M}\,\!^3$ of $\partial M^3$ whose boundary $\partial U$ has compact intersection with $M^3$. If $d:M^3\to\R:p\mapsto\dist(p,\partial M^3)$ denotes the metric distance to the boundary, then pseudo-compactness of $\partial M^3$ is equivalent to the level sets of $d$ being compact. An {\it end}\index{ind}{end} of $M^3$ then in this context is an unbounded component of $\overline{M}\,\!^3\setminus U$. It is called {\it small}\index{ind}{end ! small} if
\begin{equation*}
\int_{s_0}^\infty\left(\sigma(d^{-1}(s)\cap E)\right)^{-1}ds=\infty
\end{equation*}
where $\sigma$ denotes the induced surface measure. As above, an end is called {\it asymptotically flat}\index{ind}{end ! asymptotically flat}\index{ind}{asymptotically flat} if it possesses a structure of infinity with respect to the induced metric. The following theorem gives insight into the behavior of static vacuum solutions in an end $E$ using\footnote{For more details, especially on why this limit is well-defined, cf.~\cite{Anderson2}. The author has introduced the constants $c$ and $G$ in order to make $m_*$ a physical mass and thus comparable with other notions of mass discussed in this thesis.}
\begin{equation*}
m_*(E):=\lim_{s\to\infty}\frac{c^2}{4\pi G}\!\int_{d^{-1}(s)\cap E} g(\my{g}{\grad}\log N,\my{g}{\grad}\,d)\,d\sigma.
\end{equation*}

\begin{Thm}[Michael T.~Anderson]\label{Ander2}
Let $(M^3,g{3},N)$ be a maximal solution of the vacuum static metric equations with $N>0$ everywhere and suppose that $\partial M^3$ is pseudo-compact with corresponding tubular neighborhood $U$. Then $\overline{M}\,\!^3\setminus U$ is a union of finitely many ends. If $E$ is one of the ends and if there is a sequence $\{p_i\}\subset M^3$ with $d(p_i)\to\infty$ as $i\to\infty$ and a constant $N_0>0$ such that $N(p_i)\geq N_0$ for all $i\in\N$, $E$ is either asymptotically flat or small. If the mass $m_*(E)$ is positive and if $N$ is bounded above on $E$, $E$ must be asymptotically flat.
\end{Thm}

Thus, for the physically very reasonable class of vacuum solutions $(M^3,g,N)$ having pseudo-compact boundary $\partial M^3$, possessing a bounded lapse function $N>0$ which is not tending to $0$ in any end, and which satisfy that all ends of $M^3$ have positive $\ast$-mass $m_*$, all ends are automatically asymptotically flat. We therefore do not restrict our attention severely when focussing on asymptotically flat solutions of the static metric equations\footnote{As announced earlier, we will focus our attention on situations where the matter has compact support and thus the static spacetimes under consideration in fact satisfy the vacuum static metric equations outside some compact set. This means that we can apply Michael T.~Anderson's theorem \ref{Ander2} to our situation statically extending the spacetime to be maximal.}.

There might be even more to say. Asymptotic flatness could be a generic behavior among the class of ends for which both $g$ and the conformally transformed metric $\gamma:=N^2g$ are complete up to the boundary in the metric sense, cf.~Mart\'in Reiris \cite{Reiris}. Observe that this conformally transformed metric $\gamma$ reappears in Section \ref{sec:pseudo} where we reformulate static relativity in a pseudo-Newtonian form.

Adding to the above facts, static asymptotically flat solutions of the vacuum Einstein equation \eqref{EEq} have very specific fall-off at infinity. Well-known results on this fall-off go back to Robert Geroch \cite{GerochI,GerochII}, Rolf Hansen \cite{Hansen}, Robert Beig \cite{Beig2}, Robert Beig and Walter Simon \cite{BS1,BS2,BS3,BS4}, Prasun Kundu \cite{Kundu} as well as Daniel Kennefick and Niall \'O Murchadha \cite{KM} among others. In particular, it was asserted by Robert Beig and Walter Simon that asymptotically flat static vacuum solutions are analytic at infinity\footnote{This analyticity at infinity is defined by conformal compactification. It thus in principle differs from the analyticity at infinity we will assume in Theorems \ref{thm:Nunique} and \ref{thm:gunique}.} approximately Schwarzschildian at spatial infinity assuming specific fall-off rates at infinity.

Daniel Kennefick and Niall \'O Murchadha generalized these results to solutions satisfying weaker fall-off assumptions. We will apply their result when studying higher order terms of the expansion in Sections \ref{sec:asym}, \ref{sec:CoMasym}. Their result constitutes the reason for our assumptions that $q=2$ and $k\geq3$ in the definition of geometrostatic systems. In fact, it would allow us to drop the assumption $\tau\geq1/2$ which we have denoted in the definition to make it obvious that the ADM-mass is well-defined for geometrostatic systems.

\begin{Thm}[Daniel Kennefick \& Niall \'O Murchadha]\label{thm:KM}
Let $g$ be a Riemannian $3$-metric on some exterior set $E_R\subset \R^3$, $R>1$, such that $g_{ij}-\delta_{ij} \in W^{k,2}_{-\tau}(E_R)$ with differentiability order $k\geq3$ and decay rate $\tau>0$ and assume that $(g,N)$ solves the vacuum static metric equations for some $N:E_R\to\R^+$. Then if $N$ goes to $1$ at infinity, $(M^3,g,N)$ must be Schwarzschildian at infinity, i.~e.~such that $N$ and $g$ satisfy
\begin{equation}\label{eq:KM}
N-1+\frac{M}{r} \in W^{k+1,2}_{-(\tau+1)}(E_R)\quad\mbox{and}\quad g_{ij}-\left(1+\frac{2M}{r}\right)\delta_{ij}\in W^{k,2}_{-(\tau+1)}(E_R)
\end{equation}
for some constant $M\in\R$ with respect to wave harmonic coordinates\footnote{In \cite{KM}, the condition on the coordinates is formulated through harmonicity with respect to the conformally transformed metric $\gamma:=N^2 g$ which is equivalent to wave harmonicity, cf.~Lemma \ref{confharm} above.}.
\end{Thm}

\begin{Prop}\label{M=m}
Under the assumptions of Theorem \ref{thm:KM}, the constant $M$ in the above theorem equals $$M=\frac{m_{ADM}(E_{R},g)\,G}{c^2}.$$
\end{Prop}
\begin{Pf}
By definition of ADM-mass and by the asymptotics proven in Theorem \ref{thm:KM}, we find
\begin{eqnarray*}
m_{ADM}(E_{R},g)&=&\frac{c^2}{16\pi G}\lim_{r\to\infty}\int_{S_r^2}\left(g_{ii,j}-g_{ij,i}\right)\nu_\delta^j\,d\sigma_\delta\\
&=&\frac{c^2}{16\pi G}\lim_{r\to\infty}\int_{S_r^2}\left(\frac{2Mx_j}{r^3}\delta_{ii}-\frac{2Mx_i}{r^3}\delta_{ij}+\mathcal{O}(r^{-3})\right)\frac{x^j}{r}\,d\sigma_\delta\\
&=&\frac{c^2}{16\pi G}\lim_{r\to\infty}\left(16\pi M+\mathcal{O}(r^{-1})\right)=\frac{Mc^2}{G}.
\end{eqnarray*}\qed
\end{Pf}

Robert Geroch, Rolf Hansen, Prasun Kundu, Robert Beig and Walter Simon have also defined and studied multipole moments of asymptotically flat solutions to the vacuum static metric equations\footnote{Some of these results also continue to hold for stationary solutions to the vacuum Einstein equations.} using a conformal compactification of the asymptotically flat ends. Moreover, they asserted that these multipoles give rise to an analytic expansion of the lapse function and the $3$-metric at spatial infinity and discussed to what extent the moments determine the metric. Thomas B\"ackdahl \cite{Baeck} has considered the inverse problem of when specified moments lead to a solution of the equations. As the statement of these so-called Geroch-Hansen multipole moments is quite involved and as we will not recur to them, we refer the interested reader directly to the articles mentioned above.

\subsection*{Foliation by Levels of $N$}\label{subsec:foli}
Before we go on and study the center of mass, let us quickly state and prove a small lemma we will frequently make use of for both technical (uniqueness of photon spheres, cf.~Section \ref{sec:photon}) and physical (surfaces of equilibrium, cf.~Section \ref{sec:equi}) reasons. A more elaborate analysis of these surfaces will be presented at the end of the next section.
\begin{Lem}\label{lem:foli}
Let $(E^3,g,N,\rho,S)$ be a geometrostatic end with non-vanishing ADM-mass $m$. Then there exists a standardized compact interior $K\subset E^3$ such that $N$ foliates $E^3\setminus K$ with spherical level sets enclosing the support of the matter.
\end{Lem}
\begin{Pf}
Daniel Kennefick and Niall \'O Murchadha's theorem \ref{thm:KM} tells us that $$\frac{\partial N}{\partial x^i}-\frac{Mx_i}{r^3}\in W^{k,2}_{-\tau-2}(E^3)$$ holds in wave harmonic asymptotically flat coordinates. Therefore, by Corollary \ref{M=m} (i.~e.~by $M=mG/c^2$), $dN\neq0$ holds in a neighbourhood of infinity. By the implicit function theorem, $N$ thus locally foliates $E^3\setminus K$ for a suitable standardized compact interior $K$. The leaves of the foliation must be spherical as $N=1-M/r+\mathcal{O}(r^{-2})$ is radial up to second order again by Theorem \ref{thm:KM}.
\qed\end{Pf}

\section{The Asymptotic Center of Mass}\label{sec:CoMasym}
As we have just seen, Daniel Kennefick and Niall \'O Murchadha have used the static metric equations (and their pseudo-Newtonian equivalents which we are going to discuss in Chapter \ref{chap:mCoM}) to establish that both the Riemannian $3$-metric $g$ and the lapse function $N$ constituting a geometrostatic system can be split into a term proportional to $M/r$ and lower order terms in wave harmonic coordinates -- a fact they formulate in the language of weighted Sobolev spaces. We will now head in the same direction and expand both $N$ and $g$ one order further thereby relying on the faster fall-off trick Theorem \ref{higherreg}. Our result is summarized in Theorem \ref{thm:asym-CoM}.

The specific form the next order term displays, namely $M\vec{z}_{A}\cdot\vec{x}/r^3$ with $\vec{z}_{A}=(z_{A}^1,z_{A}^2,z_{A}^3)^t$ a vector in $\R^3$, suggests we can read off the center of mass of the system at this order of the expansion\footnote{cf.~Theorem \ref{thm:CoWu}, which, however, assumes $k\geq5$.}, though this will surely only be possible if $M\neq0$. We will see that the behavior of the moment $\vec{z}_{A}$ under Euclidean rigid body changes of coordinates endorses this interpretation and we will call $\vec{z}_{A}$ the ``asymptotic center of mass'' of the system. In fact, using the results presented, we will prove that $\vec{z}_{A}$ coincides with the ADM and the CMC centers of masses $\vec{z}_{ADM}$ and $\vec{z}_{CMC}$ of the system described in Chapter \ref{chap:iso}.

Recall that a $(k,\tau)$-geometrostatic end consists of a $3$-dimensional manifold $E^3$ diffeomorphic to an exterior domain $E_{R}\subset\R^3$, a $(k,2,\tau)$-asymptotically flat Riemannian metric $g$ on $E^3$, a lapse function $N:E^3\to\R^+$ normalized by $1$ at infinity, a mass density $\rho$, and a symmetric stress tensor $S$. Together, these fields satisfy the static metric equations \eqref{SME}. We begin by proving the following theorem. For convenience, we use the suggestive notation $f_{1}=f_{2}+W^{k,2}_{-\tau}(E^3)$ for $f_{1}-f_{2}\in W^{k,2}_{-\tau}(E^3)$.\index{sym}{$+W^{k,2}_{-\tau}$}
\begin{ThmDef}[Asymptotic Center of Mass]\label{thm:asym-CoM}
Let $\mathcal{S}:=(E^3,g,N,\rho,S)$ be a $(k,\tau)$-geometrostatic end with $\tau>1/2$, and and let $(x^{i})$ be a system of wave harmonic asymptotically flat coordinates in $E^3$. If the ADM-mass $m$ of $(E^3,g)$ is non-zero, there exists a unique vector $\vec{z}_{A}\in\R^3$\index{sym}{$\vec{z}_{A}$} such that
\begin{eqnarray}\label{eq:Nasym}
N-\my{m,S}\!{N}+\frac{M\vec{z}_{A}\cdot\vec{x}}{r^3}&\in&W^{k+1,2}_{-(\tau-\varepsilon+2)}(E^3)\\\label{eq:gasym}
g_{ij}-\myg{m,S}_{ij}-\frac{2M\vec{z}_{A}\cdot\vec{x}}{r^3}\,\delta_{ij}&\in&W^{k,2}_{-(\tau-\varepsilon+2)}(E^3)
\end{eqnarray}
for all $i,j=1,2,3$ and all $0<\varepsilon\leq\tau-1/2$. Here, $\my{m,S}\!N$ and $\myg{m,S}$ denotes the Schwarzschild lapse and metric of ADM-mass $m$ corresponding to the given system of coordinates and we have $M=mG/c^2$, as usual. We call $\vec{z}_{A}$ the {\bf asymptotic center of mass}\index{ind}{center of mass ! asymptotic} of the system.
\end{ThmDef}
\begin{Pf}
First of all, let us drop the label $m$ on the Schwarzschild notions for simplicity and let $\rho,S=0$ without loss of generality (shrinking the end to a standardized exterior domain if necessary). From Theorem \ref{thm:KM}, we know that
\begin{eqnarray}\label{eq:N}
N-\my{S}\!{N}=N-1+\frac{M}{r}&\in&W^{k+1,2}_{-(\tau+1)}(E^3)\\\label{eq:gij}
g_{ij}-\myg{S}_{ij}=g_{ij}-(1+\frac{2M}{r})\delta_{ij}&\in&W^{k,2}_{-(\tau+1)}(E^3).
\end{eqnarray}
Writing $\delta N:=N-\my{S}\!{N}$ and using the vacuum static metric equation for the Schwarzschild case, $\mylap{S}\my{S}\!{N}=0$, the vacuum static metric equations $\mylap{g}N=0$ can be rewritten as
\begin{eqnarray*}
0&=&\mylap{g}N=g^{ij}\left(N_{,ij}-\my{g}{\Gamma}^k_{ij}N_{,k}\right)=g^{ij}(\delta N_{ij}-\my{g}{\Gamma}^k_{ij}\delta N_{,k})+g^{ij}(\my{S}\!{N}_{,ij}-\my{g}{\Gamma}^k_{ij}\my{S}\!{N}_{,k})\\
&=&\mylap{g}\delta N+g^{ij}(\my{S}\!{N}_{,ij}-\my{S}{\Gamma}^k_{ij}\my{S}\!{N}_{,k})+g^{ij}(\my{S}{\Gamma}^k_{ij}-\my{g}{\Gamma}^k_{ij})\my{S}\!{N}_{,k}\\
&=&\mylap{g}\delta N+\cancelto{0}{\mylap{S}\my{S}\!{N}}+\underbrace{(g^{ij}-\inv{S}{ij})(\my{S}\!{N}_{,ij}-\my{S}{\Gamma}^k_{ij}\my{S}\!{N}_{,k})}_{\in W^{k-1,2}_{-(\tau-\varepsilon+4)}(E^3)\;\mbox{\scriptsize by Theorem }\ref{multi}}+g^{ij}(\my{S}{\Gamma}^k_{ij}-\my{g}{\Gamma}^k_{ij})\my{S}\!{N}_{,k}\\
&=&\mylap{g}\delta N+\underbrace{\inv{S}{ij}(\my{S}{\Gamma}^k_{ij}-\my{g}{\Gamma}^k_{ij})\my{S}\!{N}_{,k}+(g^{ij}-\inv{S}{ij})(\my{S}{\Gamma}^k_{ij}-\my{g}{\Gamma}^k_{ij})\my{S}\!{N}_{,k}}_{\in W^{k-1,2}_{-(\tau-\varepsilon+4)}(E^3)\;\mbox{\scriptsize by Theorem }\ref{multi}}+W^{k-1,2}_{-(\tau-\varepsilon+4)}(E^3)\\
&=&\mylap{g}\delta N+W^{k-1,2}_{-(\tau+4)}(E^3)=g^{ij}\left(\delta N_{,ij}-\my{g}{\Gamma}^k_{ij}\delta N_{,k}\right)+W^{k-1,2}_{-(\tau-\varepsilon+4)}(E^3)\\
&=&\inv{S}{ij}(\delta N_{,ij}-\my{g}{\Gamma}^k_{ij}\delta N_{,k})+(g^{ij}-\inv{S}{ij})(\delta N_{,ij}-\my{g}{\Gamma}^k_{ij}\delta N_{,k})+W^{k-1,2}_{-(\tau-\varepsilon+4)}(E^3)\\
&=&\mylap{S}\delta N+\underbrace{\inv{S}{ij}(\my{S}{\Gamma}^k_{ij}-\my{g}{\Gamma}^k_{ij})\delta N_{,k}+(g^{ij}-\inv{S}{ij})(\delta N_{,ij}-\my{S}{\Gamma}^k_{ij}\delta N_{,k})}_{\in W^{k-1,2}_{-(\tau-\varepsilon+4)}(E^3)\;\mbox{\scriptsize by Theorem }\ref{multi}}\\
&&\,\;\;\quad\quad\underbrace{+(g^{ij}-\inv{S}{ij})(\my{S}{\Gamma}^k_{ij}-\my{g}{\Gamma}^k_{ij})\delta N_{,k}}_{\in W^{k-1,2}_{-(\tau-\varepsilon+4)}(E^3)\;\mbox{\scriptsize by Theorem }\ref{multi}}+ W^{k-1,2}_{-(\tau-\varepsilon+4)}(E^3)\\
&=&\mylap{S}\delta N+W^{k-1,2}_{-(\tau-\varepsilon+4)}(E^3)
\end{eqnarray*}
In the above calculation, we have used the specific fall-off of the Schwarzschild lapse and metric \eqref{Nfall} and \eqref{gfall}, the a priori fall-off \eqref{eq:N} and \eqref{eq:gij} as well as Theorems \ref{Sobprop} and \ref{multi} in order to see that the respective terms lie in $W^{k-1,2}_{-(\tau-\varepsilon+4)}(E^3)$ (recall $\tau-\varepsilon\geq1/2$ so that $2(\tau-\varepsilon)\geq1$). The $\varepsilon$ arises from the condition $\varepsilon_{1}(=-(\tau+2))+\varepsilon_{2}(=-2)<\varepsilon(=-(\tau-\varepsilon+4))$ in the multiplication theorem \ref{multi}.

In other words, we have shown that $$\mylap{S}\delta N\in W^{k-1,2}_{-(\tau-\varepsilon+4)}(E^3).$$ By definition of the Schwarzschild metric $\myg{S}$, this is equivalent to $$\mylap{\delta}\delta N\in W^{k-1,2}_{-(\tau-\varepsilon+4)}(E^3).$$

The faster fall-off trick Theorem \ref{higherreg} now tells us that there is a harmonic polynomial $p$ of degree $d\leq\lceil\tau-\varepsilon\rceil$ such that $\delta N-\mathcal{K}\left[p\right]\in W^{k+1,2}_{-(\tau-\varepsilon+2)}(E^3)$. By Theorem \ref{thm:Bartnik}, we know that $\tau-\varepsilon>1$ implies $m=0$. Therefore, as $\tau-\varepsilon\geq1/2$ by definition of geometrostatic systems, we know that $d\leq1$. Moreover, our a priori knowledge ensures that $p$ has no constant term. $p$ must thus be a (possibly vanishing) homogeneous linear polynomial or in other words there must exist a vector $\vec{z}_{A}\in\R^3$ such that $$p(\vec{x})=-M\vec{z}_{A}\cdot\vec{x}$$ for all coordinate vectors $\vec{x}\in\R^3$. The choice of normalization will become clear in the course of this proof. The Kelvin transform of $p$ then reads $$\mathcal{K}\left[p\right](\vec{x})=-\frac{M\vec{z}_{A}\cdot\vec{x}}{r^3}$$ where as usual $r=\vert\vec{x}\vert$ so that we have proven Equation \eqref{eq:Nasym}.

Uniqueness of $\vec{z}_{A}$ directly follows from the fact that the difference between two terms of the form $M\vec{z}_{A}\cdot\vec{x}/r^3$ does not lie in $W^{k+1,2}_{-(\tau-\varepsilon+2)}(E^3)$ unless it vanishes. In particular, Equation \eqref{eq:Nasym} implies that
\begin{eqnarray}\label{eq:Ni}
N_{,i}&=&\my{S}\!{N}_{,i}-\frac{M(r^2 z_{i}-3\vec{z}\cdot\vec{x} x_{i})}{r^5}+W^{k,2}_{-(\tau-\varepsilon+3)}(E^3)\\\label{eq:Nij}
N_{,ij}&=&\my{S}\!{N}_{,ij}+\frac{3M(r^2 z_{i}x_{j}+r^2z_{j}x_{i}+\vec{z}\cdot\vec{x}\delta_{ij}-5\vec{z}\cdot\vec{x}x_{i}x_{j})}{r^7}+W^{k-1,2}_{-(\tau-\varepsilon+4)}(E^3),
\end{eqnarray}
where we have suppressed the label $A$ on $\vec{z}_{A}$ for abbreviation. Now write $$\delta g_{ij}:=g_{ij}-\myg{S}_{ij}-\frac{2M\vec{z}\cdot\vec{x}}{r^3}\delta_{ij}=:g_{ij}-\rho_{ij}.$$ Observe that $\rho_{ij}$ is an asymptotically flat metric, and that a priori $\delta g_{ij}\in W^{k,2}_{-(\tau+1)}(E^3)$ by Equation \eqref{eq:gij}. Using the multiplication theorem \ref{multi}, we compute from this that
\begin{eqnarray}\label{eq:inv}
g^{ij}&=&\rho^{ij}-\rho^{ik}\rho^{jl}\delta g_{kl}+W^{k,2}_{-(\tau-\varepsilon+2)}(E^3)\\\label{eq:gijk}
g_{ij,k}&=&\rho_{ij,k}+\delta g_{ij,k}\\\label{eq:Gamma}
\my{g}{\Gamma}^k_{ij}&=&\my{\rho}{\Gamma}^k_{ij}-\frac{1}{2}\left(\delta g_{\;i,j}^k+\delta g_{\;j,i}^k-{\delta g_{ij,}}^k\right)+W^{k-1,2}_{-(\tau-\varepsilon+3)}(E^3)\\\label{eq:Ricij}
\my{g}{\Ric}_{ij}&=&\my{\rho}{\Ric}_{ij}-\frac{1}{2}(\triangle(\delta g_{ij})+(\tr\delta g)_{,ij}-{\delta g_{ik,j}}^k-{\delta g_{jk,i}}^k)+W^{k-2,2}_{-(\tau-\varepsilon+4)}(E^3),
\end{eqnarray}
where we have raised indices, calculated traces and taken Laplacians with respect to the flat metric $\delta_{ij}$. Recalling the specific form of $\my{S}\!{N}$ and $\myg{S}$ discussed on pages \pageref{subsec:schwarz}ff, we obtain
\begin{eqnarray}\label{SNayam}
\my{S}\!{N}&=&1-\frac{M}{r}+\frac{M^2}{2r^2}+W^{k+1,2}_{-(\tau-\varepsilon+2)}(E^3)\\\label{SNi}
\my{S}\!{N}_{,i}&=&\frac{Mx_{i}}{r^3}-\frac{M^2x_{i}}{r^4}+W^{k,2}_{-(\tau-\varepsilon+3)}(E^3)\\\label{SNij}
\my{S}\!{N}_{,ij}&=&\frac{M(r^2\delta_{ij}-3x_{i}x_{j})}{r^5}-\frac{M^2(r^2\delta_{ij}-4x_{i}x_{j})}{r^6}+W^{k-1,2}_{-(\tau-\varepsilon+4)}(E^3)\\\label{Sgasym}
\myg{S}_{ij}&=&\left(1+\frac{2M}{r}+\frac{M^2}{r^2}\right)\delta_{ij}+\frac{M^2x_{i}x_{j}}{r^4}+W^{k,2}_{-(\tau-\varepsilon+3)}(E^3)\\\label{rhoij}
\rho_{ij}&=&\left(1+\frac{2M}{r}+\frac{M^2}{r^2}+\frac{2M\vec{z}\cdot\vec{x}}{r^3}\right)\delta_{ij}+\frac{M^2x_{i}x_{j}}{r^4}+W^{k,2}_{-(\tau-\varepsilon+3)}(E^3)\\\nonumber
\my{\rho}{\Gamma}^k_{ij}&=&-\left(\frac{M}{r^3}-\frac{M^2}{r^4}+\frac{3M\vec{z}\cdot\vec{x}}{r^5}\right)\left(x_{i}\delta^k_{j}+x_{j}\delta^k_{i}-x^k\delta_{ij}\right)\\\label{Gammarho}
&&+\frac{M^2x^k}{r^6}(r^2\delta_{ij}-2x_{i}x_{j})+\frac{M}{r^3}\left(z_{i}\delta^k_{j}+z_{j}\delta^k_{i}-z^k\delta_{ij}\right)\\\nonumber
&&+W^{k-1,2}_{{-(\tau-\varepsilon+3)}}(E^3)\\\nonumber
\my{\rho}{\Ric}_{ij}&=&\frac{M(r^2\delta_{ij}-3x_{i}x_{j})}{r^5}+\frac{M^2(3x_{i}x_{j}-r^2\delta_{ij})}{r^6}+\frac{3M(x_{i}z_{j}+x_{j}z_{i})}{r^5}\\\label{Ricrho}
&&+\frac{3M\vec{z}\cdot\vec{x}(r^2\delta_{ij}-5x_{i}x_{j})}{r^7}+W^{k-2,2}_{-(\tau-\varepsilon+4)}(E^3)
\end{eqnarray}
The remaining vacuum static metric equations $N\,\my{g}{\Ric}_{ij}=\my{g}{\nabla}^2_{ij}N$ then amount to
\begin{equation}\label{eq:hallo}
\triangle(\delta g_{ij})+(\tr\delta g)_{,ij}-{\delta g_{ik,j}}^k-{\delta g_{jk,i}}^k\in W^{k-2,2}_{-(\tau-\varepsilon+4)}(E^3)
\end{equation}
while the wave harmonic coordinate condition (cf.~Lemma \ref{waveharm}) and its first derivatives give us
\begin{eqnarray}\label{hallohallo}
2{\delta g_{jl,}}^{l}-\tr\delta g_{,j}&\in& W^{k-1,2}_{-(\tau-\varepsilon+3)}(E^3)\\\label{hel}
2{\delta g_{jl,i}}^{l}-\tr\delta g_{,ij}&\in& W^{k-2,2}_{-(\tau-\varepsilon+4)}(E^3)
\end{eqnarray}
which implies
\begin{equation}\label{eq:halloagain}
(\tr\delta g)_{,ij}-{\delta g_{ik,j}}^k-{\delta g_{jk,i}}^k\in W^{k-2,2}_{-(\tau-\varepsilon+4)}(E^3)
\end{equation}
for all $i,j=1,2,3$. Equations \eqref{hel} and \eqref{eq:halloagain} combine to
$\triangle(\delta g_{ij})\in W^{k-2,2}_{-(\tau-\varepsilon+4)}(E^3)$ so that the faster fall-off trick Theorem \ref{higherreg} gives us the desired $\delta g_{ij}\in W^{k,2}_{-(\tau-\varepsilon+2)}(E^3)$.
\qed\end{Pf}
Next, we prove that the asymptotic center of mass defined within the above theorem transforms adequately under changes of wave harmonic asymptotically flat coordinates.
\pagebreak
\begin{Prop}\label{prop:transfo}
Let  $\mathcal{S}:=(E^3,g,N,\rho,S)$ be a $(k,\tau)$-geometrostatic end with non-vanishing ADM-mass $m$ and $\tau>1/2$, and let $(x^{i})$, $(y^{i})$ be two systems of wave harmonic asymptotically flat coordinates\footnote{Observe that Niall \'O Murchadha and Robert Bartnik's result \ref{harmonic} ensures that two wave harmonic asymptotically flat systems of coordinates behave can only asymptotically differ by such a rigid motion.} in $E^3$ such that $y^{i}=O^{i}_{j}x^{j}+b^{i}$  for some orthogonal matrix $O\in O(\R^3)$, some vector $\vec{b}\in\R^3$, and all $i=1,2,3$. Then the centers of mass $\my{x}\!{\vec{z}_{A}}$ and $\my{y}\!{\vec{z}_{A}}$ with respect to the coordinates $(x^{i})$ and $(y^{i})$ satisfy $$\my{y}\!{z^{i}_{A}}=O^i_{j}\,\my{x}\!{z^{j}_{A}}+b^{i}.$$
\end{Prop}
\begin{Pf}
Let $\varepsilon$ be as in Theorem \ref{thm:asym-CoM}. Theorem \ref{thm:asym-CoM} holds true for $\tau-\varepsilon=:\overline{\tau}$. Dropping the bar for ease of notation, we have
\begin{eqnarray}\label{eq:Nr}
N-\left(1-\frac{M}{r}+\frac{M^2}{2r^2}-\frac{M\,\my{x}\!{\vec{z}\cdot\vec{x}}}{r^3}\right)\in W^{k+1,2}_{-(\tau+2)}(E^3)\mbox{ with respect to }(x^k)\\\label{eq:Ns}
N-\left(1-\frac{M}{s}+\frac{M^2}{2s^2}-\frac{M\,\my{y}\!{\vec{z}\cdot\vec{y}}}{s^3}\right)\in W^{k+1,2}_{-(\tau+2)}(E^3)\mbox{ with respect to }(y^k),
\end{eqnarray}
where $r:=\vert \vec{x}\vert$ and $s:=\vert \vec{y}\vert$, $M=mG/c^2$, and where we have again dropped the label $A$ in order to simplify notation. But by assumption $y^{i}=O^{i}_{j} x^j+b^j$ so that in particular
\begin{equation}\label{sr}
s-r\left(1+\frac{O\vec{x}\cdot\vec{b}}{r^2}+\frac{r^2\vert\vec{b}\vert^2-(O\vec{x}\cdot\vec{b})^2}{r^4}\right)\in W^{k+1,2}_{-(\tau+1)}(E^3)
\end{equation} with respect to $(x^k)$. This shows that the weighted Sobolev space with respect to the coordinates $(x^k)$ agree with the corresponding weighted Sobolev spaces with respect to the coordinates $(y^k)$. We can thus work in either of them.

By Equation \eqref{sr} and binomial expansion, we derive $$\frac{1}{s^p}-\frac{1}{r^p}\left(1-\frac{p\,O\vec{x}\cdot\vec{b}}{r^2}+\frac{2p^2(O\vec{x}\cdot\vec{b})^2-p\,r^2\vert\vec{b}\vert^2}{r^4}\right)\in W^{k+1,2}_{-(\tau+p+2)}(E^3)$$ for all $p\in\N$. Inserted into \eqref{eq:Ns}, this leads to
$$N-\left(1-\frac{M}{r}\left(1-\frac{O\vec{x}\cdot\vec{b}}{r^2}\right)+\left[\frac{M^2}{2r^2}-\frac{M\my{y}\!{\vec{z}}\cdot(O\vec{x}+\vec{b})}{r^3}\right]\right)\in W^{k+1,2}_{-(\tau+2)}(E^3)$$
which is equivalent to $$-\frac{M\,\my{x}\!{\vec{z}}\cdot\vec{x}}{r^3}-\left(\frac{M\,O\vec{x}\cdot\vec{b}}{r^3}-\frac{M\my{y}\!{\vec{z}}\cdot O\vec{x}}{r^3}\right)=-\frac{M(\my{x}\!{\vec{z}}+O^t\vec{b}-O^t\my{y}\!{\vec{z}})\cdot\vec{x}}{r^3}\in W^{k+1,2}_{-(\tau+2)}(E^3)$$
by \eqref{eq:Nr}. We therefore get that $\my{x}\!{\vec{z}}+O^t\vec{b}-O^t\my{y}\!{\vec{z}}=0$ or in other words that $\my{y}\!{\vec{z}}=O\,\my{x}\!{\vec{z}}+\vec{b}$.
\qed\end{Pf}
We are now going to prove the promised coincidence of the centers of mass defined in Section \ref{sec:HCoM}.
\begin{Thm}\label{thm:centers}
Let  $\mathcal{S}:=(E^3,g,N,\rho,S)$ be a $(k,\tau)$-geometrostatic end with ADM-mass $m\neq0$ and $\tau>1/2$, let $(x^{i})$ be a  wave harmonic asymptotically flat system of coordinates in $E^3$, and let $\vec{z}_{A},\vec{z}_{ADM}, \vec{z}_{CMC}, \vec{z}_{I}\in\R^3$ be the asymptotic, ADM, CMC, and intrinsic centers of mass of $g$ in these coordinates, respectively. Then
$\vec{z}_{A}=\vec{z}_{ADM}=\vec{z}_{CMC}=\vec{z}_{I}.$
\end{Thm}
\begin{Pf}
From Theorem \ref{def:Huang}, we know that both $\vec{z}_{ADM}$ and $\vec{z}_{CMC}$ agree with the intrinsic center of mass defined by Lan-Hsuan Huang as geometrostatic metrics automatically satisfy the Regge-Teiltelboim conditions \eqref{propdef:RT} as we can deduce from Theorem \ref{thm:KM}. To conclude the theorem, it thus suffices to show that $/vec{z}_{A}=\vec{z}_{ADM}$ which we will now do. To this end, we will use the well-known and easy to calculate fact that
\begin{equation}\label{S2fact}
\int_{S^2_{r}}\frac{x^{i}x^{j}}{r^4}d\sigma_{\delta}=\frac{4\pi}{3}\delta^{ij}.
\end{equation}

Inserting the expansion \eqref{eq:gij} into the definition of the ADM center of mass and using $M=mG/c^2$, the claimed result follows as follows from the definition of the asymptotic center of mass and the fall-off proven in Theorem \ref{thm:asym-CoM}: Let $\varepsilon$ be as in Theorem \ref{thm:asym-CoM} and set $\rho_{ij}:=g_{ij}-\myg{S}_{ij}-2M\vec{z}_{A}\cdot\vec{x}/r^3\in W^{k,2}_{-(\tau-\varepsilon+2)}(E^3)$. We find
\begin{eqnarray*}
z^k_{ADM}(g)&=&\frac{c^2}{16\pi mG}\lim_{r\to\infty}\int_{S^2_{r}}\sum_{i=1}^3\left[(g_{ij,i}-g_{ii,j})x^k\nu^j_{\delta}-(g_{ij}\delta^{jk}\nu^{i}_{\delta}-g_{ii}\nu^k_{\delta})\right]d\sigma_{\delta}\\
&=&\frac{1}{16\pi M}\lim_{r\to\infty}\int_{S^2_{r}}\sum_{i=1}^3\left[\left((\myg{S}+\frac{2M\vec{z}_{A}\cdot\vec{x}}{r^3}\delta+\rho)_{ij,i}\right.\right.\\
&&-\left.\left.(\myg{S}+\frac{2M\vec{z}_{A}\cdot\vec{x}}{r^3}\delta+\rho)_{ii,j}\right)x^k\nu^j_{\delta}\right.\\
&&-\left.\left((\myg{S}+\frac{2M\vec{z}_{A}\cdot\vec{x}}{r^3}\delta+\rho)_{ij}\delta^{jk}\nu^{i}_{\delta}-(\myg{S}+\frac{2M\vec{z}_{A}\cdot\vec{x}}{r^3}\delta+\rho)_{ii}\nu^k_{\delta}\right)\right]d\sigma_{\delta}\\
&\stackrel{\mbox{\scriptsize linearity}}{=}&\underbrace{z^k_{ADM}(\myg{S})}_{=0\,\mbox{\scriptsize by }\ref{prop:schwarzgeo}}+\frac{1}{8\pi}\lim_{r\to\infty}\int_{S^2_{r}}\sum_{i=1}^3\left[\left((\frac{\vec{z}_{A}\cdot\vec{x}}{r^3})_{,i}\delta_{ij}-(\frac{\vec{z}_{A}\cdot\vec{x}}{r^3}\delta)_{,j}\delta_{ii}\right)x^k\nu^j_{\delta}\right.\\
&&-\left.\left(\frac{\vec{z}_{A}\cdot\vec{x}}{r^3}\delta_{i}^{k}\nu^{i}_{\delta}-\frac{\vec{z}_{A}\cdot\vec{x}}{r^3}\delta_{ii}\nu^k_{\delta}\right)\right]d\sigma_{\delta}+\underbrace{``z^k_{ADM}(\rho)''}_{=0\,(\rho_{ij}\in W^{k,2}_{-(\tau-\varepsilon+2)}(E^3),\,\tau-\varepsilon>0)}\\
&\stackrel{\eqref{S2fact}}{=}&z^k_{A}(g).
\end{eqnarray*}
Note that we have put $z^k_{ADM}(\rho)$ in quotation marks as $\rho$ is not really an asymptotically flat metric with non-vanishing mass; however, the notation obviously suggests that we mean to apply the formula for $\vec{z}_{ADM}$ to $\rho$ (with respect to the mass of $g$).
\qed\end{Pf}

Alternatively, one could use Lan-Hsuan Huang's formulas \ref{prop:Huang-formula} or \ref{def:Huang} in order to obtain the above theorem from the asymptotics proven in Theorem \ref{thm:asym-CoM}. The assumption $\tau>1/2$ which was stipulated throughout this section can actually be relaxed with the aid of Theorem \ref{thm:KM}.
\newpage
\subsection*{Levels of $N$ as Graphs over Round Spheres around $\vec{z}$}\label{subsec:graph}
Recall that $N$ foliates a neighbourhood of infinity in each end of a geometrostatic system with non-vanishing ADM-mass (cf.~Lemma \ref{lem:foli}). Having gained a better understanding of the asymptotics of $N$ and $g$, this information can now be made quantitative by saying that its levels are in fact graphs over round (coordinate) spheres close to infinity. For the remainder of this subsection, let us relax our notation to the more suggestive $\mathcal{O}$-notation for simplicity.
\begin{Coro}[Levels of $N$ as Graphs over Round Spheres]\label{coro:graph}
Let $(E^3,g,N,\rho,S)$ be a geometrostatic end with positive ADM-mass $m$ and let $(x^i)$ be an asymptotically flat wave harmonic system of coordinates for $E^3$ as provided by Theorem \ref{harmonic}. Then each level set $\Sigma$ of $N$ outside some compact interior $K$ can be written as a graph
\begin{eqnarray*}
F:S^2_R(\vec{z})\subset\R^3\to\R^3&:&\vec{x}\mapsto \vec{x}+f(\vec{x})\frac{\vec{x}-\vec{z}}{\vert \vec{x}-\vec{z}\vert},\\
x^i(\Sigma)&=&F^i(S^2_R(\vec{z}))
\end{eqnarray*}
over suitable coordinate spheres $S^2_R(\vec{z})$ with $\vec{z}\in\R^3$ the center of mass of the system and $f:S^2_R(\vec{z})\to\R$ a smooth function. It furthermore holds that
\begin{eqnarray*}
f&=&\mathcal{O}(r^{-1}),\\
\vert df\vert_\delta&=&\mathcal{O}(1),\\
R&=&r+\mathcal{O}(r^{-1}),\\
H&=&\frac{2}{r}-\frac{4M}{r^2}+\frac{2\vec{z}\cdot\vec{x}}{r^3}+\mathcal{O}(r^{-3}),\\
\frac{\partial N}{\partial\nu}&=&\frac{M}{r^2}-\frac{2M^2}{r^3}+\frac{2M\vec{z}\cdot\vec{x}}{r^4}+\mathcal{O}(r^{-4}),\\
\frac{1}{\vert\Sigma\vert_\delta}\int_\Sigma x^i\,d\sigma_\delta&=&z^i+\mathcal{O}(r^{-1})
\end{eqnarray*}
as $r\to\infty$ where again $M=mG/c^2$.
\end{Coro}
\begin{Pf}
By Proposition \ref{prop:transfo}, we can assume that $\vec{z}=0$ without loss of generality by translating the coordinates (observing that $\lvert\vec{x}-\vec{z}\rvert=r+\mathcal{O}(1)$ and $\lvert\Sigma\rvert_{\delta}^{-1}\int_{\Sigma}z^{i}\,d\sigma_{\delta}=z^{i}$). From Theorem \ref{thm:asym-CoM}, we then know that $N=\my{S}{N}+\mathcal{O}(r^{-3})$ and $g_{ij}=\myg{S}_{ij}+\mathcal{O}(r^{-3})$, where the Schwarzschild notions refer to the mass $m$ of the end $E^3$. Using spherical symmetry of the Schwarzschild solutions and the fact that the radial derivative of $\my{S}\!{N}$ is strictly positive in a neighborhood of infinity, this fall-off ensures that for any given level set $\overline{\Sigma}$ with $N\equiv\overline{N}$ which lies suitably close to infinity there is a radius $\overline{R}$ such that $\my{S}\!{N}(\overline{R})=\overline{N}$ with a slight abuse of notation. This, the above fall-off behavior, the mean value theorem of calculus, techniques related to tubular neighborhoods and the exponential map of Riemannian geometry, well-known explicit formulae for the mean curvature of a level set as well as the transformation formula for integrals are the main ingredients for the proof of the corollary which is very technical but straightforward and which we will thus spare the reader.
\qed\end{Pf}
We will apply this corollary when treating photon spheres in Section \ref{sec:photon} but only in a lower order version ($H=2/r+\mathcal{O}(r^{-2})$ and $\partial N/\partial\nu=M/r^2+\mathcal{O}(r^{-3})$). 

\section{Asymptotic Uniqueness in Geometrostatics}\label{sec:asym}
The classic result of Theorem \ref{thm:KM} states that static solutions to the vacuum Einstein equations possess very specific fall-off properties -- a result which we have improved in the last section by extracting a higher order term in the ``expansion at infinity''. Our proof relied on methods including the theory of harmonic polynomials, the Kelvin transform\index{ind}{Kelvin transform}, and the faster fall-off trick Theorem \ref{higherreg}. In this section, we will extend these techniques and use them to prove two uniqueness results for geometrostatic systems.

The first theorem we will prove states that the lapse function $N$ of a geometrostatic system $\mathcal{S}:=(M^3,g,N,\rho,S)$ is unique given all other data. We will give an analytic proof of this theorem, here, and a geometric and/or physical proof in Chapter \ref{chap:geo}. The second theorem is in some sense complementary; it states that the metric is unique when all other data is given and a system of wave-harmonic coordinates is specified. Fixing these coordinates trivializes the isometry relating two equivalent geometrostatic systems and therefore simplifies concepts and notations.

We will make use of the theory of harmonic homogeneous polynomials $H_{lm}=r^l Y_{lm}$, $l=0,1,2,\dots$, $m=-l,\dots,l$ (where the $Y_{lm}$ denote spherical harmonics) and of the Kelvin transform which we have already met in Section \ref{sec:sobo}. We refer the interested reader to \cite{AxBourRam} for an introduction into this field.\index{sym}{$Y_{lm}$}\index{sym}{$H_{lm}$} Within this section, we say that a function $f:M^3\to\R$ on a manifold $M^3$ with asymptotically flat coordinates $(x^k)$ is {\it analytic at infinity with respect to the coordinates $(x^k)$}\index{ind}{analytic at infinity} if the Kelvin transform $\mathcal{K}[f\circ\Phi^{-1}]$ is analytic at the origin. Here, $\Phi$ denotes the diffeomorphism corresponding to the coordinates $(x^k)$.

\begin{Thm}[Uniqueness of $N$]\label{Nuniqueasym}
Let $\mathcal{S}:=(M^3,g,N,\rho,S)$ and $\widetilde{\mathcal{S}}:=(M^3,g,\widetilde{N},\rho,S)$ be geometrostatic systems. Assume that both $N$ and $\widetilde{N}$ are analytic at infinity with respect to a system of wave harmonic and asymptotically flat coordinates $(x^{i})$ outside some compact $K\subset M^3$ containing the support of the matter. Then $N=\widetilde{N}$ in $M^3\setminus K$. If, in addition, $(M^3,g)$ is geodesically complete and $\inter K$ is diffeomorphic to a bounded domain in $\R^3$ having smooth boundary\footnote{We need these conditions on $K$ in order to deduce that elliptic Dirichlet boundary value problems are uniquely solvable in $\inter K$.}, then $N=\widetilde{N}$ holds in all of $M^3$.
\end{Thm}
\begin{Rem}
Analyticity at infinity of both $N$ and $g_{ij}$ is a property geometrostatic systems are generally expected\footnote{cf.~\cite{KM} and several remarks in the series of papers by Robert Beig and Walter Simon \cite{Beig2,BS2,BS3}.} to possess in wave harmonic coordinates when the support of the matter is bounded away from infinity. Note that wave harmonic coordinates are natural candidates for analyticity statements by Theorem \ref{MzH}. We remark that this proof works whether or not the ADM-masses of the systems vanish -- other than the geometric proof of the same fact which we will present in Chapter \ref{chap:geo}. 
\end{Rem}
\begin{Pf}
By Theorem \ref{thm:KM}, we know that $N,\widetilde{N}=1-M/r+\mathcal{O}(r^{-2})$ as $r\to\infty$ in each end $E_{R}$ of $M^3$, where $M=m_{ADM}(E_{R})G/c^2$ by Proposition \ref{M=m}. This implies $N-\widetilde{N}=\mathcal{O}(r^{-2})$. We will show by induction that in fact $$N-\widetilde{N}=\mathcal{O}(r^{-2-n})\mbox{ for all }n\in\N_{0}$$ as $r\to\infty$. This then allows to give an argument based on the assumed analyticity at infinity and on ellipticity of the curvilinear Poisson equation to show that this indeed gives us $N=\widetilde{N}$.

So assume from now on that $N-\widetilde{N}=\mathcal{O}(r^{-2-n^\ast})$ for some $0\leq n^\ast$, set $l^\ast:=n^\ast+1$, and work in a fixed but arbitrary end of $M^3\setminus K$. The vacuum static metric equations \eqref{SMEvac} in the given end for both $N$ and $\widetilde{N}$ imply
\begin{eqnarray}\label{eq:a}
\mylap{g}\delta N&=&0\\\label{eq:b}
\delta N\,\my{g}{\Ric}&=&\my{g}{\nabla}^2\delta N
\end{eqnarray}
where $\delta N:=N-\widetilde{N}$. Inserting the given asymptotics $\delta N=\mathcal{O}(r^{-(l^\ast+1)})$ into the first one of these equations, we obtain
\begin{equation}\label{eq:acirc}
\mylap{\delta}(\delta N)=\mathcal{O}(r^{-(l^\ast+4)})
\end{equation}
as $r\to\infty$ in the given end. By the faster fall-off trick Theorem \ref{higherreg}, there must exist a harmonic polynomial $p$ of degree $d\leq l^\ast$ such that $\delta N=\mathcal{K}\left[p\right]+\mathcal{O}(r^{-(l^\ast+2)})$, where $\mathcal{K}\left[p\right]$ denotes the Kelvin transform\index{ind}{Kelvin transform} of $p$, cf.~page \pageref{subsec:faster}. It therefore suffices to show that $\mathcal{K}[p]=0$ or, equivalently, that  $p=0$. By harmonicity, we can expand $p$ into the canonical harmonic homogeneous polynomials $\{H_{lm}\}_{m=-l,\dots,l}$\index{sym}{$H_{lm}$} which form a basis for the space of harmonic polynomials. This means we can write $$p=\sum_{l=0}^{d}\sum_{m=-l}^l\alpha^{lm}H_{lm}\quad\mbox{and}\quad\mathcal{K}\left[p\right]=\sum_{l=0}^{d}\sum_{m=-l}^l\frac{\alpha^{lm}H_{lm}}{r^{2l+1}}$$ for some constants $\alpha^{lm}\in\R$ by definition of the Kelvin transform and homogeneity of the polynomials. It is easy to see that the induction hypothesis in fact ensures that only basis polynomials of degree $l^\ast$ appear in the linear combination. Thus, the above expressions simplify to $$p=\alpha^{m}H_{lm}\quad\mbox{and}\quad\mathcal{K}\left[p\right]=\frac{\alpha^{m}H_{lm}}{r^{2l+1}},$$ where we have dropped the asterisk and the index $l$ on the constants for convenience. Moreover, we have abused Einstein's summation convention by implicitly summing over pairs of one lower and one upper index $m$. Equation \eqref{eq:b} can then be rephrased into
\begin{equation}\label{eq:bcirc}
\alpha^m\left(r^4H_{lm,ij}+(2l+1)\left[((2l+3)x_{i}x_{j}-r^2\delta_{ij})H_{lm}-r^2(H_{lm,i}x_{j}+H_{lm,j}x_{i})\right]\right)=0
\end{equation}
for all $i,j=1,2,3$, where we have dropped the lower order terms as analytic expansions of identical functions must coincide at each order, individually. Before we go on, let us recall that homogeneity of the basis polynomials implies $H_{lm,k}x^k=lH_{lm}$ and thus by induction
\begin{equation}\label{Htrick}
H_{lm,i_{1}\dots i_{p}k}x^k=(l-p)H_{lm,i_{1}\dots i_{p}}\mbox{ for all }p\in\N.
\end{equation}
Now multiply Equation \eqref{bcirc} by $x^{i}x^j$ and obtain
$$0=\left(l(l-1)+(2l+1)\left[2(l+1)-2l\right]\right)\alpha^mH_{lm}=(l^2+l+2)\alpha^mH_{lm}.$$
In consequence, we find $\alpha^mH_{lm}=0$. By linear independence of the $H_{lm}$, we see that $\alpha^m=0$ must hold for all $m=-l,\dots,l$. By construction, this implies $p=\mathcal{K}\left[p\right]=0$ and thus the desired fall-off $\delta N=\mathcal{O}(r^{-(l^\ast+2)})$.

This fall-off translates into the fall-off $\mathcal{K}[\delta N]=\mathcal{O}(r^{l^\ast+1})$. Analyticity of $\delta N$ now allows us to deduce that $\mathcal{K}[\delta N]=0$ and thus $N=\widetilde{N}$ in $M^3\setminus K$. In case $(M^3,g)$ is geodesically complete and $K$ satisfies the regularity conditions stated in the theorem, (linear) ellipticity of the static metric Equation \eqref{lapN} in wave harmonic asymptotically flat coordinates and $N=\widetilde{N}$ on $\partial K$ ensure uniqueness on the entire manifold $M^3$ by classical PDE theory\footnote{cf.~e.~g.~theorem 6.14 on p.~107 in \cite{GT}.}.
\qed\end{Pf}

Before we go on to formulate and prove uniqueness of the metric, let us remark that although we make use of ellipticity of the static metric equations for both theorems to extend uniqueness into the interior, uniqueness in a neighborhood of infinity is actually proven without recurrence to a Dirichlet boundary problem; in fact, we do not prescribe any data at the inner boundaries of the asymptotically flat ends but obtain uniqueness directly from the structure of the equations combined with asymptotic behavior. This will become even more clear in the geometric proof of the uniqueness of $N$ which we will present in Chapter \ref{chap:geo}. Back to the analytic approach, let us continue by stating the second uniqueness theorem.

\begin{Thm}[Uniqueness of $g$]\label{thm:gunique}
Let $\mathcal{S}:=(M^3,g,N,\rho,S)$, $\widetilde{\mathcal{S}}:=(M^3,\widetilde{g},N,\rho,S)$ be geometrostatic systems with a common wave harmonic and asymptotically flat system of coordinates $(x^i)$ outside some compact $K\subset M^3$ containing the support of the matter. Assume furthermore that both $g$ and $\widetilde{g}$ are analytic at infinity with respect to these coordinates outside $K$. Then $g_{ij}=\widetilde{g}_{ij}$ in all of $M^3\setminus K$ (possibly) unless the ADM-masses of both metrics vanish. If, in addition, $g$ and $\widetilde{g}$ are geodesically complete and the coordinates $(x^{i})$ can be extended into $K$ as a global wave harmonic system of coordinates and if $\inter K$ is diffeomorphic to a bounded domain in $\R^3$ having smooth boundary\footnote{Again, we only need these conditions on $K$ in order to deduce that elliptic Dirichlet boundary value problems in $K$ are uniquely solvable.}, then $g_{ij}=\widetilde{g}_{ij}$ holds in all of $M^3$ whether or not the ADM-mass vanishes.
\end{Thm}
\begin{Rem}
The formulation that $\mathcal{S}$ and $\widetilde{\mathcal{S}}$ should have a common wave harmonic and asymptotically flat system of coordinates $(x^i)$ outside $K$, we mean that these coordinates make both systems asymptotically flat and that they satisfy $\my{ds^2}{\square}x^{i}=0=\my{\widetilde{ds^2}}{\square}x^{i}$.
\end{Rem}
\begin{Pf}
We first show that the claim holds in the case where $m:=m_{ADM}(g)=0$. By Theorems \ref{thm:KM} and \ref{M=m}, we obtain $N=1+\mathcal{O}(r^{-2})$ and thus also $m_{ADM}(\widetilde{g})=0$. By rigidity of the positive mass theorem \ref{positive mass thm} and geodesic completeness, both $(M^3,g)$ and $(M^3,\widetilde{g})$ must be isometric to $(\R^3,\delta)$, where $\delta$ denotes the Euclidean metric on $\R^3$. Here, $\Psi:(M^3,g)\to(\R^3,\delta)$ and $\widetilde{\Psi}:(M^3,\widetilde{g})\to(\R^3,\delta)$ denote the respective isometries. Moreover, as we will see in theorem\footnote{Note that Theorem \ref{thm:spm} does not logically rely on the theorem under consideration and can thus be applied, here.} \ref{thm:spm}, the lapse function must satisfy $N=1$ and therefore $\Box f=\triangle f$ and $\widetilde{\Box}f=\widetilde{\triangle}f$ for all $f:M^3\to\R$ in all of $M^3$, where the $\Box$ and $\widetilde{\Box}$ operators correspond to the associated static Lorentzian metrics $ds^2$ and $\widetilde{ds^2}$, respectively. The wave harmonic coordinates $(x^{i})$ are thus in fact harmonic coordinates with respect to both $g$ and $\widetilde{g}$ by Lemma \ref{waveharm}. The coordinate functions $y^{i}:=x^{i}\circ\Psi^{-1}:\R^3\to\R$, $\widetilde{y}^{i}:=x^{i}\circ\widetilde{\Psi}^{-1}:\R^3\to\R$ must thus be harmonic and asymptotic to each other in the Euclidean space $(\R^3,\delta)$. This implies that there must exist a translation $\vec{b}\in\R^3$ and a rotation $O\in O(\R^3)$ such that $\widetilde{y}^{i}=O^{i}_{j}y^{j}+b^{i}$ -- in complete analogy to the uniqueness statement in Theorem \ref{harmonic}. This gives us $$\widetilde{g}_{ij}=\widetilde{g}(\partial_{x^{i}},\partial_{x^{j}})=\delta(\partial_{\widetilde{y}^{i}},\partial_{\widetilde{y}^{j}})=\delta(O^{-1}\partial_{y^{i}},O^{-1}\partial_{y^{j}})=\delta(\partial_{y^{i}},\partial_{y^{j}})=g(\partial_{x^{i}},\partial_{x^{j}})=g_{ij}$$ in the case of vanishing ADM-mass.

Now assume $m>0$ for the rest of this proof. Observe that by \ref{thm:KM} and \ref{M=m}, $g$ and $\widetilde{g}$ have the same ADM-mass $m$ just as in the case where $m=0$, and thus $g_{ij}-\widetilde{g}_{ij}=\mathcal{O}(r^{-2})$ in every end of $M^3$. We now intend to argue by induction that in fact $$g_{ij}-\widetilde{g}_{ij}=\mathcal{O}(r^{-2-n})\mbox{ for all }n\in\N_{0}$$ as $r\to\infty$. Combined with analyticity at infinity of both $g_{ij}$ and $\widetilde{g}_{ij}$ and ellipticity of the static metric equations, this will give the desired result as we will see at the end of this proof.

So assume for what follows that the induction hypothesis is true for some $0\leq n^\ast\in\N$ and work in an arbitrary but fixed component of $M^3\setminus K$. By induction hypothesis, we know that the difference $g_{ij}-\widetilde{g}_{ij}$ must a priori fall of as $g_{ij}-\widetilde{g}_{ij}=\mathcal{O}(r^{-2-n^\ast})$. We set $l^\ast:=n^\ast+1$ and $\delta g_{ij}:=g_{ij}-\widetilde{g}_{ij}$. Then $\delta g_{ij}\in\mathcal{O}(r^{-(l^\ast+1)})$. We can thus compute the quantities listed below in the common wave harmonic coordinates given by assumption and using Lemma \ref{waveharm}. We obtain
\begin{eqnarray}\label{(i)}
\widetilde{g}^{ij}&=&g^{ij}+\delta^{ik}\delta^{jp}\delta g_{kp}+\mathcal{O}(r^{-(l^\ast+2)})\\\label{(ii)}
\widetilde{\Gamma}^k_{ij}&=&\Gamma_{ij}^k-\frac{1}{2}\delta^{kp}\left(\delta g_{ip,j}+\delta g_{jp,i}-\delta g_{ij,p}\right)+\mathcal{O}(r^{-(l^\ast+3)})\\\label{(iii)}
\widetilde{\Ric}_{ij}&=&\Ric_{ij}+\frac{1}{2}\delta^{kp}\left(\delta g_{kp,ij}+\delta g_{ij,kp}-\delta g_{ip,jk}-\delta g_{jk,ip}\right)+\mathcal{O}(r^{-(l^\ast+4)})\\\label{(iv)}
\widetilde{\Box}x^{k}&=&\Box x^{k}+\frac{1}{2}\delta^{ij}\delta^{kp}\left(2\delta g_{ip,j}-\delta g_{ij,p}\right)+\mathcal{O}(r^{-(l^\ast+3)})
\end{eqnarray}
where the tilde is used suggestively to distinguish the geometric quantities referring to $\widetilde{g}$ and $g$, respectively. Working in a vacuum exterior region now gives us the static metric and wave harmonic coordinates equations (cf.~Lemma \ref{waveharm} and Equations \eqref{SMEvac})
\begin{eqnarray}\label{(1)}
\widetilde{\triangle}N&=&\triangle N\\\label{(2)}
N\widetilde{\Ric}_{ij}=\widetilde{\nabla}^2_{ij} N\quad&\mbox{and}&\quad N\Ric_{ij}=\nabla^2_{ij} N\\\label{(3)}
\widetilde{\Box}x^k&=&\Box x^k.
\end{eqnarray}
Using our asymptotic considerations \eqref{(i)} through \eqref{(iv)}, these transform into
\begin{eqnarray}\label{acirc}
2{\delta g_{ij,}}^j x^{i}-(\tr\delta g)_{,i} x^{i}+2\tr\delta g-6\delta g_{ij} x^{i}x^j r^{-2}&=&\mathcal{O}(r^{-(l^\ast+2)})\\\label{ccirc}
(\tr\delta g)_{,ij}+\mylap{\delta}(\delta g_{ij})-{\delta g_{ik,j}}^k-{\delta g_{jk,i}}^k&=&\mathcal{O}(r^{-(l^\ast+4)})\\\label{bcirc}
2{\delta g_{ij,}}^j-(\tr\delta g)_{,i}&=&\mathcal{O}(r^{-(l^\ast+3)})
\end{eqnarray}
where we have used the Kennefick-\'O Murchadha result $N=1-\frac{M}{r}+\mathcal{O}(r^{-2})$ from \ref{thm:KM} and our above assumption $M\neq0$ (and where indices have been raised and lowered and traces are taken with respect to the flat background metric $\delta$). Contracting \eqref{bcirc} by $x^{i}$, inserting the result into \eqref{acirc}, and rewriting \eqref{ccirc} using \eqref{bcirc} leads to
\begin{eqnarray}\label{dcirc}
\tr\delta g-3\delta g_{ij}x^{i}x^jr^{-2}&=&\mathcal{O}(r^{-(l^\ast+2)})\\\label{ecirc}
\mylap{\delta}(\delta g_{ij})&=&\mathcal{O}(r^{-(l^\ast+4)}),
\end{eqnarray}
respectively. By the faster fall-off trick \ref{higherreg}, the latter of these implies that $\delta g_{ij}$ can be rewritten as $\delta g_{ij}=\mathcal{K}\left[p_{ij}\right]+\mathcal{O}(r^{-(l^\ast+2)})$ (with at least one derivative) for some harmonic polynomials $p_{ij}$ of degrees $d_{ij}\leq l^\ast$ and satisfying $p_{ij}=p_{ji}$. Thus, we find that
\begin{equation}\label{fall-g}
g_{ij}-\widetilde{g}_{ij}-\mathcal{K}\left[p_{ij}\right]=\mathcal{O}(r^{-(l^\ast+2)})
\end{equation}
(with at least one derivative). To finish the induction step, we need to show that $\mathcal{K}[p_{ij}]=0$ for all $i,j=1,2,3$. We do this now.

As in the proof of Theorem \ref{Nuniqueasym}, we can write $p_{ij}$ as a linear combination of the canonical basis $\{H_{lm}\}$ of harmonic homogeneous polynomials of degree $l$\index{sym}{$H_{lm}$} so that $$p_{ij}=\sum_{l=0}^{d_{ij}}\sum_{m=-l}^l\alpha_{ij}^{lm}H_{lm}\quad\mbox{and thus}\quad\mathcal{K}\left[p_{ij}\right]=\sum_{l=0}^{d_{ij}} \sum_{m=-l}^l\alpha_{ij}^{lm} \frac{H_{lm}}{r^{2l+1}},$$ for constants $\alpha_{ij}^{lm}\in\R$ with $\alpha_{ij}^{lm}=\alpha_{ji}^{lm}$ by definition of the Kelvin transform \ref{def:Kel}. By our induction hypothesis, however, $\alpha_{ij}^{lm}=0$ for all $l<\lceil \tau-1+n^\ast\rceil=n^\ast-2=l^\ast$, all associated $m$, and all indices $i,j=1,2,3$, and thus the linear combination reduces to 
\begin{equation}\label{K(p)}
\mathcal{K}\left[p_{ij}\right]=\sum_{m=-l^\ast}^{l^\ast}\alpha_{ij}^{m} \frac{H_{l^\ast m}}{r^{2l^\ast+1}}
\end{equation}
where we have dropped the index $l^\ast$ on the constants $\alpha$ in order to simplify notation. Similarly, we will also drop the asterisk in what follows. Moreover, we will abuse Einstein's summation convention as in the proof of Theorem \ref{Nuniqueasym}. In the same spirit as there, Equations \eqref{bcirc} and \eqref{dcirc} can be converted into the equations
\begin{eqnarray}\label{alpha}
2r^2\alpha_{ij}^m{H_{lm,}}^i-r^2\alpha^m H_{lm,j}&=&(2l+1)\left[2\alpha_{ij}^mH_{lm}x^{i}-\alpha^mH_{lm}x_{j}\right]\\\label{beta}
3\alpha_{ij}^mH_{lm}x^ix^j&=&r^2\alpha^mH_{lm}
\end{eqnarray}
using \eqref{fall-g} and \eqref{K(p)}. In these equations, indices were again raised and lowered by $\delta^{ij}$ and $\delta_{ij}$, respectively, and $\alpha^m:=\alpha_{ij}^m\delta^{ij}$ abbreviates the trace of the symmetric tensor $(\alpha_{ij}^m)_{i,j=1,2,3}$. Multiplying Equation \eqref{alpha} by $x^j$ then gives
$$2r^2\alpha_{ij}^m{H_{lm,}}^ix^j=2(2l+1)\alpha_{ij}^mH_{lm}x^{i}x^j-(l+1)r^2\alpha^mH_{lm}$$
which combines with \eqref{beta} to the useful equation
\begin{equation}\label{delta}
6\alpha_{ij}^m{H_{lm,}}^ix^j=(l-1)\alpha^mH_{lm}.
\end{equation}
We now take the Laplacian of Equation \eqref{beta} and calculate with the aid of \eqref{Htrick} that $$3\alpha_{ij}^m {H_{lm,}}^{i}x^j=l\alpha^m H_{lm}$$ so that, using Equation \eqref{delta} as well as linear independence of $\{H_{lm}\}_{m=-l,\dots,l}$, we obtain $\alpha^m=0$ for all $m=-l,\dots,l$. We can thus simplify equations \eqref{alpha} and \eqref{beta} to
\begin{eqnarray}\label{alphatil}
r^2\alpha_{ij}^m{H_{lm,}}^i&=&(2l+1)\alpha_{ij}^mH_{lm}x^{i}\\\label{betatil}
\alpha_{ij}^mH_{lm}x^{i}x^{j}&=&0.
\end{eqnarray}
Let us now abbreviate $a_{ij}:=\alpha_{ij}^mH_{lm}$ so that $a_{ij}$ is a a symmetric and trace-free tensor field on $\R^3$ with entries that are harmonic homogeneous polynomials of degree $l$. In the new notation, \eqref{alphatil} and \eqref{betatil} read
\begin{eqnarray}\label{1circ}
r^2 {a_{ij,}}^{i}&=&(2l+1)a_{ij} x^{i}\\\label{2circ}
a_{ij}x^{i}x^j&=&0
\end{eqnarray}
for all $j=1,2,3$ and all $x\in\R^3$. These equations imply $a_{ij}=0$ by a symmetrization technique which we will now describe. First of all, the above two equations have the simple consequences that
\begin{eqnarray}\label{3circ}
{a_{ij,}}^{i}x^j&=&0\\\label{4circ}
{a_{ij,k}}^{i}x^j+{a_{ik,}}^{i}&=&0\\\label{5circ}
{a_{ij,kp}}^{i}x^j+{a_{ik,p}}^{i}+{a_{ip,k}}^{i}&=&0
\end{eqnarray}
for all $j,k,p=1,2,3$ and all $x\in\R^3$ which arise through multiplying \eqref{1circ} by $x^j$ and using \eqref{2circ} as well as taking derivatives of the result in directions $\partial_{k}$ and $\partial_{p}$, subsequently. Taking the $k$-th derivative of both \eqref{1circ} and \eqref{2circ}, we obtain
\begin{eqnarray}\label{6circ}
2{a_{ij,}}^{i}x_{k}+r^2{a_{ij,k}}^{i}&=&(2l+1)\left(a_{ij,k}x^{i}+a_{jk}\right)\\\label{7circ}
a_{ij,k}x^{i}x^j+2a_{ik}x^{i}&=&0
\end{eqnarray}
for all $j,k=1,2,3$ and all $x\in\R^3$. Antisymmetrization of \eqref{6circ} w.~r.~t.~$j$ and $k$ implies
\begin{equation}\label{8circ}
2\left({a_{ij,}}^{i}x_{k}-{a_{ik,}}^{i}x_{j}\right)+r^2\left({a_{ij,k}}^{i}-{a_{ik,j}}^{i}\right)=(2l+1)\left(a_{ij,k}-a_{jk,i}\right)x^{i}
\end{equation}
which we derivate in direction $\partial_{p}$ and afterwards multiply by $x^j$ to obtain
\begin{equation}\label{9circ}
a_{ij,kp}x^{i}x^j=-2\left(a_{pi,k}x^{i}+a_{ki,p}x^{i}-(l-1)a_{kp}\right)
\end{equation}
where we have used \eqref{1circ} through \eqref{6circ}. Multiplying \eqref{9circ} by $x^p$ and using again \eqref{1circ} as well as \eqref{7circ}, this leads to $(2l+3)a_{ik}x^{i}=0$ so that
\begin{eqnarray}\label{10circ}
a_{ij}x^{i}&=&0\\\label{11circ}
{a_{ij,}}^{i}&=&0
\end{eqnarray}
hold for all $j=1,2,3$ and all $x\in\R^3$, where we have again made use of \eqref{1circ}. In other words, we have ``taken the gradient''  of equations \eqref{1circ} and \eqref{2circ}. To finish proving $a_{ij}=0$, observe that the $k$-th derivative of Equation \eqref{10circ}, the $p$-th derivative of the result multiplied by $x^j$, and a subsequent use of \eqref{9circ} imply that
\begin{eqnarray}\label{12circ}
a_{ij,k}x^{i}+a_{jk}&=&0\\\label{13circ}
a_{ij,kp}x^{i}x^j&=&-\left(a_{pj,k}+a_{jk,p}\right)x^j\stackrel{\eqref{12circ}}{=}2a_{pk}
\end{eqnarray}
Together with \eqref{9circ}, this gives $a_{ij}=0$ for all $i,j=1,2,3$ and all $x\in\R^3$ which in turn -- by definition of $a_{ij}$ and again linear independence of $\{H_{lm}\}_{m=-l,\dots,l}$ -- ensures also $\alpha_{ij}^m=0$ for all $i,j=1,2,3$, $m=-l,\dots,l$. In consequence, we find $\mathcal{K}\left[p_{ij}\right]=0$ and hence $g_{ij}-\widetilde{g}_{ij}=\mathcal{O}(r^{-2-(n^\ast+1)})$ in every end of $M^3$.

By induction and by the assumed analyticity of $g_{ij}-\widetilde{g}_{ij}$ at infinity, we deduce that $g_{ij}=\widetilde{g}_{ij}$ in $M^3\setminus K$. Now assume that $\inter K$ is diffeomorphic to a domain in $\R^3$ with smooth boundary $\partial K$. Then the above results can be reinterpreted to say that the Dirichlet boundary data for the functions $g_{ij}$ and $\widetilde{g}_{ij}$ on $\partial K$ must agree and it thus suffices to show that the Dirichlet problem for the static metric equations on $K$ is uniquely solvable for the variables $(g_{ij})_{i,j=1,2,3}$. Again making use of the fact that the chosen coordinates are wave harmonic, Lemma \ref{waveharm} tells us that $-g^{ij}\Gamma^k_{ij}=\mylap{g}x^k=N_{,l}g^{kl}/N$ and thus $g^{ij}\Gamma_{ij,l}^k$ can be seen to only depend on $g_{ij}$ and $g_{ij,k}$ and not on second derivatives of the ends of the metric. It is then straightforward to see that $$\my{g}{\Ric}_{ij}=-\frac{1}{2}\mylap{g}(g_{ij})+\mbox{zeroth and first order terms}$$ and that the static metric equations \eqref{SM3} become a quasi-linear elliptic system for the $g_{ij}$. The associated Dirichlet problem then has a unique $C^{2,\alpha}(\inter K)$ (and thus unique smooth) solution $(g_{ij})_{i,j=1,2,3}$ inside $\inter K$ by classical PDE theory\footnote{cf.~e.~g.~theorem 6.14 on p.~107 in \cite{GT}.} which completes our proof.
\qed\end{Pf}

\chapter{Pseudo-Newtonian Gravity}\label{chap:mCoM}\label{chap:pseudo}
In the last chapter, we have introduced and studied the theory of geometrostatics modeling static isolated gravitational systems. In this chapter, we will now introduce the so-called ``pseudo-Newtonian gravity'', a conformal variant of geometrostatics very useful for physical considerations and for the Newtonian limit as we will see in this chapter and the following ones. In Section \ref{sec:pseudo}, we will define pseudo-Newtonian systems, study their elementary properties, and translate the well-known facts on geometrostatics as well as the asymptotic considerations concerning uniqueness and the center of mass from geometrostatics to the pseudo-Newtonian setting.

In Sections \ref{sec:mpseudo} and \ref{sec:CoMpseudo}, we will prove explicit quasi-local formulae for the asymptotic quantities of ADM-mass and ADM/CMC/intrinsic/asymptotic center of mass, respectively. As they resemble the quasi-local Newtonian expressions for mass and its center, we will call them pseudo-Newtonian mass and center of mass, respectively. These notions will allow for simple addition formulae which are analog to the well-known Newtonian ones (Theorems \ref{thm:madd}, \ref{thm:CoMadd}) and many more physicogeometric insights which we will present in Chapter \ref{chap:geo}. Moreover, these formulae tremendously facilitate the Newtonian limit considerations we are aiming at and will therefore be reencountered in Chapter \ref{chap:FT}.

\section{Pseudo-Newtonian Gravity}\label{sec:pseudo}
The lapse function $N$ appearing in geometrostatics is in many respects similar to the (Newtonian) potential $U$\index{sym}{$U$}\index{ind}{potential ! Newtonian} in the Newtonian theory of gravity: It satisfies a Poisson equation relating it to the matter content of the system (cf.~Equation \eqref{lapN}), its level sets foliate the end(s) of $3$-space (cf.~Lemma \ref{lem:foli}) and are asymptotically approaching round spheres centered at the center of mass (cf.~Corollary \ref{coro:graph}), and its asymptotics allow to read off the total mass and the center of mass of the system (cf.~Theorems \ref{thm:KM} and \ref{thm:asym-CoM}). In this and the following chapters, we will encounter even more similarities in spirit between $N$ and $U$ like e.~g.~the fact that their level sets are surfaces of equilibrium (cf.~Section \ref{sec:equi}). These analogies could now lead one to the hypothesis that $N$ might be a good relativistic replacement of the Newtonian potential $U$ and ``converge'' to it in the Newtonian limit\footnote{For a more precise formulation of the Newtonian limit, cf.~Chapter \ref{chap:FT}.} -- a hypothesis that is not fully valid but will lead us to the study of so-called ``pseudo-Newtonian systems''.

As an intuitive example of why $N$ cannot converge to the Newtonian potential in the Newtonian limit consider the Schwarzschild metric with positive mass $m$ in isotropic coordinates such that $$N(p)=\left(1-\frac{mG}{2r(p)c^2}\right)\left(1+\frac{mG}{2r(p)c^2}\right)^{-1}$$ as calculated on page \pageref{subsec:schwarz}ff. Now the Newtonian limit $c\to\infty$ would imply $N\to1$ globally in the domain of the coordinates which does not contain any information and should thus not be considered as the Newtonian potential of any Newtonian system ``corresponding'' to the Schwarzschild system under consideration. In fact, a simple unit comparison already shows that the lapse function must have unit $1$ while the Newtonian potential must have unit velocity squared.

However, we will see in Chapter \ref{chap:FT} that the related function $U:=c^2\ln N$\label{U}\index{sym}{$U$} does converge to the Newtonian potential of a ``corresponding'' system under the Newtonian limit and we remark that $U$ does indeed have the appropriate unit of velocity squared. Obviously, $U$ has the same level sets as $N$ and in fact inherits all analogies to the Newtonian potential described above from $N$. In particular, the fall-off behavior of $U$ is even more similar to the one of the Newtonian potential, as we will see below.

The function $U=c^2\ln N$ is well-known in the study of static relativistic systems and is usually referred to as the ``potential'' of the system. We will follow this convention but will add the qualifier {\it pseudo-Newtonian}\index{ind}{potential ! pseudo-Newtonian}\index{ind}{pseudo-Newtonian ! potential}\index{sym}{$U$} to prevent misunderstandings. Together with this potential, both mathematicians and physicists commonly study the conformally transformed $3$-metric $\gamma$ defined by $\gamma:=N^2 g$\index{sym}{$\gamma$}\label{gamma}, which we have already encountered when dealing with wave harmonic coordinates, cf.~e.~g.~\cite{KM,Reiris} and Lemma \ref{waveharm}. We will call $\gamma$ the {\it pseudo-Newtonian metric}\index{ind}{pseudo-Newtonian ! metric}\index{ind}{metric ! pseudo-Newtonian} of the geometrostatic system in what follows. As $N\to1$ in the Newtonian limit\footnote{a fact which we have explained above for the Schwarzschild example and which will follow in general from our considerations in Chapter \ref{chap:FT}.}, we can reasonably expect that $\gamma$ and $g$ have the same Newtonian limits and thus information is neither lost nor gained by this transform. 

Returning to the geometrostatic setting, let us now study how the relevant facts and equations transform under a change from the geometrostatic perspective to the pseudo-Newtonian one. A straightforward computation tells us that if $\mathcal{S}=(M^3,g,N,\rho,S)$ is a geometrostatic system, the corresponding pseudo-Newtonian quantities $U$ and $\gamma$ satisfy a conformally transformed version of the static metric equations \eqref{SME} which reads
\begin{eqnarray}\nonumber
\mylap{\gamma}U &=& 4\pi G\left(\frac{\rho}{e^{2c^{-2}U}}+\frac{\my{\gamma}{\tr}S}{c^2}\right)\\[-2ex]\label{ConfSME}
\\\nonumber
\my{\gamma}{\Ric}&=&\frac{2}{c^4}\,dU\otimes dU+\frac{8\pi G}{c^4}\left(S-\my{\gamma}{\tr}S\,\gamma\right).
\end{eqnarray}
In vacuum, these equations reduce to
\begin{eqnarray}\nonumber
\mylap{\gamma}U &=& 0\\[-2ex]\label{ConfSMEvac}
\\\nonumber
\my{\gamma}{\Ric}&=&\frac{2}{c^4}\,dU\otimes dU.
\end{eqnarray}
We call these equations the {\it pseudo-Newtonian static metric equations}\index{ind}{static metric equations ! pseudo-Newtonian} or the {\it equations of pseudo-Newtonian gravity (pNG)}\index{ind}{pseudo-Newtonian ! gravity}\index{ind}{equations of pNG}\index{ind}{pNG} for $U$ and $\gamma$. We choose the name ``pseudo-Newtonian'' (and consider just this transform) mainly by reason of the similarity of the first of the pseudo-Newtonian equations to the Newtonian equation $\triangle U=4\pi G\rho$ and because the metric $\gamma$ singles out the wave harmonic coordinates of the Lorentzian $4$-metric as its own harmonic coordinates (cf.~Lemma \ref{waveharm}) which -- together with some facts we will prove in Chapter \ref{chap:geo} like the second pseudo-Newtonian law of motion \ref{thm:F=ma} -- also suggests that $\gamma$ has physical relevance. Moreover, the behavior of \eqref{ConfSME} under the Newtonian limit which we will discuss in Chapter \ref{chap:FT} indicates that it is worthwhile and physically significant to study this conformal transform. Heuristically speaking, the system \eqref{ConfSME} indeed reduces to the Newtonian equation $\triangle U=4\pi G\rho$ when we take its Newtonian limit anticipating that $\gamma$ and $U$ will converge to the flat metric and the Newtonian potential, respectively.

From an analytic point of view, observe that while the original static metric equations \eqref{SME} are linear in $N$ both in vacuum and in the presence of matter, the pseudo-Newtonian ones contain the quadratic expression $dU\otimes dU$ even in vacuum. In addition, the Ricci tensor, which is coupled to the Hessian of $N$ in the original equations, is only coupled to first order derivatives of $U$ after the conformal transformation.  This is another important reason for considering the pseudo-Newtonian version of the static metric equations.

Having discussed the transformation behavior of the equations, let us now for convenience define a notion of ``pseudo-Newtonian systems''.

\begin{Def}\label{DefpseudoNewton}
Let $\mathcal{S}_{PN}:=(M^3,\gamma,U,\rho,S)$ be a solution of the equations of pseudo-Newtonian gravity \eqref{ConfSME} and let $k\in\N$, $k\geq3$, and $\tau\geq1/2$ such that $-\tau$ is non-exceptional (i.~e.~$\tau\notin\Z$). We call $\mathcal{S}_{PN}$ a {\it $(k,\tau)$-pseudo-Newtonian system}\index{ind}{pseudo-Newtonian ! system}\index{ind}{system ! pseudo-Newtonian} if, in addition, the following conditions hold:
\begin{enumerate}
\item[(i)] $(M^3,\gamma)$ is a $(k,q=2,\tau)$-asymptotically flat manifold
\item[(ii)] $U(p)\to0$ as $p\to\infty$ in each end of $M^3$.
\item[(iii)] $\rho\geq0$, $S\geq0$, and the supports of $\rho$ and $S$ are bounded away from infinity.
\end{enumerate}
As always, we will call $\mathcal{S}_{PN}$ a pseudo-Newtonian system for short if $k$ and $\tau$ are either clear from context or arbitrary. Moreover, if we are concerned with a vacuum solution (i.~e.~if $\rho=0$, $S=0$), we call $(M^3,\gamma,U)$ a {\it vacuum pseudo-Newtonian system}\index{ind}{vacuum} for simplicity. Pseudo-Newtonian ends\index{ind}{end ! pseudo-Newtonian}\index{ind}{pseudo-Newtonian ! end} are defined accordingly.\index{sym}{$\mathcal{S}_{PN}$}
\end{Def}

The following proposition describes the relation between geometrostatic and pseudo-Newtonian systems and therewith justifies the above definition. It will implicitly accompany us throughout the remainder of this thesis.
\begin{Prop}\label{prop:geopseu}
Let $k\in\N$, $k\geq3$ and let $\tau\geq1/2$ such that $-\tau$ is non-exceptional. Let $(M^3,g)$ be a smooth Riemannian $3$-dimensional manifold, $N:M^3\to\R$ a smooth positive function, $\rho:M^3\to\R$ a smooth non-negative function, and $S$ a smooth and positive semi-definite $(0,2)$-tensor field on $M^3$. Assume furthermore that $0<K^{-1}\leq N\leq K<\infty$ on $M^3$ for some $K\in\R$. Set $\gamma:=N^2 g$ and $U:=c^2\ln N$. Then $\mathcal{S}=(M^3,g,N,\rho,S)$ is a $(k,\tau)$-geometrostatic system if and only if the {\bf corresponding}\index{ind}{system ! corresponding} system $\mathcal{S}_{PN}:=(M^3,\gamma,U,\rho,S)$ is a $(k,\tau)$-pseudo-Newtonian system. In particular, the asymptotic flatness is exhibited in the same systems of coordinates (if it is exhibited).
\end{Prop}
\begin{Rem}\label{rem:horizon}
By definition, the Riemannian manifold $(M^3,g)$ coming from a geometrostatic system must either be geodesically complete or an individual end and thus diffeomorphic to a cylinder $\R^3\setminus B$ where one end of the cylinder corresponds to ``infinity'' while the other one corresponds to the ``inner boundary'' $\partial B$. From the boundary values $N\to1$ as $r\to\infty$ and the requirement that $N>0$ in $M^3$, we know that $N$ is automatically bounded below if the metric is complete. Moreover, by  the maximum principle\footnote{cf.~Theorem 17.1 in \cite{GT}.} for the elliptic partial differential inequality $$\triangle N=\frac{4\pi G}{c^2}N\left(\rho+\frac{\my{g}{\tr} S}{c^2}\right)\geq0,$$ arising from the static metric equations \eqref{SME} and in view of the boundary values $N\to1$ as $r\to\infty$, $N$ is also bounded from above in that case. If, however, the system is an individual end, the assumption $0<K^{-1}\leq N\leq K<\infty$ does impose a restriction. Namely, as $N>0$ in $M^3$ and $N\to1$ as $r\to\infty$ also in that case, the assumption excludes ends where $N(p)\to0$ or $N(p)\to\infty$ as $p$ approaches the inner boundary.

This is not only technical but also has physical meaning: Recall that $N$ was defined as the length of the timelike Killing vector field $X$ providing staticity. A surface on which $N=0$ holds thus constitutes a so-called ``Killing horizon'' in the sense that the static Killing vector field becomes lightlike there.\index{ind}{horizon}\index{ind}{Killing horizon} Killing Horizons are closely related to black holes. An example for a static spacetime with a Killing horizon is provided by the Schwarzschild metrics\index{ind}{Schwarzschild}, cf.~page \pageref{subsec:schwarz}ff. We are thus assuming here that possible horizons have been cut out when we are performing the conformal transform. In other words, the pseudo-Newtonian approach breaks down in a neighborhood of a horizon.
\end{Rem}
\begin{Pf}
First of all, it follows from a straightforward computation relying on the well-known formulae for conformal transformations\footnote{cf.~e.~g.~pp.~105 in \cite{Willmore}.} that $\mathcal{S}$ satisfies the static metric equations on $M^3$ if and only if $\mathcal{S}_{PN}$ satisfies the pseudo-Newtonian ones there. Also, as $N\geq K^{-1}>0$, $(M^3,\gamma)$ is geodesically complete if and only if $(M^3,g)$ is. Furthermore, $-\infty<-\ln K\leq U\leq \ln K<\infty$ holds on $M^3$. Assume now that $(x^{i})$ is a system of coordinates for an end $E^3$ of $M^3$. Then $(g_{ij})$ is uniformly positive definite and bounded for some constant $\mu>0$ or in other words $$\mu^{-1}\vert\xi\vert^2\leq(\Phi_\ast g)\vert_x(\xi,\xi)\leq\mu\vert\xi\vert^2$$ for all $\xi\in\R^3$ holds on $E^3$ if and only if the same is true for $(\gamma_{ij})$ and the constant $\widetilde{\mu}:=K\mu$.

Moreover, we have $g_{ij}-\delta_{ij}\in W^{k,2}_{-\tau}(E^3)$ for all $i,j=1,2,3$ if and only if $\gamma_{ij}-\delta_{ij}\in W^{k,2}_{-\tau}(E^3)$ for all $i,j=1,2,3$ as $N-1\in W^{k+1,2}_{-\tau}(E^3)$ by \ref{thm:KM} and by the H\"older inequality \ref{Hoelder} and because $U\in W^{k+1,2}_{-\tau}(E^3)$ follows from the pseudo-Newtonian equations \eqref{ConfSME} and the boundary condition $U\to0$ as $r\to\infty$ just as $N-1\in W^{k+1,2}_{-\tau}(E^3)$ follows from the static metric equations \eqref{SME} and the boundary condition $N\to1$ as $r\to\infty$, a fact which is proven and explained in Daniel Kennefick and Niall \'O Murchadha's article \cite{KM}. We therefore see from Definition \ref{def:asym} that $(E^3,g)$ is a $(k,q=2,\tau)$-asymptotically flat end with respect to the coordinates $(x^{i})$ if and only if $(E^3,\gamma)$ is (where we have suppressed the associated diffeomorphisms $\Phi$ for ease of notation). Secondly, from $\ln1=0$ we see that $N\to1$ is equivalent to $U\to0$ in any end of $M^3$ and this finishes our proof of the desired equivalence.
\qed\end{Pf}

\subsection*{Example: Pseudo-Newtonian Schwarzschild Solutions}\label{subsec:schwarzpseudo}
Before we go on by carrying over the well-known facts on and the asymptotic properties of geometrostatic systems collected in Chapter \ref{chap:static}, it is useful to calculate the pseudo-Newtonian equivalent of the Schwarzschild example introduced on page \pageref{subsec:schwarz}ff which we will now do. So let us summarize the explicit expressions for the pseudo-Newtonian potential $\my{m,S}{U}=c^2\ln(\my{m,S}\!{N})$ and the associated metric $\mygam{m,S}:=\my{m,S}\!{N}^2\,\myg{m,S}$ for the Schwarzschild family\index{ind}{Schwarzschild ! solution}\index{ind}{Schwarzschild ! pseudo-Newtonian potential}\index{ind}{Schwarzschild ! pseudo-Newtonian metric}\index{sym}{$\my{m,S}{U}$}\index{sym}{$\mygam{m,S}$}\index{sym}{$\my{S}{U}$}\index{sym}{$\mygam{S}$} in wave harmonic coordinates $(y^{i})$ with $s:=\sqrt{(y^1)^2+(y^2)^2+(y^3)^2}$. Dropping the explicit reference to the mass parameter $m$, we find that
\begin{eqnarray}\label{pseudoNschwarz}
\my{S}{U}&=&\frac{c^2}{2}\left(\ln(1-\frac{M}{s})-\ln(1+\frac{M}{s})\right)\\\label{pseudogamschwarz}
\mygam{S}_{ij}&=&\frac{(1-M/r)(\varphi\circ r)^4}{1+M/r}\left(\psi^2\delta_{ij}+2s\psi\psi'\,\frac{y_iy_j}{s^2}+s^2(\psi')^2\frac{y_iy_j}{s^2}\right)
\end{eqnarray}
from \eqref{schwarzharm} through \eqref{N,yi}, where again $M=mG/c^2$, $\varphi(r)=1+M/2r$, $r(s)=s\psi(s)$, and $\psi(s)=(1+\sqrt{1-M^2/s^2)})/2$. It is now immediate that the pseudo-Newtonian versions of the members of the Schwarzschild family constitute vacuum pseudo-Newtonian systems.
\begin{Prop}
Let $m\geq0$, $k\in\N$, $k\geq3$, $\tau\geq1/2$ such that $-\tau$ is non-exceptional. Let $M^\ast>M$ where $M:=mG/c^2$ as before. Then the pseudo-Newtonian Schwarzschild system $\mathcal{S}_{PN}:=(E_{M^*}(0),\mygam{m,S},\my{m,S}{U})$ is a $(k,\tau)$-pseudo-Newtonian system with structure of infinity given by the $\mygam{m,S}$-harmonic coordinates $(y^{i})$.
\end{Prop}
\begin{Pf}
This follows from Propositions \ref{prop:schwarzgeo} and \ref{prop:geopseu} using Lemma \ref{waveharm}.
\qed\end{Pf}

As before, we will drop the label $m$ on the Schwarzschild metrics whenever no confusion can arise. In what follows, we will be happy to be able to make use of the asymptotics
\begin{eqnarray}\label{Ufall}
\my{S}{U}&=&-\frac{mG}{r}+\mathcal{O}(r^{-3})\\\label{gamfall}
\mygam{S}_{ij}&=&\left(1-\frac{M^2}{r^2}\right)\delta_{ij}+\frac{2M^2x_{i}x_{j}}{r^4}+\mathcal{O}(r^{-3})
\end{eqnarray}
which can be straightforwardly computed from \eqref{pseudoNschwarz} and \eqref{pseudogamschwarz} using harmonic coordinates and analyticity at infinity of the relevant expressions. They directly correspond to the geometrostatic expansions \eqref{Nfall} and \eqref{gfall}. Moreover, they suggest that $\mygam{S}$ has vanishing ADM-mass. The expansions are very similar to the Newtonian situation where $U=-mG/r+\mathcal{O}(r^{-3})$ (``no gravitational dipoles'') and $\gamma_{ij}=\delta_{ij}$ in Cartesian (and thus $\gamma$-harmonic) coordinates.

\subsection*{Properties of Pseudo-Newtonian Systems}\label{subsec:pseudoprop}
Let us now return our attention to general pseudo-Newtonian systems and first of all define their mass and center of mass in consistence with the ADM-mass and the ADM/CMC/ intrinsic/asymptotic center of mass discussed in Sections \ref{sec:HCoM} and \ref{sec:CoMasym}.

\begin{Def}[Mass and Center of Mass]\label{def:massasym}
Let $\mathcal{S}_{PN}=(E^3,\gamma,U,\rho,S)$ be a $(k,\tau)$-pseudo-Newtonian end. Then the {\it physical mass}\index{ind}{mass}\index{ind}{mass ! physical} $m(\mathcal{S}_{PN})$\index{sym}{$m(\mathcal{S}_{PN})$} of $\mathcal{S}_{PN}$ is defined as the ADM-mass of the corresponding geometrostatic system $\mathcal{S}=(E^3,g,N,\rho,S)$, $$m(\mathcal{S}_{PN}):=m_{ADM}(E^3,g).$$ If $m(\mathcal{S}_{PN})\neq0$, $\tau>1/2$, and $(x^{i})$ is a system of asymptotically flat harmonic coordinates for $\gamma$, then the {\it physical center of mass}\index{ind}{center of mass}\index{ind}{center of mass ! physical} with respect to these coordinates, $\vec{z}(\mathcal{S}_{PN})\in\R^3$\index{sym}{$\vec{z}(\mathcal{S}_{PN})$}, is defined as $$\vec{z}(\mathcal{S}_{PN}):=\vec{z}_{A}(E^3,g),$$ where the latter is the asymptotic center of mass defined in \ref{thm:asym-CoM}.
\end{Def}
We can now characterize the fall-off behavior of pseudo-Newtonian systems by carrying over that of geometrostatic systems to the present situation. This gives the following theorem.
\begin{Thm}\label{confasym}
Let $\mathcal{S}_{PN}=(E^3,\gamma,U,\rho,S)$ be a $(k,\tau)$-pseudo-Newtonian end, and let $(x^{i})$ be a system of $\gamma$-harmonic asymptotically flat coordinates in $E^3$. Let $m$ denote its physical mass. Then we have
\begin{eqnarray}\nonumber
U+\frac{mG}{r}&\in& W^{k+,2}_{-(\tau+1)}(E^3)\\[-2ex]\label{eq:Ugamasym}
\\\nonumber
\gamma_{ij}-\delta_{ij}&\in& W^{k,2}_{-(\tau+1)}(E^3)
\end{eqnarray}
for all $i,j=1,2,3$. Moreover, if $m\neq0$, $\tau>1/2$, and $\vec{z}$ denotes the physical center of mass of $\mathcal{S}_{PN}$, it holds that
\begin{eqnarray}\nonumber
U-\my{S}{U}+\frac{mG\vec{z}\cdot\vec{x}}{r^3}&\in& W^{k+1,2}_{-(\tau-\varepsilon+2)}(E^3)\\[-2ex]\label{eq:UgamasymCoM}
\\\nonumber
\gamma_{ij}-\mygam{S}_{ij}&\in& W^{k,2}_{-(\tau-\varepsilon+2)}(E^3)
\end{eqnarray}
for all $i,j=1,2,3$ and all $0<\varepsilon\leq\tau-1/2$, where $\mygam{S}$ and $\my{S}{U}$ are the pseudo-Newtonian Schwarzschild metric of physical mass $m$ and the associated potential, respectively.
\end{Thm}
\begin{Rems}
There are two points which we would like to remark. Firstly, the fall-off in \eqref{eq:Ugamasym} could equally well be written as $U-\my{S}{U}\in W^{k,2}_{-(\tau+1)}(E^3)$ and $\gamma_{ij}-\mygam{S}_{ij}\in W^{k,2}_{-(\tau+1)}(E^3)$ by the above expansions \eqref{Ufall} of $\my{S}{U}$ and \eqref{gamfall} of $\mygam{S}$ although the version printed above is more explicit. Secondly, it is worth noting that $\gamma_{ij}$ actually does not contain a center of mass term. Differently put, the physical center of mass only shows in the potential in the pseudo-Newtonian setting.
\end{Rems}
\begin{Pf}
The first claim follows from Theorem \ref{thm:KM}, Proposition \ref{prop:geopseu}, the Schwarzschild fall-off \eqref{gamfall}, and from analyticity of the logarithm in a neighborhood of the number $1$ with power series $$\ln(1+\frac{\alpha}{r}+\frac{\beta}{r^2}+\mathcal{O}(r^{-3}))=\frac{\alpha}{r}+\frac{2\beta-\alpha^2}{2r^2}+\mathcal{O}(r^{-3})$$ as $r\to\infty$ when we apply the H\"older inequality \eqref{Hoelder} and the embedding theorems listed in Section \ref{sec:sobo}. The second statement follows in a similar manner from Theorem \ref{thm:asym-CoM}.
\qed\end{Pf}

In analogy to the expansion \eqref{gamfall} for the Schwarzschild example, the above theorem suggests that the ADM-mass of any pseudo-Newtonian metric $\gamma$ should vanish. Intuitively, we can restate this as saying that the ADM-mass of the given end is ``transported away from infinity'' by the conformal transform. This idea is accurate as the next lemma states. Observe that the lemma does not contradict the fact that the geometrostatic (physical) metric $g$ can have non-vanishing (physical) mass which indeed shows in the $-\frac{mG}{r}$ term of the expansion of the pseudo-Newtonian potential $U$ as it would do in the Newtonian setting. In contrast, the lemma just claims that the physical mass does not show in the first order term of the expansion of $\gamma$ as it does in the first order term of the expansion of $g$.
\begin{Lem}
Let $\mathcal{S}_{PN}=(E^3,\gamma,U,\rho,S)$ be a pseudo-Newtonian end. Then its ADM-mass $m_{ADM}(E^3,\gamma)$ vanishes. 
\end{Lem}
\begin{Pf}
The ADM-mass of $(E^3,\gamma)$ is well-defined by Lemma \ref{confasym} and by Theorem \ref{thm:Bartnik}. Now use asymptotically flat $\gamma$-harmonic coordinates $(x^{i})$, set $\rho_{ij}:=\gamma_{ij}-\delta_{ij}$, and note that $\rho_{ij}\in W^{k,2}_{-(\tau+1)}(E^3)$ by Theorem \ref{confasym} for all $i,j=1,2,3$. Then by definition of the ADM-mass, we have
\begin{eqnarray*}
m_{ADM}(E^3,\gamma)&=&\frac{c^2}{16\pi G}\lim_{r\to\infty}\int_{S_r^2}\left(\gamma_{ii,j}-\gamma_{ij,i}\right)\nu_\delta^j\,d\sigma_\delta\\
&=&\frac{c^2}{16\pi G}\lim_{r\to\infty}\int_{S_r^2}\underbrace{(\rho_{ii,j}-\rho_{ij,i})\frac{x^j}{r}}_{\in W^{k-1,2}_{-(\tau+2)}(E^3)}\,d\sigma_\delta\\
&\stackrel{\mbox{\scriptsize Lemma }\ref{confasym},\,\tau>0}{=}&0.
\end{eqnarray*}
\qed\end{Pf}

\subsection*{Pseudo-Newtonian Uniqueness Properties}\label{subsec:gunique}
In Newtonian gravity, the potential $U$ of a compactly supported matter distribution is uniquely determined by the matter density and the normalization $U\to0$ as $r\to\infty$, a well-known fact which follows from elliptic PDE theory. Just as for geometrostatic systems (cf.~Theorems \ref{Nuniqueasym} and \ref{thm:gunique}), the ``same'' actually holds true in pseudo-Newtonian gravity in the following sense.

\begin{Thm}[Uniqueness of $U$]\label{thm:Uunique}
Let $\mathcal{S}_{PN}:=(M^3,\gamma,U,\rho,S)$, $\widetilde{\mathcal{S}_{PN}}:=(M^3,\gamma,\widetilde{U},\rho,S)$ be geodesically complete pseudo-Newtonian systems. Then $U=\widetilde{U}$ in all of $M^3$.
\end{Thm}
\begin{Rem}
If the systems in Theorem \ref{thm:Uunique} consisted of individual ends, only, a similar argument shows that $U=\widetilde{U}$ if we assume that both $U(p),\widetilde{U}(p)\to-\infty$ as $p$ approaches any inner boundary -- an assumption which is reasonable as it corresponds to $N(p),\widetilde{N}(p)\to0$, making the boundary a Killing horizon\footnote{Recall that $N$ is the length of the static Killing vector field and cf.~the remark on page \pageref{rem:horizon}.}.
\end{Rem}
\begin{Pf}
The equations of pNG \eqref{ConfSME} imply that $dU\otimes dU=d\widetilde{U}\otimes d\widetilde{U}$ in all of $M^3$ so that $U=\pm\widetilde{U}+\mbox{const}$. The constant must vanish as both $U$ and $\widetilde{U}$ are normalized to $0$ at infinity and thus $U=\pm\widetilde{U}$. Assume $U\neq\widetilde{U}$ (so that $U=-\widetilde{U}$) and let $m,\widetilde{m}$ denote the respective physical masses. Observe that the asymptotics of $U,\widetilde{U}$ given by Lemma \ref{confasym} imply $m=-\widetilde{m}$. The positive mass theorem \ref{positive mass thm} applied to the corresponding geometrostatic systems now gives us\footnote{It is applicable as the scalar curvatures satisfy $\my{g}{\Scal}=\widetilde{\my{g}{\Scal}}=16\pi G\rho/c^2\geq0$ by Equation \eqref{SM1}.} $m=\widetilde{m}=0$ so that by rigidity $g$ and $\widetilde{g}$ must be flat and $M^3$ must be diffeomorphic to $\R^3$. In particular, we must have $\rho=0$ by the trace of the static metric equation $0=\my{g}{\Scal}=16\pi G\rho/c^2$. By the pseudo-Newtonian equations, it follows that $\mylap{\gamma}U=-\mylap{\gamma}\widetilde{U}=\pm 4\pi G\,\my{\gamma}{\tr}S$ and therefore we know that $\my{\gamma}{\tr}S=0$ and thus $U$ (and hence also $\widetilde{U}$) must be $\gamma$-harmonic on all of $M^3$. As $M^3$ is diffeomorphic to $\R^3$, $\gamma$ is asymptotically flat, and $U\to0$ as $r\to\infty$, it follows from Section \ref{sec:sobo} that $U=0$ and thus also $U=\widetilde{U}=0$, a contradiction.
\qed\end{Pf}

Just as in the geometrostatic setting, the ``complementary'' statement also holds true.
\begin{Thm}[Uniqueness of $\gamma$]\label{thm:gammaunique}
Let $\mathcal{S}_{PN}:=(M^3,\gamma,U,\rho,S)$, $\widetilde{\mathcal{S}_{PN}}:=(M^3,\widetilde{\gamma},U,\rho,S)$ be pseudo-Newtonian systems with a common global harmonic and asymptotically flat system of coordinates $(x^i)$ outside some compact $K\subset M^3$ containing the support of the matter. Assume furthermore that $\gamma$, $\widetilde{\gamma}$, and $U$ are analytic at infinity with respect to these coordinates outside $K$. Then $\gamma_{ij}=\widetilde{\gamma}_{ij}$ in all of $M^3\setminus K$ (possibly) unless the physical masses of both metrics vanish. If, in addition, $\gamma$ and $\widetilde{\gamma}$ are geodesically complete and the coordinates $(x^{i})$ can be extended into $K$ as a global harmonic system of coordinates and if $\inter K$ is diffeomorphic to a bounded domain in $\R^3$ having smooth boundary\footnote{Again, we only need these conditions on $K$ in order to deduce that elliptic Dirichlet boundary value problems in $K$ are uniquely solvable.}, then $\gamma_{ij}=\widetilde{\gamma}_{ij}$ holds in all of $M^3$ whether or not the ADM-mass vanishes.
\end{Thm}
\begin{Pf}
We will work in the corresponding geometrostatic systems. Recall that harmonic coordinates with respect to a pseudo-Newtonian metric $\gamma$ are wave harmonic with respect to the corresponding geometrostatic metric by Lemma \ref{waveharm}. The fact that both systems $\mathcal{S}_{PN}$ and $\widetilde{\mathcal{S}}_{PN}$ have the same potential now gives us $N=\widetilde{N}$ and we are therefore back in the case of Theorem \ref{thm:gunique} as analyticity carries over and because $g$ and $\widetilde{g}$ must be complete metrics whenever $\gamma$ and $\widetilde{\gamma}$ are complete, respectively. We therefore obtain that $g_{ij}=\widetilde{g}_{ij}$ and thus again $\gamma_{ij}=\widetilde{\gamma}_{ij}$ in all of $M^3\setminus K$ or $M^3$, respectively, due to $N=\widetilde{N}$.
\qed\end{Pf}

\section{The Pseudo-Newtonian Mass}\label{sec:mpseudo}
Isolated systems or in other words asymptotically flat Riemannian $3$-manifolds have a well-defined global or ``total'' mass, namely the ADM-mass\index{ind}{mass}. As we have seen, this mass arises as a ``surface integral at infinity'' or in other words as a limit of surface integrals where the surfaces are coordinate spheres with radii tending to infinity. Although the surface integrals can already give the correct value for finite radii outside the support of the matter in special cases, e.~g.~for Schwarzschild metrics (static vacuum and spherically symmetric metrics, cf.~pp.~\pageref{subsec:schwarz}ff), it will in general not do so. Intuitively said, this is due to the curvature present even outside the support of the matter, cf.~Section 5 in \cite{Bartnik}.

This, however, is not particularly satisfying from a physical point of view as one would not expect the vacuum region outside a  nisolated star or galaxy to contribute to its mass which should be a property of the star or galaxy itself and should not depend on its exterior. On the other hand, Albert Einstein's formula $E=mc^2$ hints to a reason why mass (or rather energy) cannot in general be a property of the star or galaxy (or, as a matter of fact, of any other relativistic system) alone, since (classically put) the kinetic energies of material bodies and possibly also of black holes and gravitational waves enter the game. 

In the case of static systems, though, this kinetic energy is not present. One can therefore hope to find a notion of mass in the static realm that is purely local in the sense that it does not depend on the exterior of the isolated body $B$ the mass of which we would like to describe but allows to read off the total (ADM-)mass already in the finite regime. In analogy to Newtonian gravity, one could even hope for a volume integral expression like $\int_{B}\rho\,dV$ integrated over (the support of the matter of) the body $B$ to give its total mass. This is too optimistic in general as the body might be a black hole so that singularities and other non-classical effects can occur. But the idea of such a volume integral expression leads us to a surface integral expression that can be evaluated at any surface enclosing the body (giving the same value) and thus being fairly close to the intuitive expectations described above. 

In addition to clarifying the physical meaning of the ADM-mass, this surface integral or ``quasi-local''\index{ind}{quasi-local} definition of mass will be helpful for dealing with the Newtonian limit, cf.~Chapter \ref{chap:FT}. Let us begin by recapitulating the Newtonian situation in a heuristic manner. If $U:\R^3\to\R$ is the Newtonian potential\index{ind}{Newtonian ! potential}\index{ind}{potential ! Newtonian} of a given matter distribution $\rho:\R^3\to\R^+_{0}$\index{ind}{$\rho$} with compact support -- which means that $U$ satisfies the Poisson equation
\begin{equation}\label{eq:Newton}
\triangle U=4\pi G\rho\mbox{ in }\R^3,\quad U\to0\mbox{ as }r\to\infty,
\end{equation}
-- then the total (Newtonian) mass\index{ind}{mass}\index{ind}{mass ! Newtonian}\index{ind}{Newtonian ! mass} is usually defined as
\begin{equation*}
m_{N}:=\int_{\R^3}\rho\,d\mu=\int_{\Omega}\rho\,d\mu,\index{sym}{$m_{N}$}
\end{equation*}
where $d\mu$ is the ordinary Lebesgue measure on $\R^3$ and $\Omega\subset\R^3$ is any domain in $\R^3$ containing the support of the matter density $\rho$. With the aid of the Poisson equation \eqref{eq:Newton} and the divergence theorem, this can be converted into 
\begin{equation*}
m_{N}=\frac{1}{4\pi G}\int_{\Omega}\triangle U\,d\mu=\frac{1}{4\pi G}\int_{\partial\Omega}\frac{\partial U}{\partial\nu}d\sigma
\end{equation*}
if the boundary of $\Omega$ is sufficiently regular and $\nu$ and $d\sigma$ denote its outer unit normal and surface measure, respectively. We can turn this argument upside down and define a quasi-local Newtonian mass $m_{N}(\Sigma)$\index{sym}{$m_{N}(\Sigma)$}\index{ind}{mass ! quasi-local} by $$m_{N}(\Sigma):=\frac{1}{4\pi G}\int_{\Sigma}\frac{\partial U}{\partial\nu}d\sigma$$ on any smooth orientable surface $\Sigma\subset\R^3$ with again $\nu$ the outer normal and $d\sigma$ the surface measure of $\Sigma$. The above argument shows that $m_{N}(\Sigma)=m_{N}$ whenever $\Sigma$ {\it encloses the support of the matter}\index{ind}{enclose the support of the matter} (meaning that $\Sigma$ is a topological sphere the outside of which is a vacuum asymptotically flat end). Note that this expression is purely geometric and we can thus directly translate it into pseudo-Newtonian gravity as follows.
\begin{Def}[Quasi-Local Pseudo-Newtonian Mass]\label{pseudo-m}
Let $(E^3,\gamma,U,\rho,S)$ be a pseudo-Newtonian end. For any smooth orientable surface $\Sigma\subset E^3$ with outer unit normal $\nu$ and surface measure $d\sigma$ both induced from $\gamma$, define the {\it pseudo-Newtonian quasi-local mass}\index{ind}{mass ! pseudo-Newtonian}\index{ind}{pseudo-Newtonian ! mass}\index{sym}{$m_{PN}(\Sigma)$} by $$m_{PN}(\Sigma):=\frac{1}{4\pi G}\int_{\Sigma} \frac{\partial U}{\partial\nu}\,d\sigma.$$
\end{Def}
\begin{Rem}
The integral $\frac{1}{4\pi G}\int_{{S_{\infty}}} \frac{\partial U}{\partial\nu}\,d\sigma:=\frac{1}{4\pi G}\lim_{r\to\infty}\int_{S^2_{r}}\frac{\partial U}{\partial\nu}\,d\sigma$ is well-known as the {\it Komar mass} of the system, cf.~Arthur Komar's original work \cite{Komar2}.
\end{Rem}
And indeed we can see that this pseudo-Newtonian mass is independent of the surface if the surface encloses the support of the matter just as in the Newtonian case.
\begin{Prop}\label{prop:samesame}
Let $\mathcal{S}_{PN}=(E^3,\gamma,U,\rho,S)$ be a pseudo-Newtonian end and let $\Sigma_{i}\subset E^3,\,i=1,2$, be smooth surfaces with outer unit normals $\nu_{i}$ and surface measures $d\sigma_{i}$ both enclosing the support of the matter. Then $m_{PN}(\Sigma_{1})=m_{PN}(\Sigma_{2})$.
\end{Prop}
\begin{Pf}
Let $(x^{i})$ be a system of asymptotically flat coordinates for $E^3$ and let $R>0$ be so large that $S^2_{R}$ does intersect neither $\Sigma_{1}$ nor $\Sigma_{2}$ and does enclose the support of the matter. Let $\nu$ and $d\sigma$ denote the outer unit normal and surface measure of $S^2_{R}$, respectively, and let $\Omega_{i}$ denote the cylindrical domain enclosed by $\Sigma_{i}$ and $S^2_{R}$. We calculate
\begin{eqnarray*}
0=\int_{\Omega_{i}}\triangle U\,d\mu&\stackrel{\mbox{\scriptsize div.~thm.}}{=}&\int_{S^2_{R}}\frac{\partial U}{\partial\nu}d\sigma-\int_{\Sigma_{i}}\frac{\partial U}{\partial\nu_{i}}d\sigma_{i}\quad\mbox{ for }i=1,2\\
\Leftrightarrow\quad \int_{\Sigma_{1}}\frac{\partial U}{\partial\nu_{1}}d\sigma_{1}&=&\int_{\Sigma_{2}}\frac{\partial U}{\partial\nu_{2}}d\sigma_{2}
\end{eqnarray*}
which proves the claim.
\qed\end{Pf}

As announced above, the pseudo-Newtonian mass of a surface enclosing the support of the matter agrees with the physical mass of the system or in other words with the ADM-mass of the corresponding geometrostatic metric. This is a direct consequence of the above proposition in combination with a result by Robert Beig \cite{BeigADM} who proves that the Komar mass agrees with the ADM-mass for stationary and so in particular for static metrics. In the static case, though, an easy direct argument works, so we prefer to give our own proof of this fact.
\begin{Thm}\label{ADM=pseudo}
Let $\mathcal{S}_{PN}=(E^3,\gamma,U,\rho,S)$ be a pseudo-Newtonian end with physical mass $m=m_{ADM}(E^3,g=e^{-2U/c^2}\gamma)$ and let $\Sigma\subset E^3$ be any smooth surface enclosing the support of the matter. Then $m=m_{ADM}(E^3,g)=m_{PN}(\Sigma)$ holds irrespective of the specific position and form of $\Sigma$.
\end{Thm}
\begin{Pf}
We know from our asymptotic considerations in Theorem \ref{confasym} that $U$ and $\gamma_{ij}$ can be expanded as $U+\frac{mG}{r}\in W^{k+1,2}_{-(\tau+1)}(E^3)$ and $\gamma_{ij}-\delta_{ij}\in W^{k,2}_{-(\tau+1)}(E^3)$ in asymptotically flat $\gamma$-harmonic coordinates. On any large coordinate sphere $S^2_{R}$, this implies $\nu^k-\my{\delta}{\nu}^k\in W^{k,2}_{-(\tau-\varepsilon+1)}(E^3)$ and $d\sigma-d\sigma_{\delta}\in W^{k,2}_{-(\tau-\varepsilon+1)}(E^3)$ for all $0<\varepsilon<\tau/2$ by the multiplication theorem \ref{multi} and thus in the suggestive notation introduced in Section \ref{sec:CoMasym}
\begin{eqnarray*}
\int_{S^2_{R}}\frac{\partial U}{\partial\nu}d\sigma&=&\int_{S^2_{R}}\left(\frac{mGx_{i}}{R^3}+W^{k,2}_{-(\tau+2)}(E^3)\right)\left(\frac{x^{i}}{R}+W^{k,2}_{-(\tau-\varepsilon+1)}(E^3)\right)d\sigma\\
&\stackrel{\ref{multi}}{=}&mG\int_{S^2_{R}}\frac{1}{R^2}d\sigma+\int_{S^2_{R}}\left(W^{k,2}_{-(\tau-\varepsilon+2)}(E^3)\right)d\sigma\\
&=&4\pi mG+W^{k,2}_{-(\tau-2\varepsilon)}(E^3).
\end{eqnarray*}
In consequence, we find that
\begin{eqnarray*}
m_{PN}(\Sigma)&\stackrel{\ref{prop:samesame}}{=}&m_{PN}(S^2_{R})\stackrel{\mbox{\scriptsize by def.}}{=}\frac{1}{4\pi G}\int_{S^2_{R}}\frac{\partial U}{\partial\nu}d\sigma\stackrel{\ref{prop:samesame}}{=}\lim_{R\to\infty}\frac{1}{4\pi G}\int_{S^2_{R}}\frac{\partial U}{\partial\nu}d\sigma\\
&\stackrel{\mbox{\scriptsize see above}}{=}&\lim_{R\to\infty}\left(m+W^{k,2}_{-(\tau-2\varepsilon)}(E^3)\right)=m
\end{eqnarray*}
which proves the claim.
\qed\end{Pf}

According to the rules of conformal transformation, we can translate the definition of pseudo-Newtonian mass so that is can be stated in the language of geometrostatics, directly. The result is formulated in the following corollary. It will become useful in the proof of the static positive mass theorem and in the proof of photon sphere uniqueness, cf.~Chapter \ref{chap:FT} for both of them.
\begin{Coro}\label{formula:NU}
Let $(E^3,g,N,\rho,S)$ be a geometrostatic system and let $\gamma$ and $U$ denote the corresponding pseudo-Newtonian metric and potential, respectively. Then we have
\begin{equation*}
\int_\Sigma \frac{\partial U}{\partial\,\my{\gamma}{\nu}}\,d\sigma_\gamma=c^2\int_\Sigma \frac{\partial N}{\partial\,\my{g}{\nu}}\,d\sigma_g
\end{equation*}
and therefore $$m_{ADM}(E^3,g)=\frac{c^2}{4\pi G}\int_{\Sigma}\frac{\partial N}{\partial\,\my{g}{\nu}}d\sigma_{g}$$ for any smooth surface $\Sigma$ enclosing the support of the matter.
\end{Coro}
\begin{Pf}
Immediate from \ref{ADM=pseudo} by a direct computation.
\qed\end{Pf}

We have hence found a notion of quasi-local mass in geometrostatics -- or rather in pseudo-Newtonian gravity -- that accommodates the intuitive expectations described above. The pseudo-Newtonian mass will remain well-defined even if the matter does not have compact support but falls off suitably fast as $r$ approaches infinity, but will then clearly not deliver the full ADM-mass on any finite surface. Nevertheless, it will continue to constitute a physically relevant and analytically useful quasi-local concept of mass. In particular, it is non-negative and monotone with respect to inclusion by the divergence theorem and by $\mylap{g} N=4\pi G c^{-2}N\left(\rho+\my{g}{\tr}Sc^{-2}\right)\geq0$ or $\mylap{\gamma}U=4\pi G\left(\rho e^{-2c^{-2}U}+\my{\gamma}{\tr}S c^{-2}\right)\geq0$. It is zero for the flat spacetime and agrees with the ADM-mass in the limit of the coordinate spheres, and it is exactly the mass parameter in the spherically symmetric (Schwarzschild) setting and thus satisfies many of the criteria both Demetrios Christodoulou and Shing-Tung Yau \cite{ChrisYau} and Robert Bartnik \cite{Bartnik3} have suggested for a quasi-local notion of mass albeit only in the static asymptotically flat setting.

The pseudo-Newtonian mass has another practical property analogous to the Newtonian mass: The masses of separate bodies can be added. This is again due to the divergence theorem.
\begin{Thm}(Addition of Mass)\label{thm:madd}
Let $\mathcal{S}_{PN}=(E^3,\gamma,U,\rho,S)$ be a pseudo-Newtonian end with physical mass $m$ and let $\Sigma_{i}\subset E^3$, $i=1,2,\dots,I$ be a finite number of non-interesecting smooth orientable surfaces not contained inside each other and together enclosing the support of the matter (i.~e.~each component of the support of the matter is enclosed by one of the surfaces and none of the surfaces is enclosed (``shielded from infinity'') by any other one). Then $$\sum_{i=1}^{I}m_{i}:=\sum_{i=1}^{I}m_{PN}(\Sigma_{i})=m.$$
\end{Thm}
\begin{Pf}
By assumption, there is a domain $\Omega\subset E^3$ in which $\rho,S=0$ and the boundary of which consists of all the $\Sigma_{i}$ and of ``infinity'' (or any large coordinate sphere $S^2_{R}$) -- imagine $\Omega$ to look like a pair of trousers with $I$ legs. We have
\begin{eqnarray*}
4\pi G\sum_{i=1}^{I}m_{i}&=&\sum_{i=1}^{I}\int_{\Sigma_{i}}\frac{\partial U}{\partial\nu}d\sigma\\
&\stackrel{\mbox{\scriptsize div. thm}}{=}&-\int_{\Omega}\triangle U\,d\mu+\int_{S^2_{R}}\frac{\partial U}{\partial\nu}d\sigma\\
&\stackrel{\eqref{ConfSME},\,\ref{ADM=pseudo}}{=}&\quad\quad\;\; 0\quad\quad+\quad4\pi Gm
\end{eqnarray*}
which we proves the claim.\qed\end{Pf}

We close this section by remarking that the pseudo-Newtonian mass seems to be a good tool for attacking problems in geometrostatics. Some examples of its usefulness will be provided  in Chapter \ref{chap:geo}.

\section{The Pseudo-Newtonian Center of Mass}\label{sec:CoMpseudo}
Arguing similarly as we did above for the mass, it would be desirable to have a quasi-local (surface integral) expression for the center of mass which gives the exact answer already in a vicinity of a material body or black hole. This would not only be useful for technical considerations but also give intuitive and physical insight into and justification of the definitions of center of mass described above, cf.~Section \ref{sec:HCoM}. To define such a notion will be our goal in this section. We will again imitate the Newtonian situation and begin by describing it. As above, let $U$ be the Newtonian potential for a mass distribution $\rho$, assume that the Newtonian mass $m_{N}$ of the system does not vanish, and let $(x^{i})$ denote Cartesian (and thus in particular harmonic) coordinates on $\R^3$. Then the Newtonian center of mass\index{ind}{center of mass}\index{ind}{center of mass ! Newtonian}\index{ind}{Newtonian ! center of mass} $\vec{z}_{N}=(z^1_N,z^2_{N},z^3_{N})^t\in\R^3$ is defined by
\begin{eqnarray*}
z_{N}^k:=\frac{1}{m_{N}}\int_{\R^3}\rho x^k\,d\mu=\frac{1}{m_{N}}\int_{\Omega}\rho x^k\,d\mu,\index{sym}{$\vec{z}_{N}$}
\end{eqnarray*}
where again $d\mu$ denotes the Lebesgue measure on $\R^3$ and $\Omega\subset\R^3$ is any domain enveloping the support of the matter. By the Poisson equation for the Newtonian potential and Green's formula, this can be rephrased into the geometric expression 
\begin{equation*}
z^k_{N}=\frac{1}{4\pi Gm_{N}}\int_{\Omega}\triangle U x^k\,d\mu=\frac{1}{4\pi Gm_{N}}\int_{\partial\Omega}\left(\frac{\partial U}{\partial\nu}x^k-U\frac{\partial x^k}{\partial\nu}\right)d\sigma
\end{equation*}
as $\triangle x^k=0$, whenever $\Omega$ has smooth boundary and $\nu, d\sigma$ are as above. Observe that if $\partial\Omega$ were a regular level set of $U$, the second term on the right hand side of the above equation vanishes by as $$\int_{\partial\Omega}U\frac{\partial x^k}{\partial\nu}d\sigma=U\int_{\partial\Omega}\frac{\partial x^k}{\partial\nu}d\sigma=U\int_{\Omega}\triangle x^k\,d\sigma=0$$ holds by the divergence theorem. We put the cart before the horse and define the pseudo-Newtonian center of mass in analogy to the above quasi-local surface integral expression.
\begin{Def}[Quasi-Local Pseudo-Newtonian Center of Mass]\label{pseudo-CoM}
Let $(E^3,\gamma,U,\rho,S)$ be a pseudo-Newtonian end with non-vanishing physical mass $m$ and let $(x^{i})$ be a system of asymptotically flat $\gamma$-harmonic coordinates in $E^3$. For any smooth orientable surface $\Sigma\subset E^3$ with outer unit normal $\nu$ and surface measure $d\sigma$ both induced from $\gamma$, the {\it pseudo-Newtonian quasi-local center of mass}\index{ind}{center of mass ! pseudo-Newtonian}\index{ind}{pseudo-Newtonian ! center of mass}\index{sym}{$\vec{z}_{PN}(\Sigma)$} $\vec{z}_{PN}(\Sigma)=(z^1_{PN}(\Sigma),z^2_{PN}(\Sigma),z^3_{PN}(\Sigma))^t\in\R^3$ is defined by $$z^k_{PN}(\Sigma):=\frac{1}{4\pi Gm}\left(\int_{\Sigma} \frac{\partial U}{\partial\nu}x^k-U\frac{\partial x^k}{\partial\nu}\right)\,d\sigma.$$
\end{Def}
Just as the pseudo-Newtonian mass expression has been known before at infinity as the Komar mass, considerations similar to the pseudo-Newtonian concept of center of mass have been pursued asymptotically in the special cases of asymptotically harmonic and asymptotically conformally flat metrics, cf.~the overview article \cite{CoPo} by Justin Corvino and Daniel Pollack and the paper \cite{CoWu} by Justin Corvino and Haotian Wu. In the same sense as the pseudo-Newtonian mass, the pseudo-Newtonian center of mass is independent of a particular surface if this surface encloses the support of the matter.
\begin{Prop}\label{prop:samesameCoM}
Let $\mathcal{S}_{PN}=(E^3,\gamma,U,\rho,S)$ be a pseudo-Newtonian end with non-vanishing physical mass $m$, let $(x^{i})$ be a system of asymptotically flat $\gamma$-harmonic coordinates on $E^3$, and let $\Sigma_{i}\subset E^3,\,i=1,2$ be smooth surfaces with outer unit normals $\nu_{i}$ and surface measures $d\sigma_{i}$ both enclosing the support of the matter. Then $\vec{z}_{PN}(\Sigma_{1})=\vec{z}_{PN}(\Sigma_{2})$.
\end{Prop}
\begin{Pf}
Let again $R>0$ be so large that $S^2_{R}$ does intersect neither $\Sigma_{1}$ nor $\Sigma_{2}$ and encloses the support of the matter. Let $\nu$ and $d\sigma$ denote the outer unit normal and surface measure of $S^2_{R}$, respectively, and let $\Omega_{i}$ denote the cylindrical domain bounded by $\Sigma_{i}$ and $S^2_{R}$. Using again Green's formula and the fact that the coordinates are harmonic, we find
\begin{eqnarray*}
0=\int_{\Omega_{i}}\left(\triangle U x^k-U\triangle x^k\right)\,d\mu&=&\int_{S^2_{R}}\left(\frac{\partial U}{\partial\nu}x^k-U\frac{\partial x^k}{\partial\nu}\right)d\sigma-\int_{\Sigma_{i}}\left(\frac{\partial U}{\partial\nu_{i}}x^k-U\frac{\partial x^k}{\partial\nu_{i}}\right)d\sigma_{i}\\
\Leftrightarrow \int_{\Sigma_{1}}\left(\frac{\partial U}{\partial\nu_{1}}x^k-U\frac{\partial x^k}{\partial\nu_{1}}\right)d\sigma_{1}&=&\int_{\Sigma_{2}}\left(\frac{\partial U}{\partial\nu_{2}}x^k-U\frac{\partial x^k}{\partial\nu_{2}}\right)d\sigma_{2}.
\end{eqnarray*}
This proves the claim.
\qed\end{Pf}
\noindent The expression for the pseudo-Newtonian center of mass can be simplified on regular spherical level set surfaces just as in the Newtonian case. Such surfaces exist in a neighborhood of infinity by Lemma \ref{lem:foli} if the physical mass vanishes.
\begin{Prop}\label{prop:CoMequal}
Let $\mathcal{S}_{PN}=(E^3,\gamma,U,\rho,S)$ be a pseudo-Newtonian end with non-vanishing physical mass $m$, let $(x^{i})$ be a system of asymptotically flat $\gamma$-harmonic coordinates on $E^3$, and let $\Sigma\subset E^3$ be a regular spherical level set of $U$ enclosing the support of the matter. Then $$z^k_{PN}(\Sigma)=\frac{1}{4\pi mG}\int_{\Sigma}\frac{\partial U}{\partial\nu}x^k\,d\sigma.$$
\end{Prop}
\begin{Pf}
All we have to show is that $\int_{\Sigma}\frac{\partial x^k}{\partial\nu}d\sigma=0$. By the divergence theorem and the assumption that the coordinates are harmonic, the left hand side of this equation is independent of the surface $\Sigma$ in the sense that $\int_{\Sigma}\frac{\partial x^k}{\partial\nu}d\sigma=\int_{\widetilde{\Sigma}}\frac{\partial x^k}{\partial\nu}d\sigma$ for any smooth surface $\widetilde{\Sigma}\subset E^3$ also enclosing the support of the matter. Choosing $\widetilde{\Sigma}=S^2_{R}$ for $R>0$ suitably large and using the asymptotics of $\gamma_{ij}$ proven in Theorem \ref{confasym}, we have $$\int_{\Sigma}\frac{\partial x^k}{\partial\nu}d\sigma=\lim_{R\to\infty}\int_{S^2_{R}}\frac{\partial x^k}{\partial\nu}d\sigma=\lim_{R\to\infty}\int_{S^2_{R}}\frac{\partial x^k}{\partial\,\my{\delta}{\nu}}d\sigma_{\delta},$$ where the far right hand side vanishes by the above Newtonian considerations which finishes the proof. 
\qed\end{Pf}

Our next step will be to show that the pseudo-Newtonian center of mass on any suitably ``large'' surface (i.~e.~on any surface enclosing the support of the matter) coincides with the centers of mass described and defined above. The following theorem states that all of these centers in fact agree. Before we prove it, let us consider the following lemma stating the same fact for the Schwarzschild metrics (cf.~Proposition \ref{prop:schwarzgeo}).
\begin{Lem}\label{lem:schwarzCoM}
Let $m\neq0$, $M:=mG/c^2$, and $M^\ast>M$. Then the pseudo-Newtonian Schwarzschild system $\mathcal{S}_{PN}:=(E_{M^\ast}(0),\mygam{S},\my{S}{U})$ of mass $m$ has $\vec{z}_{PN}(\Sigma)=0$ on any smooth surface $\Sigma\subset E_{^3M^\ast}$ enclosing the support of the matter in the canonical asymptotically flat harmonic coordinates $(y^{i})$.
\end{Lem}
\begin{Pf}
Straighforward computation from \eqref{Ufall} and \eqref{gamfall}.
\qed\end{Pf}

\begin{Thm}[Centers of Mass Coincide]\label{CMC=pseudo}\label{asym=pseudo}
Let $\mathcal{S}_{PN}=(E^3,\gamma,U,\rho,S)$ be a $(k,\tau)$-pseudo-Newtonian end with non-vanishing physical mass $m$ and $\tau>1/2$, let $(x^{i})$ be a system of asymptotically flat $\gamma$-harmonic coordinates on $E^3$ and let $\Sigma\subset R^3$ be any smooth surface enclosing the support of the matter. Then
$$\vec{z}_{PN}(\Sigma)=\vec{z}_{ADM}(E^3,g)=\vec{z}_{A}(E^3,g)=\vec{z}_{CMC}(E^3,g)=\vec{z}_{I}(E^3,g),$$
where $g$ is the geometrostatic metric associated with $\mathcal{S}_{PN}$.

Differently put, if $\mathcal{S}=(E^3,g,N,\rho,S)$ is a $(k,\tau)$-geometrostatic system with $\tau>1/2$ and non-vanishing ADM-mass $m$ and $\Sigma$ is as above, then
$\vec{z}_{PN}(\Sigma)=\vec{z}_{ADM}=\vec{z}_{A}=\vec{z}_{CMC}=\vec{z}_{I}$
holds with respect to wave harmonic asymptotically flat coordinates on $E^3$, where $\vec{z}_{PN}(\Sigma)$ refers to the corresponding pseudo-Newtonian metric $\gamma$ and potential $U$.
\end{Thm}
\begin{Pf}
The two formulations in the theorem are equivalent by Lemma \ref{waveharm} and we already know from Theorem \ref{thm:centers} that all centers but the pseudo-Newtonian one coincide. Moreover, the asymptotic expansion in Theorem \ref{confasym} tells us that
\begin{eqnarray*}
U-\my{S}{U}+\frac{mG\vec{z}_{A}\cdot\vec{x}}{r^3}&\in& W^{k+1,2}_{-(\tau-\varepsilon+2)}(E^3)\\
\gamma_{ij}-\mygam{S}_{ij}&\in& W^{k,2}_{-(\tau-\varepsilon+2)}(E^3)
\end{eqnarray*}
for all $0<\varepsilon\leq\tau-1/2$, where $\mygam{S}$ and $\my{S}{U}$ are the pseudo-Newtonian Schwarzschild metric and potential of mass $m$. Using again Proposition \ref{prop:samesameCoM}, it suffices to show that $$\lim_{R\to\infty}\int_{S^2_{R}}\left(\frac{\partial U}{\partial\nu}x^k-U\frac{\partial x^k}{\partial\nu}\right)d\sigma=4\pi m Gz^k_{A}(E^3,g).$$ We will proceed as in the proof of Theorem \ref{ADM=pseudo} dropping the label $A$ on $\vec{z}_{A}$ for simplicty:
\begin{eqnarray*}
\int_{S^2_{R}}\left(\frac{\partial U}{\partial\nu}x^k-U\frac{\partial x^k}{\partial\nu}\right)d\sigma&=&\int_{S^2_{R}}\left(\my{S}{U}_{,i}-\frac{mGz_{i}}{R^3}+\frac{3mG\vec{z}\cdot\vec{x}x_{i}}{R^5}+W^{k,2}_{-(\tau-\varepsilon+3)}(E^3)\right)\nu^{i}x^k\,d\sigma\\
&&-\int_{S^2_{R}}\left(\my{S}{U}-\frac{mG\vec{z}\cdot\vec{x}}{R^3}+W^{k+1,2}_{-(\tau-\varepsilon+2)}(E^3)\right)\frac{\partial x^k}{\partial\nu}d\sigma\\
&=&\int_{S^2_{R}}\left(-\frac{mGz_{i}}{R^3}+\frac{3mG\vec{z}\cdot\vec{x}x_{i}}{R^5}+W^{k,2}_{-(\tau-\varepsilon+3)}(E^3)\right)\nu^{i}x^k\,d\sigma\\
&&-\int_{S^2_{R}}\left(-\frac{mG\vec{z}\cdot\vec{x}}{R^3}+W^{k+1,2}_{-(\tau-\varepsilon+2)}(E^3)\right)\frac{\partial x^k}{\partial\nu}d\sigma\\
&&+\int_{S^2_{R}}\left(\frac{\partial\my{S}{U}}{\partial\nu}x^k-\my{S}{U}\frac{\partial x^k}{\partial\nu}\right)\,d\sigma.
\end{eqnarray*}
The last term of the right hand side vanishes in the limit $R\to\infty$ by Lemma \ref{lem:schwarzCoM} and because of $\gamma_{ij}-\mygam{S}_{ij}\in W^{k,2}_{-(\tau-\varepsilon+2)}(E^3)$ and so do the lower order terms indicated suggestively by printing the weighted Sobolev spaces. For the same reason, and using the multiplication theorem \ref{multi}, the remaining terms induce $$\lim_{R\to\infty}\int_{S^2_{R}}\left(\frac{\partial U}{\partial\nu}x^k-U\frac{\partial x^k}{\partial\nu}\right)d\sigma=4\pi mGz^k$$ which proves our claim.
\qed\end{Pf}

We can now prove a pseudo-Newtonian center of mass addition theorem in analogy to the well-known one in Newtonian theory. For abbreviational purposes, we write $m_{PN}:=m_{PN}(\Sigma)$ and $\vec{z}_{PN}:=\vec{z}_{PN}(\Sigma)$ where $\Sigma$ is any smooth surface enclosing the support of the matter. Similarly, we drop the qualifiers ADM-, physical, pseudo-Newtonian etc.~referring to either mass or center of mass as the above theorem ensures that no confusion can thereby arise.\index{ind}{center of mass}\index{ind}{mass}\index{sym}{$m_{PN}$}\index{sym}{$\vec{z}_{PN}$}
\begin{Thm}(Addition of Centers of Mass)\label{thm:CoMadd}
Let $\mathcal{S}_{PN}=(E^3,\gamma,U,\rho,S)$ be a pseudo-Newtonian end with mass $m\neq0$ and center of mass $\vec{z}$ with respect to some system of asymptotically flat harmonic coordinates. Let $\Sigma_{i}\subset E^3$, $i=1,2,\dots,I$ be a finite number of non-interesecting smooth surfaces not contained inside each other and together enclosing the support of the matter (i.~e.~each component of the support of the matter is enclosed by one the surfaces and none of the surfaces is ``shielded from infinity'' by any other one). Then $$\sum_{i=1}^{I}m_{i}\vec{z}_{i}:=\sum_{i=1}^{I}m_{PN}(\Sigma_{i})\vec{z}_{PN}(\Sigma_{i})=m\vec{z}.$$
\end{Thm}
\begin{Pf}
We proceed as in the proof of Theorem \ref{thm:madd}. By assumption, there is a pair of trousers with $I$ legs shaped domain domain $\Omega$ in which $\rho,S=0$ and the boundary of which consists of all the $\Sigma_{i}$ and of a suitably large coordinate sphere $S^2_{R}$. We find that
\begin{eqnarray*}
4\pi G\sum_{i=1}^{I}m_{i}\vec{z}_{i}&=&\sum_{i=1}^{I}\int_{\Sigma_{i}}\left(\frac{\partial U}{\partial\nu}x^k-U\frac{\partial x^k}{\partial\nu}\right)d\sigma\\
&\stackrel{\mbox{\scriptsize Green's formula}}{=}&-\int_{\Omega}\left(\triangle U x^k-U\triangle x^k\right)d\mu+\int_{S^2_{R}}\left(\frac{\partial U}{\partial\nu}x^k-U\frac{\partial x^k}{\partial\nu}\right)d\sigma\\
&\stackrel{\eqref{ConfSME},\,\ref{CMC=pseudo}}{=}&\quad\quad\quad\quad\quad\quad\;\, 0\quad\quad\quad\quad+\quad\quad\quad4\pi Gm\vec{z}
\end{eqnarray*}
as the coordinates were chosen to be $\gamma$-harmonic. We have thus proven the formula.
\qed\end{Pf}
\chapter{Linking Physics to Geometry}\label{chap:geo}
In the last chapters, we have studied asymptotic properties of geometrostatic and pseudo-Newtonian systems as well as their masses and centers of masses. In this chapter, we will now discuss further physical properties and facts that hold in geometrostatics and/or pseudo-Newtonian gravity. In Section \ref{sec:pm}, we will prove a static version of the famous positive mass theorem. In Section \ref{sec:2nd}, we reduce the notion of test particles which is well-known in geometrodynamics to the geometrostatic setting and prove a pseudo-Newtonian version of Newton's second law of motion. In Section \eqref{sec:equi}, we will discuss constrained test particles and surfaces of equilibrium and their consequences for uniqueness of the lapse function while in Section \eqref{sec:photon}, we will prove a rigidity theorem on static photon spheres.

\section{The Static Positive Mass Theorem}\label{sec:pm}
One of the major breakthroughs in the mathematical study of general relativity is the positive mass theorem by Richard M.~Schoen and Shing-Tung Yau. Different proofs have been put forward by a number of people under more restrictive conditions, see the discussion on p.~\pageref{positive mass thm}. In geometrostatics, a much simpler proof works. The proof illustrates the usefulness of Theorem \ref{ADM=pseudo}. Moreover, the static rigidity statement also includes $N=1$ and vanishing of the matter fields and therefore reproduces parts of the results of Andr\'e Lichnerowicz and Michael T.~Anderson stated in Theorems \ref{Lich} and \ref{Anders}. 
\begin{Thm}[Static Positive Mass Theorem]\label{thm:spm}
Let $\mathcal{S}:=(M^3,g,N,\rho,S)$ be a geometrostatics system. Assume that either $g$ is a geodesically complete metric or that $N(p)\to0$ as $p$ approaches the inner boundary\footnote{i.~e.~that all inner boundaries are Killing horizons\index{ind}{horizon}\index{ind}{Killing horizon} in the sense that the static Killing vector field becomes lightlike there, cf.~the remark on page \pageref{rem:horizon}.}. Then the ADM-mass of $\mathcal{S}$ is non-negative in each end. In case $g$ is geodesically complete, the ADM-mass is zero in one end if and only if $g$ is flat in all of $M^3$, $N=1$ in $M^3$, and all matter fields vanish identically (and whence there were no inner boundaries in the first place).
\end{Thm}
\begin{Pf}
Let us prove the rigidity statement, first. If a geometrostatic system has vanishing ADM-mass, it also has vanishing pseudo-Newtonian mass by Theorem \ref{ADM=pseudo}. By \eqref{formula:NU} and using the abbreviational notation $\nu(f):=\partial f/\partial\nu$ for a smooth function $f$\index{sym}{$\nu(N)$}, this means that $$c^2\int_\Sigma \my{g}{\nu}(N)\,d\sigma_g=\int_\Sigma \my{\gamma}{\nu}(U)\,d\sigma_\gamma=0$$ on any smooth surface $\Sigma$ enclosing the support of the matter where $\gamma$ and $U$ are the associated pseudo-Newtonian metric and potential, respectively. As $g$ is geodesically complete by assumption, this leads to
$$0\leq\frac{4\pi G}{c^2}\int_\Omega N\left(\rho+\frac{\my{g}{\tr} S}{c^2}\right)\,d\mu_{g}\stackrel{\eqref{SME}}{=}\int_\Omega\mylap{g} N\,d\mu_{g}
\stackrel{\mbox{\scriptsize{div. thm}}}{=}\int_\Sigma\nu(N)\,d\sigma_{g}
\stackrel{\eqref{formula:NU}}{=}0,
$$
where $\Omega\subset M^3$ denotes the volume enclosed by $\Sigma$ and we have used that $N>0$ in $M^3$ and $\rho+\frac{\my{g}{\tr}S}{c^2}\geq0$ by definition of geometrostatics systems. This implies that $\rho=\my{g}{\tr} S=0$ and by positive semi-definitness of $S$, also $S=0$. In other words, $N$ is a solution of the curvilinear Laplace equation on a the geodesically connected asymptotically flat Riemannian manifold with $N\to1$ as $r\to\infty$ and thus $N=1$ by classical elliptic PDE theory, cf.~e.~g.~\cite{GT}. Finally, the vacuum static metric equations with $N=1$, $\rho=0$, and $S=0$ tell us that $g$ must be Ricci-flat and thus flat by \eqref{formula:3Rm} which proves rigidity.

Let us now discuss non-negativity of mass. Suppose $m:=m_{ADM}(g)<0$ and observe as above that $$m=\frac{c^2}{4\pi G}\int_\Sigma\my{g}{\nu}(N)\,d\sigma_{g}$$ for any surface $\Sigma$ enclosing the support of the matter. By Lemma \ref{lem:foli}, we know that $N$ foliates $M^3\setminus K$ with spherical level sets enclosing the support of the matter for some compact subset $K$. We can therefore deduce that $\my{g}{\nu}(N)\neq0$ and in particular $\my{g}{\nu}(N)<0$ on any of these level sets. We rephrase this into saying that $N$ must be strictly monotonically decreasing along the flowlines of the vector field $\my{g}{\grad} N$ as $r\to\infty$. But then there must exist a point $p\in M^3$ with $N(p)>1$ which contradicts the maximum principle for the elliptic partial differential inequality $$\triangle N=\frac{4\pi G}{c^2}N\left(\rho+\frac{\my{g}{\tr} S}{c^2}\right)\geq0,$$ cf.~Theorem 17.1 in \cite{GT} as $N=0$ on all inner boundaries and $N\to1$ as $r\to\infty$.
\qed\end{Pf}

\section{The Pseudo-Newtonian Second Law of Motion}\label{sec:2nd}
In Newtonian gravity, the {\it gravitational force $\vec{F}$}\index{ind}{force ! gravitational}\index{sym}{$\vec{F}$} acting on a test body is defined as $$\vec{F}:=-m\,\vec{\nabla} U$$ with $m$ the mass of the test particle and $U$ the Newtonian (gravitational) potential. Newton's second law of motion $$\vec{F}=m\vec{a}$$ then relates this force to the {\it acceleration $\vec{a}$}\index{ind}{acceleration}\index{sym}{$\vec{a}$} of the test body. The same actually holds true in pseudo-Newtonian gravity. On order to see this, let us first review the definition of test particles in general relativity and then study what restrictions are imposed on test particles by staticity of the ambient spacetime.

A {\it test body}\index{ind}{test body} or {\it test particle}\index{ind}{test particle} is modeled in general relativity by a smooth timelike geodesic with respect to the Lorentzian $4$-metric $ds^2$ of the system. Recall furthermore that the timelike geodesics are exactly the critical points of the {\it time functional}\index{ind}{time functional} $$T(\kappa):=\frac{1}{c}\int_I \sqrt{-ds^2(\dt{\kappa},\dt{\kappa})}(\tau)\,d\tau,$$ where the variation is calculated among all smooth timelike curves $\kappa:I\to L^4$ with fixed endpoints, $I\subseteq\R$ being an interval and the dot referring to the $4$-dimensional covariant derivative calculated along the curve itself. Equivalently, the timelike geodesics can be shown to agree (up to reparametrization) with the critical points of the {\it energy functional}\index{ind}{energy functional} $$E(\kappa):=-\frac{G}{2c^3}\int_I ds^2(\dt{\kappa},\dt{\kappa})(\tau)\,d\tau$$ where the variation is again restricted to smooth timelike curves $\kappa:I\to L^4$ with fixed endpoints. Recall that the critical points of $E$ are automatically parametrized proportionally to eigentime and that the Euler-Lagrange equation of $E$ is the {\it geodesic equation}\index{ind}{geodesic equation} $\my{4}{\nabla}_{\dt{\kappa}}{\dt{\kappa}}=0$. The expression $\my{4}{\nabla}_{\dt{\kappa}}{\dt{\kappa}}=0$ is usually interpreted as the acceleration of the curve $\kappa$. Now if $\kappa:I\to L^4$ is a test body in a static spacetime $(L^4,ds^2)$, we can use the canonical 3+1 decomposition $L^4=\R\times M^3$ presented in Section \ref{sec:3+1} in order to simplify and better understand the geodesic equation $\my{4}{\nabla}_{\dt{\kappa}}{\dt{\kappa}}=0$. In this decomposition, we can write $\kappa=(t,\mu)$ with $t:I\to\R$ the time component and $\mu:I\to M^3$ the spatial component of $\kappa$. In this spirit,
we now make the following definition.
\begin{Def}
Let $(M^3,\myg{3},N,\rho,S)$ be a geometrostatic system and $\kappa=(t,\mu)$ a test particle in the corresponding Lorentzian manifold. Then the vector field $\vec{a}:=\my{3}{\nabla}_{\dt{\mu}}\dt{\mu}$ along $\mu$ is called the {\it acceleration}\index{ind}{acceleration}\index{sym}{$\vec{a}$} of the test particle.
\end{Def}
Calculating the components of the geodesic equation for $\kappa$, we find that
\begin{eqnarray}\label{Fma1}
0&=&N\circ\mu\,\ddt{t}+dN(\dt{\mu})\,\dt{t}\\\label{Fma2}
\my{3}{\nabla}_{\dt{\mu}}\dt{\mu}&=&-N\circ\mu\,c^2\,\dt{t}^2(\grad N)\circ\mu,
\end{eqnarray}
where the dot denotes differentiation with respect to the eigentime parameter. Differentiating the condition that $\kappa$ is parametrized proportionally to eigentime leads to \begin{equation}\label{Fma3}g(\dt{\mu},\my{3}{\nabla}_{\dt{\mu}}\dt{\mu})=\dt{t}\,\ddt{t}\,(N\circ\mu)^2c^2+\dt{t}^2\,N\circ\mu\,dN(\dt{\mu})c^2
\end{equation}
which combines with Equation \eqref{Fma1} to give $g(\dt{\mu},\my{3}{\nabla}_{\dt{\mu}}\dt{\mu})=0$ as $N>0$ and $\dt{t}\neq0$ (which is due to $\kappa$ being timelike). Together with Equation \eqref{Fma2}, this implies $dN(\dt{\mu})=0$ so that again by Equation \eqref{Fma1}, $\ddt{t}=0$. We have thus proven the following proposition.
\begin{Prop}\label{veca}
Let $(M^3,\myg{3},N,\rho,S)$ be a geometrostatic system and $\kappa=(t,\mu)$ a test particle in the corresponding Lorentzian manifold. Then its acceleration $\vec{a}$ satisfies $$\vec{a}=-\left[N\circ\mu\,c^2\,\dt{t}^2\right](\my{3}{\grad} N)\circ\mu,$$ where the term in square brackets is constant.
\end{Prop}

We continue by making a definition of force analogous to the Newtonian setting.
\begin{Def}
Let $(M^3,\myg{3},N,\rho,S)$ be a geometrostatic system and $\kappa=(t,\mu)$ a test particle of mass $m$ in the corresponding Lorentzian manifold. Then the {\it pseudo-Newtonian force}\index{ind}{force ! pseudo-Newtonian}\index{sym}{$\vec{F}$} $\vec{F}$ exerted on $\kappa$ by ``gravitation'' is defined as $$\vec{F}:=-m\,\my{\gamma}{\grad}\,U,$$ where $\gamma, U$ are the associated pseudo-Newtonian metric and potential, respectively.
\end{Def}

\begin{Rem}
It is natural that the acceleration of the test body is induced from the Lorentzian metric $ds^2$ and thus refers to the geometrostatic metric $g$ and not to the conformally transformed metric $\gamma$ because the definition of test bodies relies on the dynamics of spacetime. On the other hand, the definition of force uses the analogy of pseuo-Newtonian and Newtonian effects and thus should be formulated in pseudo-Newtonian terms.
\end{Rem}

After these considerations, the following theorem now is immediate.

\begin{Thm}[Second Pseudo-Newtonian Law of Motion]\label{thm:F=ma}
Let $(M^3,\myg{3},N,\rho,S)$ be a geometrostatic system and $\kappa=(t,\mu)$ a test particle of mass $m$ in the corresponding Lorentzian manifold. Fix the parametrization of $\kappa$ such that $\dt{t}=1$ along $\kappa$. Then the force $\vec{F}$ acting on $\kappa$ and its acceleration $\vec{a}$ satisfy the pseudo-Newtonian second law of motion $$\vec{F}=m\vec{a}.$$
\end{Thm}
\begin{Rem}
The statement of the theorem is not true if $\dt{t}\neq\pm1$. However, it is natural that we have to fix $\dt{t}$ because we have actually fixed the ``speed'' of background time, i.~e.~time measured by the lapse function $N$ by setting $N\to1$ as $r\to\infty$. $\dt{t}=1$ then just says that $\kappa$ is future-oriented (which is implicit in the Newtonian setting) and that $\kappa$ actually uses the ``same clock'' as the observer at infinity.
\end{Rem}

\section{Surfaces of Equilibrium}\label{sec:equi}
It is well-known in Classical Mechanics that the level sets of the Newtonian potential -- or the {\it equipotential} surfaces\index{ind}{surface ! equipotential} -- can be characterized by the behavior of constrained test bodies\footnote{We tacitly assume here that the Newtonian potential foliates $\R^3$ in a neighborhood of the surface under consideration. This is not a severe restriction (unless $\rho=0$ on all of $\R^3$) as it is well-known that the potential foliates $\R^3\setminus K$ for some compact set $K\subset\R^3$ enclosing the support of the matter, $\supp\rho\subset K$. This can be derived from its asymptotics just as in Lemma \ref{lem:foli}.}: Any test body which is forced to move inside an equipotential surface and which is otherwise subject only to gravitational forces will not accelerate. On the other hand, this behavior cannot be displayed by test bodies constrained to any non-equipotential surface. Phrased differently, equipotential surfaces are exactly those surfaces for which the virtual force exerted by the constraint that a test body must move along the surface does indeed exactly compensate for the gravitational force so that the (real and virtual) forces are in equilibrium. Because of this characterizing property, we will also call them {\it (Newtonian) surfaces of equilibrium}\index{ind}{surface ! of equilibrium}\index{ind}{surface of equilibrium ! Newtonian} in what follows.

The fact that a surface in a Newtonian gravitational system is a surface of equilibrium if and only if it is a level set of the associated Newtonian potential can be derived using d'Alembert's principle of virtual forces or a Lagrangian approach. A similar variational approach will now help us to define and identify surfaces of equilibrium in geometrostatics. We will not only see that surfaces of equilibrium do exist in geometrostatic systems but also identify them as the level sets of the corresponding lapse functions. This already hints at physical relevance of the lapse function which we will discuss further at the end of this section. Moreover, we will use this characterization of the surfaces of equilibrium to prove a uniqueness result at the end of this section.

 Let us begin our study of surfaces of equilibrium in geometrostatics by giving the following definition.
\begin{Def}
Let $(M^3,\myg{3},N,\rho,S)$ be a geometrostatic system, $\Sigma\subset M^3$ a smooth surface. A {\it test body constrained to $\Sigma$}\index{ind}{test body ! constrained} is a smooth timelike curve $\overline{\kappa}:I\to\R\times M^3$ defined on a compact interval $I\subseteq\R$ which is a critical point of the energy functional
\begin{equation}\label{energy functional}
E(\kappa):=-\frac{G}{2c^3}\int_I ds^2(\dt{\kappa},\dt{\kappa})(\tau)\,d\tau
\end{equation}
with respect to the induced Lorentzian $4$-metric
$ds^2$ given by \eqref{ds^2} where the variation is confined to all smooth timelike curves $\kappa=(t,\mu):I\to\R\times M^3$ with fixed endpoints that satisfy $\mu(I)\subset\Sigma$. $\Sigma$ is called a {\it (geometrostatic) surface of equilibrium}\index{ind}{surface ! of equilibrium}\index{ind}{surface of equilibrium ! relativistic} if the spatial parts $\mu:I\to\Sigma$ of all test bodies $\kappa=(t,\mu)$ constrained to $\Sigma$ are geodesics with respect to the induced metric on $\Sigma$.
\end{Def}

\begin{Rem}
The non-acceleration of constrained test bodies from the Newtonian setting is translated into the condition that the constrained test bodies are geodesics with respect to the induced metric on the surface. This seems reasonable as the geodesic equation $\my{2}{\nabla}_{\dt{\mu}}\dt{\mu}=0$ along the curve $\mu$ can be understood to state that the curve is not intrinsically accelerating.
\end{Rem}

Let us now turn to showing that surfaces of equilibrium exist and do indeed agree with the level sets of the lapse function. To this end, we need the following lemma.
\begin{Lem}\label{lem:equili1}
Let $(M^3,\myg{3},N,\rho,S)$ be a geometrostatic system, $\Sigma\subset M^3$ a smooth surface and $\kappa=(t,\mu):I\to\R\times\Sigma$ a test body constrained to $\Sigma$. Then $\kappa$ is automatically parametrized proportionally to eigentime.
\end{Lem}
\begin{Pf}
Let $n:U\to\R$ be a smooth function on a neighborhood $U\subset M^3$ of $\Sigma$ with $dn\neq0$ on $U$ and $n\vert_\Sigma\equiv n_0$.  By definition, $\kappa$ is a critical point of the energy functional \eqref{energy functional} under the constraint that $n\equiv n_0$. By the theory of Lagrangian multipliers\footnote{cf.~e.~g.~pp.~270f in \cite{Zeidler}.}, there is a smooth function $\alpha:I\to\R$ such that $(\kappa,\alpha)$ is a critical point of the modified functional
\begin{equation*}
\tilde{F}(\kappa,\alpha):=\int_I\left(-\frac{1}{2}ds^2(\dt{\kappa},\dt{\kappa})(\tau)+\alpha\left[n\circ\mu_\kappa-n_0\right]\right)d\tau
\end{equation*}
varied among all smooth $(\kappa=(t_\kappa,\mu_\kappa),\beta):I\to(\R\times M^3)\times\R\times\R$ where $\kappa$ has fixed endpoints. Spelling out $ds^2$ in terms of $\myg{3}$ and $N$, we obtain the induced functional
\begin{eqnarray}\nonumber
F(t,\mu,\alpha)&:=&\tilde{F}((t,\mu),\alpha)\\[1ex]\label{F}
&=&\int_I\left(\frac{c^2 (N\circ\mu)^2}{2}\,\dt{t}^2-\frac{1}{2}\vert\dt{\mu}\vert_3^2+\alpha\left[n\circ\mu_\kappa-n_0\right]\right)(\tau)\,d\tau.
\end{eqnarray}
The Euler-Lagrange equations of $F$ with respect to $t$ and $\mu$ read
\begin{eqnarray}\label{ELeqt1}
0&=&-c^2\frac{d}{d\tau}\left[(N\circ\mu)^2\,\dt{t}\,\right]\\\label{ELeqmu1}
0&=&c^2 N\circ\mu\,\dt{t}^2\,dN+\frac{d}{d\tau}\left[\myg{3}(\dt{\mu},\cdot)\right]+\alpha\,dn,
\end{eqnarray}
and, using the chain rule and the Levi-Civit\`a properties of the connection, we obtain 
\begin{eqnarray*}
0&=&2 (N\circ\mu)\,dN(\dt{\mu})\,\dt{t}^2+(N\circ\mu)^2\,\ddt{t}\\\
0&=&c^2 N\circ\mu\,\dt{t}^2\,dN+\myg{3}(\ddt{\mu},\cdot)+\alpha\,dn.
\end{eqnarray*}
Combining Equation \eqref{ELeqt1} with Equation \eqref{ELeqmu1} applied to $\dt{\mu}$ we observe that $ds^2(\dt{\kappa},\dt{\kappa})\equiv\mbox{const}$ as $dn(\dt{\mu})=0$ by construction and as $\dt{t}\neq0$ because $\kappa$ is timelike. Thus $\kappa$ is parametrized proportionally to eigentime.
\qed\end{Pf}

\begin{Thm}[Surfaces of Equilibrium]\label{equili}
Let $(M^3,\myg{3},N,\rho,S)$ be a geometrostatic system and let $K\subset M^3$ be a closed subset bounded away from infinity with $\supp\rho\subset K$ such that $N$ foliates each end of $M^3\setminus K$ with connected leaves. Let $\Sigma\subset M^3\setminus K$ be a smooth connected surface (hence contained in one end). Then $\Sigma$ is a surface of equilibrium if and only if it is a level set of $N$.
\end{Thm}
\begin{Rem}
As we have seen in Lemma \ref{lem:foli}, $N$ automatically foliates appropriate standardized exteriors of all ends with non-vanishing ADM-mass. Moreover, the level sets of $N$ (in each end) outside $K$ are topological spheres (and hence connected). The assumptions of the theorem are thus automatically satisfied for appropriate $K$ if all ends have non-vanishing mass. By the rigidity statement of the positive mass theorem \ref{thm:spm}, the case of vanishing ADM-mass is not particularly interesting, anyway.
\end{Rem}
\begin{Pf}
Let $n:U\to\R$ be a smooth function on a neighborhood $U\subset M^3$ of $\Sigma$ contained in the same end as $\Sigma$ and satisfying $dn\neq0$ on $U$ and $n\vert_\Sigma\equiv n_0$. Let $\kappa=(t,\mu):I\to\R\times M^3$ be a test body constrained to $\Sigma$. Then as in the proof of the above lemma, there is a smooth Lagrangian multiplier $\alpha:I\to\R$ such that $(t,\mu,\alpha)$ is a critical point of the functional $F$ defined in \eqref{F} among all $(t,\mu):I\to\R\times\Sigma$ with fixed endpoints. Again, the Euler-Lagrange equations for $F$ read
\begin{eqnarray}\label{ELeqt}
0&=&2 (N\circ\mu)\,dN(\dt{\mu})\,\dt{t}^2+(N\circ\mu)^2\,\ddt{t}\\\label{ELeqmu}
0&=&c^2 N\circ\mu\,\dt{t}^2\,dN+\myg{3}(\ddt{\mu},\cdot)+\alpha\,dn\\
n_0&=&n\circ\mu.
\end{eqnarray}
As $\Sigma$ inherits the metric from $M^3$, we have $\ddt{\mu}=\my{3}{\nabla}_{\dt{\mu}}\dt{\mu}=\my{2}{\nabla}_{\dt{\mu}}\dt{\mu}-h(\dt{\mu},\dt{\mu})\nu$ with $\nu=\my{3}{\grad}n/\vert\my{3}{\grad}n\vert$, where $h$ denotes the second fundamental form with respect to $\nu$. Now, in order to prove the claimed identity of surfaces of equilibrium with level sets of $N$, assume first that $\Sigma$ is a level set of $N$ (and then without loss of generality choose $n=N$). In order to see that $\Sigma$ is a surface of equilibrium, we have to show that $\mu$ is a geodesic in $\Sigma$. Projecting Equation \eqref{ELeqmu} onto the tangent plane of $\Sigma$, we find that $\my{2}{\nabla}_{\dt{\mu}}\dt{\mu}=0$ using that $dN(T\Sigma)=0$ so that $\mu$ is indeed a geodesic in $\Sigma$. Thus all level sets of $N$ are surfaces of equilibrium.

For the other direction of the desired implication, assume that $\Sigma$ is a surface of equilibrium. Observe that by Lemma \ref{lem:equili1}, $\kappa$ is parametrized proportionally to eigentime, in formulae $ds^2(\dt{\kappa},\dt{\kappa})\equiv\mbox{const}$. Taking the $\tau$-derivative of this, we find that
\begin{equation}\label{eigentime}
0=-c^2(N\circ\mu)\,dN(\dt{\mu})\,\dt{t}^2-c^2(N\circ\mu)^2\,\dt{t}\,\ddt{t}+\myg{3}(\ddt{\mu},\dt{\mu}) 
\end{equation}
where the last term equals $\myg{3}(\my{2}{\nabla}_{\dt{\mu}}\dt{\mu},\dt{\mu})$ by the above. Let $\sigma:I\to I$ be a smooth orientation preserving reparametrization such that $\lambda:=\mu\circ\sigma^{-1}$ is parametrized by arclength so that $\vert\lambda'\vert_2\equiv1$, where the prime denotes differentiation with respect to $\sigma$. The chain rule and the geodesic equation $\my{2}{\nabla}_{\lambda'}{\lambda'}=0$ induce $\dt{\mu}=\lambda'\circ\sigma\,\dt{\sigma}$ as well as $\my{2}{\nabla}_{\dt{\mu}}\dt{\mu}=\dt{\sigma}\,\ddt{\sigma}\,\lambda'\circ\sigma$. Inserting this into \eqref{ELeqmu} applied to $\dt{\mu}$ and using the facts that $\lambda$ is parametrized by arclength and that $\dt{\sigma}\neq0$, $dn(\dt{\mu})=0$, we obtain
\begin{equation}\label{ELc}
0=c^2\,N\circ\lambda\circ\sigma\,\dt{t}^2\,dN(\lambda')\circ\sigma+\dt{\sigma}\,\ddt{\sigma}.
\end{equation}
On the other hand, inserting \eqref{ELeqt} into \eqref{eigentime} gives
\begin{equation}\label{ELd}
0=c^2\,N\circ\lambda\circ\sigma\,\dt{t}^2\,dN(\lambda')\circ\sigma\,\boldsymbol{\dt{\sigma}}+\dt{\sigma}\,\ddt{\sigma}
\end{equation}
so that a comparison with \eqref{ELc} leads to $\dt{\sigma}\equiv1$ and hence $\mu$ must have already been parametrized by arclength. We can hence deduce $\my{2}{\nabla}_{\dt{\mu}}\dt{\mu}=0$ which implies $g(\ddt{\mu},\cdot)=-h(\dt{\mu},\dt{\mu})dn/\vert dn\vert_2$. This shows by \eqref{ELeqmu} that $dN$ is proportional to $dn$ at any point of $\Sigma$ reached by $\mu$. As $\kappa=(t,\mu)$ was an arbitrary constrained test body -- and because the Picard-Lindel\"of theorem of ODEs\footnote{cf.~e.~g.~pp.~139f in \cite{Heuser}.} applies to constrained test bodies just as to ordinary ones and hence for each point in $\Sigma$ there is a constrained test body going through it -- $dN$ is in fact proportional to $dn$ on all of $\Sigma$ so that $N$ must in fact be constant along $\Sigma$. As $N$ has connected level sets, $\Sigma$ must be one of them which proves the theorem.
\qed\end{Pf}

\subsection*{Uniqueness of the Lapse Function}\label{subsec:Nunique}
Just as in the Newtonian setting, the vacuum region (far) outside the support of the matter thus possesses surfaces of equilibrium also in geometrostatics. These surfaces will now allow us to reprove uniqueness of the lapse function $N$ of a given geometrostatic system $(M^3,\myg{3},N,\rho,S)$. Recall that we have already discussed uniqueness of $N$ (and $\myg{3}$) in Section \ref{sec:asym} using analyticity at infinity and asymptotic considerations. Theorem \ref{equili} now allows us to give a more physical proof. It relies on the following technical lemma which shows that surfaces of equilibrium are in fact determined by ``local'' constrained test particles.

\begin{Lem}\label{lem:equili}
Let $(M^3,\myg{3},N,\rho,S)$ be a geometrostatic system, $\Sigma\subset M^3$ a smooth surface. Assume that $U^4\subset\R\times M^3$ is an open neighborhood of $S:=\{0\}\times M^3$ which is globally hyperbolic with Cauchy surface $S$ with respect to the induced $4$-metric \eqref{ds^2}. For the statement of this lemma, a smooth timelike curve $\overline{\kappa}=(\overline{t},\overline{\mu}):\left[\tau_0,\tau_1\right]\to U^4\cap(\R\times \Sigma)$ satisfying $\overline{t}(\tau_0)<0<\overline{t}(\tau_1)$ is called a ``test body within $U^4$ constrained to $\Sigma$'' if it is a critical point of the energy functional \eqref{energy functional} among all smooth timelike curves $\kappa=(t,\mu):I\to U^4\cap(\R\times\Sigma)$ with fixed endpoints that satisfy $\mu(I)\subset\Sigma$.
Then $\Sigma$ is a surface of equilibrium if and only if the spatial parts $\mu$ of all test bodies $\kappa=(t,\mu)$ within $U^4$ constrained to $\Sigma$ are geodesics with respect to the induced metric on $\Sigma$.
\end{Lem}
\begin{Pf}
Let $\overline{\kappa}=(\overline{t},\overline{\mu})$ be a test body within $U^4$ constrained to $\Sigma$. Let $\kappa=(t,\mu):\left[\tau_0,\tau_1\right]\to\R\times \Sigma$ be a smooth timelike curve parametrized by eigentime and having $\kappa(\tau_i)=\overline{\kappa}(\tau_i)$ for $i=0,1$. By continuity, there must be $\tau_*\in(\tau_0,\tau_1)$ with $t(\tau_*)=0$ and thus $\kappa(\tau_*)\in S$. This $\tau_*$ is unique as $S$ is achronal. Now assume $\kappa(I)\not\subset U^4$, then by continuity and by $\kappa(\tau_i)\in U^4$, $U^4$ open, there must be minimal and maximal parameters $\tau_{min},\tau_{max}\in\left(\tau_0,\tau_1\right)$ with $\kappa(\tau_{min})\in\partial U$. Suppose first that $\tau_*<\tau_{max}$. Then $\kappa\vert_{\left(\tau_{max},\tau_1\right]}$ is a smooth past inextendable curve through $\kappa(\tau_1)$ contained in $U^4$ which does not intersect $S$ so that $\kappa(\tau_1)\not\in D^+(S)$. On the other hand, $t(\tau_1)>0$ by assumption so that $\kappa(\tau_1)\not\in D^-(S)$ either. As $S$ is a Cauchy surface for $U^4$, this is a contradiction. Suppose then secondly that $\tau_*>\tau_{max}$. In this case, one can reverse the above argument and work on $\left[\tau_0,\tau_*\right)$ and in this way again obtain a contradiction. We can therefore conclude that $\kappa$ remains within $U^4$ automatically and thus $\overline{\kappa}$ must be a critical point of the energy \eqref{energy functional} among all smooth curves $\kappa:I\to U^4\cap(\R\times\Sigma)$ with fixed endpoints iff it is a critical point among all smooth curves $\kappa:I\to\R\times\Sigma$ with fixed endpoints. This implies the claim of the lemma by definition of surfaces of equilibrium.
\qed\end{Pf}

\begin{Thm}[Uniqueness of Lapse Function]\label{thm:Nunique}
Let $\mathcal{S}:=(M^3,\myg{3},N,\rho,S)$ and $\widetilde{\mathcal{S}}:=(M^3,\myg{3},\widetilde{N},\rho,S)$ be geometrostatic systems as in Theorem \ref{equili} (with common closed subset $K$ bounded away from the boundary) and assume that $\my{3}{\Ric}\neq0$ outside $K$. Then $N=\widetilde{N}$ in all of $M^3\setminus K$. If, in addition, $(M^3,\myg{3})$ is complete and $\inter K$ is diffeomorphic to a bounded domain in $\R^3$ having smooth boundary $\partial K$ which arises as the union of the boundaries of all ends of $M^3$, then $N=\widetilde{N}$ holds in all of $M^3$.
\end{Thm}
\begin{Rems}
We have seen in Section \ref{sec:asym} that the non-vanishing condition on the Ricci tensor in Theorem \ref{thm:Nunique} holds for any geometrostatic system with non-vanishing mass $m$ due to the asymptotic expansion $\my{3}{\Ric}_{ij}=mGc^{-2}r^{-3}\,(\delta_{ij}-3r^{-2}\,x_i\,x_j)+\mathcal{O}(r^{-4})$ for $r\to\infty$. So just as explained in the remark following Theorem \ref{equili}, any complete geometrostatic system with non-vanishing masses in all ends will have a unique lapse function. If the mass vanishes in one end and the metric is complete, then positive mass rigidity gives $N=1$ and thus also uniqueness of the lapse function.

In his paper \cite{PTod}, Paul Tod has investigated (a slight generalization of) the question whether the lapse function $N$ solving the vacuum static metric equations with respect to a given $3$-metric is unique in an arbitrary $3$-manifold (i.~e.~in a not necessarily asymptotically flat one). However, he has to make certain assumptions on the eigenvalues of the Ricci tensor and distinguishes the cases where these eigenvalues are locally pairwise distinct or where two or three of them locally coincide. As we take a geometric approach, we do not have to distinguish these cases.

Using a similar approach as \cite{PTod}, Robert Bartnik and Paul Tod \cite{BT} have derived a characterization of $3$-metrics admitting a lapse function as in the vacuum static metric equations \eqref{SMEvac}, again relying on an assumption on the eigenvalues-values of the corresponding Ricci tensor. We will not address this existence question in this thesis.
\end{Rems}

\begin{Pf}
Let $ds^2$ and $\widetilde{ds^2}$ be the corresponding Lorentzian $4$-metrics on $\R\times M^3$ with common time coordinate $t$. Then the (vacuum) energy and momentum constraints \eqref{energy constraint} and \eqref{momentum constraint} on $\{t\}\times M^3$  that hold outside $\{t\}\times K$ reduce to $\my{3}{\Scal}=0$ which is satisfied no matter what lapse function is chosen. By Theorem \ref{CHBloc}, $ds^2$ and $\widetilde{ds^2}$ must thus be extensions of a common globally hyperbolic development $(L^4,\overline{ds^2})$ of $(M^3\setminus K,\myg{3},h=0)$. By Lemma \ref{lem:equili}, $\Sigma\subset M^3\setminus K$ is a surface of equilibrium in the restrictions of either $ds^2$ or $\widetilde{ds^2}$ to $L^4$ iff it is a surface of equilibrium in $(L^4,\overline{ds^2})$ and hence their surfaces of equilibrium must coincide. Theorem \ref{equili} then tells us that $N$ and $\widetilde{N}$ have the same level sets in every end of $M^3\setminus K$ so that in each end there is a real function $f:N(M^3\setminus K)\subset\R\to\R$ such that $\widetilde{N}=f\circ{N}$. By the chain rule, we find that $\my{3}{\nabla}^2\widetilde{N}=f''\circ N\, dN\times dN+f'\circ N\,\my{3}{\nabla}^2N$. The vacuum static metric equations \eqref{SMEvac} for $(\myg{3},N)$ and $(\myg{3},\widetilde{N})$ then appear as
\begin{equation}\label{lapSME}
0=\mylap{3}\widetilde{N}=f''\circ N\,\vert dN\vert_3^2+f'\circ N\,\mylap{3}N=f''\circ N\,\vert dN\vert_3^2
\quad\Leftrightarrow\quad 0\stackrel{dN\neq0}{=}f''\circ N
\end{equation}
\begin{eqnarray}
f\circ N\,\my{3}{\Ric}&=&\my{3}{\nabla}^2\widetilde{N}\nonumber
=f''\circ N\,dN\times dN+f'\circ N\,\my{3}{\nabla}^2N\\\nonumber
&=&f''\circ N\,dN\times dN+f'\circ N\,N\,\my{3}{\Ric}\\\nonumber
&\stackrel{\eqref{lapSME}}{=}&f'\circ N\,N\,\my{3}{\Ric}\\\label{RicSME}
\Leftrightarrow f\circ N&\stackrel{\my{3}{\Ric}\neq0}{=}&f'\circ N\,N.
\end{eqnarray}
Now this means that $f$ is affine linear by \eqref{lapSME}. The boundary condition $N,\widetilde{N}\to1$ as $r\to\infty$, where $r$ is the radial coordinate of the asymptotically flat coordinate system belonging to the common metric $\myg{3}$, gives $f(n)\to1$ as $n\to1$ so that \eqref{RicSME} induces $f'(n)\to1$ as $n\to1$ where we have to use the non-vanishing assumption on the Ricci tensor. Using that $f$ is affine linear, these boundary data imply $f=\id_{N(M^3\setminus K)}$ in every end and thus $N=\widetilde{N}$ outside $K$. But then if $\inter K$ is diffeomorphic to a domain in $\R^3$ having smooth boundary $\partial K$ which arises as the union of the boundaries of all ends of $M^3$, both $N$ and $\widetilde{N}$ induce the same Dirichlet boundary data on $\partial K$. Thus, the Dirichlet problem for the elliptic equation
\begin{equation}\label{elliptic}
\mylap{3}N=\frac{4\pi G}{c^2}N\left(\rho+\frac{\my{3}{\tr}S}{c^2}\right)
 \end{equation}
coming from the static metric equations \eqref{SME} has a unique $C^{2,\alpha}(\inter K)$-solution inside $\inter K$ by classical PDE theory\footnote{cf.~e.~g.~theorem 6.14 on p.~107 in \cite{GT}.} and thus by smoothness of the lapse functions $\widetilde{N}= N$ must hold in all of $M^3$.
\qed\end{Pf}

Having proved the uniqueness of the lapse function we can now interpret the lapse function as a function telling us how we have to measure the passing of time in the corresponding $4$-dimensional Lorentzian geometry in order to ``see'' staticity of the $3$-slice just as one can only ``see'' the non-acceleration of a geodesic when the time parameter is chosen as (proportional to) its arclength parameter. This view is further supported by the following corollary.

\begin{Coro}\label{coro:Killunique}
Let $(M^3,\myg{3},N,\rho,S)$ be a geometrostatic system as in Theorem \ref{equili}. Let $\partial_t$ denote the future directed timelike hypersurface-orthogonal Killing vector field in the corresponding Lorentzian manifold $(\R\times M^3,ds^2=\myg{4})$ representing staticity and let $X$ be any other future directed timelike hypersurface-orthogonal Killing vector field in $(\R\times M^3,ds^2)$ with $ds^2(X,X)=-c^2\,N^2.$ Then $X=\partial_t$. Similarly, if $X$ is any future directed timelike hypersurface-orthogonal Killing vector field in $(\R\times M^3,ds^2)$ having the same maximal connected integral submanifolds as $\partial_t$ and $ds^2(X,X)\to -c^2$ as $r\to\infty$, then also $X=\partial_t$.
\end{Coro}
\begin{Rems}
The assumption that $ds^2(X,X)=-c^2\,N^2$ ensures that $X$ is a Killing vector field measuring the passing of time in the sense described above. In other words, this condition makes sure that $X$ is indeed a Killing vector field defining staticity of the corresponding Lorentzian metric with $N$ defined as on page \pageref{DefN}. Although we could in general reparametrize time by multiplying $\partial_t$ by a constant number -- which corresponds to a reparametrization of the parameter of a geodesic proportional to arclength in the above picture --, we prefer not to do so here as we require $N\to1$ asymptotically.

Related questions on multiple timelike Killing vector fields have been addressed e.~g.~by Robert Beig and Piotr Chr\'usciel in \cite{BC} and by Ed Ihrig and Dipak Kumar Sen in \cite{IS}. The latter paper includes a result similar to this corollary but under the condition that $X$ and $\partial_t$ must commute (and without our assumption of asymptotic flatness).
\end{Rems}
\begin{Pf}
First of all, we know that $ds^2(\partial_t,\partial_t)=-c^2\,N^2$ by definition of the lapse function $N$. Now write $X=a\partial_t+Y$ with $a:\R\times M^3\to\R$ and $ds^2(Y,\partial_t)\equiv0$. Then
\begin{eqnarray}\label{length}
0&=&-c^2N^2\left(a^2-1\right)+\vert Y\vert_4^2\\\label{lengthdifft}
0&=&-c^2N^2a\,a_{,t}+ds^2(\my{4}{\nabla}\!_tY,Y)\\\label{lengthdiffi}
0&=&-c^2NN_{,i}\left(a^2-1\right)-c^2N^2a\,a_{,i}+\myg{3}(\my{3}{\nabla}\!_iY,Y)
\end{eqnarray}
follows from the assumption on the length of $X$, where the two bottom equations are derivatives of the top one with respect to the time and space coordinates $t$ and $i$, respectively. Taking the $t$-derivative of $ds^2(Y,\partial_t)=0$, we obtain
\begin{equation}\label{Yt}
N\,a_{,t}=-Y(N). 
\end{equation}
On the other hand, the covariant derivative of $X$ can be calculated in terms of (covariant derivatives of) $a$ and $Y$ and the Killing equation for $X$ then splits into
\begin{eqnarray}\label{KEtt}
0&=&-c^2N^2a_{,t}+ds^2(\my{4}{\nabla}\!_tY,\partial_t)\\\label{KEti}
0&=&-c^2N^2a_{,i}+ds^2(\my{4}{\nabla}\!_tY,\partial_i)\\\label{KEij}
0&=&\my{3}{\nabla}\!_{\left(i\right.}Y_{\left.j\right)}
\end{eqnarray}
where we have used that the second fundamental form of $M^3$ in $(\R\times M^3,ds^2)$ vanishes by staticity. Similarly, the equation for hypersurface orthogonality $X_{\left[\alpha\right.}\my{4}{\nabla}\!_\beta X_{\left.\gamma\right]}=0$ splits into
\begin{eqnarray}\label{tij}
\my{3}{\nabla}\!_iY_j&\stackrel{\eqref{KEti},\eqref{KEij}}{=}&2Y_{\left[i\right.}(\ln(Na))_{,\left.j\right]}\\\label{ijk}
0&=&Y_{\left[i\right.}\my{3}{\nabla}\!_jY_{\left.k\right]}
\end{eqnarray}
where we have again used that the second fundamental form of $M^3$ in $(\R\times M^3,ds^2)$ vanishes and the fact that $a>0$ as $X$ is future directed. Now equations \eqref{lengthdifft} and \eqref{KEti} combine to $Y(a)=a\,a_{,t}$ while equations \eqref{lengthdiffi} and \eqref{KEij} together imply $Y(N)\,(a^2-1)+N\,a\,Y(a)=0$. Inserting \eqref{Yt} gives $a_{,t}=Y(a)=Y(N)=0$. In consequence, if we insert \eqref{tij} into \eqref{lengthdiffi} we obtain $(\ln(Na))_{,i}=(\ln N)_{,i}$ so that $a_{,i}=0$ and thus $a\equiv\mbox{const}$ and $\my{4}{\nabla}\!_tY=0$ by \eqref{KEti}. As $a=1$ implies $X=\partial_t$ by \eqref{length} let us suppose that $a\neq1$ for the following aiming to deduce a contradiction.

Surely $a$ not being equal to $1$ ensures that $Y$ vanishes nowhere. Let $\omega:=Y_b$ be the associated $1$-form. Equation \eqref{tij} can be rewritten to say that $\my{3}{\nabla}\!_jY^k=(\ln N)_{,j} Y^k+(\ln N),^kY_j$ from which we deduce that $\omega$ is closed. As $M^3\setminus K$ is homeomorphic to $\R^3\setminus B$ for some ball $B$, a set which is simply connected, a standard result from differential geometry\footnote{cf.~e.~g.~p.~401 in \cite{Lee2}, where the statement is phrased in the language of deRham cohomology.} tells us that there is a smooth function $f:M^3\setminus K\to\R$ such that $\omega=df$ or in other words $Y=\my{3}{\grad}f$. Let $\Sigma\subset M^3\setminus K$ be a level set of $N$ and hence compact. As $f$ is continuous, it must attain its maximum on $\Sigma$ at some point $p_0\in\Sigma$ and thus satisfy $df\vert_{T\Sigma}(p_0)=0$. But as $\my{3}{\grad}N$ is normal to $\Sigma$ and $df(\my{3}{\grad}N)=\myg{3}(Y=\my{3}{\grad}f,\my{3}{\grad}N)=Y(N)=0$ on $\Sigma$, we have $df\vert_{p_0}=0$ and hence $Y\vert_{p_0}=0$ which provides the desired contradiction.

Secondly, if $X$ is a Killing vector field as described above having the same maximal connected integral manifolds as $\partial_t$, then automatically $X=a\partial_t$ for some function $a:\R\times M^3\to\R$ by hypersurface-orthogonality. The Killing equations for $X$ and $\partial_t$ imply $0=\my{4}{\nabla}_{\left(\alpha\right.}(a\partial_t)_{\left.\beta\right)}=a_{,\left(\alpha\right.}\delta^t_{\left.\beta\right)}$ so that $a\equiv\mbox{const}$. The behavior of $X$ at $\infty$ then gives $a\equiv1$.
\qed\end{Pf}

The fact that $N$ is unique in the sense described above also gives insight into the so called {\it Killing initial data (KID)}\index{ind}{Killing initial data}\index{ind}{KID} of a static spacetime. The KID are defined as the elements of the kernel of the adjoint of the linearization of the vacuum Einstein constraints \eqref{energy constraint} and \eqref{momentum constraint} at a given solution, cf.~e.~g.~Justin Corvino and David Pollack's article \cite{CoPo}. Vincent Moncrief \cite{Monc} showed\footnote{We quote this from \cite{CoPo}, p.~13.} that spacetime Killing vector fields correspond precisely to the KID at a given solution. The question of uniqueness of spacetime Killing vector fields in a vacuum geometrostatic system can thus be answered by characterizing the KID of this system. In order to calculate all\footnote{or rather all KID satisfying appropriate boundary conditions at inifinity.} KID at a given vacuum geometrostatic system $\mathcal{S}=(M^3,g,N)$, we split an arbitrary spacetime Killing vector field (or in other words an arbitrary KID) $Y\in\Gamma(TL^4)$ into its lapse $n:M^3\to\R$ and shift $X\in\Gamma(TL^4), X\perp\my{4}{\nu}$, via $Y=n\,\my{4}{\nu}+X$ (recalling that $\myg{4}=ds^2=-N^2c^2dt^2+\myg{3}$). In order for our methods to be applicable, we ask that $n\to1$ and $ds^2(X,X)\leq C$ as $r\to\infty$ in all ends of $M^3$.

Calculating the linearization and its adjoint of the vacuum constraints around $\mathcal{S}$ (and using that the second fundamental form vanishes due to staticity), we find that $n$ and $X$ are characterized by the conditions that $n$ has to satisfy the vacuum static metric equations \eqref{SMEvac} with respect to the metric $\myg{3}$ and that $X$ must be a Killing vector field of $\myg{3}$. By the uniqueness theorems \ref{Nuniqueasym} and \ref{thm:Nunique}, it follows that $n=N$ and thus in particular independent of time. We have thus proven the following corollary.

\begin{Coro}[Characterization of Spacetime Killing Vector Fields (KID)]
Let $\mathcal{S}=(M^3,g,N)$ be a complete vacuum geometrostatic system. Then any Killing vector field $Y=n\,\my{4}{\nu}+X$, $n:L^4\to\R$, $X\in\Gamma(TL^4), X\perp\my{4}{\nu}$, in the associated spacetime $(L^4,ds^2)$ with asymptotic behavior $-c^2n^2=ds^2(Y-X,Y-X)\to-c^2$ and $ds^2(X,X)\leq C$ as $r\to\infty$ can be decomposed as $$Y=N\,\my{4}{\nu}+X$$ where $X(t)\in\Gamma(TM^3)$ is a Killing vector field in $(M^3,\myg{3})$ (possibly time dependent).
\end{Coro}

\section{Photon Spheres}\label{sec:photon}
The above discussion shows that the level sets of the lapse function $N$ studied are surfaces of special interest in geometrostatics. Apart from being physically relevant, they are also (technically) very useful, especially for uniqueness arguments. One famous example of the use of a foliation by level sets of the function $N$ is Werner Israel's proof \cite{Israel} of the uniqueness of static black holes. We will now proceed to prove uniqueness of metrics possessing a classical codimension 1 photon sphere thereby also relying on a foliation of $N$-levels.

\begin{Def}[Photon Sphere]
Let $(M^3,\myg{3},N,\rho,S)$ be a geometrostatic system with only one end. A smooth closed surface $\Sigma\subset M^3$ is called a {\it photon sphere}\index{Index}{photon sphere} if any null geodesic in the corresponding Lorentzian manifold $(L^4,ds^2)$ which is initially tangent to the cylinder $\R\times\Sigma$ remains tangent to it.
\end{Def}

Photon spheres have first been identified in the \schild family of solutions. They model photons spiralling around a black hole at a fixed distance. Apart from their phenomenological significance as a specifically relativistic feature, photon spheres are crucially relevant for questions of (linear) stability of special solutions, as for example in \cite{Dafermos}. In the literature, one also finds the convention that the term 'photon sphere' refers to the cylinder $\R\times\Sigma$. It is well-known that dynamical spacetimes do not usually possess classical codimension 1 photon spheres and the notion of 'photon sphere' must therefore be generalized (cf.~e.~g.~\cite{CVE}).

Nevertheless, the author is not aware of a general (non-)existence result for classical photon spheres in static relativity theory. In vacuum, this issue can now be settled through an argument similar to Werner Israel's one (as exposed in Markus Heusler's book \cite{Heusler}). We will prove that the \schild metrics are the only static asymptotically flat vacuum solutions possessing a classical photon sphere.

\begin{Thm}[Uniqueness of Photon Spheres]\label{photo}
Let $\mathcal{S}:=(M^3,\myg{3},N)$ be a vacuum geometrostatic end possessing a photon sphere $\Sigma$. Assume furthermore that $N$ regularly foliates $M^3$ and that all level sets are topological spheres. Then (the spacetime corresponding to) $\mathcal{S}$ is isometric to (the spacetime corresponding to) a member of the \schild family \eqref{schwarz}, \eqref{Nschwarz}.
\end{Thm}

\begin{Rem} We have seen in Lemma \ref{lem:foli} that $N$ regularly foliates any asymptotically flat end outside a closed set bounded away from infinity (if the total mass is positive), and that these level sets are spherical. The end in the above theorem can thus be chosen appropriately without further assumptions by cutting out such a set.
\end{Rem}

We will first prove three small lemmata:

\begin{Lem}\label{photoLem1}
Under the assumptions of Theorem \ref{photo}, $N$ must be constant on $\Sigma$. Hence, $\Sigma$ is a topological sphere and a level set of $N$.
\end{Lem}
\begin{Pf} Let $p\in\Sigma$ and pick $v_p\in T_p\Sigma$ with $\lvert v_p\rvert=1$. Set $$w_p:=\frac{1}{c N(p)}\,\left.\partial_t\right|_{(t,p)}+v_p\;\in T_p(\R\times M^3)$$ and observe that $w_p$ is null. Then there exists a unique geodesic $\gamma:\left[0,\varepsilon\right]\to \R\times M^3$, $\varepsilon>0$, $\gamma(s)=(t(s),\mu(s))$ of $ds^2$ given by \eqref{ds^2} satisfying $\dt{\gamma}(0)=w_p$ and normalized via $\lvert\dt{\mu}(s)\rvert_g\equiv1$, where the dot denotes differentiation with respect to $s$. $\gamma$ is null with $w_p$ and thus $\mu(s)\in\Sigma$ by the assumption that $\Sigma$ be a photon sphere. The time component of the geodesic equation $\my{4}{\nabla}_{\dt{\gamma}}\dt{\gamma}=0$, $$\ddt{t}+\my{4}{\Gamma}_{\alpha\beta}^t\circ\gamma\, \dt{\gamma}^{\alpha} \dt{\gamma}^{\beta}=0,$$ can easily be seen to be equivalent to $\frac{d}{ds}(N\circ\mu)=0$ as $\dt{t}=\frac{1}{c \,N\circ\mu}$ by choice of normalization. Whence $N$ is constant along $\gamma$ and $\myg{3}_p(v_p,\my{3}{\grad}_p N)=0$. As $p, v_p$ are arbitrary, this implies the desired conclusion.\qed
\end{Pf}

\begin{Lem}\label{photoLem2}
Under the assumptions of Theorem \ref{photo}, the second fundamental form $h$ of $\Sigma\subset(M^3,\myg{3})$ satisfies \begin{equation}\label{eqphotoLem2} h=\frac{\nu(N)}{N}\,\myg{2},\end{equation} where $\nu$ is the outer unit normal to $\Sigma$ in $(M^3,\myg{3})$ and $\myg{2}$ the induced metric on $\Sigma$. Conversely, any level set of $N$ satisfying \eqref{eqphotoLem2} is a photon sphere. In particular, photon spheres are extrinsically round.
\end{Lem}
\begin{Pf}
By Lemma \ref{photoLem1}, $\Sigma$ is a level set of N and hence $\nu=\frac{\my{3}{\grad} N}{\lvert\my{3}{\grad} N\rvert}$ and $\nu(N)=\lvert\my{3}{\grad N}\rvert$. In this case, the spatial components of the geodesic equation $\my{4}{\nabla}_{\dt{\gamma}}\dt{\gamma}=0$ for a curve $\gamma:\left[0,\varepsilon\right]\to \R\times M^3$, $\varepsilon>0$, $\gamma(s)=(t(s),\mu(s))$, $\lvert\dt{\mu}\rvert^2=1$, turn out to be equivalent to \begin{equation}\label{perp}\my{3}{\nabla}_{\dt{\mu}}\dt{\mu}=-\frac{\my{3}{\grad} N}{N}\circ\mu \perp T_\mu\Sigma\end{equation} irrespective of whether $\gamma$ is or is not tangential to the cylinder $\R\times\Sigma$. Now if $\Sigma$ is a photon sphere, then $\dt{\mu}\in T_\mu\Sigma$ and thus $$\my{3}{\nabla}_{\dt{\mu}}\dt{\mu}=\my{2}{\nabla}_{\dt{\mu}}\dt{\mu}-h(\dt{\mu},\dt{\mu})\,\nu\circ\mu$$ which implies \eqref{eqphotoLem2}. Conversely, splitting $\dt{\mu}$ into its normal and tangential components with respect to $\Sigma$, $\dt{\mu}=\dt{\mu}^{N}+\dt{\mu}^{T}$, and recalling $\lvert\dt{\mu}^T\rvert^2+\lvert\dt{\mu}^N\rvert^2=1$, we find from \eqref{perp} and our assumption \eqref{eqphotoLem2} that
$$\my{2}{\nabla}_{\dt{\mu}^T}\dt{\mu}^T+\my{3}{\nabla}_{\dt{\mu}^T}\dt{\mu}^N+\my{3}{\nabla}_{\dt{\mu}^N}\dt{\mu}^T+\my{3}{\nabla}_{\dt{\mu}^N}\dt{\mu}^N=-\lvert\dt{\mu}^N\rvert^2\,\,\frac{\my{3}{\grad}N}{N}\circ\mu.$$ Projecting this equation along both $\dt{\mu}^N$ and $\dt{\mu}^T$ and adding the results making use of $\lvert\dt{\mu}\rvert\equiv1$ and the fact that all used connections are Levi-Civit\`a connections leads to the desired result of $\dt{\mu}^N=0$. As $\gamma$ was an arbitrary null geodesic, this implies that $\Sigma$ indeed is a photon sphere. Finally, \eqref{eqphotoLem2} implies that $h$ is proportional to $\myg{2}$ and thus tracefree which shows that photon spheres are extrinsically round.
\qed\end{Pf}

\begin{Lem}\label{photoLem3}
Under the assumptions of Theorem \ref{photo}, the mean curvature $H$ of $\Sigma$ and the normal derivative of $N$, $\nu(N)$, are constant along $\Sigma$.
\end{Lem}
\begin{Pf}
As $H=\frac{2\,\nu(N)}{N}$ and $N$ is constant, the statements of the lemma are equivalent. Let $(y^I),\,I=1,2$ be coordinates on $\Sigma$. From the special form of $h$ derived in Lemma \ref{photoLem2}, we deduce that $(\my{2}{\nabla}\!_Kh)_{IJ}=\frac{1}{2}H\!,_K g_{IJ}$. Now the Codazzi equations \eqref{Codazzi} for $\Sigma$ read $$\myg{3}(\my{3}{\Rm}(\partial_K,\partial_I,\nu),\partial_J)=(\my{2}{\nabla}\!_Kh)_{IJ}-(\my{2}{\nabla}\!_Ih)_{KJ},$$ and thus the formula for the Riemannian curvature endomorphism in three dimensions \eqref{formula:3Rm} together with the staticity requirement of $\my{3}{\Scal}=0$ give \begin{equation}\label{-}\my{3}{\Ric}(\partial_I,\nu)=-\frac{1}{2}\, H\!,_I.\end{equation} On the other hand, staticity implies 
\begin{eqnarray*}
 N\,\my{3}{\Ric}(\partial_I,\nu)&=&\my{3}{\nabla}^2N(\partial_I,\nu)\\
&=&(\nu(N)),_I-(\my{3}{\nabla}\!_I\nu)(N)\\
&=&\frac{N H,_I}{2}-h_{IJ}\,\inv{3}{JK}\,N\!,_K\\
&=&\frac{N H,_I}{2}
\end{eqnarray*}
as $N\equiv\mbox{const}$ on $\Sigma$. Comparing this with Equation \eqref{-} shows that $H$ and thus also $\nu(N)$ are constant.
\qed\end{Pf}

Having proved these lemmata, we will now follow Werner Israel's argument (cf.~\cite{Heusler}) to induce uniqueness. To this end, we prove the following proposition which implies Theorem \ref{photo}.
\begin{Prop}\label{PropUnique}
Let $\mathcal{S}:=(M^3,\myg{3},N)$ be a vacuum geometrostatic system, $\Sigma\subset M^3$ a surface. Assume that $N$ regularly foliates $M^3$ and that all level sets are topological spheres. Then if $\Sigma$ is an extrinsically round level set of $N$ which has constant mean curvature, $(\R\times M^3,ds^2)$ must be isometric to a member of the \schild family.
\end{Prop}

\begin{Pf}
First of all, as in the proof of Lemma \ref{photoLem3}, the facts that $\Sigma$ has constant mean curvature and that its second fundamental form is pure trace imply that $\my{3}{\Ric}(\partial_I,\nu)=0$. From staticity, we deduce that $$(\nu(N)),_I=N\,\my{3}{\Ric}(\partial_I,\nu)+(\my{3}{\nabla}\!_I\nu)(N)=h_{IJ}\,\inv{3}{JK}N\!,_K=0$$ so that $\nu(N)$ is constant along $\Sigma$.
Now let $N_0:=N(\Sigma)$. Let $\Omega\subset M^3$ be the exterior of $\Sigma$. By the maximum principle for elliptic PDEs (cf.~\cite{GT}) and by the asymptotics $N=1-M/r+\mathcal{O}(r^{-2})$ as $r\to\infty$ with respect to any asymptotically flat coordinate system stated in \ref{thm:KM}, $N$ will have values in the interval $\left(N_0,1\right)$ in $\Omega$. Now by the assumption that $N$ regularly foliates $M^3$, we can extend any coordinate system $(y^I),\,I=1,2$ on $U\subset\Sigma$ to the cylinder $\left[N_0,1\right)\times U$ by letting it flow along the (nowhere vanishing) gradient of $N$. Define $\rho:M^3\to\R$ through $$\rho(p):=(\left.\nu(N)\right|_p)^{-1}=(\lvert\my{3}{\grad}_pN \rvert)^{-1}$$ and observe that $$g=\rho^2\,dN^2+\myg{2},$$ where $\myg{2}$ is the 2-metric induced on $\Sigma$. In the coordinates $(N,y^1,y^2)$, the vacuum static metric equations on any level set of $N$ can be rephrased into
\begin{eqnarray}\label{eq:A}
0&=&\frac{1}{\rho}\left(\frac{H}{N}-H\!,_N-\frac{\rho}{2}\,H^2\right)-\frac{2}{\sqrt{\rho}}\,\mylap{2}\sqrt{\rho}-\frac{1}{2}\left[ \frac{\lvert \my{2}{\grad}\rho\rvert^2}{\rho^2}+2\lvert \free\rvert^2\right]\\\label{eq:B}
0&=&\frac{1}{\rho}\left(3\frac{H}{N}-H\!,_N\right)-\my{2}{\Scal}-\mylap{2}\ln{\rho}-\left[ \frac{\lvert \my{2}{\grad}\rho\rvert^2}{\rho^2}+2\lvert \free\rvert^2\right]\\\label{eq:C}
0&=&\rho,_N-\rho^2 H.
\end{eqnarray}
Let $\mathfrak{g}:=\det(\myg{2}_{IJ})$. By definition of the second fundamental form, we have $(\sqrt{\mathfrak{g}}),_N=\sqrt{\mathfrak{g}}\,H\rho$. Using \eqref{eq:C} and non-negativeness of the terms in square brackets, we can thus reformulate equations \eqref{eq:A} and \eqref{eq:B} into the inequalities
\begin{eqnarray}\label{ineq:A}
\partial_N\left(\frac{\sqrt{\mathfrak{g}}\,H}{\sqrt{\rho}\,N}\right)&\leq&-2\frac{\sqrt{\mathfrak{g}}}{N}\,\,\mylap{2}\sqrt{\rho}\\\label{ineq:B}
\partial_N\left(\frac{\sqrt{\mathfrak{g}}}{\rho}\,\left[H N+\frac{4}{\rho}\right]\right)&\leq&-N\sqrt{\mathfrak{g}}\left(\mylap{2}\ln{\rho}+\my{2}{\Scal}\right),
\end{eqnarray}
that hold on any $N$-level. In this inequalities, equality holds if and only if the square brackets in \eqref{eq:A} and \eqref{eq:B} vanish i.~e. iff $\nu(N)\equiv\mbox{const}$ and $\free=0$ on this level set. Integrating the first of these inequalities over the interval $\left[N_0,1\right)$, we get
$$\left[\frac{H(N)}{N}\int_{\Sigma_N}\frac{1}{\sqrt{\rho}}\,\,d\sigma\!_N\right]_{N_0}^{1}\leq-2 \int_{N_0}^{1}\frac{1}{N}\int_{\Sigma_N}\mylap{2}\sqrt{\rho}\,\,d\sigma\!_N\,dN=0,$$
where the right-hand side vanishes because of the divergence theorem. As $\rho\equiv\mbox{const}$ on $\Sigma$ by assumption and using the asymptotics $H=\frac{2}{r}+\mathcal{O}(r^{-2})$, $\frac{1}{\rho(N)}=\nu(N)=\frac{M}{r^2}+\mathcal{O}(r^{-3})$ as $r\to\infty$ in any asymptotically flat wave harmonic system of coordinates (cf.~Corollary \ref{coro:graph}), this implies
$$\frac{H\,\sqrt{\nu(N)}}{N_0}\,\lvert\Sigma\rvert\geq\lim_{r\to\infty}\frac{H}{N}\int_{\Sigma_N}\sqrt{\nu(N)}\,d\sigma\!_N= 8\pi\sqrt{M},$$
Integrating inequality \eqref{ineq:B}, we get
\begin{eqnarray*}
\left[\int_{\Sigma_N}\frac{1}{\rho}\left[HN+\frac{4}{\rho}\right]d\sigma\!_N\right]_{N_0}^1&\leq&-\int_{N_0}^{1}N\int_{\Sigma_N}\left(\mylap{2}\ln{\rho}+\my{2}{\Scal}\right)\,d\sigma\!_N\,dN\\
&=&-8\pi\,\int_{N_0}^{1}N\,dN=-4\pi(1-N_0^2)
\end{eqnarray*}
again by the divergence theorem and using the theorem of Gau\ss-Bonnet \eqref{GB}. Making use of the discussed asymptotics again, we obtain $$\nu(N)\left[HN_0+4\nu(N)\right]\lvert\Sigma\rvert\geq 4\pi(1-N_0^2).$$ Finally, on our given surface $\Sigma$, the divergence theorem implies $$\nu(N)\lvert\Sigma\rvert=\int_\Sigma \nu(N)\,d\sigma=-\int_\Omega \mylap{3}{N}\,d\mu+\lim_{r\to\infty} \int_{\Sigma_N} \nu(N)\,d\sigma\!_N=4\pi M$$ where we have used the vacuum static metric equation $\mylap{3}{N}=0$ and the discussed asymptotics. This leads to
\begin{eqnarray}\label{ineq:C}
\frac{2 N_0}{\sqrt{\lvert\Sigma\rvert/4\pi}}&\leq&H\\\label{ineq:D}
M\left[HN_0+\frac{4 M}{\lvert\Sigma\rvert/4\pi}\right]&\geq&1-N_0^2.
\end{eqnarray}
Taking the difference of \eqref{eq:B} and \eqref{eq:A} on $\Sigma$ implies $K=H\nu(N)/N_0+H^2/2$ so that $K=4\pi MH/N_0\lvert\Sigma\rvert+H^2/2$ and thus in particular, $K$ is constant. Gau\ss-Bonnet \eqref{GB} now says that $K=4\pi/\lvert\Sigma\rvert$ so that $H\leq N_0(1-N_0^2)/M$ follows from \eqref{ineq:C}. Then together with \eqref{ineq:D}, we obtain $N_0\geq\sqrt{1-2M\sqrt{K}}$ so that $$N_0\leq \frac{H}{2\sqrt{K}}\leq\frac{ N_0(1-N_0^2)}{2M\sqrt{K}}\leq N_0$$ which gives us $\free\equiv0$ and $\nu(N)\equiv\mbox{const}$ on any leaf $\Sigma_N$ (recall the considerations on the equality cases of \eqref{ineq:A} and \eqref{ineq:B}). \eqref{eq:B} and \eqref{eq:A} then imply that the Gau\ss\, curvature must be constant on any level $\Sigma_N$, $N(r)=\sqrt{1-2M/r}$ for $r:=1/\sqrt{K}$, and $\nu(N)=M/r^2$ so that $$g=\frac{r^4}{M^2}N'(r)^2\,dr^2+r^2\,d\Omega^2=\frac{1}{N^2}dr^2+r^2\,d\Omega^2,$$ with $d\Omega^2$ the canonical metric on $S^2$. This proves that $g$ is isometric to the Schwarzschild metric with mass $M$.
\qed\end{Pf}

\begin{Coro}
Under the assumptions of Theorem \ref{photo} and using the notation of Proposition \ref{PropUnique}, $H\!,_N=0$ and $K\equiv\mbox{const}$ on any photon sphere.
\end{Coro}
\begin{Pf}
$K\equiv\mbox{const}$ follows for any leaf of the foliation by level sets of $N$ as discussed in the proof of Proposition \ref{PropUnique}. \eqref{eq:B} shows that on the photon sphere, we have $H\!,_N=\frac{3H}{N}-\frac{2}{M}$, \eqref{eqphotoLem2} tells us that $H=\frac{2MK}{N}$ whereas the identity $K=\frac{3H^2}{4}$ follows from the Gau\ss\, equation \eqref{Gauss3} on the photon sphere. Together, these imply $H,\!_N=0$.
\qed\end{Pf}
\chapter{The Newtonian Limit}\label{chap:FT}
This chapter is dedicated to the two last central theorems of this thesis, Theorems \ref{thm:NLm} and \ref{thm:NLCoM}. In rough terms, these theorems states that the Newtonian limits of the relativistic mass and center of mass of a geometrostatic system exist and coincide with the Newtonian mass and center of mass of the Newtonian limit of the system, respectively. As announced before, we will use the framework provided by J\"urgen Ehlers' ``frame theory'' in order to rigorously formulate and prove these claims. To this end, we will have to redefine staticity as well as the notions of pseudo-Newtonian metric and potential within frame theory in a manner that is compatible with the general relativistic definitions given in Chapters \ref{chap:static} and \ref{chap:mCoM}, respectively.

We will introduce frame theory in Section \ref{sec:FT}. In Section \ref{sec:NLStatic}, we specialize it to the realm of static isolated systems in Section \ref{sec:NLStatic} thereby relying on our preparatory generalization of Killing vector fields discussed in Section \ref{sec:LieKill}. We will formalize the immanent concept of Newtonian limit in Section \ref{sec:NLgeo}. Finally, Section \ref{sec:NLmass} will be dedicated to prove convergence of mass and its center.

As customary and convenient in frame theory, we will use abstract index notation\index{ind}{abstract index notation} throughout this chapter.

\section{J\"urgen Ehlers' Frame Theory}\label{sec:FT}
The relation between Isaac Newton's and Albert Einstein's theories of gravitation NG and GR has, of course\footnote{as J\"urgen Ehlers puts it in \cite{Ehl6}.}, been studied ever since the latter was formulated. The basic work\footnote{The historical account in this paragraph is quoted from p.~96 of J\"urgen Ehlers article \cite{Ehl6} and we refer the interested reader to this article and the references cited therein.} for elucidating this relation was done by \'Elie Cartan and Kurt Friedrichs. It was extended by Andrzej Trautman, Peter Havas, Georg Dautcourt, Hans-Peter K\"unzle, James Michael Nester, David Malament, Martin Lottermoser and others. In the 1980s, J\"urgen Ehlers has merged and consolidated the approaches taken before and devised his frame theory providing a mathematically consistent framework for the analysis of the Newtonian limit. Frame theory has been widely used ever since for many different purposes some of which we will encounter in this chapter.

As we have already discussed in the introduction, there are both physical and philosophical reasons for why it is desirable to understand the relationship of NG and GR. A mathematical reason also exists: The better we understand the Newtonian limit, the easier it becomes to transfer well-known Newtonian facts like existence or non-existence of certain configurations into GR by e.~g.~implicit function theorem type arguments or by proving that such configurations would persist under the Newtonian limit, respectively. An example for this is the existence of rotating stars asserted by Uwe Heilig \cite{Heilig} using Ehlers' frame theory and an implicit function theorem type argument.

Returning to the physical point of view, consistency and interpretational as well as modeling guidance are among the main arguments for studying the Newtonian limit. Consistency here means that -- mainly because Newtonian gravity is still used for calculations, measurements, and constructions and since it prevails in most people's intuitive views of the world -- it should be ensured that general relativity ``includes'' Newtonian gravity as a special or limit case or in other words that GR goes over to NG as $c\to\infty$, a limit relation which is generally called the ``Newtonian limit''. This applies in particular not only to the mathematical variables characterizing gravitational systems like the associated Lorentzian metrics and matter tensors but also and even more so to the physical properties of these systems like mass, momentum, etc.

The reason for this particular relevance of understanding the Newtonian limit of physical properties is twofold: First of all, the mathematical variables are in some sense only the mathematical description of a physical system while its physical properties represent the characteristic features that really matter. Secondly, relativistic notions of physical properties like mass, momentum etc.~are not automatically or canonically defined but must be specifically devised. One part of the justification of these definitions should then be to show that they converge to their Newtonian counterparts in the Newtonian limit.

This takes us back to the story line of this thesis, and we would like to continue by shortly recapitulating Ehlers' frame theory loosely following Martin Lottermoser in \cite{Lott1}. For a more extended introduction, we refer the interested reader to any of J\"urgen Ehlers' expositions \cite{Ehl5,Ehl4,Ehl6}, to \cite{Lott1} itself, or to the recent overview article by Todd Oliynyk and Bernd Schmidt \cite{OlySchmi}.

The main problem one encounters when trying to define the Newtonian limit rigorously is the fact that Newtonian gravity is a coordinate-variant\footnote{in particular allowing for inertial systems of coordinates.} (vectorial or potential) while general relativity is a coordinate-invariant (tensorial) theory which makes their comparison troublesome. One of the basic ideas underlying frame theory dates back to \'Elie Cartan \cite{Cartan} and consists in generalizing Newtonian gravity to a coordinate-invariant ``geometric'' theory which is named ``Newton-Cartan theory''\index{ind}{Newton-Cartan theory} after its inventor. It is closely related to the notion of Coriolis force as discussed in \cite{Daut}. J\"urgen Ehlers' frame theory (FT)\index{ind}{frame theory}\index{ind}{FT} brings together Newton-Cartan theory and general relativity in a common framework.

Let us now become more specific. We begin by introducing the basic notions and quantities characterizing instances of frame theory. These instances are called ``mathematical models (of frame theory)''. They consist of a smooth $4$-manifold $L^4$ together with two smooth symmetric tensor fields $s^{\alpha\beta}$, $t_{\alpha\beta}$, and a smooth torsion-free connection\footnote{It is a slight abuse of notation to denote the connection in abstract index notation since it is not a tensor; however, we follow this tradition as we believe that no confusion can arise from it.} $\Gamma^\mu_{\alpha\beta}$ defining the geometry on $L^4$. $L^4$ is called a {\it spacetime}\index{ind}{spacetime}\index{sym}{$L^4$} and $s^{\alpha\beta}$ and $t_{\alpha\beta}$ are called {\it spatial metric}\index{ind}{spatial metric}\index{sym}{$s^{}{\alpha\beta}$} and {\it temporal metric}\index{ind}{temporal metric}\index{sym}{$t_{\alpha\beta}$}, respectively. $\Gamma^\mu_{\alpha\beta}$ is called the {\it gravitational field}\index{ind}{gravitational field}.

The spatial and temporal metrics and the connection are together replacing the Lorentzian metric $ds^2$ of GR in frame theory. This is meant as follows: The Lorentzian metric of GR can be used for many purposes: to perform the musical operations\index{ind}{musical operations} of pulling indices up and pushing them down, to define space-, and timelikeness, to define the Levi-Civit\`a connection, to define the different curvature tensors etc. In frame theory, these tasks are distributed amongst the spatial and temporal metrics and the connection, respectively.

For example, the spatial metric of frame theory is used to pull indices up: If ${T^{\alpha_{1}\dots\alpha_{k}}}_{\beta_{1}\dots\beta_{l}}$ is a tensor at a point $p\in L^4$, then pulling up the index $\beta_{i}$ produces the (new) tensor $${{{T^{\alpha_{1}\dots\alpha_{k}}}_{\beta_{1}\dots\beta_{i-1}}}_{\;\;\bullet}^{\alpha_{k+1}}}_{\beta_{i+1}\dots\beta_{l}}:={T^{\alpha_{1}\dots\alpha_{k}}}_{\beta_{1}\dots\beta_{l}}s^{\alpha_{k+1}\beta_{i}}$$ at $p$, where the $\bullet$ at the former position of the index $\beta_{i}$ indicates which index we have pulled up. This indication is necessary as -- other than in GR -- pulling up indices is a potentially irreversible act in FT (because the spatial and temporal metrics are not inverses of each other). Similarly, the temporal metric can be used to define how to push an index $\alpha_{j}$ of the tensor ${T^{\alpha_{1}\dots\alpha_{k}}}_{\beta_{1}\dots\beta_{l}}$ down. This gives rise to the (equally new) tensor $${{{T^{\alpha_{1}\dots\alpha_{j-1}}}^{\;\;\bullet}_{\beta_{l+1}}}^{\alpha_{j+1}\dots\alpha_{k}}}_{\beta_{1}\dots\beta_{l}}:={T^{\alpha_{1}\dots\alpha_{k}}}_{\beta_{1}\dots\beta_{l}}t_{\alpha_{j}\beta_{l+1}}$$ at $p$. Again, the symbol $\bullet$ indicates which index we have pushed down. This procedure also is potentially irreversible (and not, as customary in GR, in any sense the 'reversion' of pulling an index up). But although the spatial and temporal metrics are not inverses of each other, they do have a very special relation, namely
\begin{equation}\label{eq:laminv}
t_{\alpha\beta}s^{\beta\mu}=-\lambda\delta^\mu_{\alpha}
\end{equation}
with $\lambda\in\R^+_{0}$ the so-called {\it causality constant}\index{ind}{causality constant}\index{sym}{$\lambda$} of the mathematical model.

The spatial metric $s^{\alpha\beta}$ can furthermore be used to define spacelikeness of $1$-forms (not tangent vectors!) at a point $p\in L^4$: $\omega_{\alpha}\in T_{p}^{\ast}\!L^4$ is called {\it spacelike}\index{ind}{spacelike} if it satisfies $s^{\alpha\beta}\omega_{\alpha}\omega_{\beta}>0$. Timelikeness is defined for tangent vectors at a point $p$ by aid of the temporal metric $t_{\alpha\beta}$: $X^\alpha\in T_{p}L^4$ is called {\it timelike}\index{ind}{timelike} if $t_{\alpha\beta}X^\alpha X^\beta>0$. As in GR, unit length ($t_{\alpha\beta}V^\alpha V^\beta=1$) timelike vectors represent ``observers''. The {\it spatial directions of an observer $X^\alpha\in T_{p}L^4$}\index{ind}{spatial directions of an observer} are those vectors $Y^\alpha\in T_{p}L^4$ that are orthogonal to $X^\alpha$ via $t_{\alpha\beta}$ or in other words those $Y^\alpha$ that satisfy $t_{\alpha\beta}X^\alpha Y^\beta=0$. We are now in a position to formulate the next two axioms the metrics $s^{\alpha\beta}$ and $t_{\alpha\beta}$ of a mathematical model of FT have to comply with:
\begin{enumerate}
\item[(i)] At each $p\in L^4$, there must exist a timelike vector $X^\alpha\in T_{p}L^4$.
\item[(ii)] At each $p\in L^4$ and for all timelike $X^\alpha\in T_{p}L^4$, $s^{\alpha\beta}$ is positive definite on the space $\{\omega_{\alpha}\in T^\ast_{p}L^4\vert \,\omega_{\alpha}X^\alpha=0\}.$
\end{enumerate}
We continue our description of how the different usages of the Lorentzian metric $ds^2$ of GR are distributed amongst the frame theoretical geometric quantities by discussing covariant derivatives and curvature. Non-surprisingly, covariant derivation is defined with respect to the connection $\Gamma^\mu_{\alpha\beta}$. As usual in the pseudo-Riemannian framework, covariant differentiation is notated by a semicolon in abstract index notation. Observe that $\Gamma^\mu_{\alpha\beta}$ was required to be torsion-free (like the Levi-Civit\`a connection of $ds^2$ in GR). As a replacement for the Riemannianness of the Levi-Civit\`a connection in GR, $\Gamma^\mu_{\alpha\beta}$, $s^{\alpha\beta}$, and $t_{\alpha\beta}$ must satisfy
\begin{equation}\label{eq:compati}
\quad{s^{\alpha\beta}}_{;\mu}=0\quad\mbox{and}\quad t_{\alpha\beta;\mu}=0.
\end{equation}
Being a torsion-free connection, $\Gamma^\mu_{\alpha\beta}$ automatically induces a Riemannian curvature endomorphism\index{ind}{Riemannian curvature endomorphism}\index{sym}{${\Rm_{\alpha\beta\gamma}}^\mu$} ${\Rm_{\alpha\beta\gamma}}^\mu$ through Formula \eqref{Rieconnection}. Note that because the musical operations are in generally irreversible in FT, the positions of the indices (i.~e.~whether they are upstairs or downstairs) in the curvature expressions is relevant. This is accounted for in the formulae and facts listed in Sections \ref{sec:subs} and \ref{sec:LieKill}. In particular, the Ricci tensor\index{ind}{Ricci tensor} can be defined by $\Ric_{\alpha\beta}:=-{\Rm_{\alpha\mu\gamma}}^\mu$\index{sym}{$\Ric_{\alpha\beta}$} without using any musical operations.

We can now formulate the last geometric compatibility requirement relating the spatial and temporal metrics of a mathematical model of frame theory with its connection. It reads
\begin{equation}\label{eq:Riesym}
\Rm_{\alpha\;\bullet\;\gamma}^{\;\;\;\nu\;\;\;\mu}=\Rm_{\alpha\;\bullet\;\gamma}^{\;\;\;\mu\;\;\;\nu}.
\end{equation}
Observe that the Riemannian curvature endomorphism of GR possesses this symmetry automatically.

Besides the geometric quantities $L^4,s^{\alpha\beta},t_{\alpha\beta},\Gamma^\mu_{\alpha\beta}$, and the causality constant $\lambda$, a mathematical model of frame theory also consists of a smooth symmetric {\it energy-momentum} or {\it matter tensor}\index{ind}{energy-momentum tensor}\index{ind}{matter tensor}\index{sym}{$T^{\alpha\beta}$} $T^{\alpha\beta}$ (just as GR does). As in GR, this matter tensor gives rise to notions of {\it matter density}\index{ind}{matter density} $\rho$\index{sym}{$\rho$}, {\it momentum density}\index{ind}{momentum density} $J$\index{sym}{$J$}, and {\it stress}\index{ind}{stress} $S$\index{sym}{$S$}. Concretely, for a given observer $X^\alpha\in TL^4$, the observed density is given by $\rho(X^\alpha)=T^{\bullet\;\bullet}_{\alpha\beta}X^\alpha X^\beta.$ The momentum observed by $X^\alpha$ in its spatial direction $Y^\alpha$ with unit length $s^{\alpha\beta}Y^\bullet_{\alpha}Y^\bullet_{\beta}=1$ is defined as $J(X^\alpha,Y^\beta):=T^{\bullet\;\bullet}_{\alpha\beta}X^\alpha Y^\beta$ and the stress observed by $X^\alpha$ in unit length spatial directions $Y^\alpha_{1},Y^\alpha_{2}$ is given by $S(Y^\alpha_{1},Y^\beta_{2}):=T^{\bullet\;\bullet}_{\alpha\beta}Y_{1}^\alpha Y_{2}^\beta$, cf.~Satz 4 in \cite{Lott1}.

The matter tensor and connection of a mathematical model of FT have to combine such that the matter tensor is {\it divergence free}\index{ind}{divergence free} (``conservation equation'')\index{ind}{conservation of matter equation}:
\begin{equation}\label{eq:divfree}
{T^{\alpha\beta}}_{;\beta}=0.
\end{equation}
Moreover, together with $t_{\alpha\beta}$, they have to satisfy the {\it generalized Einstein equations}
\begin{equation}\label{eq:genEin}
\Ric_{\alpha\beta}=8\pi G\left(T^{\,\bullet\,\bullet}_{\alpha\beta}-\frac{1}{2}T^{\mu\bullet}_{\;\;\mu}t_{\alpha\beta}\right)
\end{equation}
with $G$\index{sym}{$G$} the same gravitational constant\index{ind}{gravitational constant} as above. Note that, in contrast to the standard Einstein equations \eqref{EEq}, $\Ric_{\alpha\beta}-\frac{1}{2}\Scal g_{\alpha\beta}=\frac{8\pi G}{c^4}T_{\alpha\beta}$, the generalized Einstein equations do not explicitly contain the speed of light $c$ (nor the causality constant, which is tightly related to the speed of light as we will see in a minute). This is voluntary as it allows for taking a well-defined Newtonian limit without making the equations singular in the process.

In addition, the generalized and standard Einstein equations differ in that the first of them relates the Ricci tensor to the matter tensor and the trace of the matter tensor while the latter one relates the Ricci tensor and its trace to the matter tensor. However, taking the trace of the standard Einstein equations and inserting the result of this computation, $\Scal=-\frac{8\pi G}{c^4}\tr T$, into them shows that the standard Einstein equations are in fact equivalent to $\Ric_{\alpha\beta}=\frac{8\pi G}{c^4}(T_{\alpha\beta}-\frac{1}{2}\tr T\,g_{\alpha\beta})$ which already looks much more similar to \eqref{eq:genEin}.

Let us collect the above exposition into the following definition.
\begin{Def}[Mathematical Models of Frame Theory]\label{def:mathmodels}
A {\it mathematical model of frame theory}\index{ind}{mathematical model}\index{ind}{frame theory ! mathematical model} is a tuple $\mathcal{S}=(L^4,s^{\alpha\beta},t_{\alpha\beta},\Gamma^{\mu}_{\alpha\beta},T^{\alpha\beta},\lambda)$ consisting of a spacetime $L^4$, spatial and temporal metrics $s^{\alpha\beta}$ and $t_{\alpha\beta}$, a gravitational field (connection) $\Gamma^{\mu}_{\alpha\beta}$, a matter tensor $T^{\alpha\beta}$, and a causality constant $\lambda$ such that Equations \eqref{eq:laminv} through \eqref{eq:genEin} as well as items (i) and (ii) above are satisfied.
\end{Def}
By design, any relativistic system/solution to GR should give rise to a mathematical model of FT. Luckily enough, this holds true. The following proposition tells us how we can recover GR from FT. It is implicit in \cite{Lott1,Ehl6}.
\begin{Prop}\label{prop:GRisFT}
Let $(L^4,ds^2=g_{\alpha\beta})$ be a smooth Lorentzian spacetime, $T^{\alpha\beta}$ a smooth symmetric matter tensor field on $L^4$. Suppose that $g_{\alpha\beta}$ and $T^{\alpha\beta}$ satisfy the Einstein equations \eqref{EEq} with respect to the Levi-Civit\`a connection $\Gamma^\mu_{\alpha\beta}$ of $g_{\alpha\beta}$. Set 
\begin{equation}\label{eq:GRisFT}
\lambda:=\frac{1}{c^2}\quad\mbox{and}\quad t_{\alpha\beta}:=-\lambda g_{\alpha\beta}\quad\mbox{and}\quad s^{\alpha\beta}:=g^{\alpha\beta}.
\end{equation}
Then $(L^4,s^{\alpha\beta},t_{\alpha\beta},\Gamma^{\mu}_{\alpha\beta},T^{\alpha\beta},\lambda)$ is a mathematical model of frame theory. The relativistic and frame theoretical notions of space- and timelikeness, matter density, momentum density, and stress tensor also coincide. 
\end{Prop}
\begin{Pf}
Equations \eqref{eq:laminv} through \eqref{eq:Riesym} as well as (i) and (ii) directly follow from the fact that $\Gamma^\mu_{\alpha\beta}$ is the Levi-Civit\`a connection of the pseudo-Riemannian metric $g_{\alpha\beta}$. The generalized Einstein equations \eqref{eq:genEin} follow from the standard Einstein equations \eqref{EEq} as described above. Finally, the conservation of matter equation \eqref{eq:divfree} follows from applying the contracted second Bianchi identity (or Schur's lemma)\index{ind}{Schur's lemma}\index{ind}{Bianchi identity ! second} to the standard Einstein equations. 
\qed\end{Pf}
As before, the constant $c$ in this proposition denotes the speed of light. At first, it might seem strange to assign the value $c^{-2}$ to the causality constant as $c^{-2}$ appears to be a fixed number and thus the same one for all relativistic systems. This, however, is not the case as we have not fixed units so far. Different choices of length and time units obviously lead to different numerical values of the speed of light and thus to different values of $\lambda$, cf.~pp.~7ff and p.~14 in \cite{Lott1}. In dynamical GR, one can replace this unit argument by comparing the speed of light $c$ to ``typical'' velocities in the dynamical relativistic system. As we are interested in static systems, this does not work as there is no typical non-zero velocity in the static case.

Since FT is supposed to unite GR and NG, we also expect Newtonian gravitational systems to be mathematical models of frame theory. Sure enough, this is correct as the following proposition asserts\footnote{The Newtonian gravitational systems considered here are not the most general ones one could consider since $U$ does not even depend on time. Since we are only interested in static situations in the end, this does not concern us. More information on general Newtonian systems can be found in \cite{Lott1,OlySchmi}.}. It is proven on page 40ff in \cite{Lott1}.
\begin{Prop}\label{prop:NGisFT}
Let $\rho,U:\R^3\to\R$ be smooth functions satisfying $\triangle U=4\pi G\rho$ and $L^4:=\R\times\R^3$. Let $(x^\alpha):=(t,x^{i})$ denote Cartesian coordinates on $L^4$ and set
\begin{eqnarray}\nonumber
\lambda:&=&0\\\nonumber
t_{\mu\nu}:&=&\delta^t_{\mu}\delta^t_{\nu}\\\label{eq:NGisFT}
s^{\mu\nu}:&=&\delta^{\mu}_{i}\delta^{\nu}_{j}\delta^{ij}\\\nonumber
\Gamma^\mu_{\alpha\beta}:&=&\delta_{j}^\mu\delta_{\alpha}^t\delta_{\beta}^t\delta^{ij} U_{,i}\\\nonumber
T^{\alpha\beta}:&=&\rho\delta^{\mu}_{t}\delta^\nu_{t}.
\end{eqnarray}
Then $(L^4,s^{\alpha\beta},t_{\alpha\beta},\Gamma^{\mu}_{\alpha\beta},T^{\alpha\beta},\lambda)$ is a mathematical model of frame theory. In particular, the Newtonian and frame theoretical notions of matter density coincide.
\end{Prop}
Now that we have learned that both GR and (static) NG arise as special cases of FT it is natural to ask the opposite question of whether and if so how many and what type of other mathematical models of FT there are. The answer is simple for positive causality constant $\lambda$: Any mathematical model of FT with $\lambda>0$ corresponds to a solution of GR (with different numerical values of the speed of light $c$ depending on the value of $\lambda$). This is well-known and can be proven directly just as Proposition \ref{prop:GRisFT}.

For $\lambda=0$, the answer is more involved as all solutions to the Newton-Cartan theory mentioned above can occur. In the case of static isolated mathematical models of FT\footnote{We will give a precise definition of this notion in Section \ref{sec:NLStatic}.}, though, the Coriolis forces related to the additional freedom offered by Newton-Cartan theory vanish and static isolated solutions of NG remain the only mathematical models of FT with $\lambda=0$ (when choosing adapted coordinates), cf.~p.~5 in \cite{OlySchmi}, p.~42 in \cite{Lott1}, and Section \ref{sec:NLStatic} below.

\section{Static Isolated Systems in Frame Theory}\label{sec:NLStatic}
In this section, we will introduce staticity and asymptotic flatness concepts in FT. On our way, we will continuously compare the new frame theoretical notions with their relativistic counterparts. An analysis of what these frame theoretical notions signify in the Newtonian or Newton-Cartan sector of FT will be carried out at the end of the respective explanations, cf.~pp.~\pageref{subsec:Newton}f and Theorem \ref{thm:staticiso}.

As a first step, let us quickly discuss $3+1$ decomposition of spacetimes in FT. We cite the following definitions and lemma from pp.~14f in \cite{Lott1} adapting it to our notation and nomenclature. 
\begin{Def}
Let $(L^4,s^{\alpha\beta},t_{\alpha\beta},\Gamma^{\mu}_{\alpha\beta},T^{\alpha\beta},\lambda)$  be a mathematical model of frame theory, $X^\alpha\in\Gamma(TL^4)$ a smooth timelike vector field. We call $$\Pi^\alpha_{\beta}:=\delta^\alpha_{\beta}-\frac{X^\alpha X^\bullet_{\beta}}{t_{\mu\nu}X^\mu X^\nu}$$ the {\it projection onto the orthogonal complement $X^\perp:=\{Y^\beta\in\Gamma(TL^4)\,\vert\,t_{\alpha\beta}X^\alpha Y^\beta=0\}$} of $X^\alpha$.\index{ind}{projection onto the orthogonal complement}\index{ind}{orthogonal complement}\index{sym}{$X^\perp$}\index{sym}{$\Pi^\alpha_{\beta}$} Moreover, we set $$\inv{3}{\alpha\beta}:=s^{\alpha\beta}+\lambda\,\frac{X^{\alpha}X^\beta}{t_{\mu\nu}X^\mu X^\nu}$$\index{sym}{$\inv{3}{\alpha\beta}$} and call $\inv{3}{\alpha\beta}$ the {\it (generalized) inverse $3$-metric corresponding to $X^\alpha$}.\index{ind}{inverse metric ! generalized}\index{ind}{metric ! generalized}
\end{Def}

\begin{LemDef}\label{def:3metric}
Let $(L^4,s^{\alpha\beta},t_{\alpha\beta},\Gamma^{\mu}_{\alpha\beta},T^{\alpha\beta},\lambda)$ be a mathematical model of frame theory, $X^\alpha\in\Gamma(TL^4)$ a smooth timelike vector field. There exists a unique smooth symmetric tensor field $\myg{3}_{\alpha\beta}$ on $L^4$ satisfying
\begin{eqnarray*}
\myg{3}_{\alpha\beta}X^\alpha&=&0\\
\myg{3}_{\alpha\beta}s^{\beta\gamma}\omega_{\gamma}&=&\omega_{\alpha}
\end{eqnarray*}
for all $\omega_{\gamma}\in\Gamma(T^\ast\!L^4)$ with $X^\gamma\omega_{\gamma}=0$. We call $\myg{3}_{\alpha\beta}$ the {\bf (generalized) $3$-metric corresponding to $X^\alpha$}\index{ind}{metric ! generalized}\index{sym}{$\myg{3}_{\alpha\beta}$}. $\myg{3}_{\alpha\beta}$ is positive semi-definite when restricted to the orthogonal complement $X^\perp$. Moreover, $\myg{3}_{\alpha\beta}$, $\inv{3}{\alpha\beta}$, and $\Pi^\alpha_{\beta}$ satisfy
\begin{eqnarray*}
\Pi^\alpha_{\beta}\,\Pi^\mu_{\nu}\,\myg{3}_{\alpha\mu}&=&\Pi^\alpha_{\beta}\,\myg{3}_{\alpha\nu}=\myg{3}_{\beta\nu}\\
\lambda\,\myg{3}_{\alpha\beta}&=&\frac{X^\bullet_{\alpha}X^\bullet_{\beta}}{t_{\mu\nu}X^\mu X^\nu}-t_{\alpha\beta}\\
\inv{3}{\alpha\beta}X^\bullet_{\beta}&=&0\\
\myg{3}_{\alpha\beta}\,\inv{3}{\beta\gamma}&=&\Pi^\gamma_{\alpha}\\
\Pi^\alpha_{\beta}\,\Pi^\beta_{\gamma}&=&\Pi^\alpha_{\gamma}.
\end{eqnarray*}
\end{LemDef}

If $X^\alpha$ gives rise to an integrable tangent distribution\footnote{cf.~pp.~\pageref{subsec:foli}f.}\index{ind}{tangent distribution ! integrable}, we can interpret (the restrictions of) $\myg{3}_{\alpha\beta}$ and $\inv{3}{\alpha\beta}$ as a $3$-metric induced on the integral submanifolds and its inverse, respectively. For $\lambda>0$, i.~e.~in the case of GR, $\myg{3}_{\alpha\beta}$ indeed corresponds to the Riemannian $3$-metric induced onto the time-slice defined by $X^\alpha$ by the Lorentzian $4$-metric. Similarly, in the Newtonian case described in Proposition \ref{prop:NGisFT}, we find $\Pi^\alpha_{\beta}=\delta^\alpha_{\beta}-\delta^\alpha_{t}\delta_{\beta}^t$, $\myg{3}_{\alpha\beta}=\delta_{\alpha}^{i}\delta_{\beta}^j\delta^{ij}$ and $\inv{3}{\alpha\beta}=\delta^\alpha_{i}\delta^\beta_{i}\delta^{ij}$ when choosing $X^\alpha=\delta^\alpha_{t}$.

\subsection*{Killing Vector Fields and Staticity in Frame Theory}\label{subsec:KF}
Recall that a static spacetime in GR is a spacetime possessing a timelike Killing vector field $X^\alpha$\index{ind}{Killing vector field} which is hypersurface-orthogonal\index{ind}{hypersurface-orthogonal}\index{ind}{Killing vector field ! hypersurface-orthogonal}, i.~e.~satisfies $X_{\left[\alpha\right.}\!{\my{4}{\nabla}}_{\beta} X_{\left.\gamma\right]}=0$. In GR, hypersurface-orthogonality of the static Killing vector field allowed us to $3+1$-decompose spacetime canonically with time slices (``hypersurfaces'') orthogonal to the Killing vector field, cf.~Section \ref{sec:3+1}.

In order to define static mathematical models of FT, we need to generalize both of these notions, Killing vector fields and hypersurface-orthogonality, to FT. As both formulae implicitly refer to pulled down indices versions of the timelike vector field $X^\alpha$, this cannot be done by reinterpreting the formulae without any further thinking. To the contrary, especially the definition of hypersurface-orthogonality is fairly involved and relies on the concept of a ``defect'' tensor developed\footnote{It is attributed to Robert Geroch, there.} in \cite{Lott1}. Moreover, we will need a generalization of the concept of extrinsic curvature/second fundamental form to the realm of FT.

Let us begin by defining Killing vector fields in FT. Recall that we have extended the notion of Lie derivative $\mathcal{L}_{X^\nu}\cdot$ to include Lie derivatives of connections in Section \ref{sec:LieKill} for just this purpose.\index{ind}{Lie derivative ! of a connection}\index{sym}{$\mathfrak{L}_X\nabla$} We make the following definition.
\begin{Def}[Killing Vector Fields in FT]\label{def:KVF}
Let $\mathcal{F}:=(L^4,s^{\alpha\beta},t_{\alpha\beta},\Gamma^{\mu}_{\alpha\beta},T^{\alpha\beta},\lambda)$ be a mathematical model of frame theory, $X^\alpha\in\Gamma(TL^4)$. $X^\alpha$ is called a {\it (generalized) Killing vector field for $\mathcal{F}$}\index{ind}{Killing vector field}\index{ind}{Killing vector field ! in FT}\index{ind}{Killing vector field ! generalized} if $$\mathfrak{L}_{X^\nu} s^{\alpha\beta}=\mathfrak{L}_{X^\nu} t_{\alpha\beta}=\mathfrak{L}_{X^\nu}\Gamma^\mu_{\alpha\beta}=\mathfrak{L}_{X^\nu}{\Rm_{\alpha\beta\gamma}}^\mu=0,$$ where ${\Rm_{\alpha\beta\gamma}}^\mu$ is the induced Riemannian curvature endomorphism of the system.
\end{Def}
\begin{Rem}
Propositions \ref{Killing} and \ref{prop:GRisFT} ensure that this definition is equivalent to the classical definition of Killing vector fields in GR.
\end{Rem}

As announced above, we will need the concept of ``defect tensor'' and a generalization of extrinsic curvature/second fundamental form in FT in order to formulate hypersurface-orthogonality in FT. These notions are given by the following definitions and lemmata which correspond to Lemma 3 and Lemma 5 in \cite{Lott1}, but which we have tailored to the situation studied here in order to avoid making even more definitions.
\begin{LemDef}\label{lem:defect}
Let $(L^4,s^{\alpha\beta},t_{\alpha\beta},\Gamma^{\mu}_{\alpha\beta},T^{\alpha\beta},\lambda)$ be a mathematical model of frame theory, $X^\alpha\in\Gamma(TL^4)$ a smooth timelike vector field and $\Pi^\alpha_{\beta}$ the projection onto the orthogonal complement of $X^\alpha$. The expression $$\mathcal{D}^\gamma_{\alpha\beta}:=2\Pi^\mu_{\alpha}\,\Pi^\nu_{\beta}\,\Pi^\gamma_{\left[\nu;\mu\right]}$$ defines a tensor which we call the {\bf defect tensor of $X^\alpha$}\index{ind}{defect tensor}\index{sym}{$\mathcal{D}^\gamma_{\alpha\beta}$}. With this definition, $\mathcal{D}^\gamma_{\alpha\beta}=0$ holds if and only if the tangent distribution $D:=\{\Pi^\alpha_{\beta}Y^\beta\,\vert\,Y^\beta\in\Gamma(TL^4)\}$ is integrable.
\end{LemDef}
In simple terms, the defect tensor thus measures how far a timelike vector field is from canonically decomposing spacetime in a $3+1$ fashion. The proof of the lemma relies on Frobenius' theorem \ref{Frob}. The next lemma extends the concept of extrinsic curvature from pseudo-Riemannian geometry to FT.
\begin{LemDef}\label{lem:extcurv}
Let $(L^4,s^{\alpha\beta},t_{\alpha\beta},\Gamma^{\mu}_{\alpha\beta},T^{\alpha\beta},\lambda)$ be a mathematical model of frame theory, $X^\alpha\in\Gamma(TL^4)$ a smooth timelike vector field and $\Pi^\alpha_{\beta}$ the projection onto the orthogonal complement of $X^\alpha$. We define the {\bf associated $1$-form $\omega_{\alpha}$}\index{ind}{associated $1$-from}\index{sym}{$\omega_{\alpha}$}, the {\bf associated $\mathcal{H}$-tensor field $\mathcal{H}_{\alpha\beta}$}\index{ind}{associated $\mathcal{H}$-tensor field}\index{sym}{$\mathcal{H}_{\alpha\beta}$}, and the {\bf generalized extrinsic curvature tensor} or {\bf generalized second fundamental form $h_{\alpha\beta}$}\index{ind}{extrinsic curvature ! generalized}\index{ind}{second fundamental form ! generalized}\index{sym}{$h_{\alpha\beta}$} by
\begin{equation*}
\omega_{\alpha}:=\frac{X^\bullet_{\alpha}}{t_{\mu\nu}X^\mu X^\nu},\quad\quad \mathcal{H}_{\alpha\beta}:=-\Pi^\mu_{\beta}\,\Pi^\nu_{\alpha}\,\omega_{\mu;\nu},\quad\mbox{and}\quad h_{\alpha\beta}:=\mathcal{H}_{(\alpha\beta)},
\end{equation*}
respectively. $\omega_{\alpha}$, $\mathcal{H}_{\alpha\beta}$, and $h_{\alpha\beta}$ are smooth tensor fields. The defect tensor of $X^\alpha$ is related to $\mathcal{H}_{\alpha\beta}$ by
\begin{equation}\label{eq:extdef}
\mathcal{D}^\gamma_{\alpha\beta}=2X^\gamma\,\mathcal{H}_{\left[\alpha\beta\right]}.
\end{equation}
\end{LemDef}
\begin{Rem}
In the subcase of GR and a vector field $X^\alpha$ with vanishing defect tensor, this definition of (generalized) extrinsic curvature/second fundamental form coincides with the traditional one (if restricted to the leaves of the foliation from Lemma \ref{lem:defect}), a fact that follows from a direct computation using Proposition \ref{prop:GRisFT}.
\end{Rem}
Using these definitions and facts from \cite{Lott1}, let us now generalize the notion of hypersurface-orthogonal timelike Killing vector fields from GR to FT. Our definition is modeled on the insight that hypersurface-orthogonal timelike Killing vector fields in GR canonically $3+1$-decompose spacetime so that the time slices are orthogonal to the Killing vector field. Again, our definitions are consistent with the definitions of hypersurface-orthogonality, time slices, and staticity in GR.
\begin{Def}[Hypersurface-Orthogonality in FT]\label{def:hyportho}
Let $(L^4,s^{\alpha\beta},t_{\alpha\beta},\Gamma^{\mu}_{\alpha\beta},T^{\alpha\beta},\lambda)$ be a mathematical model of frame theory, $X^\alpha\in\Gamma(TL^4)$ a smooth timelike Killing vector field. We call $X^\alpha$ {\it hypersurface-orthogonal}\index{ind}{hypersurface-orthogonal} if the induced defect tensor $\mathcal{D}^\gamma_{\alpha\beta}$ vanishes. In that case, we call the integral submanifolds of the tangent distribution $D$ given by Lemma \ref{lem:defect} {\it time slices for the vector field $X^\alpha$}.\index{ind}{time slice}
\end{Def}
\begin{Def}[Staticity in FT]\label{def:staticFT}
Let $\mathcal{F}:=(L^4,s^{\alpha\beta},t_{\alpha\beta},\Gamma^{\mu}_{\alpha\beta},T^{\alpha\beta},\lambda)$ be a mathematical model of frame theory. $\mathcal{F}$ is called {\it static}\index{ind}{static}\index{ind}{static ! mathematical model of FT} if there exists a timelike Killing vector field $X^\alpha\in\Gamma(TL^4)$ which is hypersurface-orthogonal.\end{Def}
The following theorem is now immediate.
\begin{Thm}[Static Metric Theorem]\label{thm:staticisstatic}
Let $(L^4,ds^2=g_{\alpha\beta})$ be a smooth Lorentzian spacetime, $T^{\alpha\beta}$ a smooth symmetric matter tensor field on $L^4$. Suppose that $g_{\alpha\beta}$ and $T^{\alpha\beta}$ satisfy the Einstein equations \eqref{EEq} with respect to the Levi-Civit\`a connection $\Gamma^\mu_{\alpha\beta}$ of $g_{\alpha\beta}$. Define the associated mathematical model $\mathcal{F}$ of FT as in Proposition \ref{prop:GRisFT}. Then $\mathcal{F}$ is static in the sense of Definition \ref{def:staticFT} if and only if it is static in the sense of GR (with respect to the same field $X^\alpha$).

Moreover, if the spacetime is static, the fields $\myg{3}_{ij}$ and $\inv{3}{ij}$ arising as the restrictions of the generalized $3$-metric and inverse $3$-metric $\myg{3}_{\alpha\beta}$ and $\inv{3}{\alpha\beta}$ to the time slices coincide with the $3$-metric $\myg{3}_{ij}$  and its inverse $\inv{3}{ij}$ induced by $ds^2$.
\end{Thm}
\begin{Pf}
Follows from straightforward calculations and from Propositions \ref{Killing} and \ref{prop:GRisFT}.
\qed\end{Pf}
Just as in GR, the time slices of static mathematical models in FT have vanishing second fundamental form. This is asserted in the following proposition.
\begin{Prop}\label{prop:h0}
Let $\mathcal{F}:=(L^4,s^{\alpha\beta},t_{\alpha\beta},\Gamma^{\mu}_{\alpha\beta},T^{\alpha\beta},\lambda)$ be a mathematical model of frame theory which is static with respect to $X^\alpha\in\Gamma(TL^4)$. Then the associated $\mathcal{H}$-tensor field and generalized mean curvature tensor satisfy $$\mathcal{H}_{\alpha\beta}=0\quad\mbox{and}\quad h_{\alpha\beta}=0.$$ Moreover, the $1$-form $\omega_{\alpha}$ is closed.
\end{Prop}
\begin{Pf}
Clearly, as $\mathcal{D}^\gamma_{\alpha\beta}=0$ by staticity, $\mathcal{H}_{\alpha\beta}$ must be symmetric and thus $\mathcal{H}_{\alpha\beta}=h_{\alpha\beta}$. Let $\Pi^\alpha_{\beta}$ denote the projection onto the orthogonal complement of $X^\alpha$. By definition of $h_{\alpha\beta}$, we find
\begin{eqnarray*}
h_{\alpha\beta}&=&-\Pi^\mu_{\beta}\,\Pi^\nu_{\alpha}\,\omega_{\mu;\nu}\\
&\stackrel{\mbox{\scriptsize def. }\omega_{\mu}}{=}&-\Pi^\mu_{\beta}\,\Pi^\nu_{\alpha}\,\left(\frac{t_{\mu\rho}X^\rho_{\;;\nu}}{t_{\kappa\tau}X^\kappa X^\tau}-\frac{2\omega_{\mu}t_{\psi\phi}X^\psi X^\phi_{\;;\nu}}{t_{\kappa\tau}X^\kappa X^\tau}\right)\\
&\stackrel{\Pi^\mu_{\beta}\omega_{\mu}=0}{=}&-\Pi^\mu_{\beta}\,\Pi^\nu_{\alpha}\,\frac{t_{\mu\rho}X^\rho_{\;;\nu}}{t_{\kappa\tau}X^\kappa X^\tau}.
\end{eqnarray*}
Now the left hand side of this equation is symmetric by construction while the far right hand side is antisymmetric by the Killing equation $\mathcal{L}_{X^\nu}t_{\alpha\beta}=0$. Thus $h_{\alpha\beta}=\mathcal{H}_{\alpha\beta}=0$.

$\mathcal{H}_{\alpha\beta}=0$ implies that $\omega_{\left[\alpha;\beta\right]}$ vanishes on the product of the orthogonal complement $X^\perp$ with itself. To show that $\omega_{\alpha}$ is closed, it thus suffices to show that $\omega_{\left[\alpha;\beta\right]}X^\alpha Y^\beta=0$ for all $Y^\beta\in X^\perp$. This goes as follows: By the Leibniz rule, we find
\begin{eqnarray*}
2\omega_{\left[\alpha;\beta\right]}X^\alpha Y^\beta&\stackrel{\mbox{\scriptsize def. }\omega_{\alpha}}{=}&X^\mu_{\;;\beta}Y^\beta \omega_{\mu}-2\underbrace{\omega_{\alpha}X^\alpha}_{=1}\omega_{\mu}X^\mu_{\;;\beta}Y^\beta\\
&&-\frac{X^\mu_{\;;\beta}t_{\alpha\mu}X^\beta Y^{\alpha}}{t_{\sigma\tau}X^\sigma X^\tau}+2\underbrace{\omega_{\alpha}Y^\alpha}_{=0}\omega_{\mu}X^\mu_{\;;\beta}\\
&\stackrel{X\,\mbox{\scriptsize Killing, def. }\omega_{\alpha}}{=}&-X^\mu_{\;;\beta}Y^\beta \omega_{\mu}+X^\mu_{\;;\beta}Y^\beta \omega_{\mu}=0.
\end{eqnarray*}
This implies $\mathcal{H}_{\alpha\beta}=h_{\alpha\beta}=-\Pi^\mu_{\beta}\Pi^\nu_{\alpha}\omega_{\mu;\nu}=0$ and thus finishes the proof.
\qed\end{Pf}

The matter quantities $\rho,J^\alpha,S^{\alpha\beta}$ also behave accordingly as the following proposition asserts.
\begin{Lem}
Let $(L^4,s^{\alpha\beta},t_{\alpha\beta},\Gamma^{\mu}_{\alpha\beta},T^{\alpha\beta},\lambda)$ be a static mathematical model of frame theory with respect to some field $X^\alpha\in\Gamma(TL^4)$. Then the momentum density $J^\alpha$ observed by $X^\alpha$ vanishes and the matter density $\rho$ and stress tensor $S^{\alpha\beta}$ observed by $X^\alpha$ are constant with respect to $X^\alpha$ (in the Lie sense).
\end{Lem}

In order to close this section on staticity in FT, let us quickly review some of the above notation and suggest some suggestive abuse of it for the remainder of this thesis. Let $\mathcal{F}:=(L^4,s^{\alpha\beta},t_{\alpha\beta},\Gamma^{\mu}_{\alpha\beta},T^{\alpha\beta},\lambda)$ be a static mathematical model of FT with respect to $X^\alpha\in\Gamma(TL^4)$. Mimicking our topological assumptions described in Section \ref{sec:SME}, we call $\mathcal{F}$ {\it standard static}\index{ind}{standard static} if $L^4$ can be decomposed into $\R\times M^3$, where $M^3$ is any of the time slices of $L^4$ with respect to $X^\alpha$. Henceforth, we restrict our attention to standard static mathematical models of FT.

For any of these standard static mathematical models $\mathcal{F}$ \index{sym}{$\mathcal{F}$} of FT, let $\myg{3}_{ij}$\index{sym}{$\myg{3}$} and $\inv{3}{ij}$\index{sym}{$\inv{3}{ij}$} denote the restriction of the generalized $3$-metric and inverse $3$-metric to the tangent bundle of $M^3$, respectively. Let the {\it lapse function $N$}\index{sym}{$N$}\index{ind}{lapse function} of $\mathcal{F}$ be given by $N:=\sqrt{t_{\alpha\beta}X^\alpha X^\beta}$ and observe that this coincides with the definition given in the relativistic case as $t_{\alpha\beta}=-c^{-2}ds^2$ in that case by Proposition \ref{prop:GRisFT}. We will see below that $N=\mbox{const.}$ in the Newtonian case (i.~e.~for $\lambda=0$).

\subsubsection*{Staticity in the Case $\lambda=0$}\label{subsec:Newton}
For the remainder of this subsection, let us consider static mathematical models of FT with $\lambda=0$. In that case, any hypersurface-orthogonal timelike Killing vector field $X^\alpha$ has constant length because of the following consideration: By definition of staticity, the defect tensor vanishes. Translating this into components, we obtain $X^\alpha_{\;\,;\left[\mu\right.}\omega_{\left.\nu\right]}=0$. Multiplying this with $X^\nu$ gives $X^\alpha_{\;;\mu}=X^\alpha_{\;;\nu}X^\nu\omega_{\mu}$. The trace of this equation amounts to $X^\alpha_{\;;\alpha}=X^\alpha_{\;;\nu}X^\nu\omega_{\alpha}=-X^\alpha_{\;;\nu}X^\nu\omega_{\alpha}$ where the last step comes from $X^\alpha$ being a Killing vector field. This, however, gives us $X^\alpha_{\;;\alpha}=0$ and thus, again using the Killing property and the fact that $\omega_{\alpha}$ is closed, we obtain $(t_{\alpha\beta}X^\alpha X^\beta)_{;\mu}=-2\omega_{\mu}X^\nu_{\;;\nu}(t_{\alpha\beta}X^\alpha X^\beta)=0$. This shows that $X^\alpha$ has constant length. We henceforth assume that this constant equals $1$ without loss of generality. Then, in particular, $\omega_{\alpha}=X^\bullet_{\alpha}$ for $\lambda=0$. 

This puts us in the position to cite some results on the $\lambda=0$ case from \cite{Lott1}. There, it is shown on pp.~40ff that for $\lambda=0$ and any choice of unit length timelike vector field $\widetilde{X}^\alpha$, the associated $\widetilde{\mathcal{H}}$-tensor (which is called $\chi_{\alpha\beta}$, there) must automatically vanish. This implies that the induced defect tensor $\widetilde{\mathcal{D}}^\gamma_{\alpha\beta}$ must vanish identically for any such observer. Moreover, it is asserted there that the $3$-dimensional Ricci tensor on the time-slices with respect to this observer $\widetilde{X}^\alpha$ must vanish and thus by Formula \eqref{formula:3Rm} the time slices must be intrinsically flat. 

Now, as our particular observer $X^\alpha$ is a Killing vector field for $\myg{3}_{\alpha\beta}$, there must locally exist ``Galilei coordinates'' $(t,x^{i})$\index{ind}{Galilei coordinates} as well as possibly time dependent vector fields $\vec{g},\vec{\Omega}$ on the time slices such that $\vec{g}$ is a ``force'' and $\vec{\Omega}$ is a ``Coriolis force''. $\vec{g}$ is derived from the gravitational field $\Gamma^\mu_{\alpha\beta}$ and characterizes it completely. Since $X^\alpha$ is a Killing vector field for $s^{\alpha\beta}$ and $t_{\alpha\beta}$, neither $\vec{g}$ nor $\vec{\Omega}$ depend on time in our case. This implies that $\rot\vec{g}=0$ and thus, locally, there is a (time independent) {\it Newtonian potential $U$} for $\vec{g}$, i.~e.~$\vec{g}=-\grad U$.\index{ind}{potential ! Newtonian}\index{sym}{$U$} Observe that this Newtonian potential refers to a given system of Galilei coordinates $(t,x^{i})$ and thus in particular to a given system of Cartesian coordinates $(x^{i})$.

As it is shown in \cite{Lott1}, suitable fall-off for $\vec{\Omega}$ -- namely any fall-off that ensures a uniquely solvable Dirichlet problem for the (flat) Laplace equations $\triangle\Omega^{i}=0$ -- then gives $\vec{\Omega}=0$ (if the time slices are diffeomorphic to $\R^3$). This holds in particular if the system is ``isolated'', cf.~p.~5 in \cite{OlySchmi}. Combined with the above representation of $\vec{g}$ using the Newtonian potential $U$, we can read off the well-known (local) equation $$\triangle U=4\pi G\rho.$$ If the time slices are simply connected, the Galilei/Cartesian coordinates and the potential $U$ exist globally; we normalize $U$ by $U\to0$ as $r\to\infty$ in the case of isolated systems.

\subsection*{Asymptotic Flatness in Static Frame Theory}
In the above section, we have unified the concepts of staticity coming from GR and NT into Definition \ref{def:staticFT}, cf.~Theorem \ref{thm:staticisstatic} and the Newtonian considerations in the subsection on pp.~\pageref{subsec:Newton}f. This section is now dedicated to generalizing the concept of asymptotic flatness to FT (for static mathematical models). Again, we will model our definitions on the relativistic notions and discuss their relation to the Newtonian ones at the end of this subsection, cf.~Theorem \ref{thm:staticiso}.

In analogy to our definition of geometrostatic systems, cf.~Definition \ref{def:geomstatic}, we define ``static isolated systems in FT'' as follows.
\begin{Def}\label{def:FTasym}
Let $\mathcal{F}=(\R\times M^3,s^{\alpha\beta},t_{\alpha\beta},\Gamma^{\mu}_{\alpha\beta},T^{\alpha\beta},\lambda)$ be a static mathematical model of frame theory and let $k\in\N$, $k\geq3$, and $\tau\geq1/2$ such that $-\tau$ is non-exceptional (i.~e.~$\tau\notin\Z$). We call $\mathcal{F}$ a {\it $(k,\tau)$-static isolated system in FT}\index{ind}{static isolated system in FT}\index{ind}{system ! static isolated} if, in addition, the following conditions hold for the induced generalized $3$-metric $\myg{3}_{ij}$ and lapse function $N$:
\begin{enumerate}
\item[(ia)] $(M^3,\myg{3}_{ij})$ is a $(k,q=2,\tau)$-asymptotically flat manifold.
\item[(ib)] If $\lambda=0$, there exit global {\bf Cartesian coordinates} on $M^3$, i.~e.~coordinates $(x^k)$ such that $(\myg{3}_{ij})=(\delta_{ij})$ holds in components with respect to these coordinates.\index{ind}{Cartesian coordinates}
\item[(ii)] $N>0$ in $M^3$ and $N(p)\to1$ as $p\to\infty$ in each end of $M^3$.
\item[(iii)] $\rho\geq0$, $S_{ij}\geq0$, and the supports of $\rho$ and $S_{ij}$ are bounded away from infinity.
\end{enumerate}
As before, we will call $\mathcal{F}$ a static isolated system in FT system for short if $k$ and $\tau$ are either clear from context or arbitrary. For abbreviational purposes, we call the hypersurface-orthogonal timelike Killing vector field $X^\alpha$ making $\mathcal{F}$ a static isolated model of FT the {\it staticity field for $\mathcal{F}$}\index{ind}{staticity field}.
\end{Def}
\begin{Rems}
This definition implicitly includes the topological assumptions that $(M^3,\myg{3}_{ij})$ is either geodesically complete or an individual end diffeomorphic to an exterior domain of $\R^3$ (through (ia)). Through (ib), condition (ia) in fact reduces to a topological condition if $\lambda=0$. Observe that condition (ii) is void for $\lambda=0$.
\end{Rems}

For later convenience, we also redefine the concept of wave-harmonic coordinates in the frame theoretical setting.
\begin{Def}
If $(x^{k})$ is a system of local coordinates for the time slice $M^3$ of a given static isolated system $\mathcal{F}$ in FT, then we call $(x^{k})$ {\it wave-harmonic coordinates for $\mathcal{F}$}\index{ind}{wave harmonic}\index{ind}{coordinates ! wave harmonic} if they satisfy $$\mylap{\myg{3}_{ij}} x^k = -\frac{N_{,l}}{N}\,\inv{3}{lk}.$$
\end{Def}
\begin{Rem}
By Lemma \ref{waveharm}, this definition coincides with the relativistic one if $\lambda>0$. In the $\lambda=0$ case, $N=1$ and $\myg{3}_{ij}$ is flat and thus wave harmonic coordinates are ordinary harmonic coordinates with respect to the Euclidean metric. In particular, Cartesian coordinates are wave harmonic if $\lambda=0$.
\end{Rem}

Together, the above considerations imply the following theorem.
\begin{Thm}[Static Isolated Systems Theorem]\label{thm:staticiso}
Let $k\in\N$, $k\geq3$, and $\tau\geq1/2$ such that $-\tau$ is non-exceptional. Let $\mathcal{F}=(\R\times M^3,s^{\alpha\beta},t_{\alpha\beta},\Gamma^{\mu}_{\alpha\beta},T^{\alpha\beta},\lambda)$ be a $(k,\tau)$-static isolated system in FT. If $\lambda>0$, $\mathcal{F}$ naturally corresponds to the $(k,\tau)$-geometrostatic system $(M^3,\myg{3}_{ij},N,\rho,S_{ij})$. Here, $M^3$ denotes the time slice of $\mathcal{F}$, $\myg{3}_{ij}$ and $N$ denote the (restriction of the) generalized $3$-metric and the lapse function, respectively. $\rho$ and $S_{ij}$ denote the mass density and stress tensor induced from $T^{\alpha\beta}$, respectively.

If, on the other hand, $\lambda=0$ and $M^3$ is simply connected, then $\mathcal{F}$ naturally corresponds to a static isolated Newtonian system with a global system of asymptotically flat Cartesian coordinates on $M^3$ and a corresponding smooth Newtonian potential $U:M^3\to\R$.

Moreover, a system $(x^{k})$ of local coordinates for $M^3$ is wave harmonic for $\mathcal{F}$ if and only if it is wave harmonic in the geometrostatic sense (for $\lambda>0$) or harmonic with respect to the Euclidean metric (if $\lambda=0$).
\end{Thm}
\begin{Rem}
The role of the Newtonian potential $U$ appearing in the $\lambda=0$ case is exactly the one described in Proposition \ref{prop:NGisFT}; similarly, the $\lambda>0$ case relies on Proposition \ref{prop:GRisFT}.
\end{Rem}
\begin{Pf}
For positive causality constant, this just summarizes the results from Proposition \ref{prop:GRisFT}, Theorem \ref{thm:staticisstatic}, and the discussion before and after them. For vanishing causality constant, this is a summary of the results of \cite{Lott1} and our interpretation of them as discussed on pp.~\pageref{subsec:Newton}f using the afore-mentioned isolated systems idea formulated on p.~5 in \cite{OlySchmi} and the above considerations of wave harmonic coordinates.
\qed\end{Pf}

\subsection*{Pseudo-Newtonian Reformulation of Static Isolated Systems in FT}\label{subsec:pseudoFT}
As a final step towards our analysis of the Newtonian limit of mass and center of mass, we need to translate the notion of pseudo-Newtonian systems into FT. This is indicated because we intend to take the Newtonian limit of the quasi-local formulae of mass and center of mass which are defined in the language of pseudo-Newtonian gravity, cf.~Sections \ref{sec:mpseudo} and \ref{sec:CoMpseudo}. We make the following definition of pseudo-Newtonian metric and potential.
\begin{Def}\label{def:FTpseudo}
Let $\mathcal{F}=(\R\times M^3,s^{\alpha\beta},t_{\alpha\beta},\Gamma^{\mu}_{\alpha\beta},T^{\alpha\beta},\lambda)$ be a static isolated system in FT, $N$ its lapse function and $\myg{3}_{ij}$ its generalized $3$-metric. We define the {\it (generalized) pseudo-Newtonian metric $\gamma_{ij}$ of $\mathcal{F}$}\index{ind}{pseudo-Newtonian ! metric (generalized)}\index{ind}{metric ! pseudo-Newtonian}\index{sym}{$\gamma_{ij}$} to be the conformally transformed metric $\gamma_{ij}:=N^2\,\myg{3}_{ij}$ on $M^3$. Now, if $\lambda=0$, assume that $M^3$ is simply connected and pick a system of Cartesian coordinates for $M^3$. We define the {\it (generalized) pseudo-Newtonian potential $U$ of $\mathcal{F}$}\index{ind}{pseudo-Newtonian ! potential (generalized)}\index{ind}{potential ! pseudo-Newtonian (generalized)}\index{sym}{$U$} by $$U:M^3\to\R:p\mapsto
\begin{cases}
\lambda^{-1}\ln N(p)&\lambda\neq0\\
U(p)&\lambda=0,
\end{cases}$$ where in case $\lambda=0$ the function $U:M^3\to\R$ is the Newtonian potential given with respect to the chosen Cartesian coordinates and normalized by $U\to0$ as $r\to\infty$. 
\end{Def}
This definition of pseudo-Newtonian potential $U$ might not seem very natural as the cases $\lambda>0$ and $\lambda=0$ are treated separately. However, it is specifically tailored to ensure coincidence with the afore-defined pseudo-Newtonian and the classical Newtonian concept of a ``potential'', cf.~Theorems \ref{thm:pseudoFTispseudo} and \ref{thm:Uconv}. We can thus interpret the difference in the definitions of $U$ as being due to the systematic difference between relativistic and Newtonian systems. 

The above-mentioned coincidence is summarized in the following theorem.
\begin{Thm}[Pseudo-Newtonian Systems Theorem]\label{thm:pseudoFTispseudo}
Let $\mathcal{F}$ be a static isolated system in FT with causality constant $\lambda$ and time slice $M^3$. Assume that $M^3$ is simply connected if $\lambda=0$. The following statements hold true:
\begin{itemize}
\item If  $\lambda>0$, then the generalized pseudo-Newtonian notions of metric and potential coincide with the pseudo-Newtonian notions defined in geometrostatics.
\item If $\lambda=0$, then the generalized pseudo-Newtonian metric is flat and the generalized pseudo-Newtonian potential coincides with the classical Newtonian one (both referring to the same system of Cartesian coordinates).
\end{itemize}
Moreover, a system $(x^{k})$ of local coordinates for $M^3$ is wave harmonic for $\mathcal{F}$ if and only if it is harmonic for the associated generalized pseudo-Newtonian metric $\gamma_{ij}$.
\end{Thm}
\begin{Pf}
Let $\lambda>0$, first. The lapse function $N$ and the generalized $3$-metric $\myg{3}_{ij}$ agree with the corresponding geometrostatic variables by Theorem \ref{thm:staticiso}. As $U$ and $\gamma_{ij}$ are constructed from them in exactly the same manner in FT and in geometrostatics (cf.~p.~\pageref{U}), they must also coincide. Coincidence of wave harmonic and $\gamma_{ij}$-harmonic coordinates then follows from Lemma \ref{waveharm}.

For $\lambda=0$, the conformal change is trivial as $N=1$. Moreover, the generalized pseudo-Newtonian potential agrees with the Newtonian one by construction. By Theorem \ref{thm:staticiso}, wave harmonic coordinates are harmonic coordinates with respect to the flat metric $\myg{3}_{ij}=\gamma_{ij}$, automatically.
\qed\end{Pf}
Before we move on to define the Newtonian limit, we give another indication of why the pseudo-Newtonian potentials play similar roles in the relativistic and Newtonian settings: Recall that the Newtonian potential has arisen as a ``potential'' for the $4$-dimensional connection $\Gamma^\mu_{\alpha\beta}$ as in Proposition \ref{prop:NGisFT}, namely $\Gamma^\mu_{\alpha\beta}=\delta^\mu_{i}\delta_{\alpha}^t\delta_{\beta}^t\delta^{ij}U_{,j}$ in Cartesian coordinates. This expression is similar to the first term in the related expression for $\lambda>0$:
\begin{eqnarray}\nonumber
\Gamma^{\mu}_{\alpha\beta}&=&\delta^\mu_{i}\delta_{\alpha}^t\delta_{\beta}^te^{4\lambda U}\gamma^{ij}U_{,j}+\delta^{\mu}_s\delta^{k}_{\alpha}\delta^l_{\beta}\,\my{\gamma_{ij}}\!{\Gamma}^s_{kl}\\\label{eq:4nablaU}
&&+\lambda\left(2\delta^\mu_{t}\delta_{\left(\alpha\right.}^t\delta_{\left.\beta\right)}^{i} U_{,i}-2\delta^{\mu}_s\delta^{k}_{\left(\alpha\right.}\delta^l_{\left.\beta\right)}U_{,k}\delta^s_{l}+\delta^{\mu}_s\delta^{k}_{\alpha}\delta^l_{\beta} U_{,p}\gamma^{ps}\gamma_{kl}\right).
\end{eqnarray}
This formula is a direct consequence of Theorem \ref{thm:pseudoFTispseudo}. Anticipating that $\gamma_{ij}$ converges to $\delta_{ij}$, $\my{\gamma_{ij}}\!{\Gamma}^s_{kl}$ converges to $0$, and $\Gamma^{\mu}_{\alpha\beta}$ converges to $\Gamma^{\mu}_{\alpha\beta}$ in the Newtonian limit, a comparison of the two above expression suggests that the separate definitions of the pseudo-Newtonian potential interlock reasonably in the Newtonian limit.

This argument closes our analysis of static isolated systems in frame theory. We continue our visit to frame theory by discussing the Newtonian limit in general and in the context of static isolated systems in particular.

\section{The Newtonian Limit of Geometrostatics}\label{sec:NLgeo}
In order to enable us to properly define the Newtonian limit, let us quickly introduce notions of convergence for families of coordinates and tensor fields: Let $M^n$ be a smooth manifold and $\Omega\subset M^n$ an open subset. Let $(x^k(\lambda))$ be a family of systems of coordinates on $\Omega$ which is parametrized by $\lambda\in(0,\varepsilon)$ for some $\varepsilon>0$. Let also $(x^k(0))$ be a system of coordinates on $\Omega$. We say that $(x^k(\lambda))$ {\it converges pointwise}\index{ind}{convergence ! pointwise} to $(x^k(0))$ as $\lambda\to0$ on $\Omega$ if $x^k(\lambda)\vert_{p}\to x^k(0)\vert_{p}$ for all $p\in \Omega$ and all $k=1,\dots,n$. We say that this convergence is {\it uniform on $\Omega$}\index{ind}{convergence ! uniform} if $\lVert x^k(\lambda)-x^k(0)\rVert_{C^0(\Omega)}\to0$ as $\lambda\to0$.

Similarly, we define pointwise and uniform convergence of families of tensor fields on $M^n$. Let $T_{i_1 \dots i_s}^{j_1\dots j_t}(\lambda)$ be a family of tensor fields on an open subset $\Omega\subset M^n$ parametrized by $\lambda\in(0,\varepsilon)$. Let $(x^k(\lambda))$ be local coordinates on $\Omega$ converging pointwise to local coordinates $(x^k(0))$ on $\Omega$. We say that $T_{i_1 \dots i_s}^{j_1\dots j_t}(\lambda)$ {\it converges pointwise}\index{ind}{convergence ! pointwise} to some tensor field $T_{i_1 \dots i_s}^{j_1\dots j_t}(0)$ with respect to $(x^k(\lambda))$ as $\lambda\to0$ if for each $p\in M^n$, the family of component matrices $(T_{i_1 \dots i_s}^{j_1\dots j_t}(\lambda)\vert_{p})$ with respect to $(x^k(\lambda))$ converges to the component matrix $(T_{i_1 \dots i_s}^{j_1\dots j_t}(0)\vert_{p})$ with respect to $(x^k(0))$ as $\lambda\to0$. This convergence is measured with respect to the norm induced on the corresponding space $L_{p}(s,t)$ of multilinear maps from $(T_{p}M^n)^s\times(T_{p}^\ast\!M^n)^t$ into $\R$ by the flat metric $\delta_{ij}\vert_{p}$ in the given coordinates $(x^k(0))$.

Note that this convergence depends on the chosen local coordinates $(x^k(\lambda))$ because the coordinate transformations between two families of coordinates parametrized by $\lambda$ need not behave properly as $\lambda\to0$. The convergence does not, however, depend on the norm chosen on $L_{p}(s,t)$ as this space is finite dimensional and thus all norms on it are equivalent.

This definition can straightforwardly be extended to include pointwise convergence of families of connections parametrized by $\lambda\in(0,\varepsilon)$ as the difference of two connections is a tensor.

Moreover, we say that $T_{i_1 \dots i_s}^{j_1\dots j_t}(\lambda)$ {\it converges uniformly}\index{ind}{convergence ! uniform} to $T_{i_1 \dots i_s}^{j_1\dots j_t}(0)$ with respect to $(x^k(\lambda))$ as $\lambda\to0$ if $(x^k(\lambda))$ converges uniformly to $(x^k(0))$ and if the family of component matrices $(T_{i_1 \dots i_s}^{j_1\dots j_t}(\lambda))$ converges uniformly to the component matrix $(T_{i_1 \dots i_s}^{j_1\dots j_t}(0))$ with respect to the norm induced on the corresponding tensor bundle on $\Omega$ by the flat metric $\delta_{ij}$ in the given coordinates $(x^k(0))$. Again, this convergence does depend on the coordinates chosen.

We now make the following definition of the Newtonian limit following \cite{Lott1} but specifying the sense of convergence more precisely.
\begin{Def}\label{def:NL}
Let $\mathcal{F}(\lambda):=(L^4,s^{\alpha\beta}(\lambda),t_{\alpha\beta}(\lambda),\Gamma^{\mu}_{\alpha\beta}(\lambda),T^{\alpha\beta}(\lambda),\lambda)$ be a family of mathematical models of frame theory parametrized by $\lambda\in(0,\varepsilon)$ for some $\varepsilon>0$. We say that the family $\mathcal{F}(\lambda)$ has a {\it pointwise Newtonian limit} $\mathcal{F}(0):=\left(L^4,s^{\alpha\beta}(0),t_{\alpha\beta}(0)\right.$, $\left.\Gamma^{\mu}_{\alpha\beta}(0),T^{\alpha\beta}(0),0\right)$\index{ind}{Newtonian limit}\index{sym}{$\mathcal{F}$}\index{ind}{Newtonian limit ! pointwise} if
\begin{itemize}
\item $\mathcal{F}(0)$ is a mathematical model of FT.
\item The fields $s^{\alpha\beta}(\lambda),t_{\alpha\beta}(\lambda),\Gamma^{\mu}_{\alpha\beta}(\lambda),T^{\alpha\beta}(\lambda)$ converge to the corresponding fields of $\mathcal{F}(0)$ as $\lambda\to0$ pointwise on $L^4$, respectively.
\item The induced curvature endomorphisms ${\Rm_{\alpha\beta\gamma}}^\mu(\lambda)$ converge pointwise on $L^4$ to the induced curvature endomorphism ${\Rm_{\alpha\beta\gamma}}^\mu(0)$ as $\lambda\to0$.
\end{itemize}
If $\Omega\subset L^4$ is an open subset with uniformly convergent coordinates $(x^\alpha(\lambda))\to(x^\alpha(0))$ as $\lambda\to0$, then we say that $\mathcal{F}(\lambda)$ has the {\it uniform Newtonian limit}\index{ind}{Newtonian limit ! uniform} $\mathcal{F}(0)$  on $\Omega$ if (the restrictions to $\Omega$ of) all these fields including the Riemannian curvature endomorphisms converge uniformly to the respective fields on $\Omega$ as $\lambda\to0$.
\end{Def}
\begin{Rem}
Pointwise and uniform Newtonian limits are clearly not the only possibilities. In fact, the preferred notion of convergence might actually depend on the context. If, for example, one wants to consider limits of asymptotically flat mathematical models, weighted Sobolev spaces with respect to particular asymptotically flat systems of coordinates $(x^\alpha(\lambda))$ can be more suitable. This approach is taken in many papers proving existence of families of mathematical models of FT having a Newtonian limit.

For the purpose of this thesis, however, we will be content with the uniform Newtonian limit as uniform convergence helps to see the souls of our arguments. This is particularly true for the theorems on convergence of mass (Theorem \ref{thm:NLm}) and center of mass (Theorem \ref{thm:NLCoM}) proved below. Note that although these are formulated by with uniform convergence, they can also be applied in a weighted Sobolev space setting whenever those spaces can be embedded into $C^0$. Alternatively, it should be possible to reprove the theorems in different norms if required for applications but this would lead us to far, here.
\end{Rem}

It might be surprising at first that the Newtonian limit is not defined for an individual relativistic system but for a whole family of relativistic spacetimes. This surprise might be due to the usage of the term ``Newtonian limit'' frequently encountered in the literature. It is however implicit in the popularizing notation $c\to\infty$ that whole families parametrized by $c$ in some sense are or at least should be considered. As a matter of fact, constructing a family that converges under the Newtonian limit and models a specific physical situation (or includes an individual given relativistic system) is by no means trivial. For example, the time intervals on which solutions to the $3+1$-decomposed Einstein equations exist can shrink to zero length when the ``scaling'' with $c$ is not carefully adjusted. It is therefore important to study existence of families of (physically relevant) relativistic systems having a Newtonian limit. We refer the interested reader to the overview article \cite{OlySchmi} and references cited therein for an introduction and an overview of results on existence of such families.

But assigning a family which converges in the Newtonian limit not only is non-trivial. Neither is it unique. Besides the choice of family including a given relativistic system, this is due to the fact that the convergence is considered with respect to a given system of coordinates, e.~g.~asymptotically flat ones in the case of isolated systems. Waving our hands, we could say that this choice of coordinates actually specifies a way of ``zooming in'' to a particular region of the spacetimes considered. For example, even for the very simple case of the spherically symmetric static Schwarzschild metrics introduced on pp.~\pageref{subsec:schwarz}ff, it is possible to choose coordinates ``zooming in'' to a vacuum region outside the spherically symmetric star or black hole and thus obtain a completely empty Newtonian limit which contains no physical information whatsoever. This can e.~g.~be done by choosing coordinates $(t,x^{i}(\lambda))$ in the Schwarzschild spacetime with $x^{i}(\lambda)=x^{i}(0)+z^{i}(\lambda)$, where $z^{i}(\lambda)$ is a family of ``centers of mass'' moving away suitably fast from the original center of mass and rotation.

Another example for this non-uniqueness and its consequences for physical interpretation of the Newtonian limit is discussed by Ji\v{r}\'i Bi\v{c}\'ak and David Kofro\v{n} \cite{Jiri}. They study accelerated particles and their Newtonian limits comparing an approach that ``loses track of the particles'' to one ``riding on them''.

We interpret this as saying that there is not something like a ``physically most reasonable'' Newtonian limit of a given relativistic system in general. To the contrary, the Newtonian limit ``of a system'' very much depends on choices made by those who study it, namely the choice of family including the given system and the choice of coordinates. Differently put, what one actually studies when examining the Newtonian limit ``of a system'' is the Newtonian limit of a specific way of looking at this system as a special case of a given family parametrized by $\lambda=c^{-2}$ and looked at in a given family of coordinates $(x^\alpha(\lambda))$. Which family is adequate for studying the Newtonian limit of a given relativistic system thus depends on what aspects of the system one is actually interested in.

Having this at the back of our minds, it seems natural to consider families $\mathcal{F}(\lambda)$ only consisting of static isolated systems in FT in order to understand the Newtonian limit of geometrostatics. This means that we are restricting our attention to families of (frame theoretical versions of) geometrostatic systems and not taking into account families that include  one such a system but become dynamical or non-isolated along the way. Sure enough, we would expect that such a family of geometrostatic systems converges to a static isolated Newtonian system in the Newtonian limit. We will prove this in Theorem \ref{def:NLstaticformal}.

More concretely, we assume that the staticity fields $X^\alpha(\lambda)$ inducing staticity of $\mathcal{F}(\lambda)$ have a uniform limit $X^\alpha(0)$ as $\lambda\to0$ which is a timelike vector field in $\mathcal{F}(0)$. In addition, we assume that first covariant derivatives $X^\alpha_{\;;\beta}(\lambda)$ converge uniformly to $X^\alpha_{\;;\beta}(0)$. Note that although this looks like a requirement on the interchange of limits, this is not exactly the case as the covariant derivative indicated by the semicolon itself depends on the causality constant.

It is then automatic that $X^\alpha(0)$ is a hypersurface-orthogonal timelike Killing vector field because both the Killing equations in Definition \ref{def:KVF} and the vanishing of the defect tensor (cf.~Definition \ref{def:hyportho}) are closed\footnote{By ``closed conditions'' we mean conditions parametrized by $\lambda$ which are in some sense continuous under the limit $\lambda\to0$, i.~e.~if they are satisfied for all $0<\lambda<\varepsilon$, they must also be satisfied for $\lambda=0$,} under pointwise and thus also under uniform convergence. In particular, these conditions only involve $X^\alpha(\lambda)$ and its first covariant derivatives and the tensors and connections given by $\mathcal{F}(\lambda)$.

When we want to compare the asymptotic behavior of $X^\alpha(\lambda)$ and $X^\alpha(0)$, it is necessary to fix asymptotically flat coordinates for all systems $\mathcal{F}(\lambda)$ and for $\mathcal{F}(0)$. So let us assume that there exists an open neighborhood of infinity\footnote{To be precise, we have to assume existence of such subsets in each of the ends of $M^3$.} $\Omega\subset M^3$ and asymptotically flat coordinates $(x^k(\lambda))$ for $\mathcal{F}(\lambda)$ and $(x^k(0))$ for $\mathcal{F}(0)$ defined on all of $\Omega$ such that the family of coordinates $(x^k(\lambda))$ converges uniformly to $(x^k(0))$ on $\Omega$ as $\lambda\to0$. Moreover, we assume that the coordinates $(x^k(0))$ are Cartesian. Observe that this in particular implies the radii $r(\lambda)$ converge to the radius $r(0)$ uniformly on $\Omega$ as $\lambda\to0$.

Now assume in addition that the fields $X^\alpha(\lambda)$ and $X^\alpha_{\;;\beta}(\lambda)$ converge to their counterparts $X^\alpha(0)$ and $X^\alpha_{\;;\beta}(0)$ uniformly on $\Omega$ with respect to the asymptotically flat coordinates $(x^k(\lambda))$ and $(x^k(0))$. Recall from pp.~\pageref{subsec:Newton}ff that because $X^\alpha(0)$ is a hypersurface-orthogonal timelike Killing vector field for a system with $\lambda=0$, $N^2(0)=(t_{\alpha\beta}X^\alpha X^\beta)(0)$ is constant on $M^3$. From this and uniform convergence on $\Omega$, we obtain that $(t_{\alpha\beta}X^\alpha X^\beta)(0)\equiv1$ on $M^3$ since $N^2(\lambda)=(t_{\alpha\beta}X^\alpha X^\beta)(\lambda)\to1$ as $r(\lambda)\to\infty$ by Definition \ref{def:FTasym}. This shows that $\mathcal{F}(0)$ is indeed a static isolated system in FT with respect to $X^\alpha(0)$. Moreover, the lapse functions $N(\lambda)$ must uniformly converge to $N(0)=1$.

In the same spirit, closedness of the defining conditions $\Pi^\mu_{\alpha}=\delta^\mu_{\alpha}-X^\mu X^\nu T_{\nu\alpha}N^{-2}$ and $\inv{3}{\alpha\beta}=s^{\alpha\beta}+\lambda N^{-2} X^\alpha X^\beta$ gives us that $\inv{3}{ij}(\lambda)\to\delta^{ij}=\inv{3}{ij}(0)$ and $\inv{3}{ij}_{,k}(\lambda)\to0=\inv{3}{ij}_{,k}(0)$ uniformly on $\Omega$ as $\lambda\to0$. The desired counterpart $\myg{3}_{ij}(\lambda)\to\delta_{ij}=\myg{3}_{ij}(0)$ as $\lambda\to0$ is not a direct consequence of the definition of $\myg{3}_{ij}$, though. This is due to the fact that the definition of $\myg{3}_{ij}$ is indirect and explicitly depends on $\lambda$. However, if we assume that $\myg{3}_{ij}(\lambda)\to\chi_{ij}$ and $\myg{3}_{ij,k}(\lambda)\to\chi_{ij,k}$ uniformly on $\Omega$ as $\lambda\to0$ for some smooth symmetric tensor field $\chi_{ij}$ on $\Omega$, then a direct computation shows $\chi_{ij}=\delta_{ij}=\myg{3}_{ij}(0)$ and thus $\chi_{ij,k}=0$.

This justifies the following definition.
\begin{ThmDef}[Static Newtonian Limit]\label{def:NLstaticformal}
Let $\mathcal{F}(\lambda):=(\R\times M^3,s^{\alpha\beta}(\lambda)$, $t_{\alpha\beta}(\lambda),\Gamma^{\mu}_{\alpha\beta}(\lambda)$, $T^{\alpha\beta}(\lambda),\lambda)$ be a family of static isolated systems in frame theory parametrized by $\lambda\in\left(0,\varepsilon\right]$ for some $\varepsilon>0$ and let $\mathcal{F}(0):=(\R\times M^3,s^{\alpha\beta}(0),t_{\alpha\beta}(0),\Gamma^{\mu}_{\alpha\beta}(0),T^{\alpha\beta}(0),0)$ be a static isolated system of FT with global Cartesian coordinates $(x^k(0))$. Assume that there exist global asymptotically flat systems of coordinates $(x^k(\lambda))$ for $\mathcal{F}(\lambda)$ converging to $(x^k(0))$ uniformly on $M^3$ as $\lambda\to0$. Let $X^\alpha(\lambda)$, $X^\alpha(0)$ denote the staticity fields for $\mathcal{F}(\lambda)$, $\mathcal{F}(0)$, respectively. We say that $\mathcal{F}(\lambda)$ converges to $\mathcal{F}(0)$ in the {\bf static Newtonian limit }\index{ind}{Newtonian limit ! static} if there exist a smooth vector field $\xi^\alpha$ timelike in $\mathcal{F}(0)$ and a smooth tensor field $\chi_{ij}$ on $M^3$ such that
\begin{eqnarray*}
X^\alpha(\lambda)\to \xi^\alpha\quad\quad&&\quad\quad\myg{3}_{ij}\to\chi_{ij}\\[-1.8ex]
&\mbox{and}&\\[-1.8ex]
X^\alpha_{\;;\beta}(\lambda)\to \xi^\alpha_{\;;\beta}\quad\quad&&\quad\quad\myg{3}_{ij,k}\to\chi_{ij,k}
\end{eqnarray*}
uniformly on $M^3$ as $\lambda\to0$. We then have $\xi^\alpha=X^\alpha(0)$, $\chi_{ij}=\delta_{ij}=\myg{3}_{ij}(0)$ and $\chi_{ij,k}=\myg{3}_{ij,k}(0)=0$. Moreover, $N(\lambda)$, $N_{;k}(\lambda)$, $\myg{3}_{ij}(\lambda)$, $\my{3}{\Gamma}^k_{ij}(\lambda)$ and converge uniformly on $M^3$ to $N(0)=1$, $N_{;k}(0)=0$, $\myg{3}_{ij}(0)=\delta_{ij}$, and $\my{3}{\Gamma}^k_{ij}(0)=0$, respectively.
\end{ThmDef}
\begin{Rem}
From now on, as in GR, we suppress the abstract index notation when referencing connections, covariant derivatives, and curvature tensors. For example, $\my{3}{\Gamma}^k_{ij}$ then denotes the $3$-dimensional connection corresponding to the generalized $3$-metric $\myg{3}_{ij}$.
\end{Rem}
\begin{Pf}
It is clear that $\xi^\alpha$ is a staticity field for $\mathcal{F}(0)$ by our arguments printed above this definition. Similarly, we have seen above from asymptotic arguments that $N(\lambda)$ uniformly converges to $N(0)=1$. By construction of the Cartesian coordinates and the $3+1$ decomposition in the Newtonian case (cf.~pp.~\pageref{subsec:Newton}ff), $\mathcal{F}(0)$ can only possess one staticity field (up to reversing the direction of time which we deliberately ignore in this chapter). This proves $\xi^\alpha=X^\alpha(0)$. $\chi_{ij}=\delta_{ij}=\myg{3}_{ij}(0)$ follows directly from the characterization of $\myg{3}_{ij}$ explained in Lemma \ref{def:3metric}. $\chi_{ij,k}=\myg{3}_{ij,k}$ follows from the same formula. $N_{;k}(\lambda)\to N_{;k}(0)=0$ follows from differentiating the definition $N^2=t_{\alpha\beta}X^\alpha X^\beta$ covariantly in direction $k$ and using $N(\lambda)\to1$, $X^\alpha(\lambda)\to X^\alpha(0)$ as well as $X^\alpha_{\;;k}(\lambda)\to X^\alpha_{\;;k}(0)$ uniformly on $M^3$ as $\lambda\to0$.
\qed\end{Pf}
\begin{Rems}
We have seen in Chapter \ref{chap:geo} that staticity fields are unique in GR under suitable conditions on the time slice $M^3$ and its ADM-mass, cf.~Corollary \ref{coro:Killunique}. The condition that $X^\alpha(\lambda)\to \xi^\alpha$ uniformly as $\lambda\to0$ thus does not seem particularly restrictive.

The reason for our assumption that the coordinates $(x^k(\lambda))$ should be global lies in the coordinate dependence of the definition of uniform and pointwise convergence. We will see later that we actually only need uniform convergence in a neighborhood of infinity (in each end of $M^3$) to prove convergence of mass and center of mass (in that end). Note that the above definition applies to the situation of an individual end, in particular.
\end{Rems}
We are now in the position to prove $U(\lambda)\to U(0)$ and $U_{;k}(\lambda)\to U_{;k}(0)$ as $\lambda\to0$ for any family as in Definition \ref{def:NLstaticformal}. This will allow us to deduce $m(\mathcal{F}(\lambda))\to m_{N}(\mathcal{F}(0))$ and $\vec{z}_{N}\,(\mathcal{F}(\lambda))\to\vec{z}\,(\mathcal{F}(0))$ as $\lambda\to0$ (in suggestive notation), cf.~Section \ref{sec:NLmass}.
\begin{Thm}[Convergence of $U$ and $\gamma_{ij}$]\label{thm:Uconv}
Let $\mathcal{F}(\lambda)$ be a family of static isolated systems in FT as in Definition \ref{def:NLstaticformal} possessing a static Newtonian limit $\mathcal{F}(0)$. Let $X^\alpha(\lambda)$, $X^\alpha(0)$ be the staticity fields and let $(x^k(\lambda))$, $(x^k(0))$ be the global asymptotically flat coordinates referred to in that definition. Let moreover $\gamma_{ij}(\lambda)$, $\gamma_{ij}(0)$ and $U(\lambda)$, $U(0)$ denote the respective associated generalized pseudo-Newtonian metrics and potentials. Then $U(\lambda)$, $U_{;i}(\lambda)$, $\gamma_{ij}(\lambda)$, $\gamma^{ij}(\lambda)$, and $\my{\gamma}{\Gamma}^k_{ij}(\lambda)$ converge uniformly to $U(0)$, $U_{;i}(0)$, $\gamma_{ij}(0)=\delta_{ij}$, $\gamma^{ij}(0)=\delta^{ij}$, and $\my{\gamma}{\Gamma}^k_{ij}(0)=0$ respectively.
\end{Thm}
\begin{Rem}
This theorem finally justifies the somewhat ad hoc definition of pseudo-Newtonian potential in Chapter \ref{chap:pseudo} and in Definition \ref{def:FTpseudo} above.
\end{Rem}
\begin{Pf}
Let us begin with the convergence of $\gamma_{ij}(\lambda)$ and $\gamma^{ij}(\lambda)$. We have already proven in and above Theorem \ref{def:NLstaticformal} that $\myg{3}_{ij}(\lambda)$ and its inverse $\inv{3}{ij}(\lambda)$ converge to $\myg{3}_{ij}(0)$ and $\inv{3}{ij}(0)$ uniformly on $M^3$ as $\lambda\to0$, respectively. Moreover, we have seen that $N(\lambda)$ uniformly converges to $N(0)=1$. Thus, by definition of $\gamma_{ij}$, $\gamma_{ij}(\lambda)$ must converge uniformly to $\gamma_{ij}(0)$ and the same is true for their inverses $\gamma^{ij}(\lambda)$ and $\gamma^{ij}(0)$. Using Formula \ref{eq:4nablaU} and the fact that the frame theoretical connections $\Gamma^\mu_{\alpha\beta}(\lambda)\to\Gamma^\mu_{\alpha\beta}(0)$ converge uniformly on $M^3$, we obtain
\begin{eqnarray*}
&&\left[\delta^\mu_{i}\delta_{\alpha}^t\delta_{\beta}^te^{4\lambda U}\gamma^{ij}U_{;j}+\lambda\left(2\delta^\mu_{t}\delta_{\left(\alpha\right.}^t\delta_{\left.\beta\right)}^{i} U_{;i}-2\delta^{\mu}_s\delta^{k}_{\left(\alpha\right.}\delta^l_{\left.\beta\right)}U_{;k}\delta^s_{l}+\delta^{\mu}_s\delta^{k}_{\alpha}\delta^l_{\beta} U_{;p}\gamma^{ps}\gamma_{kl}\right)\right](\lambda)\\
&&\quad\quad\quad\quad\quad\quad\quad\to \delta_{i}^\mu\delta_{\alpha}^t\delta_{\beta}^t\delta^{ij}U_{;j}(0)
\end{eqnarray*}
uniformly on $M^3$ as $\lambda\to0$. Comparing components and substituting $e^{\lambda U(\lambda)}=N(\lambda)$, this gives in particular $U_{;j}(\lambda)\to U_{;j}(0)$ uniformly on $M^3$ as $\lambda\to0$. Normalization by $U(\lambda)\to0$ as $r(\lambda)\to\infty$ and $U(0)\to0$ as $r(0)\to\infty$ together with uniform convergence of the coordinates gives $U(\lambda)\to U(0)$ uniformly on $M^3$. This works as follows: Set $u(\lambda,p):=U(\lambda)\vert_{p}-U(0)\vert_{p}$ for all $\lambda\in(0,\varepsilon)$ and all $p\in M^3$. Then by the mean value theorem of calculus, we find
$$\lvert u(\lambda,p_{1})-u(\lambda,p_{2})\rvert\leq\sup_{q\in M^3}\lvert \partial_{q}u(\lambda,q)\rvert$$
for all $p_{1},p_{2}\in M^3$. With the help of the triangle inequality, this implies that $$\sup_{p\in M^3}\lvert u(\lambda,p)\rvert\leq\sup_{q\in M^3}\lvert \partial_{q}u(\lambda,q)\rvert+\underbrace{\lim_{r(\lambda)\vert_{q}\to\infty}\lvert u(\lambda,q)\rvert}_{=0}\to0\mbox{ as }\lambda\to0.$$ In other words, $U(\lambda)$ converges uniformly to $U(0)$ on $M^3$ as $\lambda\to0$.

By the rules for conformal transformation, we thus find
\begin{eqnarray*}
\my{\gamma}{\Gamma}^k_{ij}(\lambda)&=&\my{3}{\Gamma}^k_{ij}(\lambda)+\lambda(U_{;i}\delta^k_{j}+U_{;j}\delta^k_{i}-U_{;l}\gamma^{lk}\gamma_{ij})(\lambda)\\
&\to&\my{3}{\Gamma}^k_{ij}(0)+\quad\quad\quad\quad0\\
&=&\my{\gamma}{\Gamma}^k_{ij}(0)
\end{eqnarray*}
uniformly on $M^3$ as $\lambda\to0$ again by Theorem \ref{def:NLstaticformal}. The explicit formulae for $\gamma_{ij}(0)$ etc.~follow from Theorem \ref{thm:pseudoFTispseudo}.
\qed\end{Pf}

Before we go on to prove convergence of mass and its center under the Newtonian limit, we would like to mention that some of the above calculations have been performed in a similar manner but less rigorous on pp.~52ff in \cite{Lott1} in the context of ``time-orthogonal coordinates''. Asymptotic considerations, uniform convergence, and staticity have not been taken into account there.

\section{The Newtonian Limit of Mass and its Center}\label{sec:NLmass}
This section is dedicated to proving two of the main theorems of this thesis, namely convergence of mass and center of mass under the static Newtonian limit. We begin by discussing the Newtonian limit of mass.

\subsection*{The Newtonian Limit of Mass}\label{subsec:NLMass}
Let us quickly recall our considerations of the mass of a geometrostatic/pseudo-Newtonian system, adapting everything to the notation used in this chapter. If $\mathcal{S}=(E^3,\myg{3}_{ij},N,\rho,S_{ij})$ is a geometrostatic end, then its ADM-mass is denoted by $m_{ADM}=m_{ADM}(\myg{3}_{ij})$. From Theorem \ref{ADM=pseudo}, we know that this mass can be recovered by the quasi-local expression $$m_{ADM}=m_{PN}(\Sigma)=\frac{1}{4\pi G}\int_{\Sigma} \frac{\partial U}{\partial\nu}\,d\sigma,$$ where $U$ is the associated pseudo-Newtonian potential, $\nu$ the outer unit normal and $d\sigma$ the surface measure, both with respect to the pseudo-Newtonian metric $\gamma_{ij}$. $\Sigma\subset E^3$ is any smooth surface enclosing the support of the matter.

When we derived this expression in Section \ref{sec:mpseudo}, we also gave an argument why the Newtonian mass $m_{N}=m_{N}(U)$ of a static isolated Newtonian system $(E^3,U,\rho)$ can be read off from the equivalent quasi-local expression $$m_{N}=\frac{1}{4\pi G}\int_{\Sigma} \frac{\partial U}{\partial\nu}\,d\sigma.$$ This time, $U$ is the Newtonian potential of the system while $\nu$ and $d\sigma$ are the outer unit normal and surface element induced by the Euclidean metric. As above, $\Sigma\subset\R^3$ is any smooth surface enclosing the support of the matter.

Combining this in the frame of FT, we define the {\it generalized pseudo-Newtonian mass of a static isolated end $(\R\times E^3,s^{\alpha\beta},t_{\alpha\beta},\Gamma^\mu_{\alpha\beta},T^{\alpha\beta},\lambda)$ in FT}\index{ind}{mass ! pseudo-Newtonian}\index{ind}{pseudo-Newtonian ! mass} by $$m_{PNFT}(\gamma,U):=\frac{1}{4\pi G}\int_{\Sigma} \frac{\partial U}{\partial\my{\gamma}\!{\nu}}\,d\sigma_{\gamma},$$ with $\gamma_{ij}$ and $U$ the induced generalized pseudo-Newtonian metric and potential, respectively. As above, $\my{\gamma}\!{\nu}$ and $d\sigma_{\gamma}$ are the induced outer unit normal and surface element and $\Sigma$ is any smooth surface enclosing the support of the matter. Now, we trivially have $m_{PNFT}(\gamma,U)=m_{N}(U)$ for $\lambda=0$. $m_{PNFT}(\gamma,U)=m_{ADM}(e^{-2\lambda U}\gamma)$ for $\lambda>0$ follows from Theorem \ref{ADM=pseudo}. In particular, the frame theoretical (generalized) pseudo-Newtonian mass is well-defined and independent of the chosen surface enclosing the support of the matter. 

We now combine our results from Chapters \ref{chap:iso}, \ref{chap:static}, \ref{chap:mCoM}, and \ref{chap:FT} to prove the following theorem.
\begin{Thm}[Newtonian Limit of Mass Theorem]\label{thm:NLm}
Let $\mathcal{F}(\lambda):=(\R\times E^3,s^{\alpha\beta}(\lambda)$, $t_{\alpha\beta}(\lambda),\Gamma^{\mu}_{\alpha\beta}(\lambda)$, $T^{\alpha\beta}(\lambda),\lambda)$ be a family of static isolated ends in frame theory parametrized by $\lambda\in(0,\varepsilon)$ for some $\varepsilon>0$ and let $\mathcal{F}(0):=(\R\times E^3,s^{\alpha\beta}(0),t_{\alpha\beta}(0),\Gamma^{\mu}_{\alpha\beta}(0),T^{\alpha\beta}(0),0)$ be a static isolated system of FT with global Cartesian coordinates $(x^k(0))$. Assume that there exist global asymptotically flat systems of coordinates $(x^k(\lambda))$ for $\mathcal{F}(\lambda)$ converging to $(x^k(0))$ uniformly on $M^3$ as $\lambda\to0$. Let $\myg{3}_{ij}(\lambda)$, $\gamma_{ij}(\lambda)$, $\gamma_{ij}(0)$, $U(\lambda)$, and $U(0)$ denote the physical and pseudo-Newtonian metrics and potentials of $\mathcal{F}(\lambda)$ and $\mathcal{F}(0)$, respectively. Then $$m_{ADM}(\myg{3}(\lambda))=m_{PNFT}(\gamma(\lambda),U(\lambda))\to m_{PNFT}(\gamma(0),U(0))=m_{N}(U(0))$$
as $\lambda\to0$.
\end{Thm}
\begin{Pf}
The equation on the left hand side follows from Theorem \ref{ADM=pseudo} combined with the consistency statements in Proposition \ref{prop:GRisFT} and in Theorems \ref{thm:staticiso} and \ref{thm:pseudoFTispseudo}. The equation on the right hand side follows from the divergence theorem argument explained above Theorem \ref{ADM=pseudo} as well as from the same consistency assertions. 

So let us now discuss the remaining Newtonian limit argument. Choose any smooth surface $\Sigma\subset E^3$ enclosing the support of the matter. By definition, we have $$m_{PNFT}(\gamma,U)=\frac{1}{4\pi G}\int_{\Sigma} \frac{\partial U}{\partial\my{\gamma}\!{\nu}}\,d\sigma_{\gamma}.$$ By Theorem \ref{thm:Uconv}, we know that $U(\lambda)$ and $\gamma_{ij}(\lambda)$ converge uniformly to $U(0)$ and $\gamma_{ij}(0)$ as $\lambda\to0$ in the given coordinates and on all of $M^3$. By definition of outer unit normal and surface element, uniform convergence of $\gamma_{ij}(\lambda)$ to $\gamma_{ij}(0)$ implies uniform convergence of $\my{\gamma}\!{\nu}(\lambda)$ and $d\sigma_{\gamma}(\lambda)$ to $\my{\gamma}\!{\nu}(0)=\my{\delta}{\nu}$ and $d\sigma_{\gamma}(0)=d\sigma_{\delta}$, respectively. As $\Sigma$ is compact and the convergence is uniform, we are done with the proof of this theorem (implicitly using a partition of unity argument to cater for $d\sigma$).
\qed\end{Pf}

\subsection*{The Newtonian Limit of the Center of Mass}\label{subsec:NLCoM}
For our proof of convergence of the center of mass under the Newtonian limit, we would like to proceed just as in the proof of convergence of mass. However, there is a new difficulty, here, namely the fact that the center of mass is a vector and thus depends on the chosen system of asymptotically flat coordinates. We remind the reader that we have established the quasi-local formula for the center of mass in wave harmonic coordinates, only  (or, equivalently, $\gamma$-harmonic coordinates by Lemma \ref{waveharm}). This was due to the Green's formula trick we have used where we needed to force $\mylap{\gamma}x^k=0$. We thus have to assume more specific coordinate conditions in the theorem corresponding to the Newtonian limit of mass theorem \ref{thm:NLm}.

Moreover, recall that we made the additional assumption that the fall-off rate $\tau$ should be strictly larger than one half when dealing with the center of mass. We needed this assumption in order to ensure equivalence of the different concepts of center of mass introduced in Chapter \ref{chap:iso}.

With the same notation as in the above subsection, we recall that the pseudo-Newtonian center of mass of a geometrostatic end $\mathcal{S}=(E^3,\myg{3}_{ij},N,\rho,S_{ij})$ with non-vanishing ADM-mass $m:=m_{ADM}(\myg{3}_{ij})$ was defined as
$$z^k_{PN}(\Sigma):=\frac{1}{4\pi Gm}\left(\int_{\Sigma} \frac{\partial U}{\partial\nu}x^k-U\frac{\partial x^k}{\partial\nu}\right)\,d\sigma$$ in wave harmonic asymptotically flat coordinates $(x^k)$. As above, $U$ and $\gamma_{ij}$ are the associated pseudo-Newtonian potential and metric while $\nu$ and $d\sigma$ are the induced outer unit normal and surface element, respectively. $\Sigma\subset E^3$ is again a smooth surface enclosing the support of the matter. We have seen in Proposition \ref{prop:CoMequal} that the pseudo-Newtonian center of mass is independent of the specific surface.

We recall furthermore that Theorem \ref{asym=pseudo} states that (under the assumptions described there or above) the well-known centers of mass introduced in Chapter \ref{chap:iso} agree for geometrostatic systems. Moreover, they all coincide with the pseudo-Newtonian center of mass:
$$\vec{z}_{PN}(\Sigma)=\vec{z}_{ADM}(e^{-2\lambda U}\gamma)=\vec{z}_{A}(e^{-2\lambda U}\gamma,e^{\lambda U})=\vec{z}_{CMC}(e^{-2\lambda U}\gamma)=\vec{z}_{I}(e^{-2\lambda U}\gamma).$$

In Section \ref{sec:CoMpseudo}, we have seen that a similar statement is true for static isolated Newtonian systems $(E^3,U,\rho)$. If the Newtonian mass $m:=m_{N}(U)$ does not vanish, the Newtonian center of mass can be calculated from the quasi-local formula $$z^k_{N}(U)=\frac{1}{4\pi Gm}\left(\int_{\Sigma} \frac{\partial U}{\partial\nu}x^k-U\frac{\partial x^k}{\partial\nu}\right)\,d\sigma,$$ where now $\nu$ and $d\sigma$ correspond to the Euclidean metric and $(x^k)$ are Cartesian coordinates. We have seen in Section \ref{sec:CoMpseudo} that this formula is independent of the specific surface enclosing the support of the matter.

Unifying these definitions in frame theory, we define the {\it generalized pseudo-Newtonian center of mass of a static isolated end $(\R\times E^3,s^{\alpha\beta},t_{\alpha\beta},\Gamma^\mu_{\alpha\beta},T^{\alpha\beta},\lambda)$ in FT}\index{ind}{center of mass ! pseudo-Newtonian}\index{ind}{pseudo-Newtonian ! center of mass} with non-zero mass $m:=m_{PNFT}(\gamma,U)$ by $$z^k_{PNFT}(\gamma,U):=\frac{1}{4\pi Gm}\left(\int_{\Sigma} \frac{\partial U}{\partial\my{\gamma}\!{\nu}}x^k-U\frac{\partial x^k}{\partial\my{\gamma}\!{\nu}}\right)\,d\sigma_{\gamma}$$ with $\gamma_{ij}$ and $U$ the induced generalized pseudo-Newtonian metric and potential, respectively.  As above, $\my{\gamma}\!{\nu}$ and $d\sigma_{\gamma}$ are the induced outer unit normal and surface element and $\Sigma$ is any smooth surface enclosing the support of the matter.

We have $z^k_{PNFT}(\gamma,U)=z^k_{N}(U)$ for $\lambda=0$ by Green's formula. $z^k_{PNFT}(\gamma,U)=z^k_{CMC}(e^{-2\lambda U}\gamma)=z^k_{ADM}(e^{-2\lambda U}\gamma)=z^k_{I}(e^{-2\lambda U}\gamma)=z^k_{A}(e^{-2\lambda U}\gamma,e^{\lambda U})$ for $\lambda>0$ follows from Theorem \ref{asym=pseudo}. In particular, the frame theoretical (generalized) pseudo-Newtonian center of mass is well-defined and independent of the chosen surface enclosing the support of the matter. 

We can now formulate and prove the final theorem of this thesis.
\begin{Thm}[Newtonian Limit of Center of Mass Theorem]\label{thm:NLCoM}
Let $k\in\N$, $k\geq3$, $\tau>1/2$ such that $-\tau$ is non-exceptional. Let $\mathcal{F}(\lambda):=(\R\times E^3,s^{\alpha\beta}(\lambda)$, $t_{\alpha\beta}(\lambda),\Gamma^{\mu}_{\alpha\beta}(\lambda)$, $T^{\alpha\beta}(\lambda),\lambda)$ be a family of $(k,\tau)$-static isolated ends in frame theory parametrized by $\lambda\in(0,\varepsilon)$ for some $\varepsilon>0$ and let $\mathcal{F}(0):=(\R\times E^3,s^{\alpha\beta}(0),t_{\alpha\beta}(0),\Gamma^{\mu}_{\alpha\beta}(0),T^{\alpha\beta}(0),0)$ be a $(k,\tau)$-static isolated system of FT with global Cartesian coordinates $(x^k(0))$. Assume that there exist global wave harmonic $(k,\tau)$-asymptotically flat systems of coordinates $(x^k(\lambda))$ for $\mathcal{F}(\lambda)$ converging to $(x^k(0))$ uniformly on $M^3$ as $\lambda\to0$. Let $\myg{3}_{ij}(\lambda)$, $N(\lambda)$, $\gamma_{ij}(\lambda)$, $\gamma_{ij}(0)$, $U(\lambda)$, and $U(0)$ denote the physical and pseudo-Newtonian metrics and potentials of $\mathcal{F}(\lambda)$ and $\mathcal{F}(0)$, respectively. Finally, assume that $m_{PNFT}(\gamma(\lambda),U(\lambda))$ and $m_{PNFT}(\gamma(0),U(0))$ are non-vanishing. Then
\begin{eqnarray*}
&&\vec{z}_{ADM}(\myg{3}(\lambda))=\vec{z}_{A}(\myg{3}(\lambda),N(\lambda))=\vec{z}_{CMC}(\myg{3}(\lambda))=\vec{z}_{I}(\myg{3}(\lambda))=\vec{z}_{PNFT}(\gamma(\lambda),U(\lambda))\\[1ex]
&&\quad\quad\quad\quad\quad\quad\to \vec{z}_{PNFT}(\gamma(0),U(0))=\vec{z}_{N}(U(0))\in\R^3
\end{eqnarray*}
as $\lambda\to0$.
\end{Thm}
\begin{Pf}
The equations on the left hand side follow from Theorem \ref{asym=pseudo} combined with the consistency statements in \ref{prop:GRisFT}, \ref{thm:staticiso}, and \ref{thm:pseudoFTispseudo}. The equation on the right hand side follows from the Green's formula argument explained before Theorem \ref{asym=pseudo} as well as from the same consistency statements.

The interesting part is Newtonian limit claim. To prove this claim, choose any smooth surface $\Sigma\subset E^3$ enclosing the support of the matter. By definition, we have $$z^k_{PNFT}(\gamma,U)=\frac{1}{4\pi Gm_{PNFT}(\gamma,U)}\left(\int_{\Sigma} \frac{\partial U}{\partial\my{\gamma}\!{\nu}}x^k-U\frac{\partial x^k}{\partial\my{\gamma}\!{\nu}}\right)\,d\sigma_{\gamma}$$ as all involved coordinates are wave harmonic and asymptotically flat. We proceed as in the proof of the Newtonian limit of mass theorem \ref{thm:NLm}. There are four ``new'' steps here: The first one is to observe that the coordinate functions $x^k(\lambda)$ uniformly converge to $x^k(0)$ by assumption. Secondly, the expression $\partial x^k/\partial\my{\gamma}\!{\nu}$ can be reformulated to give $$\frac{\partial x^k}{\partial\my{\gamma}\!{\nu}}(\lambda)=\my{\gamma}\!{\nu}^k(\lambda)\to\my{\gamma}\!{\nu}^k(0)=\frac{\partial x^k}{\partial\my{\gamma}\!{\nu}}(0)$$ uniformly on $M^3$ as $\lambda\to0$ by uniform convergence of $\gamma_{ij}$. From Theorem \ref{thm:Uconv}, we know directly that $U(\lambda)$ converges uniformly to $U(0)$ on $M^3$ and the only remaining step is to quote Theorem \ref{thm:NLm} to ensure convergence of $m_{PNFT}(\gamma(\lambda),U(\lambda))\to m_{PNFT}(\gamma(0),U(0))$ as $\lambda\to0$. Now conclude as in Theorem \ref{thm:NLm} that compactness of $\Sigma$ and uniformity of the convergence ensure the desired convergence.
\qed\end{Pf}

\chapter*{}
{\thispagestyle{plain}
\begin{center}
\vspace{25ex}
\begin{minipage}[h]{10cm}
\begin{center}
{\large ``Si enim fallor, sum. Nam qui non est, utique nec falli potest.''}\\[1ex]
{\selectlanguage{german} ãSelbst wenn ich mich t\"ausche, bin ich. Denn wer nicht ist, kann sich auch nicht t\"auschen.Ò}\\
Aurelius Augustinus (354-430), Vom Gottesstaat 11,26
\end{center}
\end{minipage}
\vfill
\end{center}
\clearpage}

\selectlanguage{english}
\bibliographystyle{amsalpha}
\bibliography{Diss-final}
\addcontentsline{toc}{chapter}{\numberline{}\refname}
\nocite{AF}
\nocite{BS1}
\nocite{BS2}
\nocite{BS3}
\nocite{BS4}
\nocite{BI}
\nocite{Bray}
\nocite{Chrus}
\nocite{Eva}
\nocite{Heb}
\nocite{Jara}
\nocite{Lee2}
\nocite{DLee}
\nocite{LP}
\nocite{List}
\nocite{Lott2}
\nocite{Beig}
\nocite{Bray2}
\nocite{BL}
\nocite{CK}
\nocite{Chrus2}
\nocite{EH}
\nocite{Ehl2}
\nocite{Ehl3}
\nocite{Ehl4}
\nocite{Ehl5}
\nocite{HI}
\nocite{Mur}
\nocite{Jan}
\nocite{Jan2}
\nocite{McO}
\nocite{Corvi}
\nocite{Rend2}
\nocite{Rend3}
\nocite{Rend4}
\nocite{RR}
\nocite{SchoenYau2}
\nocite{Scia}
\nocite{OlySchmi}

\printindex{ind}{Index}
\printindex{sym}{List of Symbols}

\end{document}